\documentclass[cernpreprint,texlive=2011,txfonts,UKenglish,texmf,siunitx=false]{atlasdoc}
\pdfoutput=1

\usepackage{atlaspackage}
\usepackage{atlasbiblatex}
\usepackage{atlasphysics}

\graphicspath{{logos/}{{figures/}}{figure_}}
\usepackage{Coupling2015-defs}

\usepackage{multirow}
\usepackage{rotating}
\usepackage[italic]{heppennames2}
\usepackage{amssymb}

\DeclareMathVersion{normal2}
\DeclareMathVersion{normal3}

\hypersetup{pdftitle={Measurements of the Higgs boson production and decay rates and coupling strengths using pp collision data at 7 and 8 TeV in the ATLAS experiment},pdfauthor={The ATLAS Collaboration}}

\AtlasTitle{Measurements of the Higgs boson production and decay rates and coupling strengths using $pp$ collision data at $\sqrt{s}=7$ and $8$~TeV in the ATLAS experiment}

\author{The ATLAS Collaboration}

\date{\today}

\PreprintIdNumber{CERN-PH-EP-2015-125}

\AtlasJournalRef{Eur. Phys. J. C76 (2016) 6}
\AtlasDOI{10.1140/epjc/s10052-015-3769-y}

\AtlasAbstract{%
Combined analyses of the Higgs boson production and decay rates as well as its coupling strengths to vector bosons and fermions are presented. The combinations include the results of the analyses of the $H\to\gamma\gamma,\, ZZ^*,\, WW^*,\, Z\gamma,\, b\bar{b},\, \tau\tau$ and $\mu\mu$ decay modes, and the constraints on the associated production with a pair of top quarks and on the off-shell coupling strengths of the Higgs boson. The results are based on the LHC proton-proton collision datasets, with integrated luminosities of up to 4.7 $\mathrm{fb}^{-1}$  at $\sqrt{s}=7$~TeV and 20.3 $\mathrm{fb}^{-1}$ at $\sqrt{s}=8$~TeV, recorded by the ATLAS detector in 2011 and 2012.  Combining all production modes and decay channels, the measured signal yield, normalised to the Standard Model expectation, is $1.18^{+0.15}_{-0.14}$.  The observed Higgs boson production and decay rates are interpreted in a leading-order coupling framework, exploring a wide range of benchmark coupling models both with and without assumptions on the Higgs boson width and on the Standard Model particle content in loop processes.  The data are found to be compatible with the Standard Model expectations for a Higgs boson at a mass of 125.36~GeV for all models considered.%
}

\begin{document}
\maketitle
\newpage
\section{Introduction}
\label{sec:Introduction}

In 2012, the ATLAS and CMS Collaborations at the Large Hadron Collider (LHC) reported the observation of a new particle at a mass of approximately 125~GeV~\cite{ATLAS:2012obs,CMS:2012obs}. The discovery made in the search for the Standard Model~(SM) Higgs boson ($H$), is a milestone in the quest to understand electroweak symmetry breaking~(EWSB). Within the SM, EWSB is achieved through the Brout--Englert--Higgs mechanism~\cite{Englert:1964et,Higgs:1964ia,Higgs:1964pj,Guralnik:1964eu,Higgs:1966ev,Kibble:1967sv} which predicts the existence of a neutral scalar particle, commonly known as the Higgs boson. While the SM does not predict the value of its mass~($m_H$), the production cross sections and decay branching ratios~(BR) of the Higgs boson can be precisely calculated once the mass is known. Therefore, precision measurements of the properties of the new particle are critical in ascertaining whether the newly discovered particle is fully responsible for EWSB and whether there are potential deviations from SM predictions.

At the LHC, SM production of the Higgs boson is dominated by the gluon fusion process $gg\to H$ (ggF), followed by the vector-boson fusion process $qq'\to qq'H$ (VBF). Associated production with a $W$ boson $q\bar{q}'\to WH$ ($WH$), a $Z$ boson $q\bar{q}/gg\to ZH$ ($ZH$) or with a pair of top quarks $q\bar{q}/gg\to t\bar{t}H$ ($ttH$) have sizeable contributions as well. The $WH$ and $ZH$ production processes are collectively referred to as the $VH$ process. Contributions are also expected from $b\bar{b}\to H$ ($bbH$) and production in association with a single top quark ($tH$). The latter proceeds through either the $qb\to tHq'$ or $gb\to WtH$ process.  With the present dataset, the LHC is expected to be most sensitive to the Higgs boson decays of $H\to\gamma\gamma,\, ZZ^*,\, WW^*,\, \tau\tau$ and $b\bar{b}$. Together they account for approximately 88\% of all decays of a SM Higgs boson at $m_H\sim 125$~GeV. 

The discovery of the Higgs boson was made through analyses of the bosonic decay modes in $\Hyy$, $\Hllll$ and $\Hlvlv$ ($\ell=e,\,\mu$) events. Since the discovery, these analyses have been improved and updated with more data~\cite{Aad:2014eha,Aad:2014eva,ATLAS:2014aga}. The $\Hlvlv$ analysis has been supplemented with a dedicated $VH$ analysis targeting $\Hww$~\cite{ATLAS:VHWW}. The ATLAS Collaboration has measured the Higgs boson mass from the $\Hyy$ and $\Hllll$ decays to be $m_H=125.36\pm 0.41$~GeV~\cite{Aad:2014aba}, reported results in the $\Htt$~\cite{Aad:2015vsa} and $\Hbb$~\cite{Aad:2014xzb} fermionic decay modes, and published upper limits on the rare decays  $\Hzg$~\cite{Aad:2014fia} and $\Hmm$~\cite{Aad:2014xva}. Furthermore, constraints have been set on the $ttH$ production rate~\cite{ATLAS:ttHhad,ATLAS:ttHlep,Aad:2014lma} and on the off-shell coupling strengths of the Higgs boson~\cite{Aad:2015xua}. These results are based on the full proton-proton collision data with integrated luminosities of up to 4.7~\ifb\ at a centre-of-mass energy $\sqrt{s}=7$~TeV recorded in 2011 and  20.3~\ifb\ at $\sqrt{s}=8$~TeV recorded in 2012 by the ATLAS detector at the LHC. A detailed description of the ATLAS detector can be found in Ref.~\cite{Aad:2008zzm}.

This paper presents the combined results of the analyses mentioned above. These analyses are designed for maximum sensitivities to SM Higgs boson production from different processes, exploiting in particular the differences in kinematics through categorisation of the selected events. Thus the yields of different Higgs boson production processes and decays can be extracted. The Higgs boson coupling strengths to SM vector bosons and fermions in different benchmark models are probed for the measured Higgs boson mass of $m_H=125.36$~GeV. All results are obtained assuming the Higgs boson has a small total decay width such that its production and decay factorise. The ATLAS Collaboration has previously published combined studies of Higgs boson production and decay rates~\cite{Aad:2013wqa} and of spin-parity properties~\cite{Aad:2013xqa,Aad:2015mxa} using diboson final states. The results are found to be consistent with expectations from the SM Higgs boson. Compared with the previous publication, the current results are based on the improved analysis sensitivities and the addition of information from other decay modes. A similar combination has been published by the CMS Collaboration~\cite{Khachatryan:2014jba}.

The paper is organised as follows. Section~\ref{sec:InputChannels} briefly summarises the individual analyses that are included in the combinations and Section~\ref{sec:StatisticalProcedure} outlines the statistical method and the treatment of systematic uncertainties used in the combinations. In Section~\ref{sec:SignalStrength}, the measured Higgs boson yields are compared with the SM predictions for different production processes and decay modes. In Section~\ref{sec:CouplingFits}, the coupling strengths of the Higgs boson are tested through fits to the observed data. These studies probe possible deviations from the SM predictions under various assumptions, motivated in many cases by beyond-the-SM (BSM) physics scenarios. An upper limit on the branching ratio to invisible or undetected decay modes of the Higgs boson is also set. Finally, a brief summary is presented in Section~\ref{sec:Conclusion}.

\section{Input analyses to the combinations}
\label{sec:InputChannels}

The inputs to the combinations are
the results from the analyses of $H\to\gamma\gamma,\, ZZ^*,\, WW^*,\, \tau\tau,\, b\bar{b},\, \mu\mu$ and $Z\gamma$ decay modes, and of the constraints on $ttH$ and off-shell Higgs boson production. These analyses and changes made for the combinations are briefly discussed in this section.
The ATLAS Collaboration has also performed a search for the rare $H\to J/\psi\gamma$ decay~\cite{Aad:2015sda} which has the potential to constrain the Higgs boson coupling strength to the charm quark. However, the current result does not add sensitivity and is therefore omitted from the combinations. Furthermore, the inclusion of the results from direct searches for Higgs boson decays to invisible particles, such as those reported in Ref.~\cite{Aad:2014iia,Aad:2015uga}, is beyond the scope of the combinations presented in this paper.

The theoretical calculations of the Higgs boson production cross sections and decay branching ratios have been compiled in Refs.~\cite{Dittmaier:2011ti,Dittmaier:2012vm,Heinemeyer:2013tqa} and are summarised in Table~\ref{tab:SMPrediction}. For the ggF process, the cross section is computed at up to NNLO in QCD corrections~\cite{Djouadi:1991tka,Dawson:1990zj,Spira:1995rr,Harlander:2002wh,Anastasiou:2002yz,Ravindran:2003um} and NLO in electroweak~(EW) corrections~\cite{Aglietti:2004nj,Degrassi:2004mx,Actis:2008ug}. The effects of QCD soft-gluon resummations at up to NNLL~\cite{Catani:2003zt} are also applied. These calculations are described in Refs.~\cite{Anastasiou:2008tj,deFlorian:2009hc,Baglio:2010ae,Anastasiou:2012hx,deFlorian:2012yg}. For the VBF process, full QCD and EW corrections up to NLO~\cite{Ciccolini:2007jr,Ciccolini:2007ec,Arnold:2008rz} and approximate NNLO~\cite{Bolzoni:2010xr,Bolzoni:2011cu} QCD corrections are used to calculate the cross section. The cross sections of the $WH$ and $ZH$ ($q\bar{q}\to ZH$) are calculated including QCD corrections up to NNLO~\cite{Han:1991ia,Brein:2003wg} and EW corrections up to NLO~\cite{Ciccolini:2003jy,Denner:2011id} whereas the cross section of the $gg\to ZH$ process is calculated up to NLO in QCD corrections~\cite{Altenkamp:2012sx,Englert:2013vua}. The cross section for $ttH$ is computed up to NLO in QCD~\cite{Beenakker:2001rj,Beenakker:2002nc,Dawson:2002tg,Dawson:2003zu}. For the $bbH$ process, the cross section is calculated in QCD corrections up to NLO~\cite{Dawson:2003kb,Dittmaier:2003ej,Dawson:2005vi} in the four-flavour scheme and up to NNLO~\cite{Harlander:2003ai} in the five-flavour scheme with the Santander matching scheme~\cite{Harlander:2011aa}. The cross sections of the $tH$ processes used are calculated at up to NLO in QCD corrections~\cite{Maltoni:2001hu,Farina:2012xp}.  
The PDF sets used in these calculations are CT10~\cite{Lai:2010vv,Gao:2013xoa}, MSTW2008~\cite{Martin:2009iq}, NNPDF2.1~\cite{Ball:2011mu,Ball:2011uy} and NNPDF2.3~\cite{Ball:2012cx} following the prescription of Ref.~\cite{Botje:2011sn}. The decay branching ratios of the Higgs boson are calculated using the {\sc Hdecay}~\cite{Djouadi:1997yw,hdecay2} and Prophecy4f~\cite{Bredenstein:2006rh,Bredenstein:2006ha} programs, compiled in Ref.~\cite{Denner:2011mq}.

\begin{table}[htbp]
\caption{SM predictions of the Higgs boson production cross sections and decay branching ratios and their uncertainties for $m_H=125.36$~GeV, obtained by linear interpolations from those at 125.3 and 125.4~GeV from Ref.~\cite{Heinemeyer:2013tqa} except for the $tH$ production cross section which is obtained from Refs.~\cite{Alwall:2014hca,Aad:2014lma}. 
The uncertainties of the cross sections are the sum in quadrature of the uncertainties resulting from variations of QCD scales, parton distribution functions and $\alpha_{\rm s}$. The uncertainty on the $tH$ cross section is calculated following the procedure in Refs.~\cite{Heinemeyer:2013tqa,Aad:2014lma}. 
}\vspace*{-0.4cm}
\begin{center}
\begin{tabular}{lcl} \\ \hline\hline

\begin{tabular}{crr} 
   Production  & \multicolumn{2}{c}{Cross section [pb]} \\ \cline{2-3}
   process     &   $\sqrt{s}=7$ TeV   &  $\sqrt{s}=8$ TeV \tsp  \\ \hline
   ggF         &   $15.0\pm 1.6$      &  $19.2\pm 2.0$     \\
   VBF         &   $1.22\pm 0.03$     &  $1.57\pm 0.04$    \\
   $WH$        &   $0.573\pm 0.016$   &  $0.698\pm 0.018$    \\
   $ZH$        &   $0.332\pm 0.013$   &  $0.412\pm 0.013$    \\ 
   $bbH$       &   $0.155\pm 0.021$   &  $0.202\pm 0.028$    \\
   $ttH$       &   $0.086\pm 0.009$   &  $0.128\pm 0.014$    \\ 
   $tH$        &   $0.012\pm 0.001$   &  $0.018\pm 0.001$     \\ \hline
   Total       &   $17.4 \pm 1.6$     &  $22.3\pm 2.0$     \\
\end{tabular} 

& \hsa &

\begin{tabular}{lr}
  Decay channel &  Branching ratio [\%]    \\ \hline
   $\Hbb$       &  $57.1\pm 1.9$    \\
   $\Hww$       &  $22.0\pm 0.9$   \\
   $\Hgg$       &  $8.53\pm 0.85$  \\
   $\Htt$       &  $6.26\pm 0.35$  \\
   $\Hcc$       &  $2.88\pm 0.35$  \\
   $\Hzz$       &  $2.73\pm 0.11$  \\
   $\Hyy$       &  $0.228\pm 0.011$  \\
   $\Hzg$       &  $0.157\pm 0.014$  \\
   $\Hmm$       &  $0.022\pm 0.001$  \\

\end{tabular} \\

\hline\hline

\end{tabular}
\end{center}

\label{tab:SMPrediction}
\end{table}

All analyses use Monte Carlo~(MC) samples to model the acceptances of the Higgs boson events. Table~\ref{tab:generator} summarises the event generators and parton distribution functions~(PDF) used for the analyses of the $\sqrt{s}=8$~TeV data. The modelling at $\sqrt{s}=7$~TeV is similar, with one notable difference of {\sc Pythia6}~\cite{Sjostrand:2006za} replacing {\sc Pythia8}~\cite{Sjostrand:2007gs}. The ggF and VBF production of the Higgs boson are simulated with the next-to-leading order~(NLO) matrix-element {\sc Powheg} program~\cite{Nason:2004rx,Frixione:2007vw,Alioli:2008tz,Alioli:2010xd,Bagnaschi:2011tu} interfaced to either {\sc Pythia6} or {\sc Pythia8} for the simulation of the underlying event, parton showering and hadronisation (referred to as the showering program). The Higgs boson transverse momentum distribution from ggF production is reweighted to match the calculation of {\sc HRes2.1}~\cite{deFlorian:2012mx,Grazzini:2013mca}, which includes QCD corrections up to the next-to-next-to-leading order~(NNLO) and next-to-next-to-leading logarithm~(NNLL) in perturbative expansions. Furthermore, ggF events with two or more jets are reweighted to match the transverse momentum distribution from {\sc MiNLO} HJJ predictions~\cite{Campbell:2006xx}. The $WH$ and $ZH$ ($q\bar{q}\to ZH$) production processes are simulated with the leading-order~(LO) {\sc Pythia8} program. The $gg\to ZH$ process contributes approximately 8\% to the total $ZH$ production cross section in the SM. For most of the analyses, the process is modelled using $q\bar{q}\to ZH$ of {\sc Pythia8}. Only the $VH$ analysis in the $\Hbb$ decay mode specifically models  $gg\to ZH$ production using  {\sc Powheg}~\cite{Nason:2004rx,Frixione:2007vw,Alioli:2008tz} interfaced to {\sc Pythia8}. The $ttH$ process is modelled using the NLO calculation in the HELAC-Oneloop package~\cite{Bevilacqua:2011xh} interfaced to {\sc Powheg} and {\sc Pythia8} for the subsequent simulation. The $tH$ production process is simulated using {\sc MadGraph}~\cite{Maltoni:2002qb} interfaced to {\sc Pythia8} for $qb\to tHq'$ and using {\sc MadGraph5\_aMC@NLO}~\cite{Alwall:2014hca} interfaced to {\sc Herwig++}~\cite{Bahr:2008pv} for $gb\to WtH$. The $bbH$ production process contributes approximately 1\%~\cite{Wiesemann:2014ioa} to the total Higgs boson cross section in the SM. It is simulated with the {\sc MadGraph5\_aMC@NLO} program for some analyses. The event kinematics of ggF and $bbH$ production are found to be similar for analysis categories that are most important for $bbH$. Thus the acceptance times efficiency for $bbH$ is assumed to be the same as for ggF for all analyses. The PDF sets used in the event generations are CT10~\cite{Lai:2010vv} and CTEQ6L1~\cite{Nadolsky:2008zw}. All Higgs boson decays are simulated by the showering programs.

\begin{table}[htb]
\caption{Summary of event generators, showering programs and PDF sets used to model the Higgs boson production and decays at $\sqrt{s}=8$~TeV.}
\begin{center}
\begin{tabular}{clccc}\hline\hline
\multicolumn{2}{c}{Production\hsb}         &  Event       &   Showering        &   PDF          \\ 
\multicolumn{2}{c}{process\hsb}            & generator    &   program          & \hsd  set \hsd \\ \hline  
& ggF                   & {\sc Powheg}        &  {\sc Pythia6}/{\sc Pythia8} &   CT10     \\  
& VBF                   & {\sc Powheg}        &  {\sc Pythia6}/{\sc Pythia8} &   CT10     \\  
& $WH$                  & {\sc Pythia8}           &  {\sc Pythia8}              &  CTEQ6L1  \\
& $ZH:\ q\bar{q}\to ZH$ & {\sc Pythia8}           &  {\sc Pythia8}              &  CTEQ6L1  \\
& $ZH:\ gg\to ZH$       & {\sc Powheg}            &  {\sc Pythia8}           &   CT10     \\
& $ttH$                 & {\sc Powheg}            &  {\sc Pythia8}              &   CT10     \\
& $bbH$                 & {\small\sc MadGraph5\_aMC@NLO} & {\sc Herwig++}       &   CT10     \\
& $tH:\ qb\to tHq'$     & {\sc MadGraph}          & {\sc Pythia8}               &   CT10     \\
& $tH:\ gb\to WtH$      & {\small\sc MadGraph5\_aMC@NLO} & {\sc Herwig++}       &   CT10     \\
\hline\hline
\end{tabular}
\end{center}
\label{tab:generator}
\end{table}

Throughout this paper, the signal-strength parameter $\mu$ is defined as the ratio of the measured Higgs boson yield to its SM expectation:
\begin{eqnarray}
\label{eq:mu}
\mu = \frac{\sigma\times {\rm BR}}{(\sigma\times {\rm BR})_{\rm SM}}\,.
\end{eqnarray}
Here $\sigma$ is the production cross section of the Higgs boson.  For a specific production process $i$ and decay channel $f$, i.e., $i\to H\to f$, the signal-strength parameter is labelled as $\mu_i^f$ and can be factorised in terms of the signal strengths of production ($\mu_i$) and decay ($\mu_f$):
\begin{eqnarray}
\label{eq:muif}
\mu_i^f = \frac{\sigma_i\times {\rm BR}_f}{(\sigma_i\times {\rm BR}_f)_{\rm SM}}\equiv \mu_i\times \mu_f,\hsa {\rm with}\hsa \mu_i=\frac{\sigma_i}{(\sigma_i)_{\rm SM}}\ \ {\rm and}\ \ \mu_f= \frac{{\rm BR}_f}{({\rm BR}_f)_{\rm SM}}\,.
\end{eqnarray}
Thus for each analysis category $(c)$ as discussed later in this section, the number of signal events ($n^c_s$) can be written as:

\begin{equation}
n^c_s
=\sum_i \sum_f  \mu_i (\sigma_i)_{\rm SM} \times \mu_f  ({\rm BR}_f)_{\rm SM}  \times A^c_{if} \times \varepsilon^c_{if} \times  \mathcal{L}^c
\label{eq:nsig}
\end{equation}

where the indices $i$ and $f$ indicate the production processes and decays contributing to the  category, $A^c_{if}$ represents the detector acceptance derived from simulation of the SM process, $\varepsilon^c_{if}$ is the reconstruction efficiency within the acceptance and $ \mathcal{L}^c$ the integrated luminosity for the given category $c$ of the given channel.

However, the experimental data do not allow to separately determine $\mu_i$ and $\mu_f$ for any given process since only their product is measured. All combined fits of signal strengths presented in this paper make assumptions about the relationship between $\mu_i$ of different production processes or similarly between $\mu_f$ of different decay modes. Thus the meaning of the signal strength depends on the assumptions made. Nevertheless, the production and decays can be factorised using the ratios of cross sections and of branching ratios as discussed in Section~\ref{sec:ratio}.

Leptons ($\ell$) refer to electrons or muons unless specified otherwise; the symbols $\tlep$ and $\thad$ refer to $\tau$ leptons identified through their decays to leptons or hadrons; and variables $\pT$, $\ET$ and $\met$ refer to transverse momentum, transverse energy and  missing transverse momentum, respectively. Notations indicating particle charges or antiparticles are generally omitted.

The ATLAS experiment uses a right-handed coordinate system with its origin at the nominal interaction point (IP) in the centre of the detector and the $z$-axis along the beam pipe. The $x$-axis points from the IP to the centre of the LHC ring, and the $y$-axis points upward. Cylindrical coordinates $(r,\phi)$ are used in the transverse plane, $\phi$ being the azimuthal angle around the beam pipe. The pseudorapidity is defined in terms of the polar angle $\theta$ as $\eta=-\ln\tan(\theta/2)$.

Table~\ref{tab:inputs} gives an overview of the analyses that are inputs to the combinations and their main results, as published.  An essential feature of these analyses is the extensive application of exclusive categorisation, i.e., classifying candidate events based on the expected kinematics of the different Higgs boson production processes. The categorisation not only improves the analysis sensitivity, but also allows for the discrimination among different production processes. Figure~\ref{fig:inputs} summarises the signal-strength measurements of different production processes that are used as inputs to the combinations.

\begin{table}[htbp]
\caption{Overview of the individual analyses that are included in the combinations described in this paper. The signal strengths, the statistical significances of a Higgs boson signal, or the 95\% CL upper limits on the Higgs boson production rates or properties are also shown wherever appropriate. A range is quoted for the upper limit on the off-shell signal strength, depending on the assumption for the continuum $gg\to WW/ZZ$ cross section. These results are taken directly from the individual publications. Results of the on-shell analyses are quoted for $m_H=125.36$~GeV except that $m_H=125.5$~GeV is assumed for the $\Hzg$ and $\Hmm$ analyses and that $m_H=125$~GeV is used for the $ttH$ searches with $\Hbb$ and $ttH\to{\rm multileptons}$. The luminosity used for the $\sqrt{s}=7$~TeV $VH(\to\bb)$ analysis differs slightly from the values used for other analyses because a previous version of the luminosity calibration was applied.
The significance is given in  units of standard deviations (s.d.). The numbers in parentheses are the expected values for the SM Higgs boson. The $ttH$ analysis in the $H\to\gamma\gamma$ decay is part of the $H\to\gamma\gamma$ analysis. It is included separately under the $ttH$ production for completeness. The checkmark (\checkmark) indicates whether the analysis is performed for the respective $\sqrt{s}=7$ and 8~TeV dataset.}
\label{tab:inputs}
\begin{center}
\small
\begin{tabular}{lccllc} \hline\hline
 Analysis   &     \multicolumn{2}{c}{Signal}  && \multicolumn{2}{c}{$\int{\cal L} dt$ [\ifb]} \tsp \\ \cline{2-3}\cline{5-6}
  \hsc Categorisation or final states             & Strength $\mu$  & Significance [s.d.] &&  7 TeV   &   8 TeV \tsp \\ \hline
$\Hyy$~\cite{Aad:2014eha}                              &  $1.17\pm 0.27$ & 5.2 (4.6) && 4.5 & 20.3 \tsp \\
  \multicolumn{3}{l}{\hsc $ttH$: leptonic, hadronic}                     && \checkmark & \checkmark  \\
  \multicolumn{3}{l}{\hsc $VH$:  one-lepton, dilepton,  $\met$, hadronic} && \checkmark & \checkmark  \\
  \multicolumn{3}{l}{\hsc VBF: tight, loose}                           && \checkmark & \checkmark  \\
  \multicolumn{3}{l}{\hsc ggF: 4 \pTt\ categories }                && \checkmark & \checkmark  \\[1mm] \hline
$\Hllll$~\cite{Aad:2014eva}                           & $1.44^{+0.40}_{-0.33}$  & 8.1 (6.2) && 4.5 & 20.3 \tsp \\
  \multicolumn{3}{l}{\hsc VBF}                             && \checkmark & \checkmark \\
  \multicolumn{3}{l}{\hsc $VH$: hadronic, leptonic}         && \checkmark & \checkmark \\
  \multicolumn{3}{l}{\hsc ggF}                             && \checkmark & \checkmark  \\ \hline
$H\to WW^*$~\cite{ATLAS:2014aga,ATLAS:VHWW}                          &  $1.16^{+0.24}_{-0.21}$  & 6.5 (5.9) && 4.5 & 20.3 \tsp \\
   \multicolumn{3}{l}{\hsc ggF: (0-jet, 1-jet) $\otimes$ ($ee+\mu\mu,\ e\mu$)}     && \checkmark & \checkmark \\
   \multicolumn{3}{l}{\hsc ggF: $\ge 2$-jet and $e\mu$}                            &&            & \checkmark \\
   \multicolumn{3}{l}{\hsc VBF: $\ge 2$-jet $\otimes$ ($ee+\mu\mu,\ e\mu$)}        && \checkmark & \checkmark \\
   \multicolumn{3}{l}{\hsc $VH$: opposite-charge dilepton, three-lepton, four-lepton} && \checkmark & \checkmark \\
   \multicolumn{3}{l}{\hsc $VH$: same-charge dilepton}                                &&            & \checkmark \\ \hline
$\Htt$~\cite{Aad:2015vsa}                                  & $1.43^{+0.43}_{-0.37}$ & 4.5 (3.4)  && 4.5 & 20.3  \tsp \\
  \multicolumn{3}{l}{\hsc Boosted: $\ttll, \ttlh, \tthh$} && \checkmark & \checkmark   \\
  \multicolumn{3}{l}{\hsc VBF: $\ttll, \ttlh, \tthh$}   && \checkmark & \checkmark  \\ \hline
$VH\to V\bb$~\cite{Aad:2014xzb}                               & $0.52\pm 0.40$ & 1.4 (2.6)  && 4.7 & 20.3 \tsp  \\
  \multicolumn{3}{l}{\hsc\small $0\ell\ (ZH\to \nu\nu \bb)$: $\Njet=2,3$, $\Nbjet=1,2$, $\pT^V\in$ 100-120 and $>\!120$~GeV}  && \checkmark & \checkmark \\
  \multicolumn{3}{l}{\hsc\small $1\ell\ (WH\to \ell\nu \bb)$: $\Njet=2,3$, $\Nbjet=1,2$,  $\pT^V<$ and $>\!120$~GeV} && \checkmark & \checkmark \\
  \multicolumn{3}{l}{\hsc\small $2\ell\ (ZH\to \ell\ell \bb)$: $\Njet=2,3$, $\Nbjet=1,2$, $\pT^V<$ and $>\!120$~GeV} && \checkmark & \checkmark \\ \hline\hline \\

                    &     &      95\% CL limit  \\ \hline
$\Hzg$~\cite{Aad:2014fia}   &     & $\mu<11\ (9)$    && 4.5 & 20.3 \\
  \multicolumn{3}{l}{\hsc 10 categories based on $\Delta\eta_{Z\gamma}$ and \pTt} && \checkmark & \checkmark \\ \hline
$\Hmm$~\cite{Aad:2014xva}   &     & $\mu<7.0\ (7.2)$  &&  4.5  &  20.3   \\
  \multicolumn{3}{l}{\hsc VBF and 6 other categories based on $\eta_\mu$ and $\pT^{\mu\mu}$} && \checkmark & \checkmark \\ \hline

$ttH$ production~\cite{ATLAS:ttHhad,ATLAS:ttHlep,Aad:2014lma}    &    &      &&  4.5  & 20.3 \\
  \multicolumn{2}{l}{\hsc $H\to \bb$: single-lepton, dilepton}                                 & $\mu<3.4\ (2.2)$  &&            & \checkmark \\
  \multicolumn{2}{l}{\hsc $ttH\to$multileptons: categories on lepton multiplicity}  & $\mu<4.7\ (2.4)$  &&            & \checkmark \\
  \multicolumn{2}{l}{\hsc $H\to \gamma\gamma$: leptonic, hadronic}  & $\mu<6.7\ (4.9)$  && \checkmark & \checkmark \\ \hline
Off-shell $H^*$ production~\cite{Aad:2015xua}    &    & $\mu<$ 5.1 -- 8.6 (6.7 -- 11.0)   &&   &  20.3  \tsp \\
  \multicolumn{3}{l}{\hsc $H^*\to ZZ\to 4\ell$}      &&            & \checkmark \\
  \multicolumn{3}{l}{\hsc $H^*\to ZZ\to 2\ell 2\nu$} &&            & \checkmark \\
  \multicolumn{3}{l}{\hsc $H^*\to WW\to e\nu\mu\nu$}  &&            & \checkmark \\
\hline\hline
\end{tabular}
\end{center}
\end{table}

\begin{figure}[htb]
\begin{center}
   \includegraphics[width=0.6\textwidth]{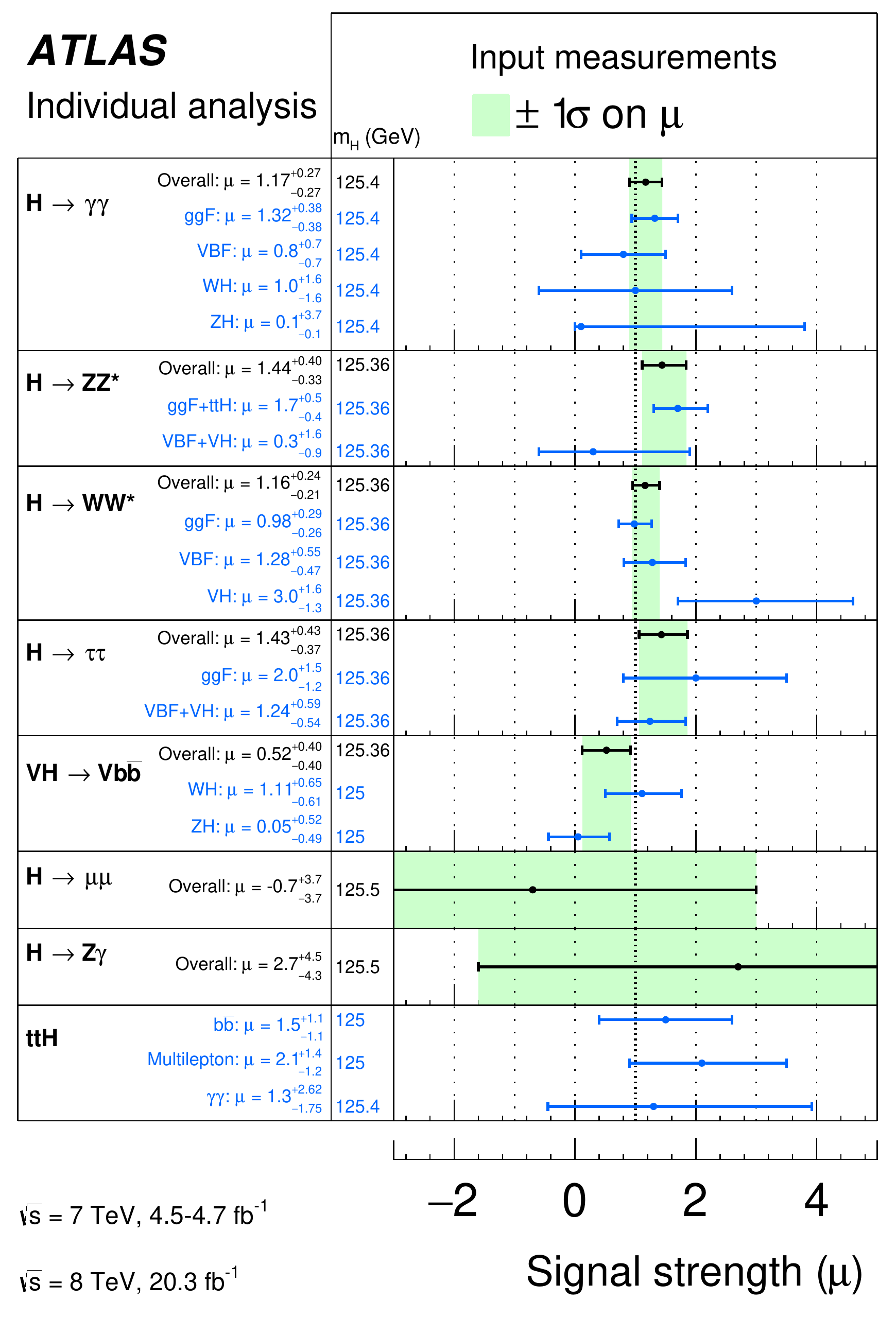}
   \caption{Summary of the signal-strength measurements, as published, from individual analyses that are inputs to the combinations. The Higgs boson mass column indicates the $m_H$ value at which the result is quoted. The overall signal strength of each analysis (black) is the combined result of the measurements for different production processes (blue) assuming SM values for their cross-section ratios.  The error bars represent $\pm 1\sigma$ total uncertainties, combining statistical and systematic contributions.  The green shaded bands indicate the uncertainty on the overall signal strength obtained by each analysis. The combined signal strength of the $\Hyy$ analysis also includes the $ttH$ contribution which is listed separately under $ttH$ production.}
   \label{fig:inputs}
\end{center}
\end{figure}

\subsection{$H\to\gamma\gamma$}
\label{sec:hgg}

In the $H\to\gamma\gamma$ analysis, described in detail in Ref.~\cite{Aad:2014eha}, the Higgs boson signal is measured in events with at least two isolated and well-identified photon candidates. The leading and subleading photon candidates are required to have $\ET /m_{\gamma\gamma} > 0.35$ and 0.25, respectively, where $m_{\gamma\gamma}$ is the invariant mass of the two selected photons.
The diphoton candidate events are grouped into twelve exclusive categories separately for the $\sqrt{s}=7$ and 8~TeV datasets; the order of categorisation is chosen to give precedence to production modes with the most distinct signatures. Each category is optimised by adjusting the event selection criteria to minimise the expected uncertainty on the signal yield of the targeted production mode. 

The first two categories are designed for $ttH$ production based on the topology of leptonic and hadronic decays of the associated $t\bar{t}$ pair. They are described in Section~\ref{sec:tth}. The next four categories are optimised for $VH$ production, targeting one-lepton, dilepton, $\met$, and hadronic signatures of $W$ and $Z$ boson decays.  Events from VBF production are identified by requiring two well-separated and high-$\pT$ 
jets and little hadronic activity between them.
A boosted decision tree (BDT)~\cite{BDT,Speckmayer:2010zz} algorithm is employed to maximise the VBF signal and background separation.
Events are sorted into two categories with different VBF purities according to 
the output value of the BDT. Finally, the remaining events  are separated into four categories based on the pseudorapidities of the photons and the \pTt\ of the diphoton system~\cite{Aad:2014eha}, the diphoton momentum transverse to its thrust axis in the transverse plane. 

For most of the categories, the background is composed of a mixture of  $\gamma$--jet and jet--jet events, where one or two jets are misidentified as photons, and $\gamma\gamma$ events. In particular the $\gamma\gamma$ background is dominant and irreducible. 
The Higgs boson signal is extracted from maximum-likelihood fits of a narrow resonance plus continuum background models to unbinned diphoton invariant-mass  distributions observed in the different event categories. In the fit, the signal is modelled by the sum of a Crystal Ball function~\cite{Oreglia:1980cs} and a smaller but wider Gaussian component while the backgrounds are modelled by category-dependent exponential functions of first- or second-order polynomials.

\subsection{$\Hllll$}
\label{sec:hzz}

The $\Hllll$ analysis, described in detail in Ref.~\cite{Aad:2014eva}, has a high signal-to-background
ratio, which is about two for each of the four final states considered: $4\mu$, $2e2\mu$, $2\mu2e$, and $4e$,
where the first lepton pair has an invariant mass closer to the $Z$ boson mass. The analysis selects Higgs boson candidates by requiring two pairs of isolated, same-flavour 
and opposite-charge leptons with one of the two pairs having a dilepton invariant mass in the range 50 -- 106~GeV.

To measure the rates of different production processes, each $H\to ZZ^*\to 4\ell$ candidate is assigned to one of four categories depending on event characteristics beyond the four selected leptons. The VBF category consists of candidates with two additional jets with dijet mass $\mjj>130$~GeV. The events failing this selection are considered for the $VH$-hadronic category, where the dijet mass is required to be $40\,\GeV <\mjj<130\,\GeV$. Events failing the VBF and $VH$-hadronic categorisation criteria are considered for the $VH$-leptonic category with the requirement of an additional lepton. Finally, the remaining events are assigned to the ggF category. The separation of VBF and $VH$ production from the dominant ggF production mode is improved by exploiting two BDT discriminants trained on the jet kinematics, one for the VBF category and the other for the $VH$-hadronic category. A third BDT discriminant based on the four-lepton kinematics is used to improve the separation between the ggF signal and its main background.

The largest background comes from continuum $ZZ^*$ production and is estimated using simulation
normalised to the SM next-to-leading-order cross-section calculation.  For the four-lepton events with an invariant
mass, $m_{4\ell}$, below about $160$\,\gev, there are also important background contributions from
$Z$+jets and $t\bar{t}$ production with two prompt leptons, where the additional charged lepton
candidates arise from decays of hadrons with $b$- or $c$-quark content, from photon conversions or
from misidentified jets.  
Their contributions are estimated with data-driven methods.

For each category, the signal is extracted from a maximum-likelihood fit to either the $m_{4\ell}$ distribution ($VH$ categories) or the combined two-dimensional distributions of $m_{4\ell}$ and a BDT discriminant (ggF and VBF categories). The four-lepton mass range of $110\,\GeV<m_{4\ell}<140\,\GeV$ is included in the fits.

\subsection{$H\to WW^*$}
\label{sec:hww}

Analyses targeting the ggF, VBF, and $VH$ production modes~\cite{ATLAS:2014aga,ATLAS:VHWW} are performed for the $H\to WW^*$ decay channel. The ggF and VBF production processes are explored through the $\Hlvlv$ decay and the $VH$ process is studied in final states with two or more leptons.

The analysis of the ggF and VBF production processes~\cite{ATLAS:2014aga} selects the signal candidate events by requiring two oppositely charged leptons. Candidates are categorised according to the number of jets ($\Njet$) and to the flavours of the leptons. The $\Njet$ categorisation separates the large top-quark production background from the ggF signal while the categorisation by lepton flavours isolates the challenging Drell--Yan background in the same-flavour categories. The categories targeting  ggF production include $\Njet=0,\: 1$ and $\ge 2$ and are further divided into the same- and different-flavour leptons for $\Njet=0,\: 1$. Only the different-flavour leptons are considered for $\Njet\ge 2$.  The categories targeting VBF production require $\Njet\ge 2$, separately for the same- or different-flavour leptons.
The primary background processes are $WW$, top quark ($t\bar{t}$ and $Wt$), $W$+jets, Drell--Yan, and other diboson ($WZ$, $W\gamma$, $W\gamma^{*}$, and $ZZ$) production. Most of the background contributions are estimated using data. 
For the ggF categories, the final signal region  is selected by requiring the dilepton mass $m_{\ell\ell} < 55$~GeV and their azimuthal angular separation $\Delta\phi_{\ell\ell} < 1.8$ and the signal is extracted through a combined fit to the transverse mass distributions of the dilepton plus $\met$ system in both the signal and control regions of different categories and lepton flavours. For the VBF categories, a BDT combining information such as rapidity separation and mass of the two leading jets and the dilepton angular separation, is used as the final discriminant, from which the signal is extracted.

The $VH$ analysis~\cite{ATLAS:VHWW} is optimised for different lepton multiplicities: opposite-charge dileptons, same-charge dileptons, three and four leptons. Most final states are required to have $\met$ and events with a $b$-tagged jet are vetoed.  Dilepton final states target $VH$ production with the $H\to WW^*$ decay with two bosons decaying leptonically and the other hadronically.
The opposite-charge dilepton final state selects events with two or more jets, with the value of $m_{jj}$ required to be close to the $W$ and $Z$ boson masses.  Similar to the ggF $\Njet\ge 2$ category, the dominant background is from top quark production. The same-charge dilepton category accepts events with either one or two jets. The dominant backgrounds are from $WZ$, $W\gamma^{(*)}$, and $W$+jets production. The three-lepton final state targets $WH$ with $H\to WW^*$ and has the highest sensitivity of the four final states.  The three leptons are required to have a net charge of $\pm 1$ and the event can have at most one jet.   The dominant background process is $WZ$ production and is reduced with a $Z\to\ell\ell$ veto. The four-lepton category is designed to accept events from $ZH$ production with the $H\to WW^*$ decay. The net charge of the leptons is required to be zero and at least one pair of  leptons is required to have the same flavour, opposite charge, and an invariant mass close to the $Z$ boson mass. The dominant background is SM $ZZ^*$ production. 
In the three-lepton category, the signal yield is extracted through  fits to distributions of a BDT  or the minimum separation in the $\eta-\phi$ plane between opposite-charge leptons depending on the lepton flavours. For other categories, the event yields are used, without exploiting information on the shapes of distributions.

\subsection{$H\to\tau\tau$}
\label{sec:htautau}

The $\Htt$ analysis~\cite{Aad:2015vsa} considers both the leptonic ($\tlep$) and hadronic ($\thad$) decays of the $\tau$ lepton.  Three
sub-channels  ($\ttll$, $\ttlh$ and $\tthh$) are defined by orthogonal requirements on the number of reconstructed hadronic $\tau$ decays and  leptons (electrons or muons) in the event.\footnote{For events with two leptons, a requirement on the invariant mass of the $\tau\tau$ system reconstructed via the collinear approximation also ensures orthogonality with the $\Hlvlv$ analysis.}

Candidate events are divided into boosted and VBF categories.
The boosted category targets signal events where the Higgs boson is produced with a large boost, primarily from the gluon fusion process, and requires the  transverse momentum of the reconstructed Higgs boson candidate to be greater than 100~GeV. The VBF category contains events with two jets separated in pseudorapidity and targets signal events produced through the vector boson fusion process.
A separate BDT  is then employed in each category and sub-channel to discriminate signal from background, utilising between five and nine input variables, chosen in order to exploit discriminating features such as Higgs boson decay properties, event activity, and the VBF topology in the corresponding category. One of the most important input variables is the mass of the $\tau\tau$ system, which is quite challenging to reconstruct due to the presence of at least two neutrinos in the final state; the Missing Mass Calculator~\cite{Elagin:2010aw} is used for this purpose.

In all three sub-channels, the most important backgrounds are  irreducible $Z\to\tau\tau$ events, and events  with one or two jets misidentified as $\tau$ lepton decay products (primarily from multijet and $W$+jets production). To estimate the $Z\to\tau\tau$ background the embedding technique~\cite{Aad:2015kxa} is used, where $Z\to\mu\mu$ events are selected in data and the reconstructed muons are replaced by simulated $\tau$ lepton decays.  Fully data-driven techniques are used for the estimation of backgrounds from misidentified $\tau$ decay products, while Monte Carlo simulation corrected to data is used for other backgrounds, such as the top quark and $Z\to\ell\ell$ production.

The signal is extracted by fitting the shape of the BDT discriminant with signal and background templates simultaneously in all signal regions.
The fit also includes dedicated control regions enriched with top quark, $Z\to\ell\ell$ and multijet events. These control regions are used to constrain normalisations of the corresponding backgrounds.

\subsection{$VH$ with $H\to \bb$}
\label{sec:hbb}

The $H\to\bb$ decay mode is predicted in the SM to have the largest branching ratio (see Table~\ref{tab:SMPrediction}). In spite of this large branching ratio, an inclusive search for $H\to\bb$ is not feasible because of the overwhelming background from multijet production. Associated production of a Higgs boson with a vector boson $V$
($W$ or $Z$), offers a viable alternative because leptonic decays of the vector boson, $W\to\ell\nu$, $Z\to\ell\ell$, and $Z\to\nu\nu$, can be efficiently used for triggering and background reduction. 

The search for associated $VH$ production with $H\to\bb$~\cite{Aad:2014xzb} is performed for events containing zero, one, or two charged leptons. Contributions from $W\to\tau\nu$ and $Z\to\tau\tau$ decays in which the $\tau$ leptons subsequently decay to electrons or muons are also included. A $b$-tagging algorithm is used to identify jets from  $H\to \bb$ decays. To improve the sensitivity, the three channels are each split into categories according to the vector-boson transverse momentum, $\pT^V$, the number of jets, and the number and quality of the $b$-tagged jets. Topological and kinematic selection criteria are applied within each of the resulting categories. The categories providing most of the sensitivity are those requiring two $b$-tagged jets and large $\pT^V$. The categories with low sensitivity are used to constrain the contributions of the dominant background processes.

A binned profile maximum-likelihood fit to all categories simultaneously 
is used to extract the signal yield and the background normalisations.
The most significant background sources are $V$+heavy-flavour-jet production
and $t\bar{t}$ production. The normalisations of these backgrounds are fully determined
by the likelihood fit. Other significant background sources are single-top-quark and
diboson ($WZ$ and $ZZ$) production, with normalisations from theory,
as well as multijet events. The shapes of all backgrounds are estimated from simulation,
except for the multijet background for which the shape and normalisation are obtained using
multijet-enriched control samples.

Two versions of the analysis are performed. In
the dijet-mass analysis, the mass of the dijet system of $b$-tagged jets is the final
discriminating variable used in the statistical analysis.
In the  multivariate analysis (MVA), which incorporates various kinematic
variables in addition to the dijet mass and the $b$-tagging information, the outputs
of boosted decision trees provide the final discriminating variable.
Since the MVA has higher expected sensitivity,
it is chosen as the nominal analysis for the $\sqrt{s}=8$~TeV dataset to extract the final results.
For the $\sqrt{s}=7$~TeV dataset, only a dijet-mass analysis is performed.

The $\sqrt{s}=7$~TeV $VH(\to\bb)$ analysis uses a previous version of the luminosity calibration and therefore has a slightly different luminosity value compared with those quoted for other analyses. However, this small difference is expected to have negligible effects on the results presented in this paper.

\subsection{$H\to Z\gamma$}
\label{sec:hzg}

The $H\to Z\gamma$ analysis~\cite{Aad:2014fia} with $Z\to\ell\ell$ searches for a narrow peak in the reconstructed $\ell\ell\gamma$ invariant-mass distribution around 125~GeV over a smooth background.  The $Z{+}\gamma$ production, $Z\to \ell\ell\gamma$ radiative decays and $Z$+jets events where a jet is misidentified as a photon dominate the background contributions.

The analysis selects  two isolated leptons of same flavour and opposite charge and one isolated photon. 
Due to the kinematics of the decay, low $\pT$ thresholds are applied to the leptons and the photon.
The invariant mass of the dilepton system must satisfy $m_{\ell\ell}>m_Z-10$ GeV and the three-body invariant mass must be consistent with 
the mass of the Higgs boson.
To enhance the sensitivity of the analysis, events are classified into categories with different signal-to-background ratios and invariant-mass resolutions, based on the pseudorapidity difference $\Delta \eta_{Z\gamma}$ between the photon and the $Z$ boson and  $p_{\rm Tt}$, the component of the Higgs boson candidate $\pT$ that is orthogonal to the $Z\gamma$ thrust axis in the transverse plane.

The final discrimination between signal and background events is based on a simultaneous likelihood fit to the $m_{\ell\ell\gamma}$ spectra in each category, separately for the $\sqrt{s}=7$ and 8~TeV datasets. Similar to the $H\to\gamma\gamma$ analysis (Section~\ref{sec:hgg}), the signal is modelled with the sum of a Crystal Ball function and a smaller but wider Gaussian component while the backgrounds are modelled with polynomials, or exponentiated polynomials depending on categories.

\subsection{$H\to\mu\mu$}
\label{sec:hmumu}

The $H\rightarrow\mu\mu$ analysis~\cite{Aad:2014xva} searches for a 
narrow peak in the dimuon invariant mass $m_{\mu\mu}$ distribution over a smooth background, 
where the width of the  signal  is dominated by the experimental resolution. 
The mass  spectrum is dominated by the continuously falling background due to 
$Z/\gamma^*$ production, with smaller contributions from top quark and diboson production.

The selected events containing a pair of oppositely charged muons  are separated into seven mutually 
exclusive categories based on the VBF dijet signature, the muon pseudorapidity $\eta_\mu$, and the transverse momentum 
of the dimuon system $p_{\rm T}^{\mu\mu}$. The events with two or more jets that match selections designed 
for the VBF process are accepted in the VBF signal region. 
All other selected events are split up into six categories based 
on the values of $\eta_{\mu}$ and $p_{\rm T}^{\mu\mu}$. This categorisation takes advantage of the 
higher momentum resolution for muons reconstructed in the central part of the
detector, and high  $p_{\rm T}^{\mu\mu}$ for the expected SM signal.

The $m_{\mu\mu}$ distribution in the 110--160 GeV region is fitted with an analytic signal-plus-background model separately for the $\sqrt{s}=7$ and 8~TeV datasets, setting a limit  on the dimuon decay of the SM Higgs boson with a mass of 125.5 GeV. In the fit, the signal is modelled as the sum of a Crystal Ball function and a Gaussian function in all
regions while the backgrounds are modelled using exponentials or polynomials.

\subsection{$ttH$ production}
\label{sec:tth}

Searches for $q\bar{q}/gg\to t\bar{t}H$ production have been performed with three analyses targeting the Higgs boson decays $H\to b\bar{b}$, $H\to (WW^*,\,\tau\tau,\,ZZ^*)\to {\rm leptons}$, and $H\to \gamma\gamma$. The  search in the $H\to\gamma\gamma$ decay mode uses both $\sqrt{s}=7$ and 8~TeV data, while the other two use only the $\sqrt{s}=8$~TeV data.

The search for $ttH$ production with $H\to b\bar{b}$~\cite{ATLAS:ttHhad} considers two separate selections optimised for  single-lepton and dilepton final states of $t\bar{t}$ decays. In the single-lepton channel, events are required to have one isolated electron or muon and at least four jets. In the dilepton channel, events are required to have two opposite-charge leptons ($ee$, $\mu\mu$ or $e\mu$) and at least two jets; events consistent with originating from a $Z\to \ell\ell$  decay are rejected.  In both cases at least two $b$-tagged jets are required.  Candidate events are categorised according to the jet and $b$-jet multiplicities with a total of nine~(six) categories for the single-lepton (dilepton) final states. The background is dominated by
$t\bar{t}$+jets events, with increasing fractions of $t\bar{t}b\bar{b}$ and $t\bar{t}c\bar{c}$ at the higher $b$-jet multiplicities characteristic of signal events. The analysis uses a neural network to discriminate signal from background in the most signal-like categories. Simpler kinematic discriminants are used in background-like categories.

The $ttH$ search with $H\to WW^*,\ \tau\tau$ and $ZZ^*$ decays~\cite{ATLAS:ttHlep} exploits several multilepton signatures resulting from leptonic decays of vector bosons and/or the presence of $\tau$ leptons. The events are categorised by the number of reconstructed electrons or muons and hadronic $\tau$ candidates.  The five channels used in this combination are: one lepton with two hadronic $\tau$ candidates, two same-charge leptons with zero or one hadronic $\tau$ candidate,  three leptons, and four leptons. The largest backgrounds in the analysis are non-prompt leptons, primarily arising from semileptonic $b$-hadron decays in $t\bar{t}$ events; electron charge misreconstruction in events where opposite-charge leptons are produced; and the production of $\ttbar W$ and $\ttbar Z$ ($\ttbar V$). The potential signal is determined from the numbers of observed events in data and of the estimated number of background events.

The $ttH$ search in the $H\to\gamma\gamma$ channel~\cite{Aad:2014lma} is part of the $H\to\gamma\gamma$ analysis (see Section~\ref{sec:hgg}) and employs the same diphoton selection. The leptonic as well as fully hadronic decay signatures of the $t\bar{t}$ system are considered. The leptonic selection requires at least one lepton and one $b$-tagged jet as well as $\met$. In the hadronic selection, different combinations of jet and $b$-tagging multiplicities are applied to improve the signal sensitivity.
The small contribution from ggF, VBF and $VH$ production is estimated from Monte Carlo simulation. The $ttH$ signal is extracted from a fit to the observed diphoton mass distribution.

\subsection{Off-shell Higgs boson production}
\label{sec:width}

Measurements of the $H^*\to ZZ$ and $H^*\to WW$ final states in the mass range above the $2m_Z$ and $2m_W$ thresholds (off-shell region) provide a unique opportunity to measure the off-shell coupling strengths of the observed Higgs boson, as discussed in Refs.~\cite{Kauer:2012hd,Caola:2013yja,Campbell:2013una,Campbell:2013wga}.
The $ZZ \to 4\ell$, $ZZ\to2\ell2\nu$ and $WW\to e\nu\mu\nu$ final states in the $\sqrt{s}=8$~TeV dataset are used in these measurements, detailed in Ref.~\cite{Aad:2015xua}. Assuming the relevant Higgs boson coupling strengths are independent of the energy scale of Higgs boson production, a combination with the on-shell measurements can be interpreted as a constraint on the total width of the Higgs boson.

The analysis in the $ZZ\to 4\ell$ final state follows closely the Higgs boson measurements in the same final state, described in Section~\ref{sec:hzz}, with the same  object definitions, event selections and background estimation methods. The off-peak region is defined to include the range $220\,\GeV < m_{4\ell} <1000\,\GeV$. Like the $\Hllll$ analysis, the background is dominated by $q\bar{q}/gg\to ZZ$ production.
A matrix-element-based discriminant~\cite{Aad:2015xua} is constructed to enhance the $gg\to H^*\to ZZ$ signal and is used in a binned maximum-likelihood fit for the final result.

The analysis in the $ZZ\to 2\ell 2\nu$ channel follows closely the $ZH$  analysis with the Higgs boson decaying to weakly interacting particles~\cite{Aad:2014iia}, with the same object definitions.
As the analysis is performed inclusively in the number of jets in the final states,
kinematic cuts are optimised accordingly. SM $ZZ$ and $WZ$ production are the major backgrounds.
The transverse mass ($m_{\mathrm{T}}^{ZZ}$)~\cite{Aad:2015xua}, reconstructed from the momentum of the 
dilepton system and the missing transverse momentum, is
chosen as the discriminating variable. Events in the range of $380\,\GeV < m_{\mathrm{T}}^{ZZ} <1000\,\GeV$ are used in a binned maximum likelihood fit for the final result.

The analysis in the $WW\to e\nu\mu\nu$ channel follows closely the Higgs boson measurements in the oppositely charged electron--muon pair final state, described in Section~\ref{sec:hww}, with the same  object definitions.
The analysis is performed inclusively in the number of jets in the final state, 
and selections are optimised for the off-shell region with revised background estimation methods. Top quark pairs and $WW$ events constitute the major backgrounds. 
In order to isolate the off-shell Higgs boson production while minimising sensitivity to higher-order QCD effects on $gg\rightarrow WW$ kinematics,
a new variable $R_8$~\cite{ATLAS:VHWW}, defined as the weighted combination of the dilepton mass and the transverse mass of the dilepton and $\met$ system, is constructed to select the signal region.
The final results are obtained from the numbers of events observed in the data and expected from background processes in the signal region of $R_8>450$~GeV.

\subsection{Modifications of analyses}
\label{sec:modifications}

To ensure a consistent interpretation of all inputs in terms of Higgs boson coupling
strengths, several minor modifications were made to the inputs of these combinations with respect
to their previously published versions:

\begin{itemize}
\item The upper limits on the $\Hzg$ and $\Hmm$ decays and the results
  of the $ttH$ searches in $\Hbb$ and $ttH\to{\rm multilepton}$ decays have
  been updated to assume a Higgs boson mass of 125.36~GeV.
\item In some individual analyses, cross-feed of other Higgs boson
  decays occurs: in the $VH,\,\hww$ selection cross-feed of \htt\ and \hzz\ 
  occurs (whereas this cross-feed is negligible in the ggF and VBF
  \hww analyses where a veto on the reconstructed $\tau\tau$ mass is
  applied). Similarly, there is cross-feed from \hww\ in the \htt\ analysis.
  In such cases, this cross-feed was treated as background in the
  relevant individual channel analyses.  For the combinations described in this paper, such
  events are interpreted as signal from the corresponding Higgs boson
  decay.
\item The rate of \ggZH\ events in the $VH$ channels is
  parameterised in terms of Higgs boson coupling strengths to $Z$
  bosons and top quarks, following the calculations of
  Ref.~\cite{Englert:2013vua} for $\sqrt{s}=7$ and 8~TeV.
\item The rate of $tH$ events in all the $ttH$ channels is
  parameterised in terms of Higgs boson coupling strengths to $W$
  bosons and top quarks.
\item In the standalone analysis of the $ttH$ channels, small contributions from Higgs boson decays to the $c\bar{c}$ and $gg$ final states
  are explicitly modelled. To avoid spurious sensitivity due to these very small
  components in the combined analyses presented in this paper, both aforementioned decays are treated 
 like \hbb\ in the fits for the Higgs boson signal strength. 
  In fits for Higgs boson coupling strengths, it is assumed that the coupling strengths of the $\Hcc$ 
  and $\Hgg$ decays scale as the $t\bar{t}\to H$ and $gg\to H$ couplings, respectively. 
\item Theoretical uncertainties from QCD scales in Higgs boson signal
  processes have been updated to be consistent with the latest
  recommendations~\cite{Heinemeyer:2013tqa} for $H \to WW^*,\,
  b\bar{b},\, \tau\tau$ and $Z\gamma$. No modifications were needed
  for the \hgg\ and \hzz\ channels.
\item In channels where $bbH$ production was not explicitly modelled,
  the signal strength of ggF is redefined to include this process. In
  channels where $bbH$ was modelled explicitly ($H \to
  \gamma\gamma,ZZ^*$), ggF and $bbH$ production are correlated with
  their ratio fixed to the SM value, allowing a consistent treatment
  of $bbH$ production across all channels. The impact of this average
  scaling on the results is negligible since, as can be seen in
  Table~\ref{tab:inputs}, the $bbH$ production process has a cross
  section which is only 1\% of the ggF production in the SM.
\item The off-shell analysis depends on the unknown $K$-factor from higher-order QCD corrections for the $gg \to VV$ background process. In the
  case of the very similar Higgs boson signal $gg\to H^*\to VV$ production process, a  $K$-factor between 0.5 
  and 2 is expected, as discussed in Ref.~\cite{Aad:2015xua}. The results are given as a function of the 
  unknown ratio of the $K$-factors for $gg\to VV$ background and $gg\to H^*\to VV$ signal, $R^B_{H^*}$. 
  The range 0.5--2.0 is chosen as a systematic uncertainty on $R^B_{H^*}$.

\end{itemize}

\section{Statistical procedure}
\label{sec:StatisticalProcedure}

\newcommand{\mT}{\ensuremath{m_{\rm T}}}
The statistical treatment of the data is described in
Refs.~\cite{paper2012prd,LHC-HCG,Moneta:2010pm,HistFactory,ROOFIT}.
Hypothesis testing and confidence intervals are based on the  $\Lambda(\vec\alpha)$ profile
likelihood ratio~\cite{Cowan:2010st} test statistic. The test statistic depends on
one or more parameters of interest $\vec\alpha$,
such as the Higgs boson signal strength $\mu$ normalised to the
SM expectation (Eq.~(\ref{eq:mu})), Higgs boson mass $m_H$,
coupling strength scale factors $\vec\Cc$ and their ratios
 $\vec\Rr$, as well as on additional parameters $\vec\theta$ that are not of interest,
\begin{equation}
  \Lambda(\vec\alpha) = \frac{L\big(\vec\alpha\,,\,\hat{\hat{\vec\theta}}(\vec\alpha)\big)}
{L(\hat{\vec\alpha},\hat{\vec\theta})\label{eq:LH}} .
\end{equation}

The likelihood functions in the numerator and denominator of the above equation
are built using sums of signal and background probability density functions (pdfs) of the discriminating variables, introduced in Section~\ref{sec:InputChannels}.
The pdfs are derived from MC simulation
for the signal and from both data and simulation for the background. Likelihood fits to the observed data are done for the
parameters of interest.
 The single circumflex in Eq.~(\ref{eq:LH}) denotes the unconditional maximum-likelihood
estimate of a parameter, i.e. both the parameters of interest and
the nuisance parameters are varied to maximise the likelihood function. The double circumflex
denotes a conditional maximum-likelihood estimate, i.e. an estimate for given
fixed values of the parameters of interest $\vec\alpha$.

Systematic uncertainties and their correlations~\cite{paper2012prd}
are modelled by introducing nuisance parameters $\vec\theta$ described
by likelihood functions associated with the estimate of the
corresponding effect. Systematic uncertainties that affect multiple
measurements are modelled with common nuisance parameters to propagate
the effect of these uncertainties coherently to all measurements. Most
experimental systematic uncertainties are modelled independently for
the $\sqrt{s}=7$ and 8~TeV data samples, reflecting independent
assessments of these uncertainties, but a subset of these
uncertainties, e.g.  material effects and some components of the jet
energy scale, are considered common to the two data taking
periods and are correspondingly described by a common set of nuisance
parameters.

Components of theoretical uncertainties, scale
uncertainties of a given Higgs boson production process as well as PDF-induced
uncertainties, that affect the inclusive signal rate are described with
common nuisance parameters in all channels, whereas components of
theoretical uncertainties that affect the acceptance of individual channels
are modelled with separate nuisance parameters for each decay
channel. 
Specifically, since PDF-induced uncertainties and scale uncertainties are described by separate
nuisance parameters, these uncertainties are effectively treated as uncorrelated. 
The PDF uncertainties of the inclusive rates are treated as correlated for $WH$, $ZH$
and VBF production, as anti-correlated for $gg\to ZH$ and $qq\to ZH$ production and as uncorrelated 
for ggF and $ttH$ production. A cross check with the full
correlation matrix as given in Ref.~\cite{Heinemeyer:2013tqa} show no differences larger
than 1\% for the most generic model (Section~\ref{sec:gen3}).
Similarly, the effects of correlations between Higgs boson branching ratios and partial
decay widths have been determined to be negligible, and are ignored in the
combinations, except for the branching ratios to $WW^*$ and $ZZ^*$ which are treated
as fully correlated. When results are provided
with a breakdown of the
systematic uncertainties in experimental and theoretical uncertainties,
the theoretical uncertainties correspond to the influence of all nuisance
parameters that can affect Higgs boson signal distributions, e.g.
parton density functions related to Higgs boson production, QCD
scale uncertainties related to Higgs boson production processes and uncertainties on
the Higgs boson branching ratios. Theoretical uncertainties that
exclusively affect background samples are included in the 
systematic uncertainty components.

The choice of the parameters of interest depends on the test under
consideration, with the remaining parameters being ``profiled", i.e.,
similarly to nuisance parameters they are set to the values that
maximise the likelihood function for the given fixed values of the
parameters of interest.

Asymptotically, a test statistic $-2\ln \Lambda(\vec{\alpha})$ of
several parameters of interest $\vec{\alpha}$ is distributed as a
$\chi^2$ distribution with $n$ degrees of freedom, where $n$ is the
dimensionality of the vector $\vec{\alpha}$. In particular, the
$100(1-\beta)\%$ confidence level (CL) contours are defined by
$-2\ln\Lambda(\vec{\alpha})<k_\beta$, where $k_\beta$ satisfies
$P(\chi^2_n > k_\beta) = \beta$.
 For one degree of freedom the 68\%
and 95\%~CL intervals are given by $-2\ln\Lambda(\vec{\alpha})=1.0$
and $4.0$, respectively.
 For two degrees of freedom the 68\%
and 95\%~CL contours are given by $-2\ln\Lambda(\vec{\alpha})=2.3$
and 6.0, respectively. All results presented in the following sections
are based on likelihood evaluations and give 
CL intervals under asymptotic approximation.%
\footnote {Whenever probabilities are translated into the number of
  Gaussian standard deviations the two-sided convention is chosen.}
For selected parameters of interest a physical boundary on the
parameter values is included in the statistical interpretation. For
example, branching ratio parameters can conceptually not be
smaller than zero. The 95\% confidence interval quoted for such
parameters is then based on the profile likelihood ratio restricted
to the allowed region of parameter space; the confidence interval is
defined by the standard $\chi^2$ cutoff, which leads to some over-coverage
near the boundaries.

For the measurements in the following sections the compatibility with
the Standard Model, $p_\text{SM}$, is quantified using the
$p$-value\footnote{The $p$-value is defined as the probability to
  obtain a value of the test statistic that is at least as high as the
  observed value, under the hypothesis that is being tested.} obtained
from the profile likelihood ratio
$\Lambda(\vec\alpha=\vec\alpha_{\rm SM})$, where $\vec\alpha$ is the set
of parameters of interest and $\vec\alpha_{\rm SM}$ are their Standard
Model values.  For a given benchmark coupling model, $\vec\alpha$ is
the set of Higgs boson coupling scale factors $\Cc_i$ and ratios of
coupling scale factors $\Rr_{ij}$ probed by that model, where the
indices $i,j$ refer to the parameters of interest of the model (see
Section~\ref{sec:CouplingFits}).  All other parameters are treated as
independent nuisance parameters.

\mathversion{normal2}
\newcommand{\RESULTGlobalMupValue}{\ensuremath{76\%}}
\newcommand{\RESULTGlobalMuSMpValue}{\ensuremath{18\%}}
\newcommand{\RESULTGlobalMuShort}{1.18\,^{+0.15}_{-0.14}}
\newcommand{\RESULTGlobalMuLong}{1.18 \pm 0.10\,({\rm stat.})\pm 0.07\,({\rm syst.})\,^{+0.08}_{-0.07}\,({\rm theo.})}

\newcommand{\RESULTGlobalAMu}{0.75\,^{+0.32}_{-0.29}=0.75\,^{+0.28}_{-0.26}\,({\rm stat.})\,^{+0.13}_{-0.11}\,({\rm syst.})\,^{+0.08}_{-0.05}\,({\rm theo.})}
\newcommand{\RESULTGlobalBMu}{1.28\,^{+0.17}_{-0.15}=1.28 \pm 0.11\,({\rm stat.})\,^{+0.08}_{-0.07}\,({\rm syst.})\,^{+0.10}_{-0.08}\,({\rm theo.})}

\newcommand{\RESULTMuggF}{1.23\,^{+0.23}_{-0.20}}
\newcommand{\RESULTMuVBF}{1.23\pm 0.32}
\newcommand{\RESULTMuVH}{ 0.80\pm 0.36}
\newcommand{\RESULTMuttH}{1.81\pm 0.80}

\newcommand{\RESULTMuggFErr}{{^{+0.14}_{-0.14}}\: {^{+0.09}_{-0.08}}\: {^{+0.16}_{-0.12}}}
\newcommand{\RESULTMuVBFErr}{{^{+0.28}_{-0.27}}\: {^{+0.13}_{-0.12}}\: {^{+0.11}_{-0.09}}}
\newcommand{\RESULTMuVHErr} {{^{+0.31}_{-0.30}}\: {^{+0.17}_{-0.17}}\: {^{+0.10}_{-0.05}}}
\newcommand{\RESULTMuttHErr}{{^{+0.52}_{-0.50}}\: {^{+0.58}_{-0.55}}\: {^{+0.31}_{-0.12}}}

\newcommand{\RESULTAXSAllShort}{22.1\,^{+7.4}_{-6.0}}
\newcommand{\RESULTAXSAllLong}{22.1\,^{+6.7}_{-5.3}\,({\rm stat.})\,^{+2.7}_{-2.3}\,({\rm syst.})\,^{+1.9}_{-1.4}\,({\rm theo.})}

\newcommand{\RESULTBMuggF}{1.23\,^{+0.25}_{-0.21}}
\newcommand{\RESULTBMuVBF}{1.55^{+0.39}_{-0.35}}
\newcommand{\RESULTBMuVH}{ 0.93\pm 0.39}
\newcommand{\RESULTBMuttH}{1.62\pm 0.78}

\newcommand{\RESULTBMuggFErr}{{^{+0.16}_{-0.16}}\: {^{+0.10}_{-0.08}}\: {^{+0.16}_{-0.11}}}
\newcommand{\RESULTBMuVBFErr}{{^{+0.32}_{-0.31}}\: {^{+0.17}_{-0.13}}\: {^{+0.13}_{-0.11}}}
\newcommand{\RESULTBMuVHErr} {{^{+0.37}_{-0.33}}\: {^{+0.20}_{-0.18}}\: {^{+0.12}_{-0.06}}}
\newcommand{\RESULTBMuttHErr}{{^{+0.51}_{-0.50}}\: {^{+0.58}_{-0.54}}\: {^{+0.28}_{-0.10}}}

\newcommand{\RESULTBXSggF}{23.9\pm 3.6}
\newcommand{\RESULTBXSVBF}{2.43\pm 0.58}
\newcommand{\RESULTBXSVH}{1.03\pm 0.53}
\newcommand{\RESULTBXSttH}{0.24\pm 0.11}

\newcommand{\RESULTBXSggFErr}{^{+3.1}_{-3.1}\:   {^{+1.9}_{-1.6}}\:   {^{+1.0}_{-1.0}}}
\newcommand{\RESULTBXSVBFErr}{^{+0.50}_{-0.49}\: {^{+0.27}_{-0.20}}\: {^{+0.19}_{-0.16}}}
\newcommand{\RESULTBXSVHErr} {^{+0.37}_{-0.36}\: {^{+0.22}_{-0.20}}\: {^{+0.13}_{-0.06}}}
\newcommand{\RESULTBXSttHErr}{^{+0.07}_{-0.07}\: {^{+0.08}_{-0.08}}\: {^{+0.01}_{-0.01}}}

\newcommand{\RESULTBXSAllShort}{27.7\pm 3.7}
\newcommand{\RESULTBXSAllLong}{27.7\pm 3.0\, ({\rm stat.})\,^{+2.0}_{-1.7}\,({\rm syst.})\,^{+1.2}_{-0.9}\,({\rm theo.})}

\newcommand{\RESULTMuttHLimitObs}{3.2}
\newcommand{\RESULTMuttHLimitExp}{1.4}
\newcommand{\RESULTMuVHLimitObs}{1.6}
\newcommand{\RESULTMuVHLimitExp}{1.8}

\newcommand{\RESULTTableProductionRates}{
\begin{tabular}{cccccc} \hline\hline

Production   &  \multicolumn{2}{c}{Signal strength $\mu$}   &&  \multicolumn{2}{c}{Cross section [pb]} \\ \cline{2-3} \cline{5-6}
process      & \hsa  $\sqrt{s}=8$ TeV \hsa  &   combined $\sqrt{s}=7$ and 8 TeV   && \hsa $\sqrt{s}=7$ TeV \hsa  & \hsa  $\sqrt{s}=8$ TeV \hsa     \tsp  \\ \hline

ggF          & $\RESULTBMuggF$  &  $\RESULTMuggF$  && $-$ &    $\RESULTBXSggF$    \tsp \\
VBF          & $\RESULTBMuVBF$  &  $\RESULTMuVBF$  && $-$ &    $\RESULTBXSVBF$    \tsp \\
$VH$         & $\RESULTBMuVH$   &  $\RESULTMuVH$   && $-$ &    $\RESULTBXSVH$     \tsp \\
$ttH$        & $\RESULTBMuttH$  &  $\RESULTMuttH$  && $-$ &    $\RESULTBXSttH$    \tsp \\[1mm] \hline
Total        &                  &                  &&  $\RESULTAXSAllShort$  &  $\RESULTBXSAllShort$ \tsp \\
   \hline\hline
\end{tabular}}

\newcommand{\RESULTTableMuProcess}{
\begin{tabular}{crlcrl} \hline\hline
Production     & \multicolumn{5}{c}{Signal strength $\mu$ at $m_H=125.36$~GeV} \\ \cline{2-6}
process         & \multicolumn{2}{c}{$\sqrt{s}=8$ TeV} &\hsb& \multicolumn{2}{c}{\hsb Combined $\sqrt{s}=7$ and 8 TeV \hsb} \tspp \\ \hline
ggF          & $\RESULTBMuggF$   & $\left[\RESULTBMuggFErr\right]$  &&  $\RESULTMuggF$ & $\left[\RESULTMuggFErr\right]$ \tspp  \\
VBF          & $\RESULTBMuVBF$   & $\left[\RESULTBMuVBFErr\right]$  &&  $\RESULTMuVBF$ & $\left[\RESULTMuVBFErr\right]$ \tspp \\
$VH$         & $\RESULTBMuVH$    & $\left[\RESULTBMuVHErr\right]$   &&  $\RESULTMuVH$  & $\left[\RESULTMuVHErr\right]$  \tspp \\
$ttH$        & $\RESULTBMuttH$   & $\left[\RESULTBMuttHErr\right]$  &&  $\RESULTMuttH$ & $\left[\RESULTMuttHErr\right]$ \tspp \\
\hline\hline
\end{tabular}}

\newcommand{\RESULTTableXSProcess}{
\begin{tabular}{ccrl} \hline\hline
\hsb Production process \hsb  && \multicolumn{2}{l}{Cross section [pb] at $\sqrt{s}=8$ TeV \hspace*{0.5cm}} \\ \hline
ggF            &&  $\RESULTBXSggF$  & $\left[\RESULTBXSggFErr\right]$ \tspp \\
VBF            &&  $\RESULTBXSVBF$  & $\left[\RESULTBXSVBFErr\right]$ \tspp \\
$VH$           &&  $\RESULTBXSVH$   & $\left[\RESULTBXSVHErr\right]$  \tspp \\
$ttH$          &&  $\RESULTBXSttH$  & $\left[\RESULTBXSttHErr\right]$ \tspp \\
\hline \hline
\end{tabular}}

\newcommand{\RESULTRatioVBFVHoverggFttH}{0.96\,^{+0.43}_{-0.31}=0.96\,^{+0.33}_{-0.26}\,({\rm stat.})\,^{+0.20}_{-0.13}\,({\rm syst.})\,^{+0.18}_{-0.10}\,({\rm theo.})}

\newcommand{\RESULTRatiogg}{0.56\,^{+0.66}_{-0.45}}
\newcommand{\RESULTRatioZZ}{0.18\,^{+1.20}_{-0.52}}
\newcommand{\RESULTRatioWW}{1.47\,^{+0.80}_{-0.54}}
\newcommand{\RESULTRatiott}{0.81\,^{+2.19}_{-0.49}}
\newcommand{\RESULTRatiobb}{0.33\,^{+1.03}_{-0.25}}
\newcommand{\RESULTRatioCombined}{0.96\,^{+0.43}_{-0.31}}

\newcommand{\RESULTRatioggErr}{{^{+0.62}_{-0.42}}\: {^{+0.15}_{-0.09}}\: {^{+0.18}_{-0.15}}}
\newcommand{\RESULTRatioZZErr}{{^{+1.16}_{-0.50}}\: {^{+0.23}_{-0.05}}\: {^{+0.23}_{-0.16}}}
\newcommand{\RESULTRatioWWErr}{{^{+0.63}_{-0.47}}\: {^{+0.37}_{-0.19}}\: {^{+0.31}_{-0.18}}}
\newcommand{\RESULTRatiottErr}{{^{+1.36}_{-0.41}}\: {^{+1.68}_{-0.15}}\: {^{+0.39}_{-0.23}}}
\newcommand{\RESULTRatiobbErr}{{^{+0.39}_{-0.20}}\: {^{+0.94}_{-0.14}}\: {^{+0.18}_{-0.06}}}
\newcommand{\RESULTRatioCombinedErr}{{^{+0.33}_{-0.26}}\: {^{+0.20}_{-0.13}}\: {^{+0.18}_{-0.10}}}

\newcommand{\RESULTMediationTable}{
\begin{tabular}{clcl}\hline\hline
& Decay channel         &\hsc&  \multicolumn{1}{c}{\hsc Cross-section ratio $R_{ff}$ \hsc}  \\ \hline
& $H\to\gamma\gamma$    &&  $\RESULTRatiogg\ \ \left[\RESULTRatioggErr\right]$ \tspp \\
& $H\to ZZ^*$           &&  $\RESULTRatioZZ\ \ \left[\RESULTRatioZZErr\right]$ \tspp \\
& $H\to WW^*$           &&  $\RESULTRatioWW\ \ \left[\RESULTRatioWWErr\right]$ \tspp \\
& $H\to \tau\tau$       &&  $\RESULTRatiott\ \ \left[\RESULTRatiottErr\right]$ \tspp \\
& $H\to b\bar{b}$       &&  $\RESULTRatiobb\ \ \left[\RESULTRatiobbErr\right]$ \tspp \\ \hline
& Combined              &&  $\RESULTRatioCombined\ \ \left[\RESULTRatioCombinedErr\right]$ \tspp \\
\hline\hline
\end{tabular}}

\newcommand{\RESULTggFWWmu}{0.98\,^{+0.29}_{-0.26}}

\newcommand{\RESULTggFWW}{4.86\,^{+0.95}_{-0.90}}
\newcommand{\RESULTggFWWErr}{^{+0.76}_{-0.74}\,^{+0.52}_{-0.48}\,^{+0.26}_{-0.18}}
\newcommand{\RESULTggFWWSM}{4.22\pm 0.47}

\newcommand{\RESULTNewRVBF}{0.081\,^{+0.035}_{-0.026}}
\newcommand{\RESULTNewRWH} {0.053\,^{+0.037}_{-0.026}}
\newcommand{\RESULTNewRZH} {0.013\,^{+0.030}_{-0.014}}
\newcommand{\RESULTNewRttH}{0.012\,^{+0.007}_{-0.005}}

\newcommand{\RESULTNewRVBFErr}{^{+0.031}_{-0.024}\,^{+0.016}_{-0.010}\,^{+0.008}_{-0.005}}
\newcommand{\RESULTNewRWHErr} {^{+0.032}_{-0.023}\,^{+0.018}_{-0.012}\,^{+0.008}_{-0.004}}
\newcommand{\RESULTNewRZHErr} {^{+0.021}_{-0.013}\,^{+0.020}_{-0.005}\,^{+0.005}_{-0.002}}
\newcommand{\RESULTNewRttHErr}{^{+0.005}_{-0.004}\,^{+0.004}_{-0.003}\,^{+0.0014}_{-0.0005}}

\newcommand{\RESULTNewRyy}{0.010\,^{+0.003}_{-0.003}}
\newcommand{\RESULTNewRZZ}{0.15\,^{+0.05}_{-0.04}}
\newcommand{\RESULTNewRtt}{0.34\,^{+0.14}_{-0.11}}
\newcommand{\RESULTNewRbb}{1.53\,^{+1.64}_{-0.94}}

\newcommand{\RESULTNewRyyErr}{^{+0.003}_{-0.002}\,^{+0.002}_{-0.001}\,^{+0.0006}_{-0.0004}}
\newcommand{\RESULTNewRZZErr}{^{+0.046}_{-0.036}\,^{+0.022}_{-0.013}\,^{+0.008}_{-0.005}}
\newcommand{\RESULTNewRttErr}{^{+0.112}_{-0.090}\,^{+0.084}_{-0.053}\,^{+0.032}_{-0.017}}
\newcommand{\RESULTNewRbbErr}{^{+1.17}_{-0.69}\,^{+1.11}_{-0.63}\,^{+0.30}_{-0.12}}

\newcommand{\RESULTNewRVBFSM}{0.082\pm 0.009}
\newcommand{\RESULTNewRWHSM} {0.036\pm 0.004}
\newcommand{\RESULTNewRZHSM} {0.021\pm 0.002}
\newcommand{\RESULTNewRttHSM}{0.007\pm 0.001}

\newcommand{\RESULTNewRyySM}{0.01036\pm 0.00011}
\newcommand{\RESULTNewRZZSM}{0.124\,\pm <\!0.001}
\newcommand{\RESULTNewRttSM}{0.285\pm 0.006}
\newcommand{\RESULTNewRbbSM}{2.60\pm 0.12}

\newcommand{\RESULTNewRmmLimit}{0.006}
\newcommand{\RESULTNewRzgLimit}{0.078}

\newcommand{\RESULTNewggFSigmaObs}{8.9}
\newcommand{\RESULTNewVBFSigmaObs}{4.3}
\newcommand{\RESULTNewWHSigmaObs} {2.1}
\newcommand{\RESULTNewZHSigmaObs} {0.9}
\newcommand{\RESULTNewVHSigmaObs} {2.6}
\newcommand{\RESULTNewttHSigmaObs}{2.5}

\newcommand{\RESULTNewggFSigmaExp}{6.8}
\newcommand{\RESULTNewVBFSigmaExp}{3.8}
\newcommand{\RESULTNewWHSigmaExp} {2.0}
\newcommand{\RESULTNewZHSigmaExp} {2.1}
\newcommand{\RESULTNewVHSigmaExp} {3.1}
\newcommand{\RESULTNewttHSigmaExp}{1.5}

\newcommand{\RESULTNewSMpValue}{80\%}

\newcommand{\RESULTNewRatioTable}{
\begin{tabular}{crlcr} \hline\hline
Parameter   & \multicolumn{2}{c}{Best-fit value}  &&  SM prediction \\ \hline
$\sigma(gg\to H\to WW^*)$ (pb)        &  $\RESULTggFWW$    & $\left[\RESULTggFWWErr\right]$   && $\RESULTggFWWSM$  \tspp \\[3mm]

$\sigma_{\rm VBF}/\sigma_{\rm ggF}$   &  $\RESULTNewRVBF$  & $\left[\RESULTNewRVBFErr\right]$ && $\RESULTNewRVBFSM$ \tspp \\
$\sigma_{WH}/\sigma_{\rm ggF}$        &  $\RESULTNewRWH$   & $\left[\RESULTNewRWHErr\right]$  && $\RESULTNewRWHSM$  \tspp \\
$\sigma_{ZH}/\sigma_{\rm ggF}$        &  $\RESULTNewRZH$   & $\left[\RESULTNewRZHErr\right]$  && $\RESULTNewRZHSM$  \tspp \\
$\sigma_{ttH}/\sigma_{\rm ggF}$       &  $\RESULTNewRttH$  & $\left[\RESULTNewRttHErr\right]$ && $\RESULTNewRttHSM$ \tspp \\[3mm]

$\Gamma_{\gamma\gamma}/\Gamma_{WW^*}$ &  $\RESULTNewRyy$   & $\left[\RESULTNewRyyErr\right]$ && $\RESULTNewRyySM$ \tspp \\
$\Gamma_{ZZ^*}/\Gamma_{WW^*}$         &  $\RESULTNewRZZ$   & $\left[\RESULTNewRZZErr\right]$ && $\RESULTNewRZZSM$ \tspp \\
$\Gamma_{\tau\tau}/\Gamma_{WW^*}$     &  $\RESULTNewRtt$   & $\left[\RESULTNewRttErr\right]$ && $\RESULTNewRttSM$ \tspp \\
$\Gamma_{bb}/\Gamma_{WW^*}$           &  $\RESULTNewRbb$   & $\left[\RESULTNewRbbErr\right]$ && $\RESULTNewRbbSM$ \tspp \\

\hline \hline
\end{tabular}}

\newcommand{\RESULTNewSigmaTable}{
\begin{tabular}{cccccc} \hline\hline
\hsb Process  \hsb  &     VBF   &  \hsb  $ttH$ \hsb    &   $WH$     & \hsa  $ZH$ \hsa   &  \hsa  $VH$ \hsa \\ \hline
Observed &  \RESULTNewVBFSigmaObs & \RESULTNewttHSigmaObs &  \RESULTNewWHSigmaObs & \RESULTNewZHSigmaObs & \RESULTNewVHSigmaObs \\
Expected &  \RESULTNewVBFSigmaExp & \RESULTNewttHSigmaExp &  \RESULTNewWHSigmaExp & \RESULTNewZHSigmaExp &  \RESULTNewVHSigmaExp \\
\hline\hline
\end{tabular}}

\newcommand{\RESULTMuggFWW}{1.15\,^{+0.28}_{-0.24}}
\newcommand{\RESULTMuggFWWErr}{{^{+0.18}_{-0.18}}\: {^{+0.12}_{-0.11}}\: {^{+0.17}_{-0.12}}}

\newcommand{\RESULTRVBFoverggF}{0.99\,^{+0.46}_{-0.33}}
\newcommand{\RESULTRWHoverggF} {1.47\,^{+1.06}_{-0.74}}
\newcommand{\RESULTRZHoverggF} {0.60\,^{+1.39}_{-0.66}}
\newcommand{\RESULTRttHoverggF}{1.81\,^{+1.10}_{-0.81}}
\newcommand{\RESULTRVHoverggF} {1.33\,^{+0.94}_{-0.68}}

\newcommand{\RESULTRVBFoverggFErr}{{^{+0.37}_{-0.29}}\: {^{+0.20}_{-0.12}}\: {^{+0.18}_{-0.10}}}
\newcommand{\RESULTRWHoverggFErr} {{^{+0.87}_{-0.65}}\: {^{+0.49}_{-0.32}}\: {^{+0.34}_{-0.15}}}
\newcommand{\RESULTRZHoverggFErr} {{^{+0.99}_{-0.60}}\: {^{+0.93}_{-0.25}}\: {^{+0.30}_{-0.07}}}
\newcommand{\RESULTRttHoverggFErr}{{^{+0.79}_{-0.64}}\: {^{+0.61}_{-0.48}}\: {^{+0.46}_{-0.17}}}
\newcommand{\RESULTRVHoverggFErr} {{^{+0.77}_{-0.60}}\: {^{+0.43}_{-0.30}}\: {^{+0.32}_{-0.15}}}

\newcommand{\RESULTRVBFSigmaObs}{\ensuremath{4.3}}
\newcommand{\RESULTRVHSigmaObs} {\ensuremath{2.6}}
\newcommand{\RESULTRttHSigmaObs}{\ensuremath{2.4}}

\newcommand{\RESULTRVBFSigmaExp}{\ensuremath{3.8}}
\newcommand{\RESULTRVHSigmaExp} {\ensuremath{3.1}}
\newcommand{\RESULTRttHSigmaExp}{\ensuremath{1.5}}

\newcommand{\RESULTRhogg}{0.97\,^{+0.32}_{-0.25}}
\newcommand{\RESULTRhoZZ}{1.24\,^{+0.42}_{-0.31}}
\newcommand{\RESULTRhott}{1.20\,^{+0.52}_{-0.38}}
\newcommand{\RESULTRhobb}{0.59\,^{+0.63}_{-0.37}}

\newcommand{\RESULTRhoggErr}{{^{+0.26}_{-0.22}}\: {^{+0.15}_{-0.10}}\: {^{+0.10}_{-0.06}}}
\newcommand{\RESULTRhoZZErr}{{^{+0.37}_{-0.29}}\: {^{+0.18}_{-0.10}}\: {^{+0.07}_{-0.04}}}
\newcommand{\RESULTRhottErr}{{^{+0.40}_{-0.32}}\: {^{+0.29}_{-0.18}}\: {^{+0.17}_{-0.09}}}
\newcommand{\RESULTRhobbErr}{{^{+0.45}_{-0.27}}\: {^{+0.43}_{-0.24}}\: {^{+0.12}_{-0.05}}}

\newcommand{\RESULTRhommLimit}{5.9}
\newcommand{\RESULTRhoZgLimit}{11.0}

\newcommand{\RESULTRatioTable}{
\begin{tabular}{crl} \hline\hline
 \hsc  Parameter \hsc     & \multicolumn{2}{c}{\hsd Best-fit value \hsd}   \\  \hline
$\mu_{\rm ggF}^{WW^*}$     &  $\RESULTMuggFWW$      & $\left[\RESULTMuggFWWErr\right]$      \tspp \\[3mm]

$R_{{\rm VBF}/{\rm ggF}}$  &  $\RESULTRVBFoverggF$  & $\left[\RESULTRVBFoverggFErr\right]$  \tspp\\
$R_{WH/{\rm ggF}}$         &  $\RESULTRWHoverggF$   & $\left[\RESULTRWHoverggFErr\right]$   \tspp \\
$R_{ZH/{\rm ggF}}$         &  $\RESULTRZHoverggF$   & $\left[\RESULTRZHoverggFErr\right]$   \tspp \\
$R_{ttH/{\rm ggF}}$        &  $\RESULTRttHoverggF$  & $\left[\RESULTRttHoverggFErr\right]$  \tspp \\[3mm]

$\rho_{\gamma\gamma/WW^*}$  & $\RESULTRhogg$   & $\left[\RESULTRhoggErr\right]$  \tspp \\
$\rho_{ZZ^*/WW^*}$          & $\RESULTRhoZZ$   & $\left[\RESULTRhoZZErr\right]$  \tspp \\
$\rho_{\tau\tau/WW^*}$      & $\RESULTRhott$   & $\left[\RESULTRhottErr\right]$  \tspp \\
$\rho_{bb/WW^*}$            & $\RESULTRhobb$   & $\left[\RESULTRhobbErr\right]$  \tspp \\

\hline\hline
\end{tabular}}

\newcommand{\RESULTRatioLongTable}{
\begin{tabular}{ccccc} \hline\hline
\hsc Parameter \hsc  &  Best-fit  &&  \multicolumn{2}{c}{Significance ($\sigma$)} \\ \cline{4-5}
                     & \hsb  value \hsb      &&   Observed   & \hsb Expected \hsb   \\ \hline
$\mu_{\rm ggF}^{WW^*}$      & $\RESULTMuggFWW\ \ \left[\RESULTMuggFWWErr\right]$ \tspp \\ \\

$R_{{\rm VBF}/{\rm ggF}}$  &  $\RESULTRVBFoverggF\ \ \left[\RESULTRVBFoverggFErr\right]$   &&  $\RESULTRVBFSigmaObs$   &  $\RESULTRVBFSigmaExp$  \tspp \\
$R_{VH/{\rm ggF}}$         &  $\RESULTRVHoverggF \ \ \left[\RESULTRVHoverggFErr\right] $   &&  $\RESULTRVHSigmaObs$    &  $\RESULTRVHSigmaExp$   \tspp  \\
$R_{ttH/{\rm ggF}}$        &  $\RESULTRttHoverggF\ \ \left[\RESULTRttHoverggFErr\right]$   &&  $\RESULTRttHSigmaObs$   &  $\RESULTRttHSigmaExp$  \tspp \\ \\

$\rho_{\gamma\gamma/WW^*}$  & $\RESULTRhogg\ \ \left[\RESULTRhoggErr\right]$ \tspp \\
$\rho_{ZZ^*/WW^*}$          & $\RESULTRhoZZ\ \ \left[\RESULTRhoZZErr\right]$ \tspp \\
$\rho_{\tau\tau/WW^*}$      & $\RESULTRhott\ \ \left[\RESULTRhottErr\right]$ \tspp \\
$\rho_{bb/WW^*}$            & $\RESULTRhobb\ \ \left[\RESULTRhobbErr\right]$ \tspp \\
\hline\hline
\end{tabular}}

\newcommand{\RESULTMuV}{1.22\,^{+0.18}_{-0.16}=1.22 \pm 0.12\,({\rm stat.})\,^{+0.08}_{-0.07}\,({\rm syst.})\,^{+0.11}_{-0.09}\,({\rm theo.})}
\newcommand{\RESULTMuF}{1.07\,^{+0.27}_{-0.26}=1.07^{+0.21}_{-0.20}\,({\rm stat.})\,^{+0.14}_{-0.15}\,({\rm syst.})\,^{+0.10}_{-0.05}\,({\rm theo.})}

\newcommand{\RESULTMuFSigmaObs}{\ensuremath{4.5}}
\newcommand{\RESULTMuFSigmaExp}{\ensuremath{4.4}}

\section{Signal-strength measurements}
\label{sec:SignalStrength}

This section discusses the measurements of the signal-strength parameter $\mu$ of different production modes and decay channels as well as their ratios for a fixed Higgs boson mass hypothesis of $m_H = 125.36$~\GeV~\cite{Aad:2013wqa}. The signal-strength parameter is a measure of potential deviations from the SM prediction under the assumption that the Higgs boson production and decay kinematics do not change appreciably from the SM expectations. In particular, the transverse momentum and rapidity distributions of the Higgs boson are assumed to be those predicted for the SM Higgs boson by state-of-the-art event generators and calculations of each production process. This assumption is corroborated by studies such as the measurements of differential production cross sections~\cite{Aad:2014lwa,Aad:2014tca} and tests of spin and CP properties of the Higgs boson~\cite{Aad:2013xqa,Khachatryan:2014kca}.

For the discussion in this section, $bbH$ is assumed to have the same signal strength as ggF, $tH$ the same as $ttH$, and $gg\to ZH$ the same as $q\bar{q}\to ZH$, unless noted otherwise. The ggF and $bbH$ processes lead to similar event signatures and no attempt is made to separate them in the analyses, thus the assumption of equal signal strength implies that the observed ggF signal is interpreted as a mixture of $bbH$ and ggF events following their SM ratio of cross sections. The $ttH$ and $tH$ events have similar topologies. The $gg\to ZH$ process leads to the same final state as the $q\bar{q}\to ZH$ process. Whenever $WH$ and $ZH$ are combined into $VH$, their signal strengths are assumed to be the same.

\subsection{Global signal strength}
\label{sec:global}

In Section~\ref{sec:InputChannels}, the published ATLAS measurements on Higgs boson production and decay modes based on individual final states as well as the changes since their publication are summarised. Figure~\ref{fig:muChannels} shows the updated measurements of the signal-strength parameter $\mu$ from a simultaneous fit to all decay channels analysed, assuming SM values for the cross-section ratios of different Higgs boson production processes (or equivalently all $\mu_i$'s of Eq.~(\ref{eq:muif}) are set to be equal). In the fit, the SM predictions of the signal yields are scaled by decay-dependent signal-strength parameters, independent of production processes.  
Compared to the separate measurements shown in Fig.~\ref{fig:inputs}, small changes are observed, resulting from the assignment of the Higgs boson yields in the $ttH$ searches to appropriate decay channels, namely $\Hww$, $\Htt$ and $\Hbb$.\footnote{The measurement of the $\ttH$ signal strength in the multiple-lepton decay mode contributes to all final states with leptons in Fig. \ref{fig:muChannels}, according to the prediction of MC simulation, i.e. predominantly to the $\Hww$ and $\Hzz$ final states.} The central values all increase slightly due to the high observed signal-strength values of the $ttH$ searches, but the uncertainties are barely improved because of the limited significance obtained for the $ttH$ production process with the current dataset. The most significant change in the signal strength is observed for the $\Hbb$ decay. The combination of the $VH(\to \bb)$ analysis and the $ttH(\to\bb)$ search leads to an observed (expected) significance of 1.8~(2.8) standard deviations for the $\Hbb$ decay channel.

\begin{figure}[htb!]
\begin{center}
   \includegraphics[width=.60\textwidth]{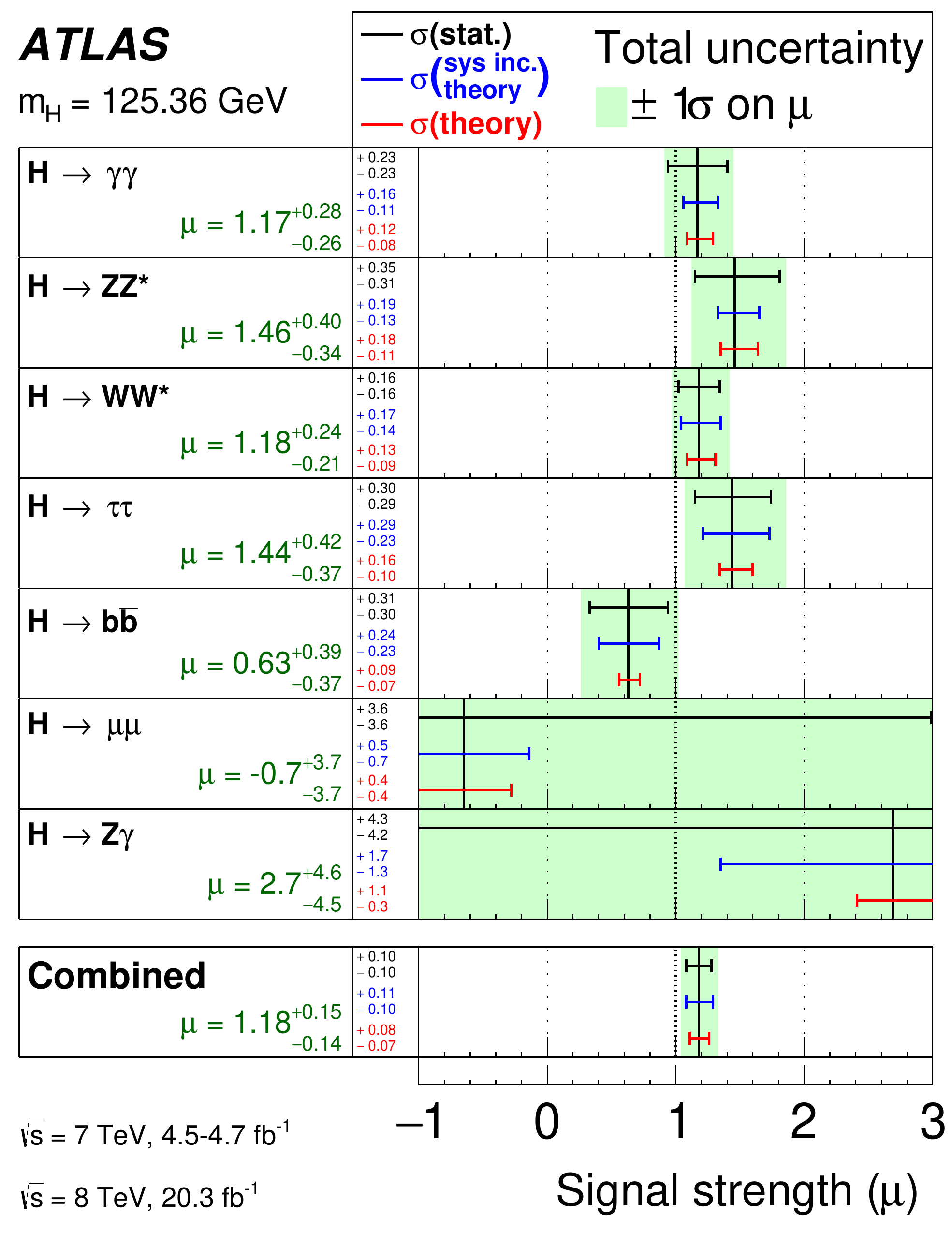}
   \caption{The observed signal strengths and uncertainties for different Higgs boson decay channels and 
     their combination for $m_H=125.36$~GeV. Higgs boson signals corresponding to the same decay channel 
     are combined together for all analyses, assuming SM values for the cross-section ratios of different
     production processes. The best-fit values are shown by
     the solid vertical lines. The total $\pm1\sigma$ uncertainties are indicated by green shaded bands,
     with the individual contributions from the statistical uncertainty (top),
     the total (experimental and theoretical) systematic uncertainty (middle), and the signal theoretical 
     uncertainty (bottom) on the signal strength shown as horizontal error bars.}
   \label{fig:muChannels}
\end{center}
\end{figure}

Assuming a multiplier common to all decay modes, signal-strength measurements of individual decay modes can be combined to give a global and more precise measurement, providing the simplest consistency test with the SM expectation.  Combining all measurements using the profile likelihood ratio $\Lambda(\mu)$ results in a global signal-strength value of 
$$\mu = \RESULTGlobalMuShort=\RESULTGlobalMuLong,$$
where the labels stat., syst. and theo. refer to statistical, systematic, and signal theoretical uncertainties, respectively. The signal theoretical uncertainty includes contributions from uncertainties in SM cross sections and branching ratios as well as in the modelling of the production and decays of the Higgs boson, as discussed in Section~\ref{sec:StatisticalProcedure}. The theoretical uncertainties of background processes are included in the uncertainty labelled as systematic uncertainty. 

The uncertainty on the global signal strength has comparable statistical and systematic components and is significantly reduced compared to the individual measurements, as illustrated in Fig.~\ref{fig:muChannels}. Here, the largest source of experimental systematic uncertainty is from background estimates in the analyses of individual channels. 
This result is consistent with the SM expectation of $\mu=1$, with a $p$-value of \RESULTGlobalMuSMpValue, All individual measurements of the signal-strength parameters are consistent and compatible with the combined value, with a $p$-value of \RESULTGlobalMupValue. 

Performing independent combinations of measurements at $\sqrt{s}=7$ and 8~TeV independently lead to signal-strength values of
\begin{eqnarray*}
  \mu(7\,{\rm TeV}) & = & \RESULTGlobalAMu,\ {\rm and} \\[2mm]
  \mu(8\,{\rm TeV}) & = & \RESULTGlobalBMu
\end{eqnarray*}
at these two energies. The relative theoretical uncertainty of $\sim 7\%$ on the measured $\mu$ value at $\sqrt{s}=8$~TeV arises predominantly from the uncertainty on the total cross section, but is nevertheless smaller than the corresponding uncertainty of $\sim 9\%$ on the total SM cross section shown in Table~\ref{tab:SMPrediction}, because $\mu$ 
is effectively a weighted average of the signal-strength measurements in all categories: the contributions
from VBF and $VH$ production, which have comparatively small theoretical uncertainties, have larger weights 
in this average than in the total cross section.

\subsection{Individual production processes}
\label{sec:process}

In addition to the signal strengths of different decay channels, the signal strengths of different production modes are also determined, exploiting the sensitivity offered by the use of event categories in the analyses of all channels. 

The Higgs boson production modes can be probed with four signal-strength parameters: $\mu_{\rm ggF}$, $\mu_{\rm VBF}$, $\mu_{VH}$ and $\mu_{ttH}$,  one for each main production mode, combining Higgs boson signals from different decay channels under the assumption of SM values for the ratios of the branching ratios of different Higgs boson decays. This assumption is equivalent to set all $\mu_f$'s in Eq.~(\ref{eq:muif}) to be equal. The SM predictions of the signal yields are scaled by these four production-dependent parameters. The best-fit values of these parameters for the $\sqrt{s}=8$~TeV data separately and in combination with the $\sqrt{s}=7$~TeV data are shown in Table~\ref{tab:muProcessErr}.
 Uncertainty components from statistics, systematics, and signal theory are also shown.
The accuracy with which the uncertainties are broken down  is limited by the precision of the fit and more importantly by the approximations made in individual analyses when neglecting uncertainties which are small with respect to, e.g., the statistical uncertainty.
The $\sqrt{s}=7$ and 8~TeV combined values with their total uncertainties are also illustrated in Fig.~\ref{fig:muProcess}. The $\sqrt{s}=7$~TeV data are included in the combinations only, as they have limited statistical power to distinguish between different production modes. The signal-strength measurements are in reasonable agreement with the SM predictions of unity. Although the results support the SM prediction of the $ttH$ production (see Section~\ref{sec:ratio}), 
this production process remains to be firmly established in future LHC runs. Thus, a 95\%~CL upper limit on its signal strength is also derived. Combining the results from various analyses with sensitivity to $ttH$ production, the observed and expected limits are $\mu_{ttH}<\RESULTMuttHLimitObs$ and $\RESULTMuttHLimitExp$, respectively.

\begin{table}[htb]
\caption{Measured signal strengths $\mu$ at $m_H=125.36$~GeV and their total $\pm 1\sigma$ uncertainties for different production modes for the $\sqrt{s}=8$~TeV data and the combination with the $\sqrt{s}=7$~TeV data.  The $\sqrt{s}=7$~TeV data do not have sufficient statistical power to yield meaningful measurements for individual production modes, but are included in the combination. Shown in the square brackets are uncertainty components: statistical (first), systematic (second) and signal theoretical (third) uncertainties. These results are derived using SM values for the ratios of branching ratios of different Higgs boson decay channels.}
\label{tab:muProcessErr}
\begin{center}
\RESULTTableMuProcess
\end{center}
\end{table}

\begin{table}[htb]
\caption{Measured cross sections of different Higgs boson production processes at $\sqrt{s}=8$~TeV for $m_H=125.36$~GeV obtained from the signal-strength values of Table~\ref{tab:muProcessErr}. Their SM predictions can be found in Table~\ref{tab:SMPrediction}.
Shown in the square brackets are uncertainty components: statistical (first), systematic (second) and signal theoretical (third) uncertainties. The theoretical uncertainties here arise from the modelling of Higgs boson production and decays. These results are derived using the SM values of the Higgs boson decay branching ratios.}
\label{tab:XSProcess}
\begin{center}
\RESULTTableXSProcess
\end{center}
\end{table}

\begin{figure}[htb!]
\begin{center}
\includegraphics[width=0.8\textwidth]{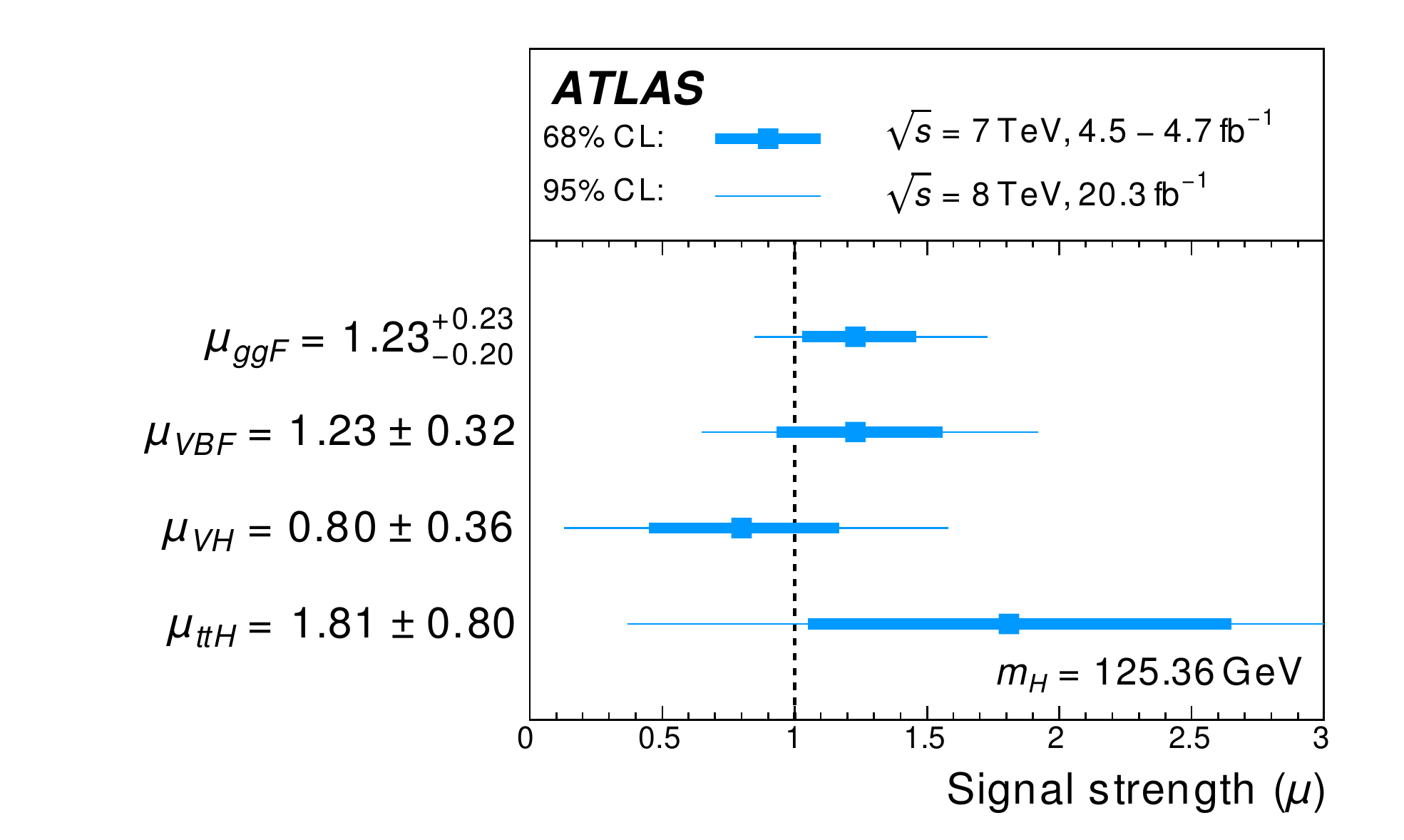}
\end{center}
\caption{The best-fit signal-strength values of different production modes determined from the combined fit to the $\sqrt{s}=7$ and 8~TeV data. Higgs boson signals corresponding to the same production process but from different decay channels are combined together, assuming SM values for the ratios of the branching ratios of different Higgs boson decay channels. The inner and outer error bars correspond to 68\%~CL and 95\%~CL intervals. Total uncertainties combining statistical, experimental and theoretical systematic uncertainties are shown.}
\label{fig:muProcess}
\end{figure} 

The signal-strength measurements shown in Table~\ref{tab:muProcessErr} are extrapolated to total cross-section measurements for each production process, as shown in Table~\ref{tab:XSProcess} for $\sqrt{s}=8$~TeV, with the further assumption of SM values for the Higgs boson decay branching ratios. The theoretical uncertainties on the absolute values of the SM Higgs boson production cross sections are thereby removed, but significant theoretical uncertainties remain, related to the modelling of the Higgs boson production and of the acceptance of the event selection. One can sum the different cross sections to obtain an overall extrapolated cross section for Higgs boson production. The measurement is performed at $\sqrt{s}=7$~TeV as well despite of the limited statistical power of the dataset. The resulting total Higgs boson production cross sections at the two energies are
\begin{eqnarray*}
   \sigma_H(7\,{\rm TeV}) & = \RESULTAXSAllShort\:{\rm pb} &= \RESULTAXSAllLong\: {\rm pb},\ {\rm and} \\[2mm]
   \sigma_H(8\,{\rm TeV}) & = \RESULTBXSAllShort\:{\rm pb} & = \RESULTBXSAllLong\: {\rm pb}\,,
\end{eqnarray*}
to be compared with the theoretical predictions of  $17.4\pm 1.6$~pb at $\sqrt{s}=7$~TeV and $22.3\pm 2.0$~pb at $\sqrt{s}=8$~TeV, as shown in Table~\ref{tab:SMPrediction}.

These cross sections are different from what one would naively expect from the global signal-strength values discussed in Section~\ref{sec:global}, particularly for $\sqrt{s}=7$~TeV. The differences are largely the result of analysis categorisation. Categories often explore production processes or phase-space regions with distinct signal-event topologies. The resulting high signal-to-background ratios can significantly improve the precision of the signal-strength measurements. However, these categories often account for small fractions of the production cross section and thus have limited impact on the total cross-section measurement, which is dominated by processes with larger expected cross sections. One good example is the VBF category. It contributes significantly to the global signal-strength measurement, but has a relatively minor impact on the total cross-section measurement.

\subsection{Boson and fermion-mediated production processes}
\label{sec:mediation}

The Higgs boson production processes can be categorised into two groups according to the Higgs boson couplings to fermions (ggF and $ttH$) or vector bosons (VBF and $VH$). 
Potential deviations from the SM  can be tested with two signal-strength parameters, $\mu^f_{\text{ggF}+ttH}\equiv(\mu^f_{\text{ggF}}=\mu^f_{ttH}$) and $\mu^f_{\text{VBF}+VH}\equiv(\mu^f_{\text{VBF}}=\mu^f_{VH})$ for each decay channel $f$, assuming SM values for the ratio of ggF and $ttH$ cross sections and the ratio of VBF and VH cross sections.
Signal contaminations from one group to another, e.g. ggF events with two jets passing the VBF selection, are taken
into account in the simultaneous fit.
The 68\% and 95\%~CL two-dimensional contours of $\mu^f_{\text{ggF}+ttH}$ and $\mu^f_{\text{VBF}+VH}$ of the five main decay channels are shown in Fig.~\ref{fig:muContour}. The measurements of $\Hmm$ and $\Hzg$ decays have relatively poor sensitivities and are therefore not included in the figure. 
The cutoff in the contours of the \hgg\ and \hzz\ decays is caused by the expected sum of signal and background yields in one of the contributing measurements going below zero in some regions of the parameter space shown in Fig.~\ref{fig:muContour}.
The SM expectation of $\mu^f_{\text{ggF}+ttH}=1$ and $\mu^f_{\text{VBF}+VH}=1$ is within the 68\%~CL contour of most of these measurements. 

\begin{figure}[htb]
 \centering
  \includegraphics[width=.85\textwidth]{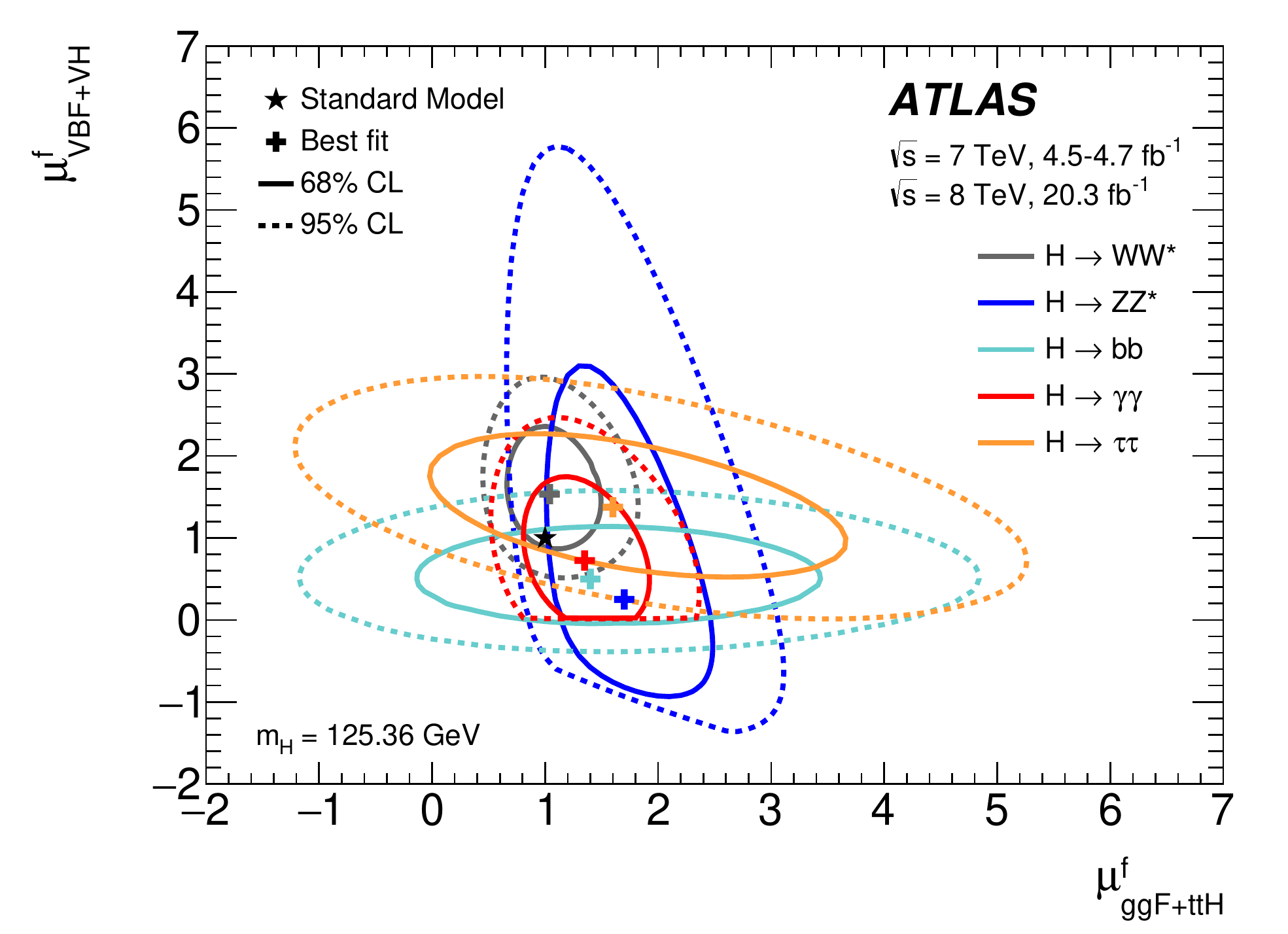}
\caption{Likelihood contours in the $(\mu^f_{\mathrm{ggF}+ttH}, \mu^f_{\mathrm{VBF}+VH})$ plane for 
        a Higgs boson mass $m_H=125.36$~\GeV\ measured separately for $H\to WW^*,\, ZZ^*,\, \bb,\, \gamma\gamma$ and  
        $\tau\tau$ decays. SM values are assumed for the relative contributions between ggF and $ttH$ 
        and between VBF and $VH$ production. 
        The straight lower portions of the $\Hyy$ and $\Hllll$ contours are due to 
        the small numbers of events in these channels and the requirement of a positive probability density
        function. The best-fit values to the data (+) and the $68\%$ (full) and $95\%$ (dashed)~CL 
        contours are indicated, as well as the SM expectation ($\star$).}
\label{fig:muContour}
\end{figure}

The relative production cross sections of the processes mediated by vector bosons and by fermions can be tested using the ratio $\mu^f_{\text{VBF}+VH}/\mu^f_{\text{ggF}+ttH}$. When measured separately for each decay channel, this ratio reduces to the ratio of production cross sections because the Higgs boson decay branching ratios cancel and is equivalent to the ratio of $\mu_i$ defined in Section~\ref{sec:global}, i.e., 
\begin{equation}
   \frac{\mu^f_{\text{VBF}+VH}}{\mu^f_{\text{ggF}+ttH}} = \frac{\sigma_{\text{VBF}+VH}/\sigma_{\text{ggF}+ttH}}{\left[\sigma_{\text{VBF}+VH}/\sigma_{\text{ggF}+ttH}\right]}_{\rm SM}=\frac{\mu_{{\rm VBF}+VH}}{\mu_{{\rm ggF}+ttH}} \equiv R_{ff}\,.
\label{eq:Rff}
\end{equation}
The observed ratios are shown in Table~\ref{tab:BFratio} and illustrated in Fig.~\ref{fig:BFratio} for the five main decay channels. The signal-strength parameter $\mu^f_{{\rm ggF}+ttH}$ of each decay channel is profiled in the fit.
The combination of these measurements yields an overall value of the ratio of cross sections for the vector-boson- and fermion-mediated processes (relative to its SM prediction):
\begin{equation*}
R_{\rm Combined}  =\RESULTRatioVBFVHoverggFttH\label{eq:VBFVHratio} .
\end{equation*}

\begin{table}[htb]
\caption{The best-fit values and their uncertainties for the ratio $R_{ff}$ of cross sections for the vector-boson- and fermion-mediated production processes relative to their SM values at $m_H=125.36$~GeV for the individual decay channels and their combination. Shown in the square brackets are uncertainty components: statistical (first),  systematic (second) and signal theoretical (third) uncertainties. These results are independent of the Higgs boson decay branching ratios.}
\begin{center}
\RESULTMediationTable
\end{center}
\label{tab:BFratio}
\end{table}

\begin{figure}[htb]
\begin{center}
\includegraphics[width=0.70\textwidth]{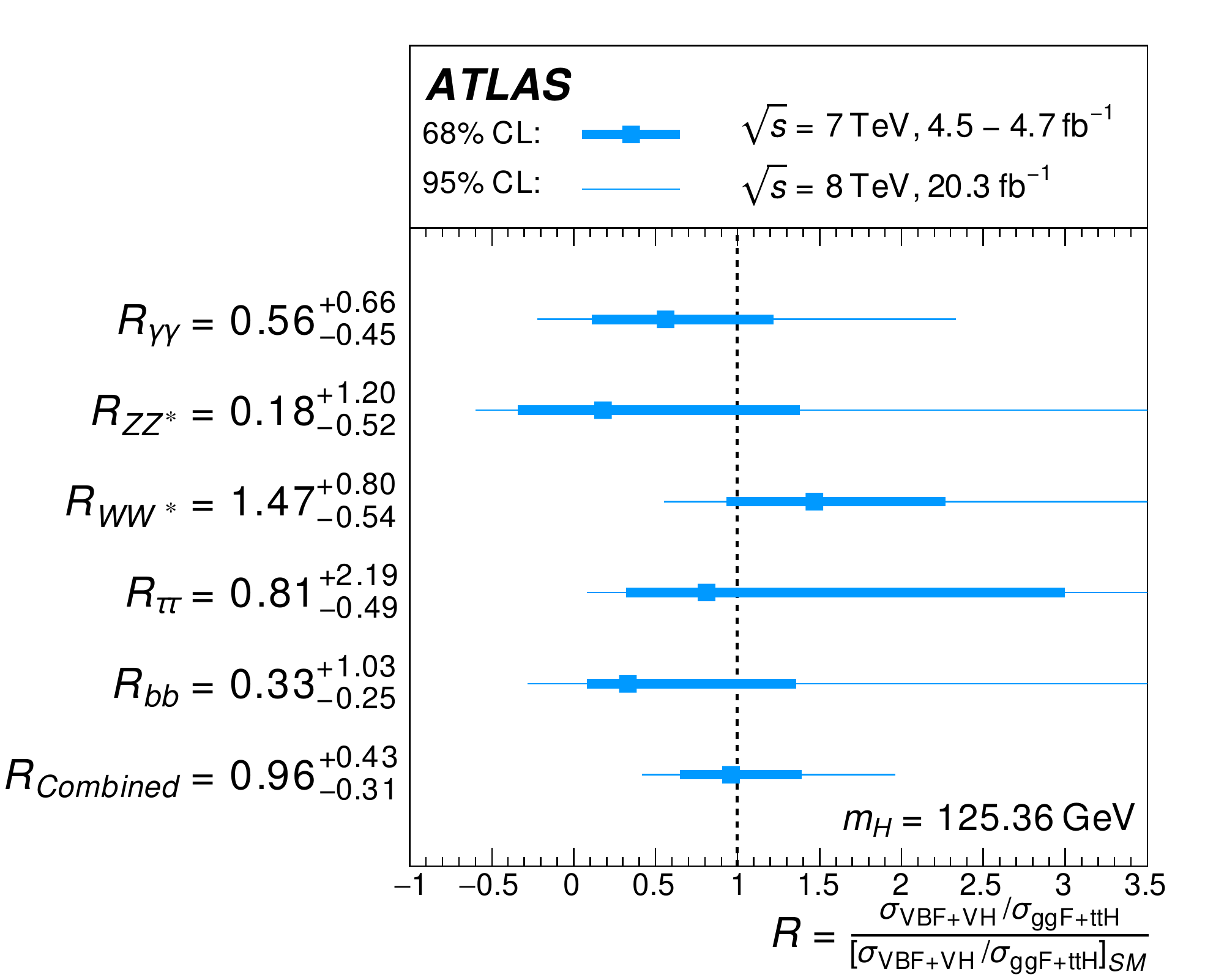}
\caption{The ratios of cross sections for the vector-boson- and fermion-mediated processes relative to their SM values at $m_H=125.36$~GeV, measured in the individual Higgs boson decay final states and their combination, $R_{\rm Combined}$ (see text). The inner and outer error bars represent 68\% CL and 95\% CL intervals, combining statistical and systematic uncertainties. These measurements are independent of Higgs boson decay branching ratios.}
\label{fig:BFratio}
\end{center}
\end{figure}

\clearpage
\subsection{Ratios of production cross sections and partial decay widths}
\label{sec:ratio}

At the LHC, the Higgs boson production cross sections and decay branching ratios cannot be separately determined in a model-independent way as only their products are measured. However, the ratios of cross sections and ratios of branching ratios can be disentangled without any assumptions, within the validity of the narrow width approximation of the Higgs boson.
By normalising to the cross section of the $gg\to H\to WW^*$ production process, $\sigma(gg\to H\to WW^*)$,  the yields of other Higgs boson production modes and decay channels can be parameterised using the ratios of cross sections and ratios of branching ratios. For the production and decay $i\to H\to f$, the yield is then
\begin{equation}
\sigma_i\cdot {\rm BR}_f = \left(\sigma_{\rm ggF}\cdot {\rm BR}_{WW^*}\right) \times \left(\frac{\sigma_i}{\sigma_{\rm ggF}}\right)\times \left(\frac{{\rm BR}_f}{{\rm BR}_{WW^*}}\right)= \sigma(gg\to H\to WW^*)\times  \left(\frac{\sigma_i}{\sigma_{\rm ggF}}\right)\times \left(\frac{\Gamma_f}{\Gamma_{WW^*}}\right).
\label{eq:ratio}
\end{equation}

The ratio of branching ratios in the above equation is substituted by the equivalent ratio of partial decay widths. The ratios extracted from the measured yields are independent of theoretical predictions on the inclusive cross sections and partial decay widths (and thus branching ratios). Furthermore, many experimental systematic uncertainties cancel in the ratios. The residual theoretical uncertainties are related to the modelling of the Higgs boson production and decay, which impacts the signal acceptance calculations. 
The $gg\to H\to WW^*$ process is chosen as the reference because it has both the smallest statistical and overall uncertainties, as shown in Fig.~\ref{fig:muChannels}.

The $\sqrt{s}=7$ and 8~TeV data are fitted with $\sigma(gg\to H\to WW^*)$, $\sigma_i/\sigma_{\rm ggF}$ and $\Gamma_f/\Gamma_{WW^*}$ as parameters of interest and the results are listed in Table~\ref{tab:ratio}, together with the SM predictions~\cite{Heinemeyer:2013tqa}. The results after normalising to their SM values are illustrated in Fig.~\ref{fig:ratio}. The results of $\sigma(gg\to H\to WW^*)$ and $\sigma_i/\sigma_{\rm ggF}$ from the combined analysis of the $\sqrt{s}=7$ and 8~TeV data are shown for $\sqrt{s}=8$~TeV, assuming the SM values for $\sigma_i(7~\TeV)/\sigma_i(8~\TeV)$. The $WH$ and $ZH$ production processes are treated independently in the fit to allow for direct comparisons with theoretical predictions. The searches for $\Hmm$ and $\Hzg$ decays are included in the fit, but the current datasets do not result in sensitive measurements for these two decays. Therefore only 95\% CL upper limits are derived, namely $\RESULTNewRmmLimit$ for $\Gamma_{\mu\mu}/\Gamma_{WW^*}$ and $\RESULTNewRzgLimit$ for $\Gamma_{Z\gamma}/\Gamma_{WW^*}$.
The $p$-value of the compatibility between the data and the SM predictions is found to be $\RESULTNewSMpValue$.

\begin{table}[htbp!]
\caption{Best-fit values of $\sigma(gg\to H\to WW^*)$, $\sigma_i/\sigma_{\rm ggF}$ and $\Gamma_f/\Gamma_{WW^*}$ for a Higgs boson with $m_H=125.36$~GeV from the combined analysis of the $\sqrt{s}=7$ and 8~TeV data. The cross-section ratios are given for $\sqrt{s}=8$~TeV assuming the SM values for $\sigma_i(7~\TeV)/\sigma_i(8~\TeV)$. Shown in square brackets are uncertainty components: statistical (first),  systematic (second) and signal theoretical (third) uncertainties. The SM predictions~\cite{Heinemeyer:2013tqa} are shown in the last column.}
\label{tab:ratio}
\begin{center}
\RESULTNewRatioTable
\end{center}
\end{table}

The results exhibit a few interesting features that are worth mentioning. As a multiplicative factor common to all rates in this parameterisation, $\sigma(gg\to H\to WW^*)$ is pulled up in the fit to accommodate the observed large global signal-strength value (Section~\ref{sec:global}). The best-fit value of $\sigma(gg\to H\to WW^*)$  is approximately 15\% above the SM prediction, to be compared to the significantly lower value of $\RESULTggFWWmu$, found from the stand-alone measurement from the $\Hww$ decay (see Fig.~\ref{fig:inputs}). Moreover, there are by construction large anti-correlations between $\sigma(gg\to H\to WW^*)$, $\sigma_i/\sigma_{\rm ggF}$ and $\Gamma_f/\Gamma_{WW^*}$.

\begin{figure}[htb!]
\begin{center}
\includegraphics[width=0.9\textwidth]{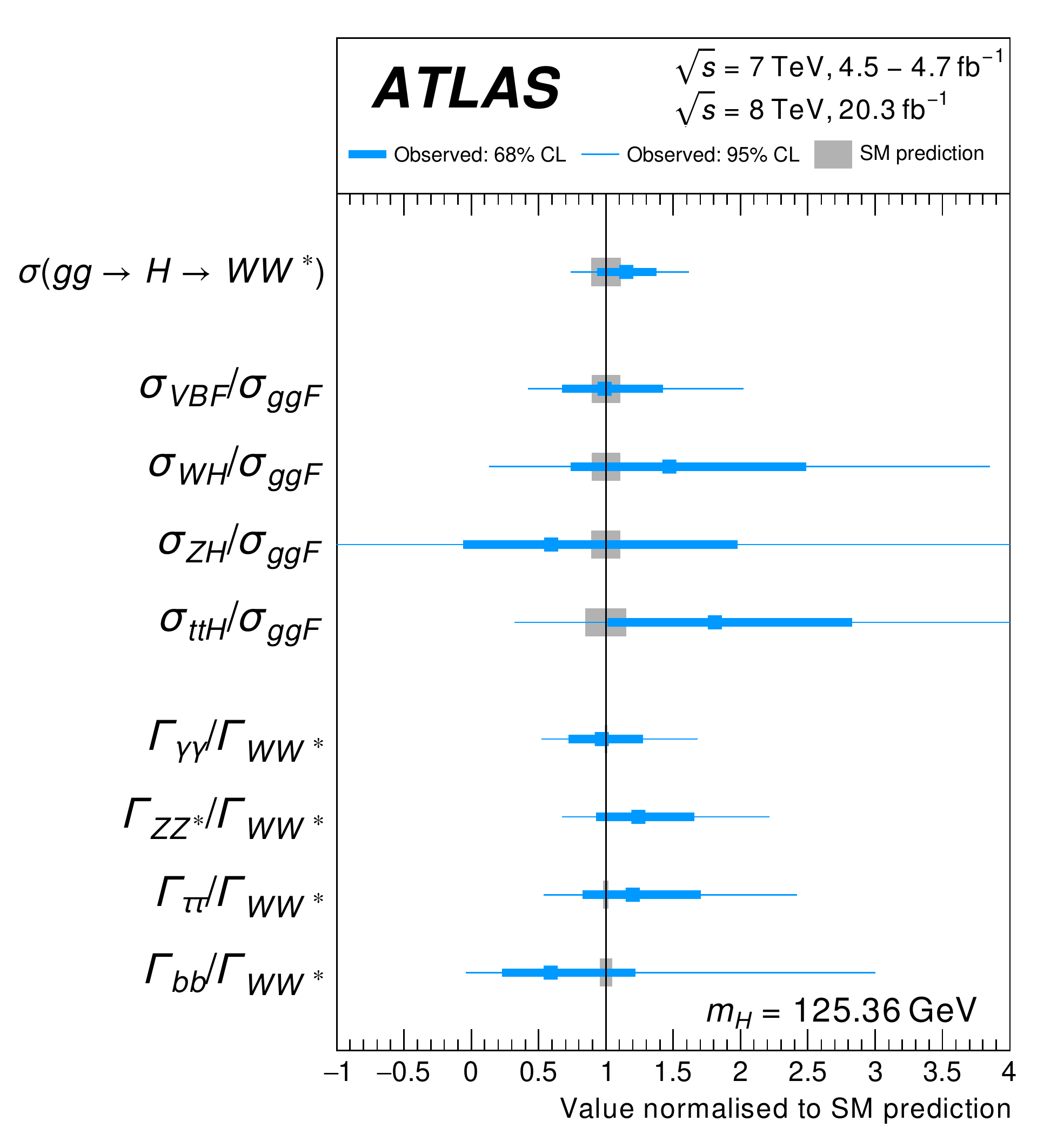}
\caption{The $gg\to H\to WW^*$ cross section, ratios of cross sections and of partial decay widths relative to their SM values at $m_H=125.36$~GeV from the combined analyses of the $\sqrt{s}=7$ and 8~TeV data. 
The inner and outer error bars on the measurements are 68\% CL and 95\% CL intervals. The SM predictions are shown as the vertical line at unity with grey bands representing theoretical uncertainties on the ratios of inclusive cross sections and of partial decay widths. 
}
\label{fig:ratio}
\end{center}
\end{figure}

\begin{table}[htbp]
\caption{The observed and expected significances in units of standard deviations for different Higgs boson production processes except ggF production which is well established (see text). The significances of $VH$ production are obtained by combining the $WH$ and $ZH$ processes, assuming the SM value for their relative cross sections. All significances are calculated under the asymptotic approximation~\cite{Cowan:2010st}.}
\label{tab:sigma}
\begin{center}
\RESULTNewSigmaTable
\end{center}
\end{table}

Table~\ref{tab:sigma} shows the observed and expected significances in units of standard deviations of the VBF, $WH$, $ZH$ and $ttH$ production processes. Listed under $VH$ are the combined significances of $WH$ and $ZH$ production, assuming the SM value for their relative cross sections. 
The significance is calculated from a likelihood scan,  where the contributions from other processes are fixed at their best-fit values. As the $gg\to H\to WW^*$ process is chosen as the reference, the significances are calculated using the observable $\sigma(gg\to H\to WW^*)$ for the ggF process and the cross-section ratios $\sigma_i/\sigma_{\rm ggF}$ for all other processes. The cross-section ratios are independent of the Higgs boson decay branching ratios and have the advantage of the cancellation of many experimental uncertainties. The result provides an unequivocal confirmation of the gluon fusion production of the Higgs boson with its significance exceeding well above five standard deviations. Furthermore, the result also offers strong evidence, at  $\RESULTRVBFSigmaObs$ standard deviations, of vector-boson fusion production and supports the SM assumptions of production in association with vector bosons or a pair of top quarks. 

An alternative parameterisation normalising the ratios of cross sections and of branching ratios to their SM values is presented in Appendix~\ref{sec:oldRatio}.

\newcommand{\RESULTHiggsMass}{\ensuremath{125.36}\xspace}

\newcommand{\RESULTHiggsEffectiveCouplings}{
\begin{eqnarray}
\Cc_{\PGg}^2    &\sim& 1.59 \cdot \Cc_{\PW}^2 -0.66 \cdot \Cc_{\PW} \Cc_{\PQt} + 0.07 \cdot \Cc_{\PQt}^2 \label{eqn:kH2:simple:a}\\
\Cc_{\Pg}^2     &\sim& 1.06 \cdot\Cc_{\PQt}^2  - 0.07\cdot\Cc_{\PQt}\Cc_{\PQb} + 0.01 \cdot \Cc_{\PQb}^2 \label{eqn:kH2:simple:b}\\
\Cc_{\PZ\PGg}^2 &\sim& 1.12 \cdot\Cc_{\PW}^2  - 0.12\cdot\Cc_{\PQt}\Cc_{\PW} \label{eqn:kH2:simple:d}\\
\Cc_\text{VBF}^2&\sim& 0.74 \cdot \Cc_{\PW}^2 + 0.26 \cdot \Cc_{\PZ}^2 \label{eqn:kH2:simple:c}\\
\Cc_{\PH}^2     &\sim& 0.57 \cdot \Cc_{\PQb}^2 + 0.22 \cdot \Cc_{\PW}^2 + 0.09 \cdot \Cc_{\Pg}^2 + 0.06 \cdot \Cc_{\PGt}^2 + 0.03 \cdot \Cc_{\PZ}^2 + 0.03 \cdot \Cc_{\PQc}^2 \label{eqn:kH2:simple} \\
\Cc_{\Pg\Pg\PZ\PH}^2  &\sim& 0.35 \cdot\Cc_{\PQt}^2  + 2.23\cdot\Cc_{\PZ}^2 - 2.23 \cdot \Cc_{\PZ}\Cc_{\PQt} \hspace{0.5cm}{\rm (7~TeV)} \label{eqn:kH2:simple:g}\\
\Cc_{\Pg\Pg\PZ\PH}^2  &\sim& 0.37 \cdot\Cc_{\PQt}^2  + 2.27\cdot\Cc_{\PZ}^2 - 1.61 \cdot \Cc_{\PZ}\Cc_{\PQt} \hspace{0.5cm}{\rm (8~TeV)} \label{eqn:kH2:simple:h} .
\end{eqnarray}
}

\newcommand{\RESULTNegativeCVCFZValue}{4.0\sigma}

\newcommand{\RESULTmodelCVCFvalueCV}{1.09\pm 0.07  \left[ ^{+0.05}_{-0.05}({\rm stat.})\:^{+0.03}_{-0.03}({\rm syst.})\:^{+0.04}_{-0.03}({\rm theo.})\right]\xspace}
\newcommand{\RESULTmodelCVCFvalueCF}{1.11\pm 0.16  \left[ ^{+0.12}_{-0.11}({\rm stat.})\:^{+0.10}_{-0.09}({\rm syst.})\:^{+0.06}_{-0.05}({\rm theo.})\right]\xspace}

\newcommand{\RESULTmodelCVCFpvalueSM}{41\%}

\newcommand{\RESULTmodelRFVCVVvalueRFV}{1.02^{+0.15}_{-0.13} \left[ ^{+0.11}_{-0.11} ({\rm stat.})\:^{+0.08}_{-0.07}({\rm syst.})\:^{+0.04}_{-0.03}({\rm theo.})\right]\xspace}
\newcommand{\RESULTmodelRFVCVVvalueCVV}{1.07^{+0.14}_{-0.13} \left[ ^{+0.11}_{-0.11} ({\rm stat.})\:^{+0.06}_{-0.06}({\rm syst.})\:^{+0.04}_{-0.04}({\rm theo.})\right]\xspace}

\newcommand{\RESULTmodelRFVCVVpvalueSM}{41\%}

\newcommand{\RESULTmodelCUSTvalueRWZ}{1.00^{+0.15}_{-0.11}}
\newcommand{\RESULTmodelCUSTvalueRFZ}{1.03^{+0.21}_{-0.18}}
\newcommand{\RESULTmodelCUSTvalueCZZ}{1.06^{+0.29}_{-0.24}}

\newcommand{\RESULTmodelCUSTpvalueSM}{62\%}

\newcommand{\RESULTmodelUPDOWNvalueRDU}{[-1.08,-0.81] \cup [0.75,1.04]}
\newcommand{\RESULTmodelUPDOWNvalueRVU}{0.92^{+0.18}_{-0.16}}
\newcommand{\RESULTmodelUPDOWNvalueCUU}{1.25^{+0.33}_{-0.33}}

\newcommand{\RESULTmodelUPDOWNvalueRDUupper}{0.90^{+0.14}_{-0.15}}

\newcommand{\RESULTmodelUPDOWNpvalueDownType}{4.5\sigma}
\newcommand{\RESULTmodelUPDOWNpvalueUpType}{8\sigma}
\newcommand{\RESULTmodelUPDOWNpvalueVType}{10.2\sigma}

\newcommand{\RESULTmodelUPDOWNpvalueSM}{51\%}

\newcommand{\RESULTmodelQLvalueRLQ}{[-1.34, -0.94] \cup [0.94, 1.34]}
\newcommand{\RESULTmodelQLvalueRVQ}{1.03^{+0.18}_{-0.15}}
\newcommand{\RESULTmodelQLvalueCQQ}{1.03^{+0.24}_{-0.20}}

\newcommand{\RESULTmodelQLvalueRLQupper}{1.12^{+0.22}_{-0.18}}

\newcommand{\RESULTmodelQLpvalueLepton}{\sim 4.4\sigma}
\newcommand{\RESULTmodelQLpvalueV}{\sim 9.8\sigma}
\newcommand{\RESULTmodelQLpvalueQuark}{\sim 8\sigma}

\newcommand{\RESULTmodelQLpvalueSM}{53\%}

\newcommand{\RESULTmodelELOOPvalueCGL}{1.13 ^{+0.11}_{-0.11}}
\newcommand{\RESULTmodelELOOPvalueCGA}{0.99^{+0.14}_{-0.12}}
\newcommand{\RESULTmodelELOOPvalueCZGA}{1.59^{+0.94}_{-4.91}}
\newcommand{\RESULTmodelELOOPvalueCGGZH}{0.00^{+3.24}_{-3.24}}

\newcommand{\RESULTmodelELOOPpvalueSM}{69\%}

\newcommand{\RESULTmodelELOOPBRUIvalueCGL}{1.12 ^{+0.14}_{-0.11}   \left[ ^{+0.10}_{-0.08}({\rm stat.})\:^{+0.05}_{-0.05}({\rm syst.})\:^{+0.07}_{-0.07}({\rm theo.})\right]}
\newcommand{\RESULTmodelELOOPBRUIvalueCGA}{1.00 \pm 0.12    \left[ ^{+0.11}_{-0.11}({\rm stat.})\:^{+0.05}_{-0.05}({\rm syst.})\:^{+0.04}_{-0.03}({\rm theo.})\right]}
\newcommand{\RESULTmodelELOOPBRUIvalueCZGA}{ 3.3~~(95\%~{\rm C.L.})}
\newcommand{\RESULTmodelELOOPBRUIvalueBRINVlimit}{ 0.27~~(95\%~{\rm C.L.})}
\newcommand{\RESULTmodelELOOPBRUIvalueBRINV}{-0.15 ^{+0.21}_{-0.22} \left[ ^{+0.17}_{-0.17}({\rm stat.})\:^{+0.11}_{-0.11}({\rm syst.})\:^{+0.06}_{-0.07}({\rm theo.})\right]\xspace}

\newcommand{\RESULTmodelELOOPBRUIpvalueSM}{74\%}

\newcommand{\RESULTmodelELOOPBRUIlimitFC}{0.27}

\newcommand{\RESULTmodelELOOPBRUIlimitFCSM}{0.37}

\newcommand{\RESULTmodelELOOPBRUIvalueGammaH}{1.03^{+0.13}_{-0.03}}

\newcommand{\RESULTmodelELOOPBRUICVCFAvalueCV}{1.00^{+0}_{-0.05})}
\newcommand{\RESULTmodelELOOPBRUICVCFAvalueCF}{1.10^{+0.19}_{-0.21}}
\newcommand{\RESULTmodelELOOPBRUICVCFAvalueCGL}{1.20^{+0.21}_{-0.18}}
\newcommand{\RESULTmodelELOOPBRUICVCFAvalueCGA}{1.00^{+0.14}_{-0.12}}
\newcommand{\RESULTmodelELOOPBRUICVCFAvalueCZGA}{1.60^{+0.93}_{-4.13}}
\newcommand{\RESULTmodelELOOPBRUICVCFAvalueCGGZH}{x.xx^{+0.xx}_{-0.xx}}
\newcommand{\RESULTmodelELOOPBRUICVCFAvalueBRINV}{0^{+0.10}_{-0.00}}

\newcommand{\RESULTmodelELOOPBRUICVCFBvalueCV}{1.00^{+0.23}_{-0.16}}
\newcommand{\RESULTmodelELOOPBRUICVCFBvalueCF}{1.00^{+0.23}_{-0.16}}
\newcommand{\RESULTmodelELOOPBRUICVCFBvalueCGL}{1.00^{+0.23}_{-0.16}}
\newcommand{\RESULTmodelELOOPBRUICVCFBvalueCGA}{1.17^{+0.16}_{-0.13}}
\newcommand{\RESULTmodelELOOPBRUICVCFBvalueCZGA}{x.xx^{+0.xx}_{-0.xx}}
\newcommand{\RESULTmodelELOOPBRUICVCFBvalueCGGZH}{x.xx^{+0.xx}_{-0.xx}}
\newcommand{\RESULTmodelELOOPBRUICVCFBvalueBRINV}{-0.16^{+0.29}_{-0.30}}

\newcommand{\RESULTmodelELOOPBRUICVCFApvalueSM}{96\%}
\newcommand{\RESULTmodelELOOPBRUICVCFBpvalueSM}{64\%}

\newcommand{\RESULTmodelELOOPBRUICVCFAlimitFC}{0.27}
\newcommand{\RESULTmodelELOOPBRUICVCFBlimitFC}{0.54}

\newcommand{\RESULTmodelELOOPBRUICVCFAlimitFCSM}{0.39}
\newcommand{\RESULTmodelELOOPBRUICVCFBlimitFCSM}{0.72}

\newcommand{\RESULTmodelBRUICVCFAvalueCV}{1.00^{+0.00}_{-0.04}}
\newcommand{\RESULTmodelBRUICVCFAvalueCF}{1.10^{+0.17}_{-0.15}}
\newcommand{\RESULTmodelBRUICVCFAvalueBRINV}{0^{+0.03}_{-0.00}}

\newcommand{\RESULTmodelBRUICVCFApvalueSM}{99\%}
\newcommand{\RESULTmodelBRUICVCFBpvalueSM}{29\%}

\newcommand{\RESULTmodelBRUICVCFAlimitFC}{0.13}
\newcommand{\RESULTmodelBRUICVCFBlimitFC}{0.52}

\newcommand{\RESULTmodelBRUICVCFAlimitFCSM}{0.24}
\newcommand{\RESULTmodelBRUICVCFBlimitFCSM}{0.71}

\newcommand{\RESULTmodelGENIpvalueSM}{57\%}

\newcommand{\RESULTmodelGENIIpvalueSMkV}{80\%}
\newcommand{\RESULTmodelGENIIpvalueSMoffshell}{57\%}
\newcommand{\RESULTmodelGENIIpvalueSMnoBRui}{73\%}

\newcommand{\RESULTmodelGENIIIpvalueSM}{73\%}

\newcommand{\RESULTmodelGENIIBRuiA}{0.49}
\newcommand{\RESULTmodelGENIIBRuiB}{0.68}
\newcommand{\RESULTTableGenericModel}{
\begin{tabular}{ccccccccc} \hline\hline
\multirow{2}{*}{\hsa Parameter \hsa} && \multirow{2}{*}{\hsa  $\kappa_V <1$  \hsa}            && \multirow{2}{*}{\hsa   $\Cc_{\rm on}=\Cc_{\rm off}$ \hsa }   && \multicolumn{2}{c}{  \hsa   $\BRinv = 0$ \hsa} \\ \cline{7-8}
&&                                                       &&                                                              &&    Fitted value         & Uncertainty breakdown \\ \hline
$\Cc_{\PW}$        &         &  $>0.64$ (95\% CL)                  &&  $=0.96\pm^{0.35}_{0.16}$                  &&   $=0.92^{+0.14}_{-0.15}$  \tsp             & $\left[ ^{+0.11}_{-0.11}({\rm stat.})\:^{+0.07}_{-0.08}({\rm syst.})\:^{+0.03}_{-0.03}({\rm theo.})\right]$ \\
  $\Cc_{\PZ}$        &         &  $> 0.71$ (95\% CL)                 &&  $=1.05\pm^{0.38}_{0.17}$                  &&   $\in [-1.08,-0.84]\cup [0.86,1.14]$  \tsp & $\left[ ^{+0.13}_{-0.13}({\rm stat.})\:^{+0.05}_{-0.07}({\rm syst.})\:^{+0.03}_{-0.02}({\rm theo.})\right]$  \\
  $\Cc_{\PQt}$       &         &  $=1.28^{+0.32}_{-0.35}$                   &&  $=1.35^{+0.61}_{-0.39}$                   &&  $\in [-1.12,-1.00]\cup [0.93,1.60]$ \tsp   & $\left[ ^{+0.20}_{-0.22}({\rm stat.})\:^{+0.22}_{-0.26}({\rm syst.})\:^{+0.12}_{-0.06}({\rm theo.})\right]$  \\
  $|\Cc_{\PQb}|$     & $=$     &  $0.62 \pm 0.28$                    &&  $0.64^{+0.34}_{-0.28}$                    &&  $0.62^{+0.31}_{-0.27}$ \tsp                & $\left[ ^{+0.21}_{-0.20}({\rm stat.})\:^{+0.17}_{-0.18}({\rm syst.})\:^{+0.06}_{-0.03}({\rm theo.})\right]$  \\
  $|\Cc_\tau|$       & $=$     &  $0.99^{+0.22}_{-0.18}$               &&  $1.03^{+0.21}_{-0.40}$                    &&  $1.00\pm 0.20$ \tsp                      & $\left[ ^{+0.15}_{-0.14}({\rm stat.})\:^{+0.12}_{-0.11}({\rm syst.})\:^{+0.06}_{-0.04}({\rm theo.})\right]$  \\
  $|\Cc_\mu|$        & $<$     &  $2.3$ (95\% CL)                    &&  $2.8$ (95\% CL)                        &&  $2.3$ (95\% CL)   \\ \\[-2mm]

  $\Cc_{\PGg}$       & =       & $0.90^{+0.16}_{-0.14}$                &&  $0.93\pm^{0.36}_{0.17}$                   &&  $0.90 \pm 0.15$  \tsp & $\left[ ^{+0.13}_{-0.12}({\rm stat.})\:^{+0.07}_{-0.07}({\rm syst.})\:^{+0.04}_{-0.03}({\rm theo.})\right]$ \\
  $\Cc_{\Pg}$        & =       & $0.92^{+0.23}_{-0.16}$                &&  $1.02\pm^{0.37}_{0.19}$                   &&  $0.92 \pm 0.17$  \tsp & $\left[ ^{+0.14}_{-0.12}({\rm stat.})\:^{+0.10}_{-0.09}({\rm syst.})\:^{+0.07}_{-0.05}({\rm theo.})\right]$  \\
  $\Cc_{\PZ\PGg}$    & $<$      & $3.15$ (95\% CL)                    &&  $4.03$ (95\% CL)                        &&  $3.18$ (95\% CL)    \tsp \\

  $\BRinv$                   & $<$ & $0.49$ (95\% CL)                &&  $0.68$ (95\% CL) && -    \\
  $\Gamma_H/\Gamma_H^{\rm SM}$ & =   & $0.64^{+0.40}_{-0.25}$           &&  $0.74^{+1.57}_{-0.21}$ && $0.64 ^{+0.31}_{-0.25}$ & $\left[ ^{+0.24}_{-0.21}({\rm stat.})\:^{+0.19}_{-0.15}({\rm syst.})\:^{+0.06}_{-0.05}({\rm theo.})\right]$ \tsp \\
  \hline\hline

  \end{tabular}}

  \newcommand{\RESULTTableGenericModelThree}{
  \begin{tabular}{ccccccccc} \hline\hline
  \hsa Parameter \hsa & & \hsa  Measurement \hsa  & Uncertainty breakdown \\ \hline
  $\Cc_{\Pg\PZ}$       & $=$   &  $1.18 \pm 0.16$ \tsp &                  $\left[ ^{+0.14}_{-0.14}({\rm stat.})\:^{+0.04}_{-0.04}({\rm syst.})\:^{+0.08}_{-0.06}({\rm theo.})\right]$ \\
  $\Rr_{\PZ\Pg}$       & $=$   &  $1.09^{+0.26}_{-0.22}$ \tsp &           $\left[ ^{+0.21}_{-0.20}({\rm stat.})\:^{+0.12}_{-0.10}({\rm syst.})\:^{+0.08}_{-0.06}({\rm theo.})\right]$ \\
    $\Rr_{\PW\PZ}$       & $\in$ &  $[-1.04,-0.81] \cup [0.80,1.06]$ \tsp & $\left[ ^{+0.13}_{-0.11}({\rm stat.})\:^{+0.05}_{-0.05}({\rm syst.})\:^{+0.02}_{-0.02}({\rm theo.})\right]$ \\
    $\Rr_{\PQt\Pg}$      & $\in$ &  $[-1.70, -1.07] \cup [1.03,1.73]$ &     $\left[ ^{+0.26}_{-0.25}({\rm stat.})\:^{+0.20}_{-0.24}({\rm syst.})\:^{+0.14}_{-0.08}({\rm theo.})\right]$ \\
    $\Rr_{\PQb\PZ}$      & $=$   &  $0.60 \pm 0.27$ \tsp &                  $\left[ ^{+0.21}_{-0.19}({\rm stat.})\:^{+0.14}_{-0.16}({\rm syst.})\:^{+0.05}_{-0.03}({\rm theo.})\right]$ \\
    $\Rr_{\PGt\PZ}$      & $=$   &  $0.99^{+0.23}_{-0.19}$ \tsp &           $\left[ ^{+0.19}_{-0.16}({\rm stat.})\:^{+0.11}_{-0.09}({\rm syst.})\:^{+0.06}_{-0.04}({\rm theo.})\right]$ \\
    $|\Rr_{\PGm\PZ}|$    & $<$   &  $2.3$ (95\% CL)                     \\
    $\Rr_{\PGg\PZ}$      & $=$       & $0.90 \pm 0.15$ \tsp &               $\left[ ^{+0.15}_{-0.13}({\rm stat.})\:^{+0.05}_{-0.04}({\rm syst.})\:^{+0.03}_{-0.03}({\rm theo.})\right]$\\
    $|\Rr_{(\PZ\PGg)\PZ}|$  & $<$       & $3.2$ (95\% CL) \\
    \hline\hline
    \end{tabular}}

\section{Coupling-strength fits}
\label{sec:CouplingFits}

In the previous section signal-strength parameter $\mu_{i}^{f}$
for a given Higgs boson production or decay mode is discussed. For a
measurement of Higgs boson coupling strengths, production and decay
modes cannot be treated independently, as each observed process
involves at least two Higgs boson coupling strengths.  Scenarios with
a consistent treatment of coupling strengths in production and decay
modes are studied in this section. All uncertainties on the best-fit
values shown take into account both the experimental and theoretical
systematic uncertainties.  For selected benchmark models a breakdown
of parameter uncertainties in statistical uncertainties and in experimental and theoretical
systematic uncertainties is presented.

\subsection{Framework for coupling-strength measurements}
\label{sec:framework}
Following the leading-order (LO) tree-level-motivated framework and
benchmark models recommended in Ref.~\cite{Heinemeyer:2013tqa}, measurements
of Higgs boson coupling-strength scale factors $\Cc_{j}$ are implemented for the combination of all
analyses and channels summarised in Table~\ref{tab:inputs}.

\subsubsection{Structure and assumptions of the framework for benchmark models}
The framework is based on the assumption that the signals observed in
the different channels originate from a single narrow resonance
with a mass near $\RESULTHiggsMass\GeV$.  The case of several,
possibly overlapping, resonances in this mass region is not
considered.
Unless otherwise noted, the Higgs boson production and decay
kinematics are assumed to be compatible with those expected for a
SM Higgs boson, similar to what was assumed for the signal-strength
measurements of Section~\ref{sec:SignalStrength}.

The width of the assumed Higgs boson near $\RESULTHiggsMass\GeV$ is neglected in the Higgs boson propagator, i.e.\ the zero-width
approximation is used. In this approximation, the cross section $\sigma(\mathit{i}\to H\to\mathit{f})$ for on-shell measurements
can always be decomposed as follows:
\begin{equation}
\label{eq:zwa}
\sigma(\mathit{i}\to H \to\mathit{f}) = \frac{\sigma_{\mathit{i}}(\Cc_{j})\cdot\Gamma_{\mathit{f}}(\Cc_{j})}{\Gamma_{\PH}(\Cc_{j})}
\end{equation}
where $\sigma_{\mathit{i}}$ is the Higgs boson production cross section through
the initial state $\mathit{i}$, $\Gamma_{\mathit{f}}$ its the partial
decay width into the final state $\mathit{f}$ and $\Gamma_{\PH}$ the
total width of the Higgs boson. The index $j$ runs over all Higgs
boson couplings.  The components of $\sigma_{\mathit{i}}$,
$\Gamma_{\mathit{f}}$, and $\Gamma_{\PH}$ of Eq.~(\ref{eq:zwa}) are
expressed in scale factors $\Cc_{j}$ of the Higgs boson coupling
strengths to other particles $j$ that are motivated by the leading-order
processes that contribute to production or decay, and are
detailed in Section~\ref{sec:charact}. All scale factors are defined
such that a value of $\Cc_{j}=1$ corresponds to the best available SM
prediction, including higher-order QCD and EW corrections. This
higher-order accuracy is generally lost for $\Cc_{j}\ne 1$, nevertheless
higher-order QCD corrections approximately factorise with respect to
coupling rescaling and are accounted for wherever possible.

Modifications of the coupling scale factors change the Higgs boson
width $\Gamma_{\PH}(\Cc_j)$ by a factor $\Cc_{\PH}^2(\Cc_{j})$ with
respect to the SM Higgs boson $\Gamma_{\PH}^{\rm SM}$,
\begin{equation*}
    \Gamma_{\PH}(\Cc_j) =  \Cc_{\PH}^2(\Cc_{j})\cdot \Gamma_{\PH}^{\rm SM}  ,
\end{equation*}
where $\Cc_{\PH}^2(\Cc_{j})$ is the sum
of the scale factors $\Cc_{j}^{2}$ weighted by the corresponding SM branching
ratios. The total width of the Higgs boson increases beyond
modifications of $\Cc_{j}$ if invisible or undetected Higgs boson
decays\footnote{Invisible final states can be directly searched for
through the \met\ signature~\cite{Aad:2014iia}.  An example of an
undetected mode would be a decay mode to multiple light jets, which
presently cannot be distinguished from multijet backgrounds.} occur
that are not present in the SM. Including a Higgs boson branching
fraction \BRinv to such invisible or undetected decays, the full
expression for the assumed Higgs boson width becomes
\begin{equation}
    \Gamma_{\PH}(\Cc_j,\BRinv) =  \frac{\Cc_{\PH}^2(\Cc_{j})}{(1-\BRinv)} \Gamma_{\PH}^{\rm SM}.
     \label{eq:CH:1}
\end{equation}
\noindent As \BRinv scales all observed cross-sections of on-shell Higgs boson
production $\sigma(\mathit{i}\to H\to\mathit{f})$, some assumption about
invisible decays must be made to be able to interpret these
measurements in terms of absolute coupling-strength scale factors $\Cc_{j}$.
The signal-strength measurements of off-shell Higgs boson production~\cite{Aad:2015xua}, on the other hand, is assumed
to only depend on the coupling-strength scale factors and not on the total width ~\cite{Kauer:2012hd,Caola:2013yja}, i.e.
\begin{equation}
  \label{eq:sigma_off}
  \sigma^{\rm off}(\mathit{i}\to H^{*}\to\mathit{f}) \sim \Cc_{i,{\rm off}}^2 \cdot \Cc_{f,{\rm off}}^2
\end{equation}
where the additional assumption of non-running coupling-strength scale factors, $\Cc_{j,{\rm off}} = \Cc_{j,{\rm on}}$
allows $\Gamma_{H}$ to be constrained using using Eq.~(\ref{eq:CH:1}), from a simultaneous
measurement of on-shell and off-shell measurements.  While this
assumption of non-running coupling-strength scale factors cannot hold universally
for ggF and
VBF production without violating unitarity, it is
assumed to hold in the region of phase space of the off-shell
{$H^\ast \to WW$} and {$H^\ast \to ZZ$} measurements described in
Section~\ref{sec:width} which is relatively close to the on-shell
regime~\cite{Englert:2014aca}. Alternatively, ratios of coupling-strength scale factors can be measured
without assumptions on the Higgs boson total width, as the identical contributions of $\Gamma_{\PH}$ to
each coupling strength cancel in any ratio of these.

Finally, only modifications of coupling strengths, i.e.\ of absolute
values of coupling strengths, are taken into account, while the tensor
structure of the couplings is assumed to be the same as in the
SM. This means in particular that the observed state is assumed to be
a CP-even scalar as in the SM. This assumption was tested by both the
ATLAS~\cite{Aad:2013xqa} and CMS~\cite{Khachatryan:2014kca}
Collaborations.

\subsubsection{Characterisation of the input measurements in terms of coupling strengths}
\label{sec:charact}
The combined input channels described in Table~\ref{tab:inputs} probe eight
different production processes:
$\sigma(\text{ggF})$, $\sigma(\text{VBF})$, $\sigma(WH)$,
$\sigma(q\bar{q}\to ZH)$, $\sigma(gg\to ZH)$, $\sigma{(bbH)}$,
$\sigma(ttH)$, and $\sigma(tH)$ whose SM cross sections are listed in
Table~\ref{tab:SMPrediction}.\footnote{The $ZH$ production cross section
  quoted in Table~\ref{tab:SMPrediction} comprises both the
  $q\bar{q}\to ZH$ and $gg\to ZH$ processes.} Table
\ref{tab:kexpr} summarises the Higgs boson coupling-strength
characteristics of all production processes and lists the rate scaling
behaviour in terms of Higgs boson coupling-strength scale factors.

\begin{figure}[hbt]
  \center
  \begin{minipage}{4.5cm}
    \subfloat[]{
      \resizebox{4.5cm}{!}{\includegraphics[width=.99\textwidth]{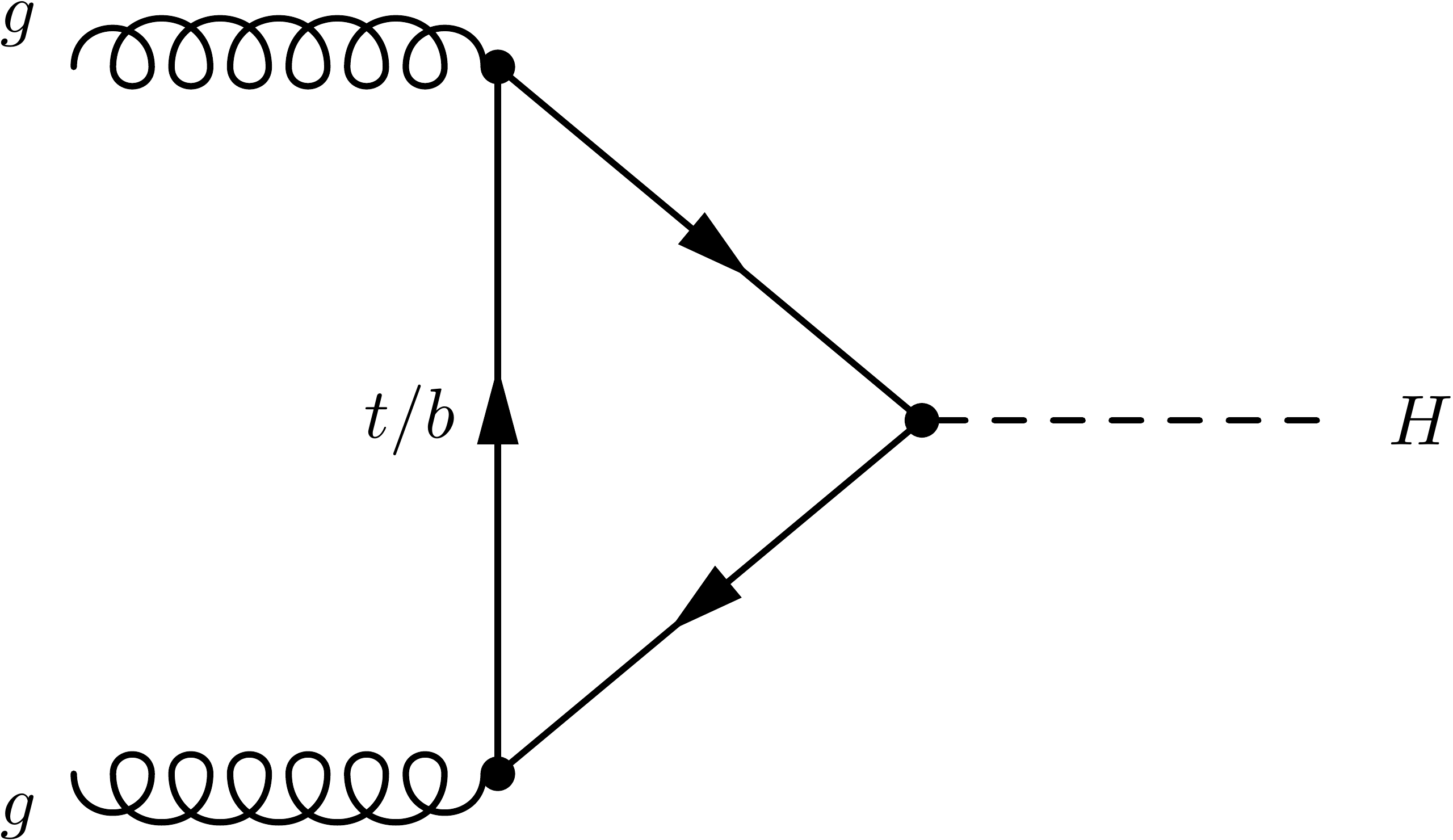}}
      \label{fig:feyn_ggF}
    }
  \end{minipage}
  \qquad
  \begin{minipage}{4.5cm}
    \subfloat[]{
      \resizebox{4.5cm}{!}{\includegraphics[width=.99\textwidth]{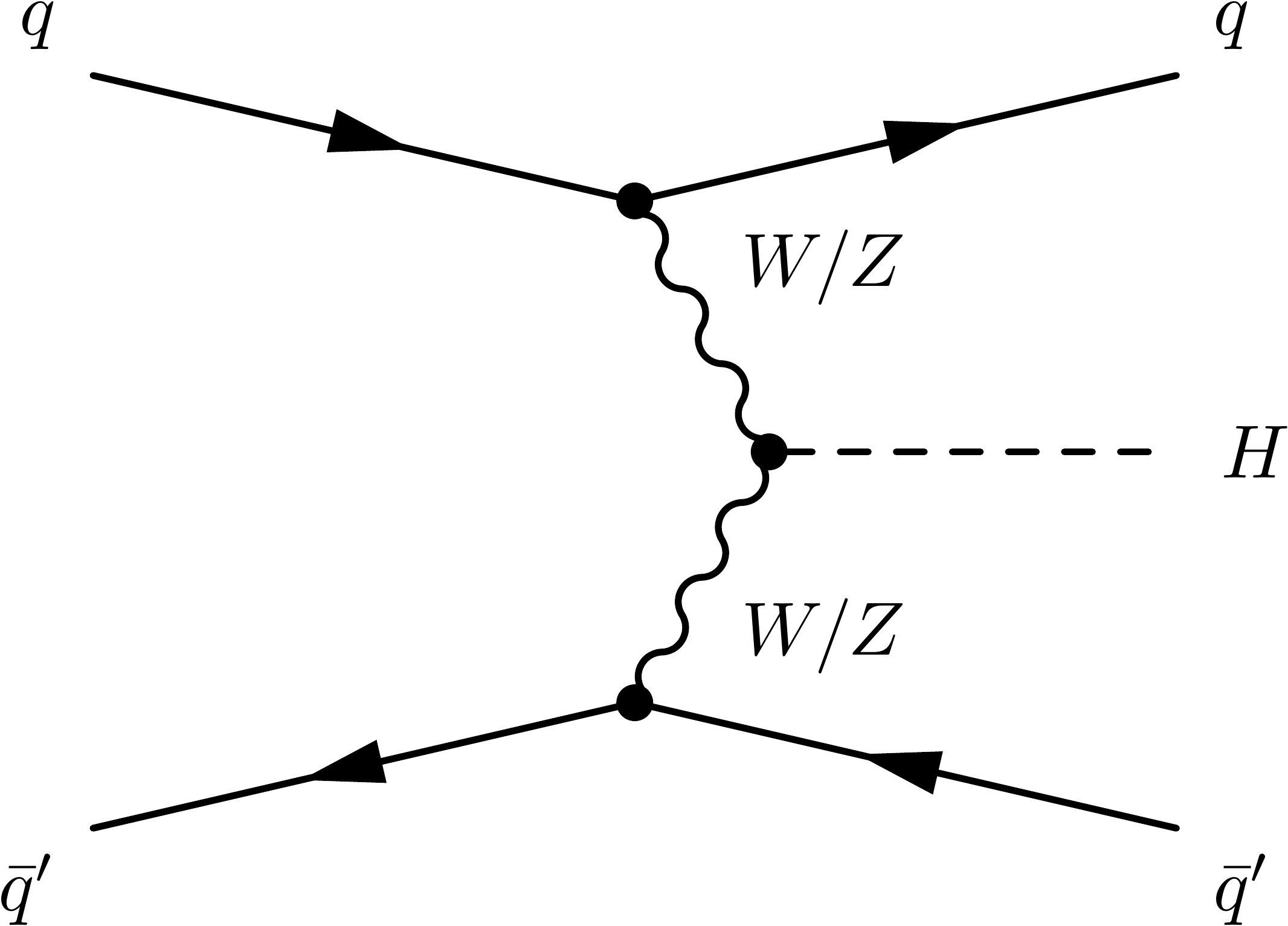}}
      \label{fig:feyn_VBF}
    }
  \end{minipage}
  \caption{Feynman diagrams of Higgs boson production via (a) the ggF and (b) VBF production processes.}
\end{figure}

The ggF production process (Fig.~\ref{fig:feyn_ggF}) involves a loop process at lowest order, with
contributions from $t$- and $b$-quark loops and a small interference between
them. The VBF production (Fig.~\ref{fig:feyn_VBF}) process probes a combination of $\Cc_{\PW}$ and $\Cc_{\PZ}$
coupling-strength scale factors, with a negligible amount ($\ll 0.1\%$) of interference between these
tree-level contributions.

\begin{figure}[hbt]
  \center
  \begin{minipage}{4.5cm}
    \subfloat[]{
      \resizebox{4.5cm}{!}{\includegraphics[width=.99\textwidth]{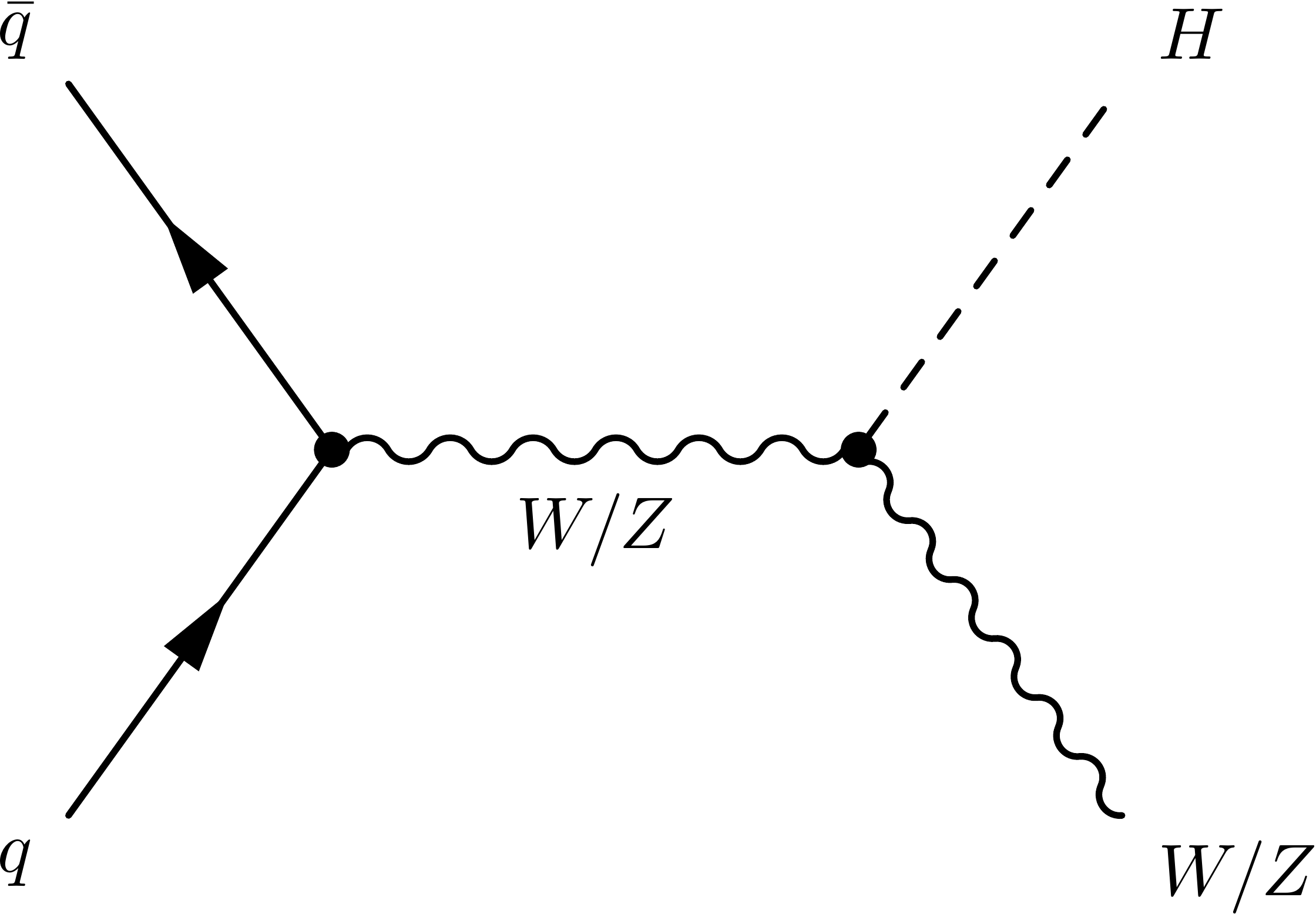}}
      \label{fig:feyn_vH}
    }
  \end{minipage}
  \qquad
  \begin{minipage}{4.5cm}
    \subfloat[]{
      \resizebox{4.5cm}{!}{\includegraphics[width=.99\textwidth]{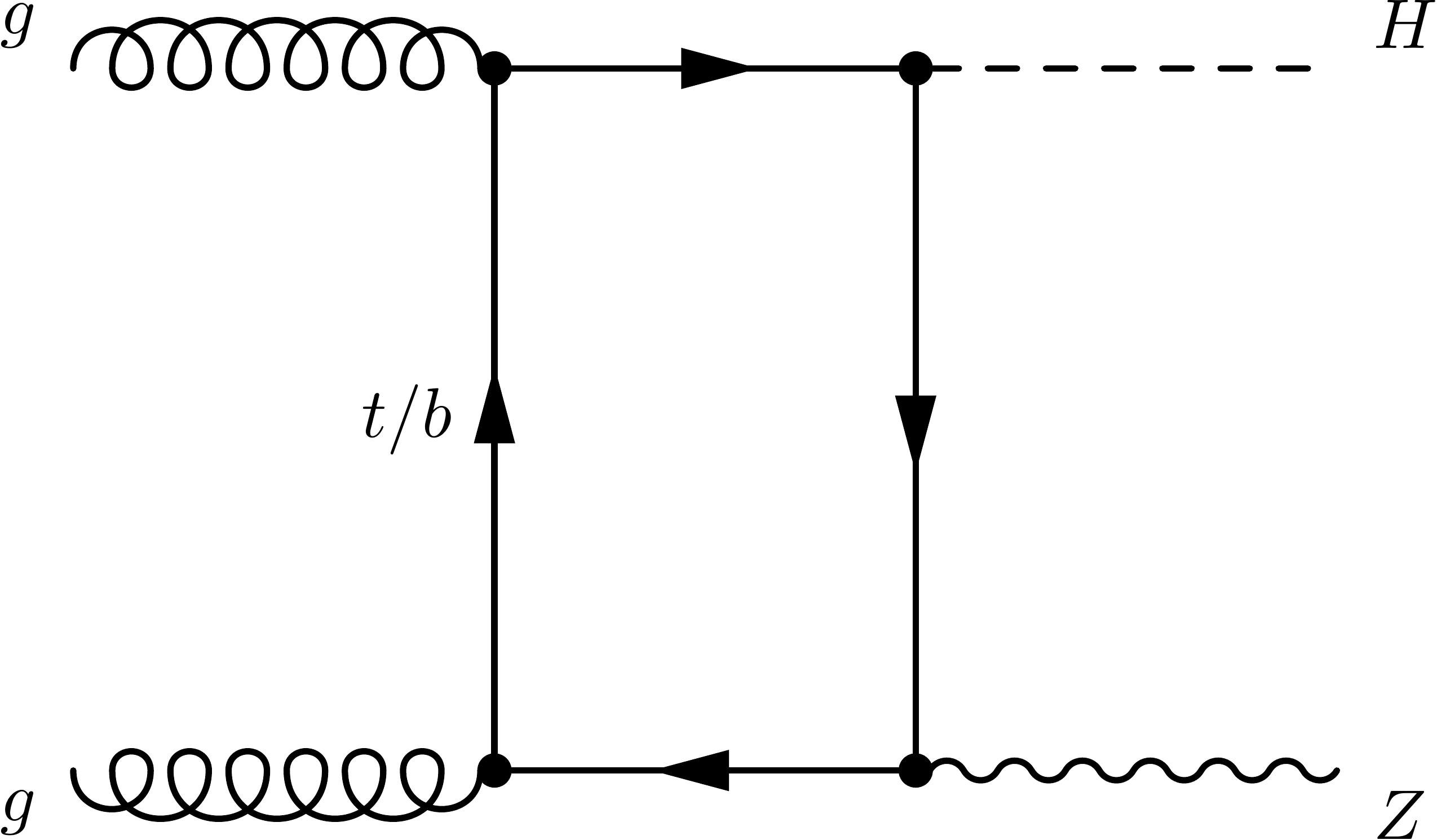}}
      \label{fig:feyn_ggZH2}
    }
  \end{minipage}
  \qquad
  \begin{minipage}{4.5cm}
    \subfloat[]{
      \resizebox{4.5cm}{!}{\includegraphics[width=.99\textwidth]{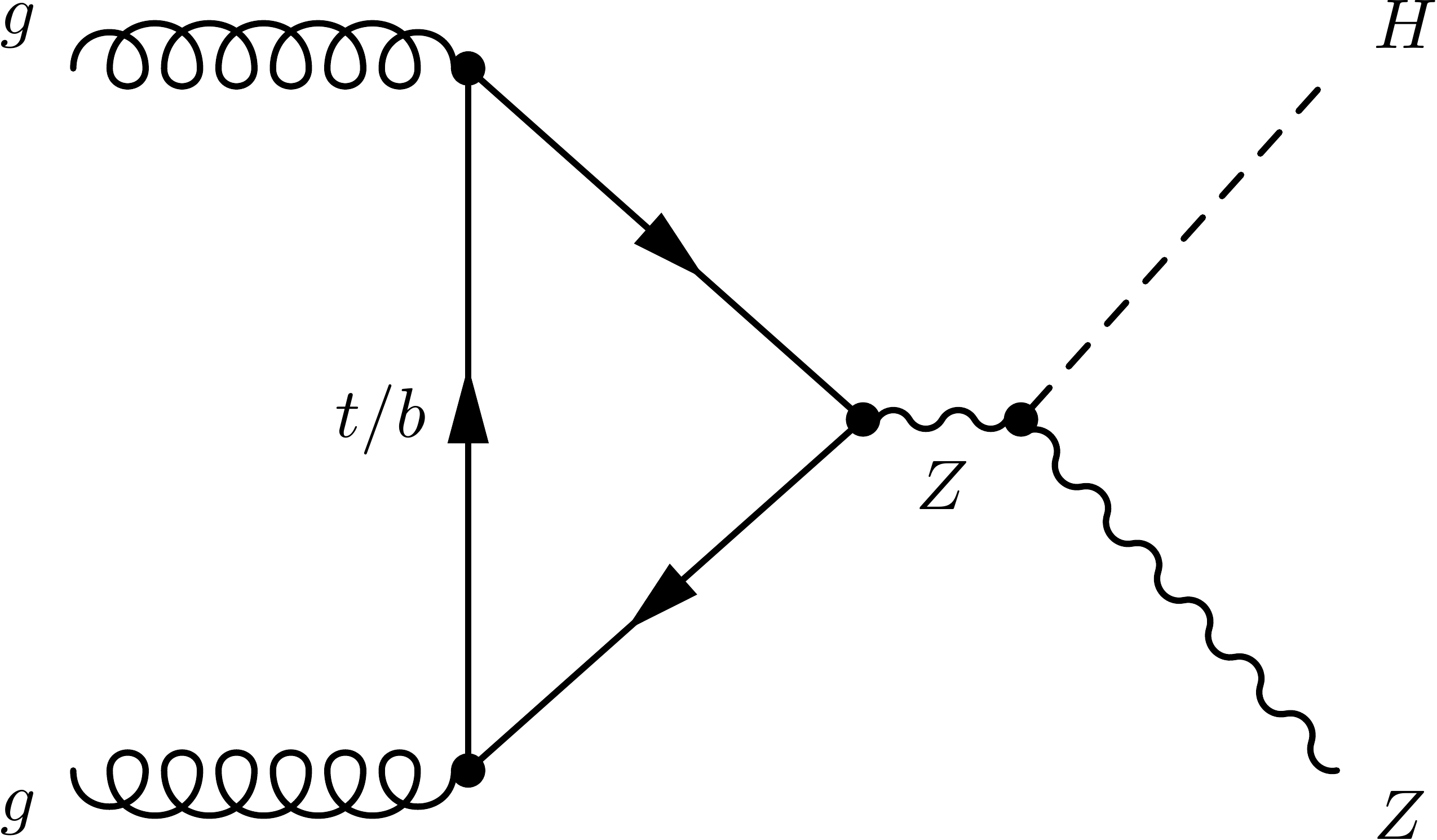}}
      \label{fig:feyn_ggZH1}
    }
  \end{minipage}
  \caption{Feynman diagrams of Higgs boson production via (a) the $q\bar{q} \to VH$ and (b,c) \ggZH\ production processes.}
\end{figure}

The $q\bar{q}\to WH$ and $q\bar{q}\to ZH$ processes (Fig.~\ref{fig:feyn_vH}) each probe a single coupling strength, with scale factors $\Cc_{\PW}$ and
$\Cc_{\PZ}$, respectively. The gluon-initiated associated production of a Higgs boson with a $Z$
boson, $\sigma(gg\to ZH)$, is characterised by gluon-fusion-style production
involving $t,b$-quark loops where the $Z$ boson is always radiated from the
fermion loop and the Higgs boson is either radiated directly from
the fermion loop (Fig.~\ref{fig:feyn_ggZH2}), or is radiated from the outgoing $Z$ boson (Fig.~\ref{fig:feyn_ggZH1}). The cross
section of $gg\to ZH$ production is sensitive to the relative sign
between $\Cc_{\PQt}$ and $\Cc_{\PZ}$ due to interference between these
contributions. This separate treatment of $gg\to ZH$ production is not present in the framework described in Ref.~\cite{Heinemeyer:2013tqa}.

\begin{figure}[hbt]
  \center
  \begin{minipage}{4.5cm}
    \subfloat[]{
      \resizebox{4.5cm}{!}{\includegraphics[width=.99\textwidth]{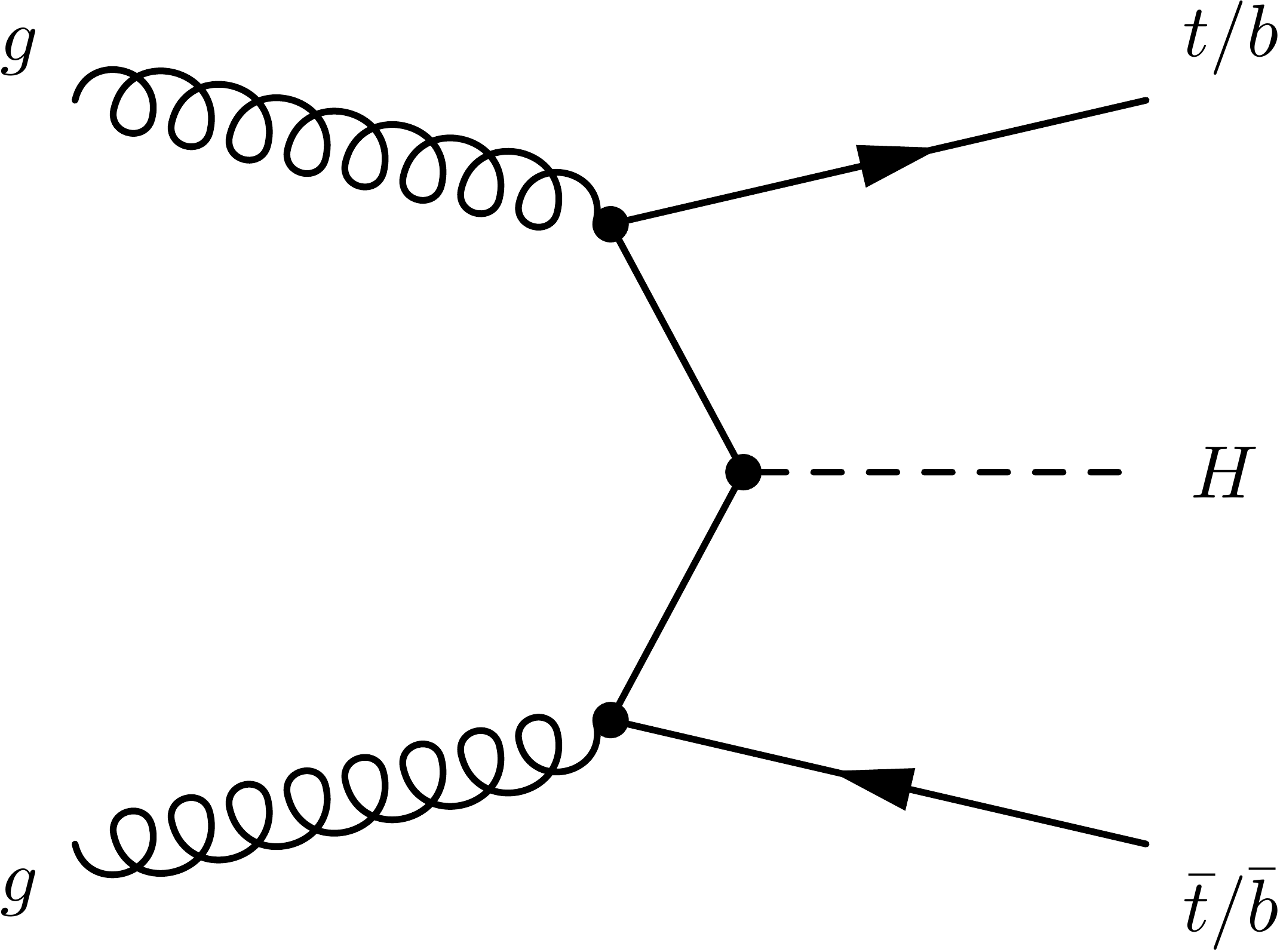}}
      \label{fig:feyn_ttH}
    }
  \end{minipage}
  \qquad
  \begin{minipage}{4.5cm}
    \subfloat[]{
      \resizebox{4.5cm}{!}{\includegraphics[width=.99\textwidth]{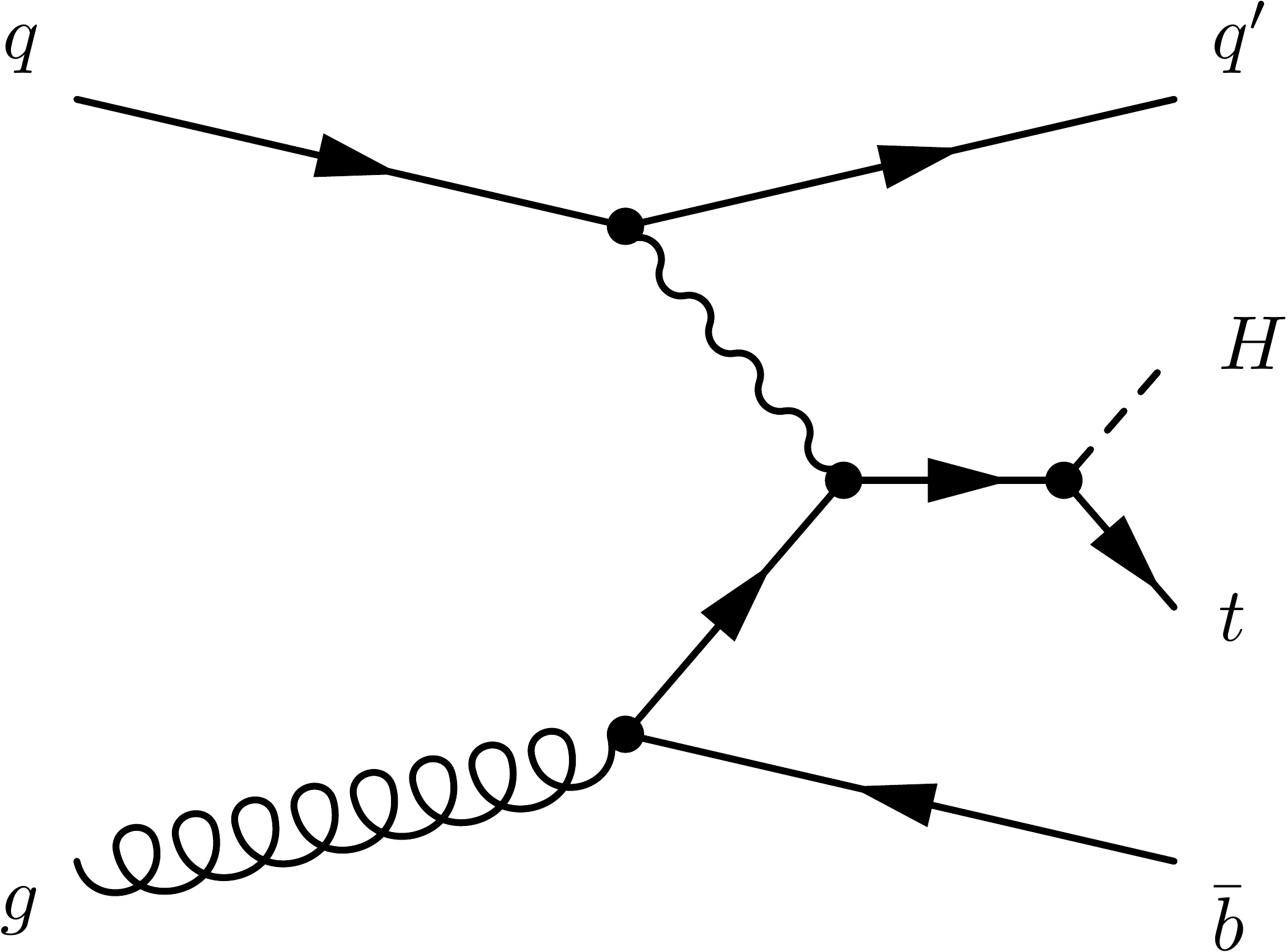}}
      \label{fig:feyn_tHjb1}
    }
  \end{minipage}
  \qquad
  \begin{minipage}{4.5cm}
    \subfloat[]{
      \resizebox{4.5cm}{!}{\includegraphics[width=.99\textwidth]{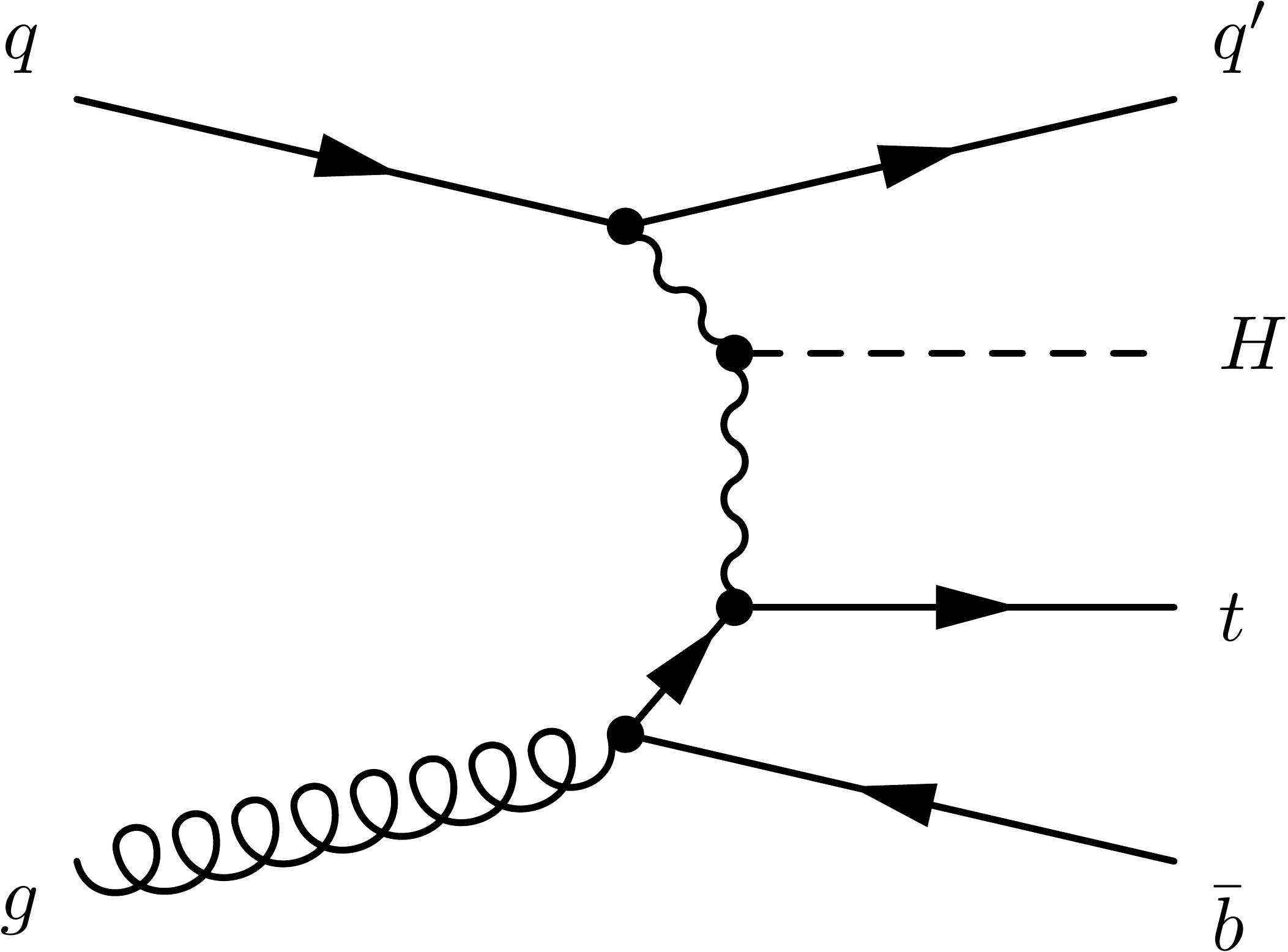}}
      \label{fig:feyn_tHjb2}
    }
  \end{minipage}

  \vspace{0.5cm}

  \begin{minipage}{4.5cm}
  \resizebox{4.5cm}{!}{\qquad}
  \end{minipage}
  \qquad
  \begin{minipage}{4.5cm}
    \subfloat[]{
      \resizebox{4.5cm}{!}{\includegraphics[width=.99\textwidth]{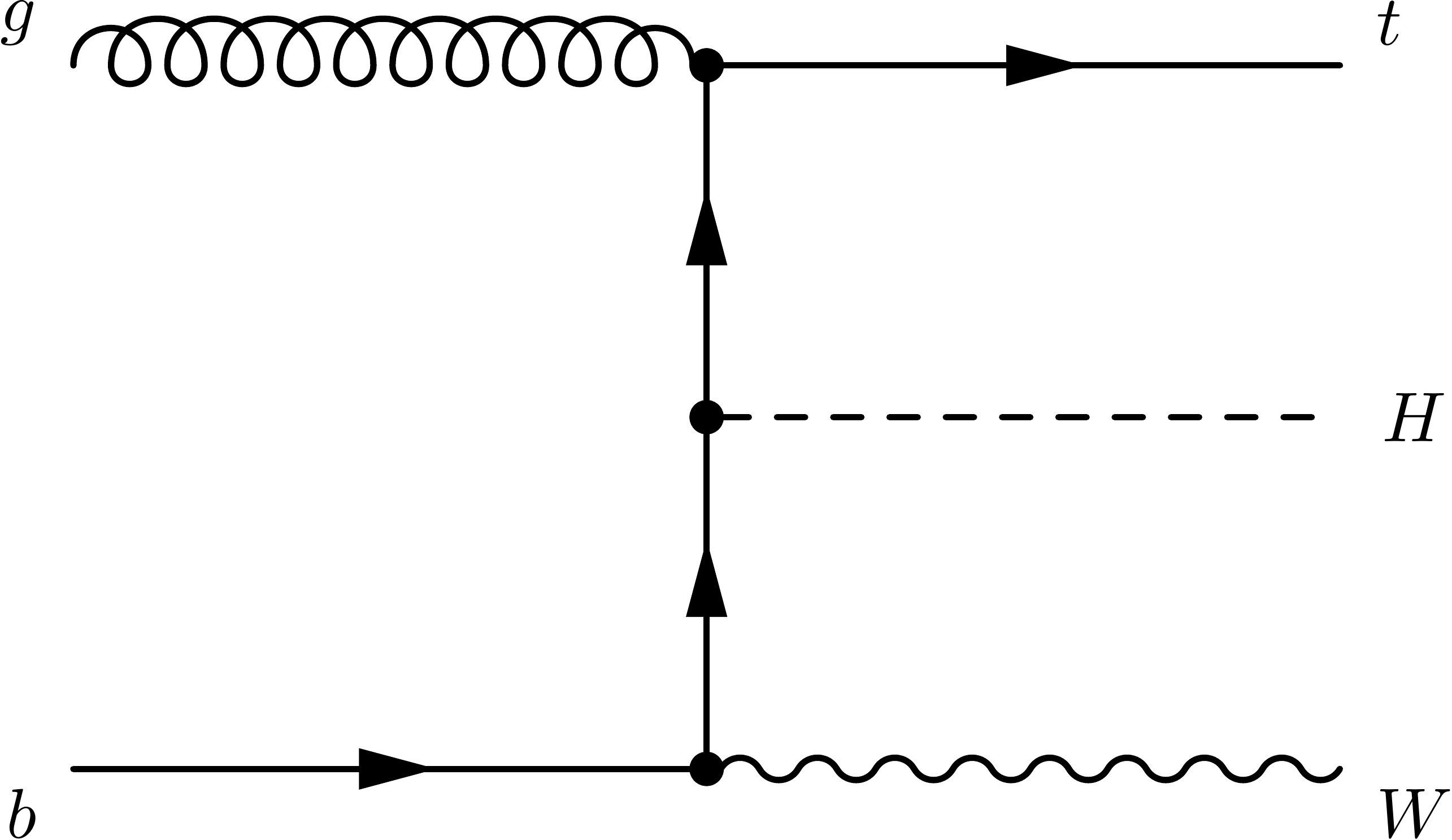}}
      \label{fig:feyn_WtH1}
    }
  \end{minipage}
  \qquad
  \begin{minipage}{4.5cm}
    \subfloat[]{
      \resizebox{4.5cm}{!}{\includegraphics[width=.99\textwidth]{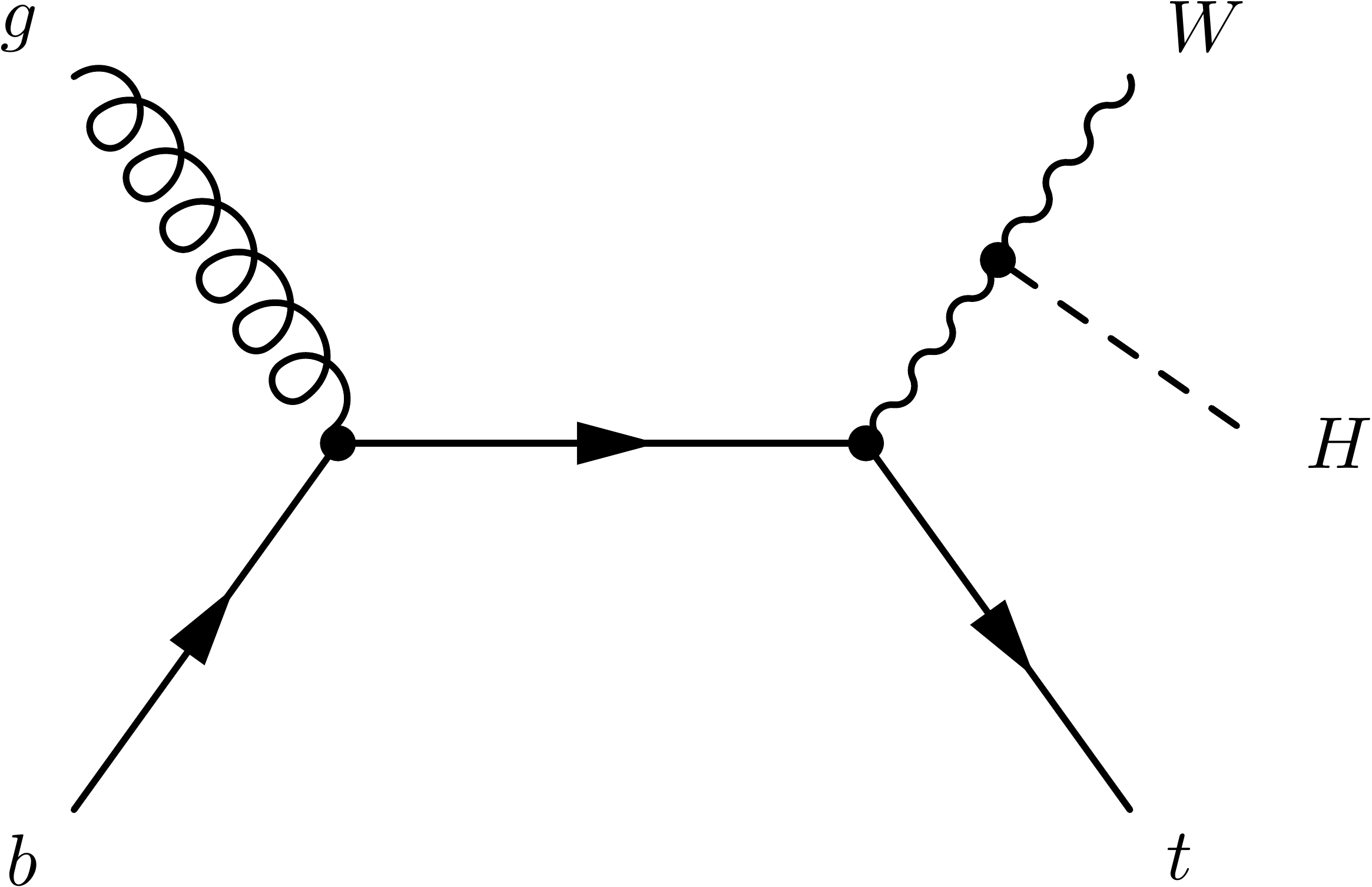}}
      \label{fig:feyn_WtH2}
    }
  \end{minipage}
  \caption{Feynman diagrams of Higgs boson production via (a) the $ttH$ ($bbH$), (b,c) $tHq^{\prime}b$ and (d,e) $WtH$ processes.}
\end{figure}

The $ttH$ production process (Fig.~\ref{fig:feyn_ttH}) directly probes
the Higgs boson coupling strength to top quarks, parameterised in the
framework with the scale factor $\Cc_{\PQt}$.  Tree-level $tH$
production, comprising the processes $qg \to tHbq^{\prime}$
(Fig.~\ref{fig:feyn_tHjb1},~\ref{fig:feyn_tHjb2}) and $gb \to WtH$
(Fig.~\ref{fig:feyn_WtH1},~\ref{fig:feyn_WtH2}), is included as
background to events in all reconstructed $ttH$ categories, and has
for SM Higgs boson coupling strengths a large destructive
interference~\cite{Farina:2012xp} between contributions where the
Higgs boson is radiated from the $W$ boson and from the top quark. The
SM cross section for $tH$ production is consequently small, about 14\%
of the $ttH$ cross section. However, for negative $\Cc_{\PQt}$ the
interference becomes constructive and, following
Table~\ref{tab:kexpr}, the cross section increases by a factor of 6
(13) for $\abs{\Cc_{\PQt}} = \abs{\Cc_{\PW}} = 1$ for the $gb \to WtH$
($qg \to tHbq^\prime$) process, making the $tH$ process sensitive to
the relative sign of the $W$ and top-quark coupling strength, despite
its small SM cross section. The modelling of $tH$ production is not
present in the framework described in Ref.~\cite{Heinemeyer:2013tqa}.

The $bbH$ (Fig.~\ref{fig:feyn_ttH}) production process directly probes
the Higgs boson coupling strength to $b$-quarks, with scale factor
$\Cc_{\PQb}$. Simulation studies using $bbH$ samples produced in the
four-flavour scheme~\cite{Alwall:2014hca,Wiesemann:2014ioa} have shown that the ggF samples are a good
approximation for $bbH$ production for the most important analysis
categories, therefore $bbH$ production is always modelled using
simulated ggF events (see Section~\ref{sec:modifications}).

\begin{figure}[hbt]
  \center
  \begin{minipage}{4.5cm}
    \subfloat[]{
      \resizebox{4.5cm}{!}{\includegraphics[width=.99\textwidth]{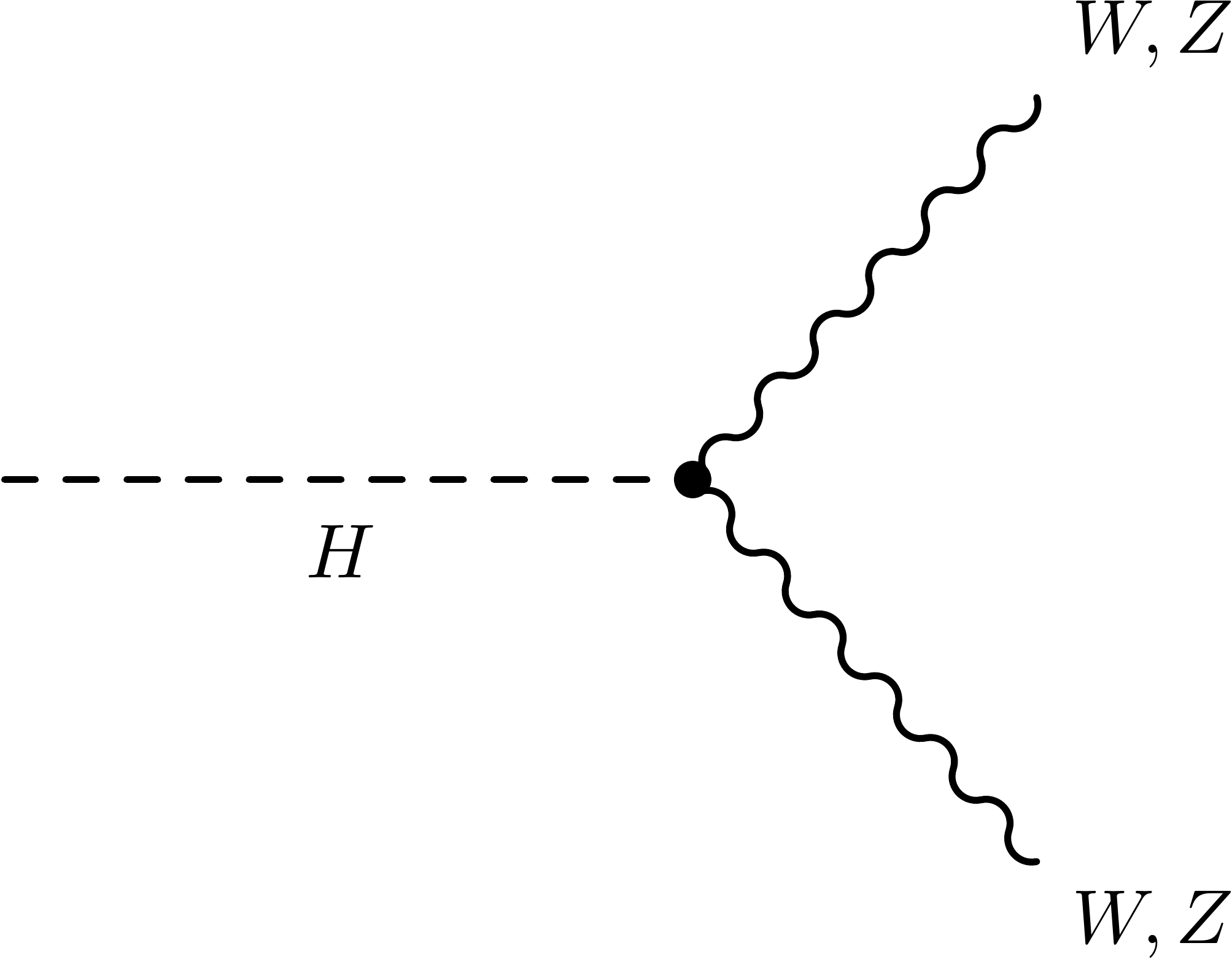}}
      \label{fig:feyn_hVV}
    }
    \end{minipage}
    \qquad
    \begin{minipage}{4.5cm}
      \subfloat[]{
        \resizebox{4.5cm}{!}{\includegraphics[width=.99\textwidth]{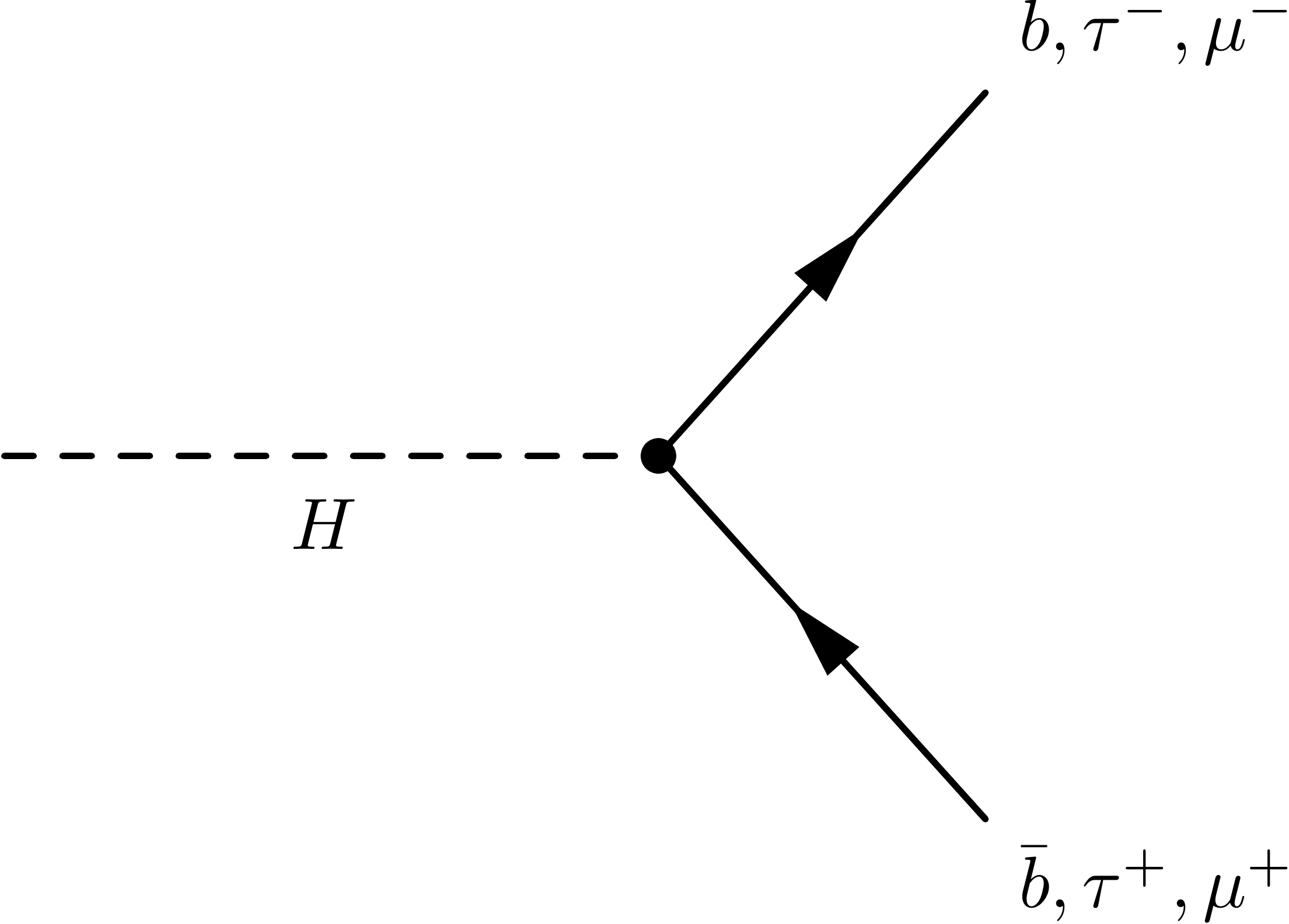}}
        \label{fig:feyn_hff}
      }
    \end{minipage}
  \caption{Feynman diagrams of Higgs boson decays (a) to $W$ and $Z$ bosons and (b) to fermions.}
\end{figure}

The combined input channels probe seven Higgs boson decay modes. Five
of these decay modes, $H\to WW^*$, $H\to ZZ^*$, $H\to\bb$, $H\to{\tau\tau}$, and $H\to\mu\mu$ 
each probe a single coupling-strength scale factor to either a gauge boson
(Fig.~\ref{fig:feyn_hVV}) or to a fermion (Fig.~\ref{fig:feyn_hff}). The
remaining two decay modes, $H\to \gamma\gamma$ and
$H\to Z\gamma$ are characterised by the interference between $W$
boson or top-quark loop diagrams (Fig.~\ref{fig:feyn_hgg}). These modes probe the $W$ and $t$
coupling strengths as well as their relative sign through interference
effects.

\begin{figure}[hbt]
  \center
  \begin{minipage}{4.5cm}
    \subfloat[]{
      \resizebox{4.5cm}{!}{\includegraphics[width=.99\textwidth]{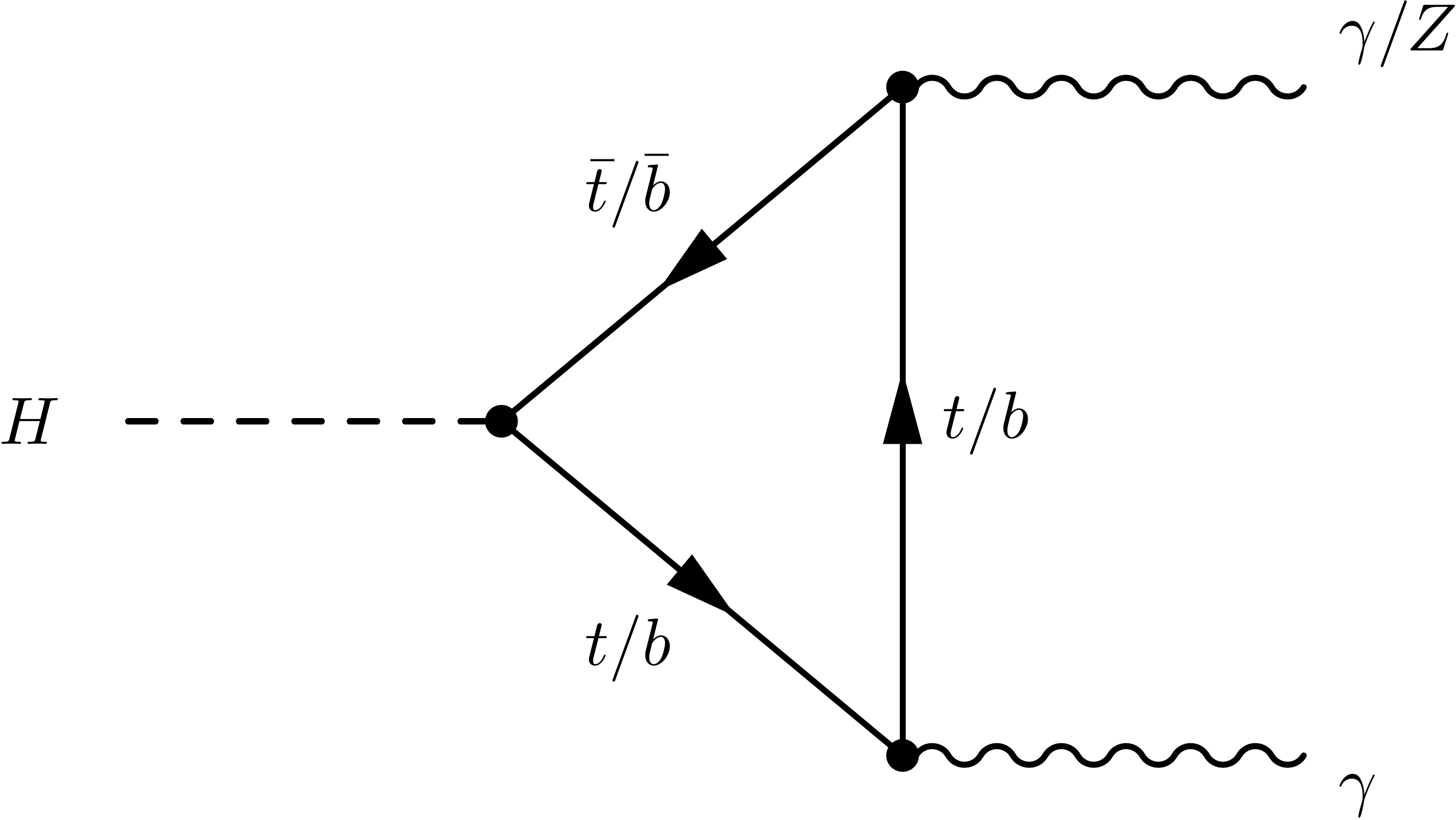}}
      \label{fig:feyn_hggf}
    }
  \end{minipage}
  \qquad
  \begin{minipage}{4.5cm}
    \subfloat[]{
      \resizebox{4.5cm}{!}{\includegraphics[width=.99\textwidth]{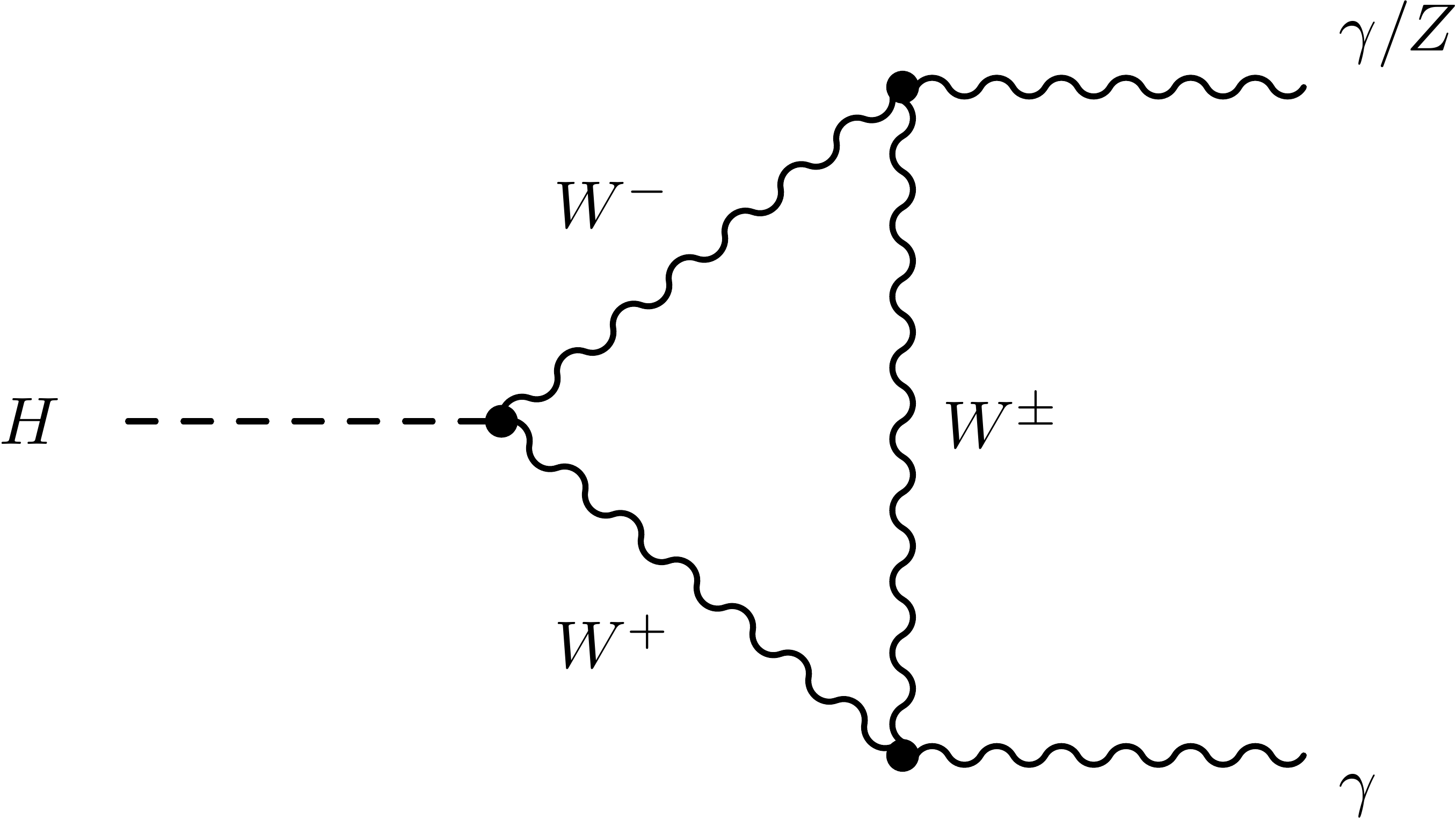}}
      \label{fig:feyn_hggW1}
    }
  \end{minipage}
  \qquad
  \begin{minipage}{4.5cm}
    \subfloat[]{
      \resizebox{4.5cm}{!}{\includegraphics[width=.99\textwidth]{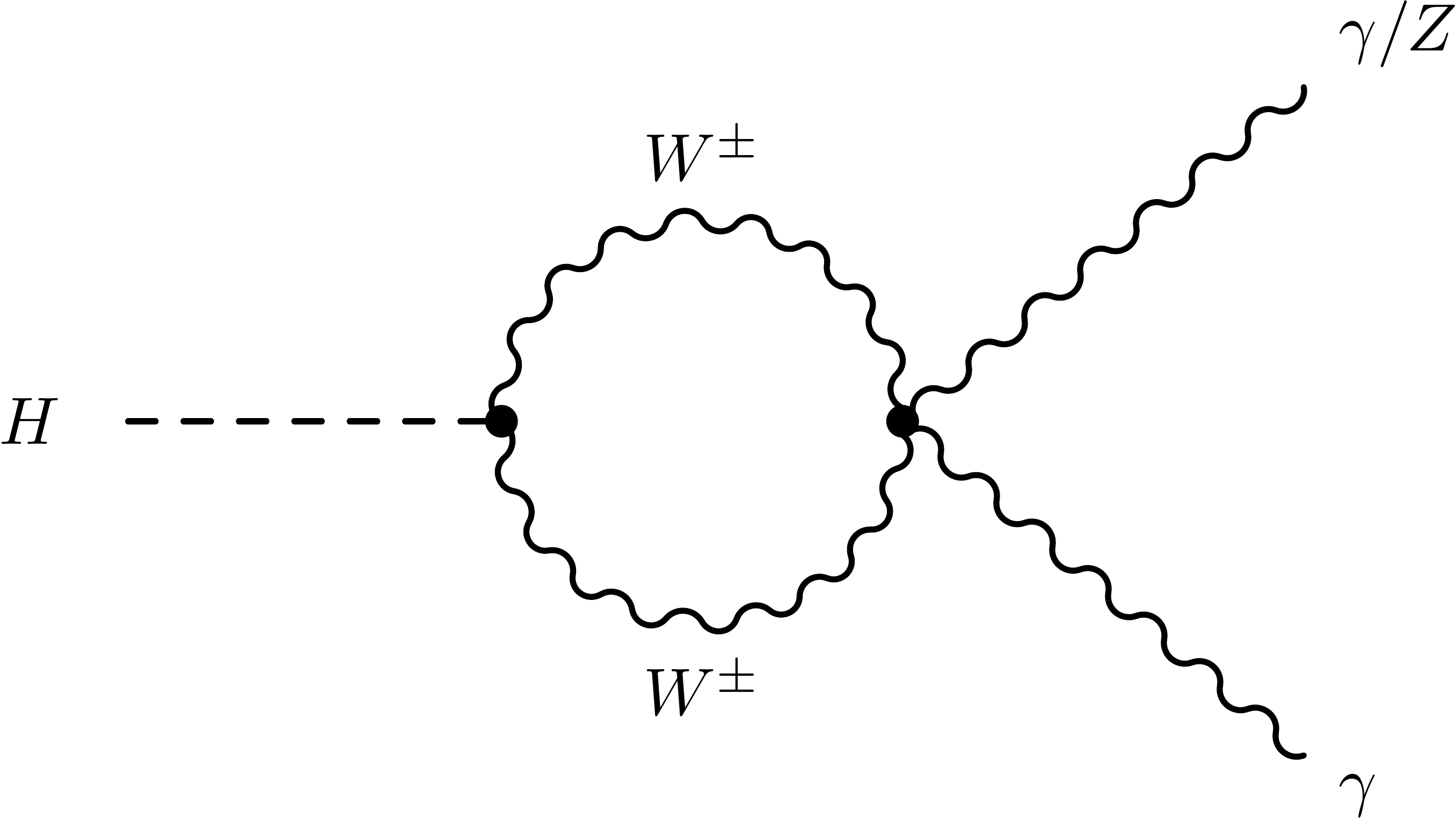}}
      \label{fig:feyn_hggW2}
    }
  \end{minipage}
  \caption{Feynman diagrams of Higgs boson decays to a pair of photons, or to a photon and a $Z$ boson.}
  \label{fig:feyn_hgg}
\end{figure}

For completeness it should be noted also that the ggF, $tH$ and
\ggZH\ cross sections expressed in Higgs boson coupling strengths
depend on the kinematic selection criteria used. The $b$--$t$
interference expression quoted in Table~\ref{tab:kexpr} for ggF is
valid for the inclusive cross section, but in events with additional
jets the top-quark loop dominates, and the observed interference is
somewhat smaller. For \ggZH\ production the effect of phase-space
dependence was estimated for $H\to \bb$ decays with a variant of the
coupling model that introduces separate coupling-dependent
cross-section expressions for each of the $Z$ boson $p_{\rm T}$ bins of the
$H\to \bb$ analysis. The effect on coupling strength measurements of
approximating the \ggZH\ production cross section with an inclusive
expression instead of using the set of $p_{\rm T}$-dependent expressions was
determined to be negligible at the current experimental precision,
with the largest effect being a $\sim 0.1\sigma$ reduction of the
expected sensitivity in the determination of the relative sign of the
$W/Z$ couplings.  Neither this phase-space dependence, nor that of ggF
are considered in this paper.  For the $tH$ process on the other hand,
which features a comparatively large $W$--$t$ interference term, the
effect of phase-space dependence is taken into account, even though
Table~\ref{tab:kexpr} only lists the inclusive expression. 

\subsubsection{Effective coupling-strength scale factors}
In some of the fits, effective scale factors $\Cc_{\Pg}$, $\Cc_{\PGg}$
and $\Cc_{\PZ\PGg}$ are introduced to describe the processes
$gg \to H$, $H \to \gamma\gamma$ and $H\to Z\gamma$, which are
loop-induced in the SM, as shown in Figures~\ref{fig:feyn_ggF}
and \ref{fig:feyn_hgg}, respectively.  In other fits they are treated
as a function of the more fundamental coupling-strength scale factors
$\Cc_{\PQt}$, $\Cc_{\PQb}$, $\Cc_{\PW}$, and similarly for all other
particles that contribute to these SM loop processes. In these cases,
the loop contributions are expressed in terms of the fundamental
coupling strengths, including all interference effects, as listed for
the SM in Table~\ref{tab:kexpr}. 
The loop process \ggZH\ is never treated as an effective scale factor, as unlike 
in the other loop processes, a $ggHZ$ contact interaction from new physics would 
likely show a kinematic structure very different from the SM \ggZH\ 
process~\cite{Englert:2013vua} assumed in the current study and is expected to be 
suppressed. What then remains of BSM effects on  the \ggZH\ process are modifications 
of the Higgs boson couplings to the top quark~(Fig.~\ref{fig:feyn_ggZH2}) 
and the $Z$ boson~(Fig.~\ref{fig:feyn_ggZH1}), which are taken into account within the limitation 
of the framework by the coupling-strength scale factors $\Cc_{\PQt}$ and $\Cc_{\PZ}$.

\begin{table}[hbt]
\caption{Overview of Higgs boson production cross sections
  $\sigma_{i}$, the Higgs boson partial decay widths $\Gamma_{f}$ and
  the Higgs boson total width $\Gamma_{\PH}$. For each production or
  decay mode the scaling of the corresponding rate in terms of Higgs
  boson coupling-strength scale factors is given. For processes where
  multiple amplitudes contribute, the rate may depend on multiple
  Higgs boson coupling-strength scale factors, and interference terms
  may give rise to scalar product terms $\Cc_{i}\Cc_{j}$ that allow
  the relative sign of the coupling-strength scale factors $\Cc_{i}$
  and $\Cc_{j}$\label{tab:kexpr} to be determined. Expressions
  originate from Ref.~\cite{Heinemeyer:2013tqa}, except for
  $\sigma(gg\to ZH$) (from Ref.~\cite{Englert:2013vua}) and $\sigma(gb
  \to WtH)$ and $\sigma(qb \to tHq^\prime)$ (calculated using
  Ref.~\cite{Alwall:2014hca}). The expressions are given for $\rts =
  8\TeV$ and $m_H = 125.36\GeV$ and are similar for $\rts =
  7\TeV$. Interference contributions with negligible magnitudes have
  been omitted in this table.}  \center
\scalebox{0.95}{
\begin{tabular}{rccrl}
\hline
\hline
Production & Loops & Interference & \multicolumn{2}{l}{Expression in fundamental coupling-strength scale factors} \\
\hline
$\sigma(\text{ggF})$ & $\checkmark$ & $b$--$t$ &  $\Cc_{\Pg}^2 \sim$&$ 1.06 \cdot\Cc_{\PQt}^2 + 0.01 \cdot \Cc_{\PQb}^2 - 0.07\cdot\Cc_{\PQt}\Cc_{\PQb}$ \\
$\sigma(\text{VBF})$     & -            &  -    &  $\sim$ & $ 0.74 \cdot \Cc_{\PW}^2 + 0.26 \cdot \Cc_{\PZ}^2$ \\
$\sigma(WH)$             & -            &  -    & $\sim$ & $\Cc_{\PW}^2$\\
$\sigma(q\bar{q}\to ZH)$ & -            &  -    & $\sim$ & $\Cc_{\PZ}^2$\\
$\sigma(gg\to ZH)$       & $\checkmark$ & $Z$--$t$ &  $\Cc_{\Pg\Pg\PZ\PH}^2  \sim$&$  2.27\cdot\Cc_{\PZ}^2 + 0.37 \cdot\Cc_{\PQt}^2 - 1.64 \cdot \Cc_{\PZ}\Cc_{\PQt} $\\
$\sigma(bbH)$            & -            &  -    & $\sim$ & $\Cc_{\PQb}^2$ \\
$\sigma(ttH)$            & -            &  -    & $\sim$ & $\Cc_{\PQt}^2$ \\
$\sigma(gb \to WtH)$     & -            & $W$--$t$ &   $\sim$ & $ 1.84 \cdot \Cc_{\PQt}^2 + 1.57 \cdot \Cc_{\PW}^2 - 2.41 \cdot \Cc_{\PQt}\Cc_{\PW}$ \\
$\sigma(qb \to tHq^\prime)$ & -          & $W$--$t$ &   $\sim$ & $ 3.4 \cdot \Cc_{\PQt}^2 + 3.56 \cdot \Cc_{\PW}^2 - 5.96 \cdot \Cc_{\PQt}\Cc_{\PW}$ \\
\hline
Partial decay width \\
\hline
$\Gamma_{b\bar{b}}$      & -             &  -    & $\sim$ & $\Cc_{\PQb}^2$ \\
$\Gamma_{WW}$            & -             &  -    & $\sim$ & $\Cc_{\PW}^2$ \\
$\Gamma_{ZZ}$            & -             &  -    & $\sim$ & $\Cc_{\PZ}^2$ \\
$\Gamma_{\tau\tau}$      & -             &  -    & $\sim$ & $\Cc_{\tau}^2$ \\
$\Gamma_{\mu\mu}$        & -             &  -    & $\sim$ & $\Cc_{\mu}^2$ \\
$\Gamma_{\gamma\gamma}$  & $\checkmark$  & $W$--$t$ &  $\Cc_{\PGg}^2 \sim$&$ 1.59 \cdot \Cc_{\PW}^2 + 0.07 \cdot \Cc_{\PQt}^2 -0.66 \cdot \Cc_{\PW} \Cc_{\PQt}$ \\
$\Gamma_{Z\gamma}$       & $\checkmark$  & $W$--$t$ &  $\Cc_{\PZ\PGg}^2 \sim$&$ 1.12 \cdot\Cc_{\PW}^2 + 0.00035  \cdot\Cc_{\PQt}^2 - 0.12\cdot\Cc_{\PW}\Cc_{\PQt}$ \\
\hline
Total decay width \\
\hline
$\Gamma_{\PH}$           & $\checkmark$  & $\begin{array}{cc}W-t\\b-t\end{array}$ & $\Cc_{\PH}^2 \sim$ & $\begin{array}{lll}0.57 \cdot \Cc_{\PQb}^2 + 0.22 \cdot \Cc_{\PW}^2 + 0.09 \cdot \Cc_{\Pg}^2 + \\ 0.06 \cdot \Cc_{\PGt}^2 + 0.03 \cdot \Cc_{\PZ}^2 + 0.03 \cdot \Cc_{\PQc}^2 + \\ 0.0023 \cdot \Cc_{\PGg}^2 + 0.0016 \cdot \Cc_{\PZ\PGg}^2 + 0.00022 \cdot \Cc_{\PGm}^2\end{array}$ \\
\hline
\hline
\end{tabular}}
\end{table}

\subsubsection{Strategies for measurements of absolute coupling strengths}
As all observed Higgs boson cross sections in the LO framework are inversely proportional to the Higgs boson width (Eq.~(\ref{eq:zwa})),
which is not experimentally constrained to a meaningful precision at the LHC, only ratios of coupling strengths can be measured
at the LHC without assumptions about the Higgs boson width. To make measurements of absolute coupling strengths, an assumption about the Higgs  boson width
must be introduced.

The simplest assumption is that there are no invisible or undetected
Higgs boson decays, i.e. $\BRinv=0$ is assumed in Eq.~(\ref{eq:CH:1}). An
alternative, less strong assumption, is that $\Cc_{\PW} \le 1$ and $
\Cc_{\PZ} \le 1$\cite{Heinemeyer:2013tqa}. This assumption is
theoretically motivated by the premise that the Higgs boson should solve
the unitarity problem in vector boson scattering and also holds in a wide
class of BSM models. In particular, it is valid in any model with an
arbitrary number of Higgs doublets, with and without additional Higgs
singlets. The assumption is also justified in certain classes of
composite Higgs boson models.
A second alternative is to assume that the coupling strengths in off-shell Higgs boson production are identical to those
for on-shell Higgs boson production. Under the assumption that the
off-shell signal strength and coupling-strength scale factors are independent
of the energy scale of Higgs boson production, the total Higgs
boson decay width can be determined from the ratio of off-shell to on-shell
signal strengths~\cite{Aad:2015xua}.
The constraint $\BRinv \ge 0$, motivated by the basic assumption that
the total width of the Higgs boson must be greater or equal to the sum
of the measured partial widths, always introduces a lower bound on the Higgs
boson width. The difference in effect of these assumptions is therefore mostly
in the resulting upper limit on the Higgs boson width.
The assumptions made for the various measurements are summarised in
Table~\ref{tab:coupling_fits} and discussed in the next sections
together with the results.

\begin{sidewaystable}[tbp]
\caption{
  Summary of benchmark coupling models considered in this paper, where
  $\Rr_{ij}\equiv\Cc_i/\Cc_j$, $\Cc_{ii}\equiv\Cc_i\Cc_i/\Cc_{\PH}$, and the functional
  dependence assumptions are:
  $\Cc_{V}=\Cc_{\PW}= \Cc_{\PZ}$, $\Cc_F = \Cc_{\PQt} = \Cc_{\PQb} = \Cc_{\PGt} = \Cc_{\PGm}$ (and similarly for the
  other fermions), $\Cc_{\Pg} =
  \Cc_{\Pg}(\Cc_{\PQb}, \Cc_{\PQt})$,  $\Cc_{\PGg}=\Cc_{\PGg}(\Cc_{\PQb}, \Cc_{\PQt}, \Cc_{\PGt},
  \Cc_{\PW})$, and $\Cc_{\PH}=\Cc_{\PH}(\Cc_{i})$. The tick marks indicate which assumptions are made
  in each case. The last column shows, as an example,  the relative coupling strengths
  involved in the $gg\to\hgg$ process. 
\label{tab:coupling_fits}}
\begin{center}
\scalebox{0.85}{
\begin{tabular}{@{}l|c|p{0.2\linewidth}|@{}c@{}|c|c|c|c|c|c}
\hline\hline
Section in & Corresponding  & Probed  & Parameters of & \multicolumn{5}{c|}{Functional assumptions} & Example: $gg\to\hgg$ \\
this paper       &  table in Ref.\cite{Heinemeyer:2013tqa}  & couplings     & interest      & $\Cc_{V}$ & $\Cc_{F}$ & $\Cc_{\Pg}$ & $\Cc_{\PGg}$ & $\Cc_{\PH}$ &  \\ \hline
\ref{sec:bm:CF,CV}& 43.1 & \multirow{4}{0.99\linewidth}{Couplings to fermions and bosons} & $\Cc_{V}$, $\Cc_{F}$ & $\checkmark$ & $\checkmark$ & $\checkmark$ &
$\checkmark$ & $\checkmark$ & $\Cc_{F}^2 \cdot \Cc_{\PGg}^2(\Cc_{F},\Cc_{V})/ \Cc_{\PH}^2(\Cc_{F},\Cc_{V})$ \\[0.1em]\cline{1-2}\cline{4-9}
\ref{sec:CV,CF,BRinv}& 43.2 &                              & $\Cc_{F}$, $\Cc_{V}$, $\BRinv$ & $\begin{array}{cc}\le 1\\-\end{array}$ &  $\begin{array}{cc}-\\-\end{array}$ & $\begin{array}{cc}\checkmark\\\checkmark\end{array}$ &  $\begin{array}{cc}\checkmark\\\checkmark\end{array}$ &$\begin{array}{cc}\checkmark\\\Cc_{\rm on}=\Cc_{\rm off}\end{array}$ & $\frac{\Cc_{F}^2 \cdot \Cc_{\PGg}(\Cc_F,\Cc_V)^2}{\Cc_{\PH}^2(\Cc_{F},\Cc_{V})} \cdot (1-\BRinv)$ \\[0.1em]\cline{1-2}\cline{4-9}
\ref{sec:bm:CF/CV,CV2/CH}& 43.3 & & $\Rr_{FV}$, $\Cc_{VV}$ & $\checkmark$ & $\checkmark$ & $\checkmark$ & $\checkmark$ & $-$ & $\Cc_{VV}^2 \cdot \Rr_{FV}^2 \cdot \Cc_{\PGg}^2(\Rr_{FV},\Rr_{FV},\Rr_{FV},1)$ \\[0.1em]\hline
\ref{sec:Cd/Cu,CV/Cu,Cu2/CH}& 46 &Up-/down-type fermions & $\Rr_{du}$, $\Rr_{Vu}$, $\Cc_{uu}$ & $\checkmark$ & $\Cc_u$, $\Cc_d$ & $\checkmark$ & $\checkmark$ & $-$ & $\Cc_{uu}^2 \cdot \Cc_{\Pg}^2(\Rr_{du},1) \cdot \Cc_{\PGg}^2(\Rr_{du},1,\Rr_{du},\Rr_{Vu})$ \\[0.1em]\hline
\ref{sec:Cl/Cq,CV/Cq,Cq2/CH}& 47 &Leptons/quarks & $\Rr_{\ell q}$, $\Rr_{Vq}$, $\Cc_{qq}$ & $\checkmark$ & $\Cc_\ell$, $\Cc_q$ & $\checkmark$ & $\checkmark$ & $-$ & $\Cc_{qq}^2 \cdot \Cc_{\PGg}^2(1,1,\Rr_{\ell q},\Rr_{Vq})$ \\[0.1em]\hline
\ref{sec:Cg,Cam}& 48.1 &\multirow{2}{0.99\linewidth}{Vertex loops + $\PH\!\to$invisible/undetected decays}& $\begin{array}{cc}\Cc_{\Pg}, \Cc_{\PGg},\\ \Cc_{\PZ\PGg}\end{array}$ & =1 & =1 & $-$ & $-$ & $\checkmark$ & $\Cc_{\Pg}^2 \cdot \Cc_{\PGg}^2/ \Cc_{\PH}^2(\Cc_{\Pg},\Cc_{\PGg})$ \\[0.1em]\cline{1-2}\cline{4-9}
\ref{sec:Cg,Cam,BRinv}& 48.2 & & $\begin{array}{cc}\Cc_{\Pg}, \Cc_{\PGg}, \\ \Cc_{\PZ\PGg}, \BRinv\end{array}$ & =1 & =1 & $-$ & $-$ &$\checkmark$ & $\Cc_{\Pg}^2 \cdot \Cc_{\PGg}^2/ \Cc_{\PH}^2(\Cc_{\Pg},\Cc_{\PGg}) \cdot (1-\BRinv)$ \\[0.1em]\cline{1-2}\cline{4-9}
\ref{sec:CV,CF,Cg,Cam,BRinv}& 49 &                       &$\begin{array}{cc}\Cc_{F}, \Cc_{V}, \Cc_{\Pg}, \Cc_{\PGg}, \\ \Cc_{\PZ\PGg}, \BRinv\end{array}$ & $\begin{array}{cc}\le 1\\-\end{array}$ & $\begin{array}{cc}-\\-\end{array}$ &  $\begin{array}{cc}-\\-\end{array}$ &  $\begin{array}{cc}-\\-\end{array}$ &$\begin{array}{cc}\checkmark\\\Cc_{\rm on}=\Cc_{\rm off}\end{array}$ & $\frac{\Cc_{F}^2 \cdot \Cc_{\PGg}(\Cc_F,\Cc_V)^2}{\Cc_{\PH}^2(\Cc_F,\Cc_V,\Cc_{\Pg},\Cc_{\PGg})} \cdot (1-\BRinv)$ \\[0.1em]\hline
\ref{sec:gen1}& 51 &\multirow{2}{0.99\linewidth}{Generic models with and without assumptions on vertex loops and $\Gamma_{\PH}$} & $\Cc_{\PW}$, $\Cc_{\PZ}$, $\Cc_{\PQt}$, $\Cc_{\PQb}$, $\Cc_{\PGt}$, $\Cc_{\PGm}$ & $-$ & $-$ & $\checkmark$ &
$\checkmark$ & $\checkmark$ & $\frac{\Cc_{\Pg}^2(\Cc_{\PQb},\Cc_{\PQt}) \cdot \Cc_{\PGg}^2(\Cc_{\PQb},\Cc_{\PQt},\Cc_{\PGt},\Cc_{\PGm},\Cc_{\PW})}{\Cc_{\PH}^2(\Cc_{\PQb},\Cc_{\PQt},\Cc_{\PGt},\Cc_{\PGm},\Cc_{\PW},\Cc_{\PZ})}$ \\[0.1em]\cline{1-2}\cline{4-9}
\ref{sec:gen2}& 50.2 & & $\begin{array}{cc}\Cc_{\PW}, \Cc_{\PZ}, \Cc_{\PQt}, \Cc_{\PQb},\\ \Cc_{\PGt}, \Cc_{\PGm}, \Cc_{\Pg}, \Cc_{\PGg}, \\ \Cc_{\PZ\PGg}, \BRinv \end{array}$ & $\begin{array}{cc}\le 1\\-\\-\end{array}$ & $\begin{array}{ccc}-\\-\\-\end{array}$ &  $\begin{array}{ccc}-\\-\\-\end{array}$ &
 $\begin{array}{ccc}-\\-\\-\end{array}$ & $\begin{array}{ccc}\checkmark\\\checkmark\\\Cc_{\rm on}=\Cc_{\rm off}\end{array}$ & $\frac{\Cc_{\Pg}^2 \cdot \Cc_{\PGg}^2}{\Cc_{\PH}^2(\Cc_{\PQb},\Cc_{\PQt},\Cc_{\PGt},\Cc_{\PGm},\Cc_{\PW},\Cc_{\PZ})} \cdot (1-\BRinv)$ \\[0.1em]\cline{1-2}\cline{4-9}

\ref{sec:gen3}& 50.3 & & $\begin{array}{ccc}\Rr_{\PW\PZ}, \Rr_{\PQt\Pg}, \Rr_{\PQb\PZ}\\ \Rr_{\tau\PZ}, \Rr_{\Pg\PZ}, \Rr_{\PGg\PZ},\\ \Rr_{(\PZ\PGg)\PZ}, \Cc_{\Pg\PZ} \end{array}$  & $-$ & $-$ & $-$ &
$-$ & $-$ & $\Cc_{\Pg\PZ}^2 \cdot \Rr_{\PGg\PZ}^2$ \\[0.1em]\hline
\hline
\end{tabular}
}
\end{center}
\end{sidewaystable}

\subsection{Fermion versus vector (gauge) coupling strengths}
\label{sec:CFCV}
Benchmark coupling models in this section allow for different Higgs
boson coupling strengths to fermions and bosons, reflecting the different
structure of the interactions of the SM Higgs sector with gauge bosons
and fermions.  It is always assumed that only SM particles contribute
to the {\ggF}, {\hgg}, {\Hzg} and {\ggZH} vertex loops, and
modifications of the coupling-strength scale factors for fermions and vector
bosons are propagated through the loop calculations. Models with and
without assumptions about the total width are presented.

\subsubsection{Assuming only SM contributions to the total width}
\label{sec:bm:CF,CV}
In the first benchmark model no undetected or invisible Higgs boson decays are assumed to exist, i.e. $\BRinv=0$.
The universal coupling-strength scale factors $\Cc_F$ for all fermions and $\Cc_V$ 
for all vector bosons are defined in this model as:
\begin{eqnarray*}
  \Cc_V &=& \Cc_{\PW} = \Cc_{\PZ} \label{kVWZ} \\
  \Cc_F &=& \Cc_{\PQt} = \Cc_{\PQb} = \Cc_{\PGt} = \Cc_{\Pg} = \Cc_{\PGm} .
\label{kftb}
\end{eqnarray*}

As only SM particles are assumed to contribute to the {\ggF}
loop in this benchmark model, the gluon fusion process depends directly on
the fermion scale factor $\Cc_F^2$.
Only the relative sign between $\Cc_F$ and $\Cc_V$ is physical and hence in the
following only $\Cc_V>0$ is considered, without loss of generality.
Sensitivity to this relative sign is gained from the negative interference
between the loop contributions of the $W$ boson and the $t$-quark in {\hgg} and
{\Hzg} decays and in \ggZH\ production, as well as from the $tH$ processes (see the corresponding expressions in Table \ref{tab:kexpr}).

Figure~\ref{fig:spbm:CVCF} shows the results of the fits for this
benchmark model.  Figure~\ref{fig:spbm:CVCF:CVCF_overlay} illustrates
how the decays \hgg, \hzz, \hww, \htt\ and \hbb\ contribute to the
combined measurement. The slight asymmetry in $\Cc_{F}$ for 
\hww\ and \hbb\ decays is introduced by the small contributions of the $tH$
and \ggZH\ production processes that contribute to these decay modes, and which are
sensitive to the sign of $\Cc_{F}$ due to interference effects. The
strong constraint on $\Cc_F$ from \hww\ decays is related to the
$3.2\sigma$ observation of the VBF production process in this
channel\cite{ATLAS:2014aga}. Outside the range shown in
Fig.~\ref{fig:spbm:CVCF:CVCF_overlay} there are two additional minima
for \hgg. The long tails in the \hbb\ contour towards high values of
$|\Cc_{V}|$ are the result of an asymptotically disappearing
sensitivity of the observed signal strength in the $\bb$ final states
to $\Cc_{V}$ at large values of $\Cc_{V}$.
The combined measurement without overlays is also shown in Figure~\ref{fig:spbm:CVCF:CVCF}.

Figures~\ref{fig:spbm:CVCF:CVCF_overlay} and \ref{fig:spbm:CVCF:CVCF}
only show the SM-like minimum with a positive relative sign, as
the local minimum with negative relative sign is disfavoured at the
$\RESULTNegativeCVCFZValue$ level, which can been seen in the wider
scan of $\Cc_{F}$, where $\Cc_V$ is profiled, shown in
Fig.~\ref{fig:spbm:CVCF:CF}.  The likelihood as a function of $\Cc_V$,
profiling $\Cc_F$, is given in Fig.~\ref{fig:spbm:CVCF:CV}. Around
$\Cc_{V}=0.8$ the sign of the chosen profiled solution for $\Cc_{F}$
changes, causing a kink in the likelihood. The profile likelihood
curves restricting $\Cc_{F}$ to either positive or negative values are
also shown in Fig.~\ref{fig:spbm:CVCF:CV} as thin curves, and
illustrate that this sign change in the unrestricted profile
likelihood is the origin of the kink.

Both $\Cc_{F}$ and $\Cc_{V}$ are measured to be
compatible with their SM expectation and the two-dimensional
compatibility of the SM hypothesis with the best-fit point is
$\RESULTmodelCVCFpvalueSM$. The best-fit values and uncertainties are:
\begin{eqnarray*}
  \Cc_V &=& \RESULTmodelCVCFvalueCV\\
  \Cc_F &=& \RESULTmodelCVCFvalueCF.
\end{eqnarray*}

\begin{figure}[ptb]
  \center
  \subfloat[]{
    \includegraphics[width=.90\textwidth]{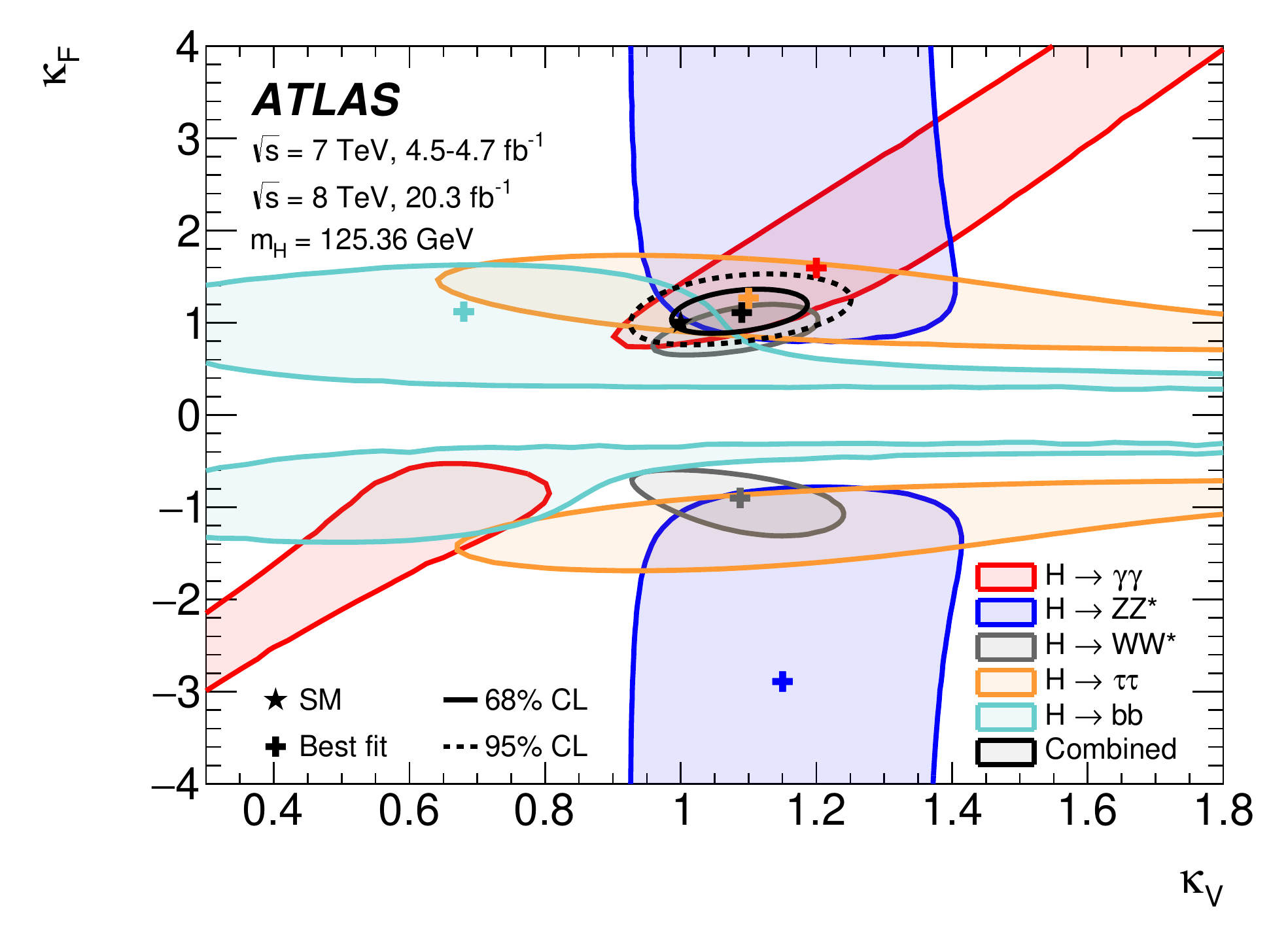}
    \label{fig:spbm:CVCF:CVCF_overlay}
   }

   \subfloat[]{
     \includegraphics[width=.33\textwidth]{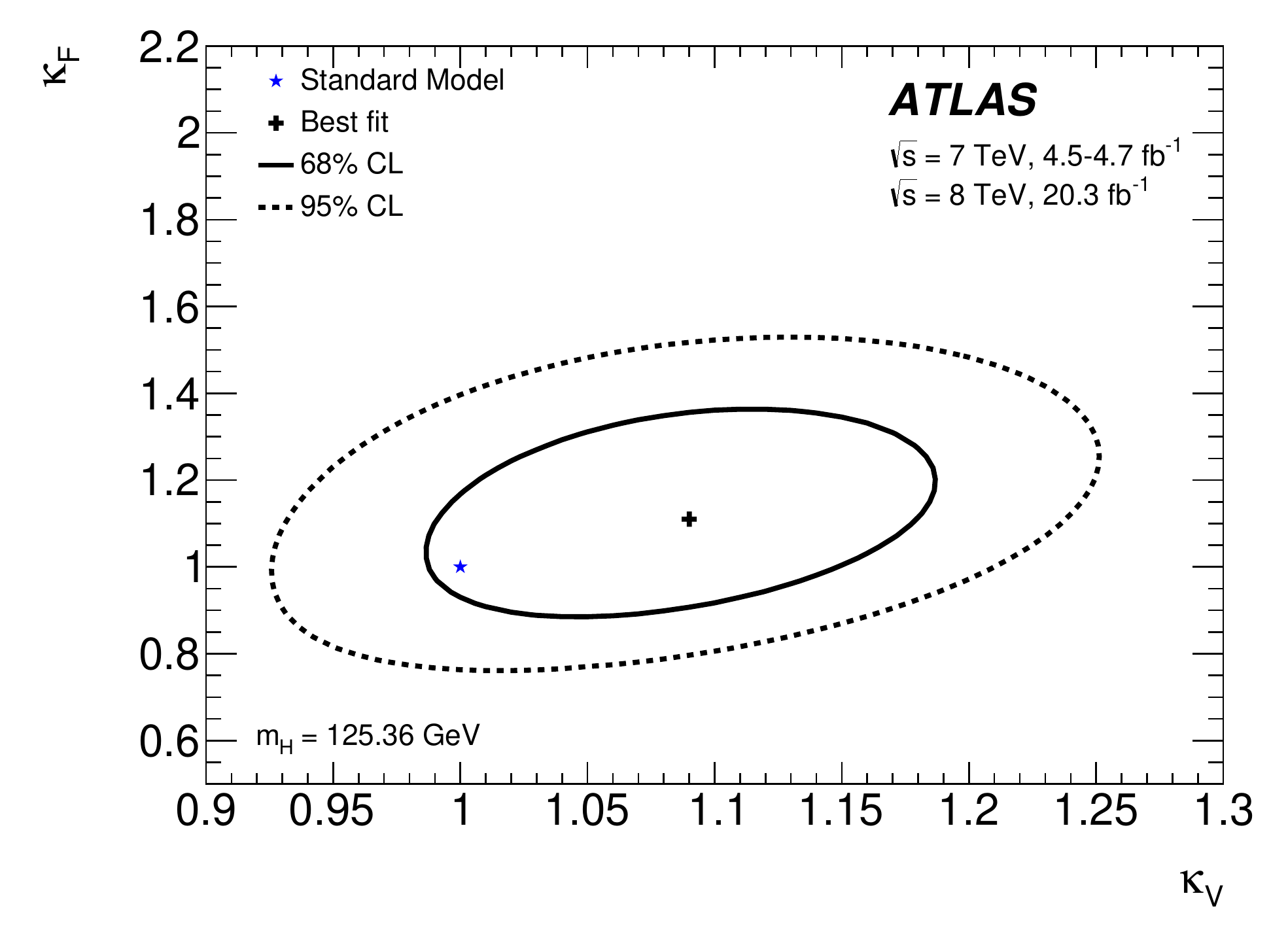}
     \label{fig:spbm:CVCF:CVCF}
   }
  \subfloat[]{
    \includegraphics[width=.33\textwidth]{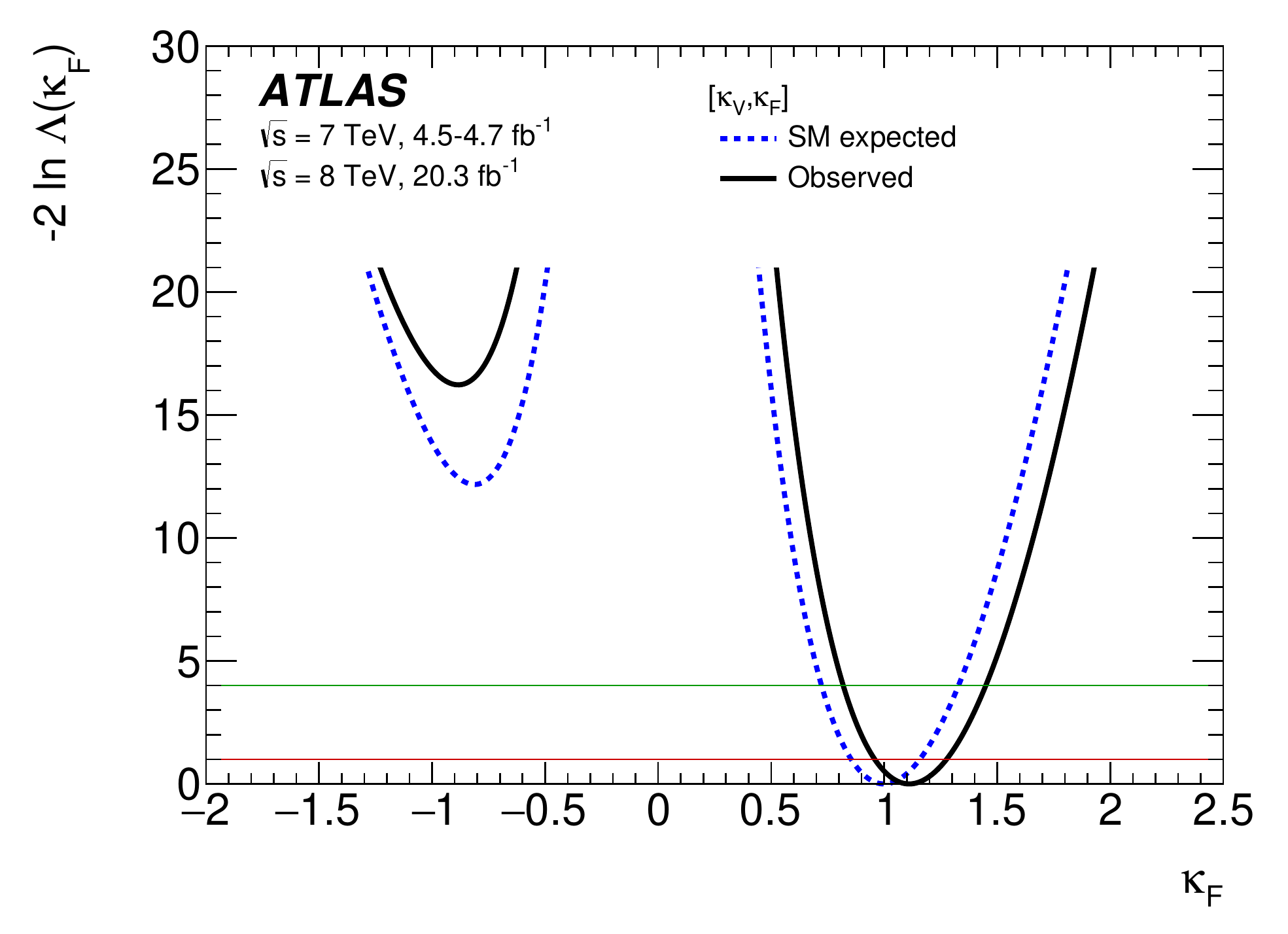} 
    \label{fig:spbm:CVCF:CF}
  }
  \subfloat[]{
    \includegraphics[width=.33\textwidth]{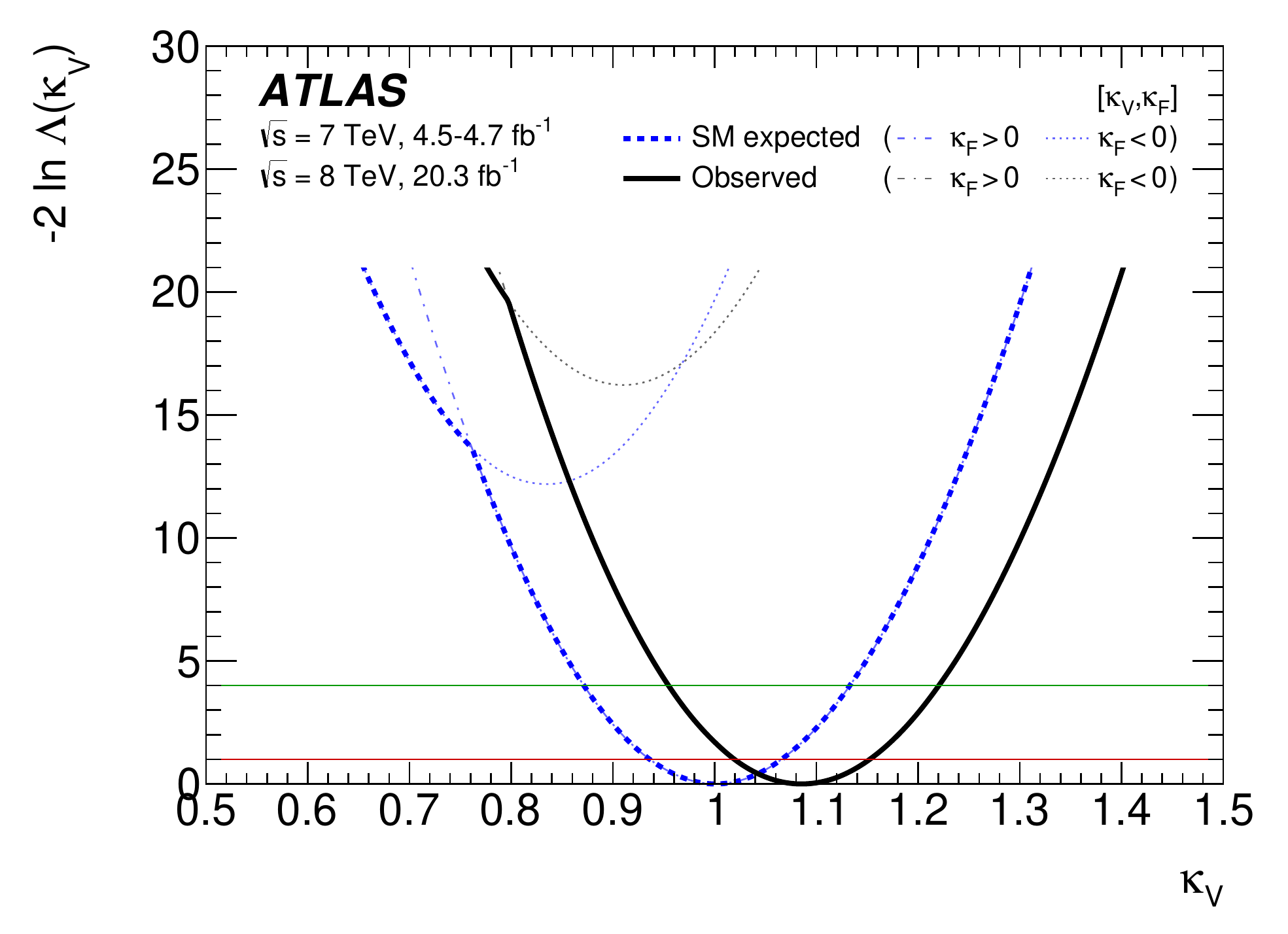}
    \label{fig:spbm:CVCF:CV}
    }

  \caption{Results of fits for the two-parameter benchmark model
    defined in Section~\ref{sec:bm:CF,CV} that probes different
    coupling-strength scale factors for fermions and vector bosons,
    assuming only SM contributions to the total width: (a)~results of the two-dimensional fit to $\Cc_{F}$ and $\Cc_{V}$, including $68\%$ and $95\%$ CL contours;
    overlaying the $68\%$ CL contours derived from
    the individual channels and their combination; 
    (b) ~the same measurement, without the overlays of the individual channels; (c)~the profile likelihood ratio as a function of the coupling-strength scale factors 
    $\Cc_{F}$ ($\Cc_{V}$ is profiled) and (d)~as a function of ~$\Cc_{V}$ ($\Cc_{F}$ is profiled).
    The dashed curves in (c) and (d) show the SM expectations. In (d) the sign of the
    chosen profiled solution for $\Cc_{F}$ changes at $\Cc_{V} \approx 0.8$ , causing a kink in the likelihood. 
    The profile likelihood curves restricting $\Cc_{F}$ to be either positive or negative are also shown
    to illustrate that this sign change in the unrestricted profile likelihood is the origin of the kink.
    The red (green) horizontal line indicates the value of the profile likelihood ratio corresponding to a 68\% (95\%) confidence interval for the parameter of interest, assuming the asymptotic $\chi^2$ distribution for the test statistic.
    \label{fig:spbm:CVCF}}
\end{figure}

\subsubsection{Allowing for invisible or undetected Higgs boson decays in the total width}
\label{sec:CV,CF,BRinv}

The second benchmark model of this section allows for the presence of invisible or undetected Higgs boson decays by introducing \BRinv as a free parameter
in the expression of Eq.~(\ref{eq:CH:1}) for the Higgs boson total width. The free parameters of this model thus are $\Cc_F$, $\Cc_V$ and \BRinv. Loop processes
are still assumed to have only SM content.

With the introduction of \BRinv as a free parameter, the assumed Higgs
boson width has no intrinsic upper bound and an additional constraint must
be imposed on the model that infers an upper bound on
$\Gamma_{\PH}$. Both choices of constraints on the total width discussed
in Section~\ref{sec:framework} are studied: $\Cc_{V}<1$ and $\Cc_{\rm
  on}=\Cc_{\rm off}$.

\begin{figure}[htb!]
  \centering
  \includegraphics[width=0.8\textwidth]{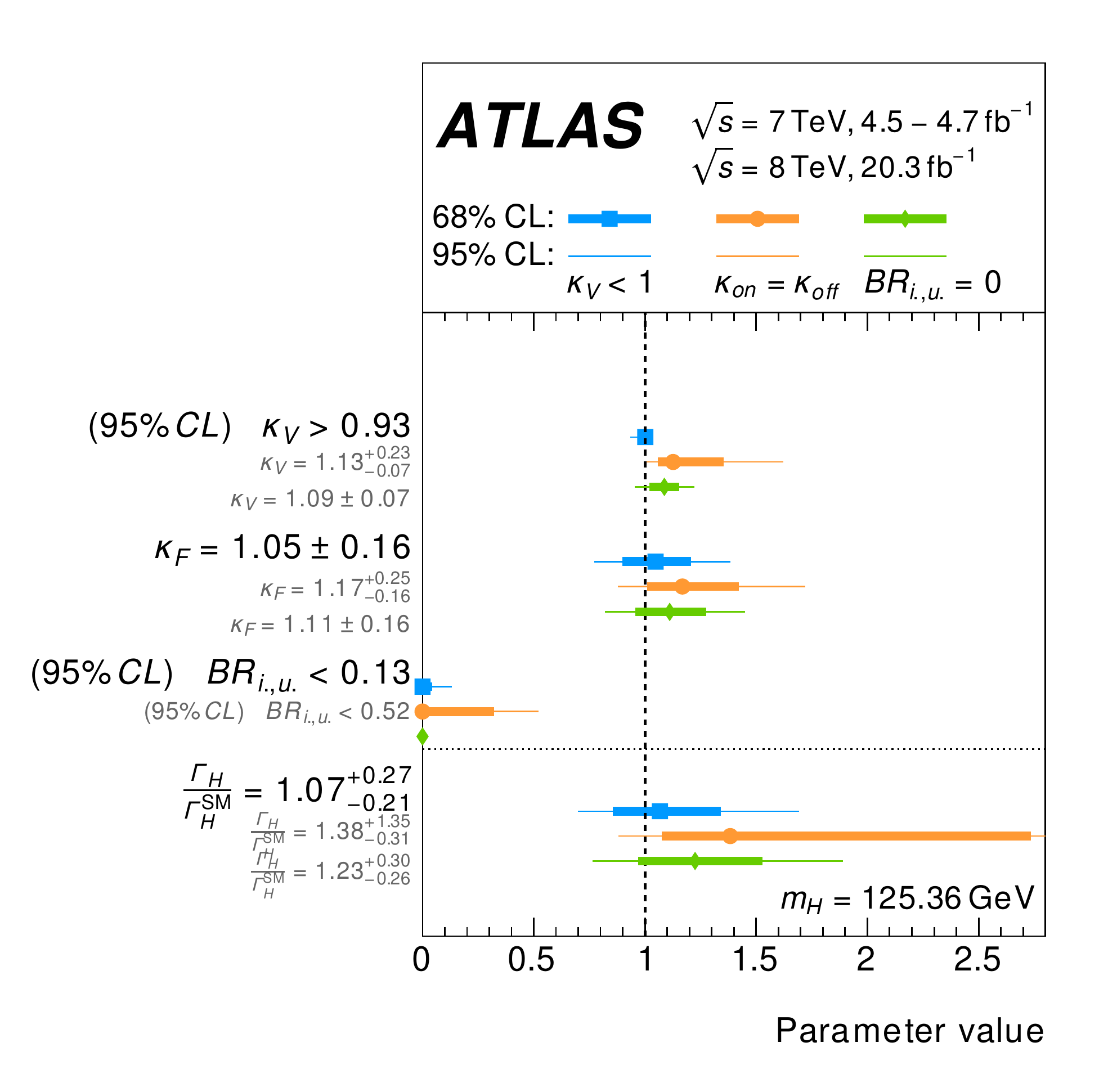}
  \caption{Results of fits for benchmark models that probe for potential extra contributions to the total width, but do not allow contributions
    from non-SM particles in the \hgg, \ggF\ and \hzg\ loops,
    with free gauge and fermion coupling-strength scale factors $\Cc_V,\Cc_F$.
    The estimated values of each parameter under the constraint $\Cc_{V}<1$, $\Cc_{\rm on}=\Cc_{\rm off}$ or $\BRinv=0$ are shown with markers in the shape of box, circle, or diamond, respectively.
    The inner and outer bars correspond to 68\%~CL and 95\%~CL intervals.
    The confidence intervals of $\BRinv$ and, in the benchmark model with the constraint $\Cc_{V}<1$, also $\Cc_{V}$, are estimated with respect to their physical bounds, as described in the text. The numerical values of the fit under the constraint $\Cc_{V}<1$ are shown on the left. Values for the two alternative constraints are also shown (in a reduced font size due to space constraints).
    \label{fig:bm:BRUICVCF}
  }

\end{figure}

Figure~\ref{fig:bm:BRUICVCF} shows the results of fits for this
benchmark scenario. For comparison the results of the benchmark
model of Section~\ref{sec:bm:CF,CV} are included, corresponding to
the condition $\BRinv=0$. The coupling-strength scale factors $\Cc_{F}$ and
$\Cc_{V}$ are measured to be compatible with the SM values and a limit is set
on the fraction of Higgs boson decays to invisible or undetected final
states. The three-dimensional compatibility of the SM hypothesis with
the best-fit point is $\RESULTmodelBRUICVCFApvalueSM$ $(\RESULTmodelBRUICVCFBpvalueSM)$, when applying the
$\Cc_V<1$ (off-shell) constraint, respectively.  When imposing the physical
constraint $\BRinv\ge 0$, the $95\%$ CL upper limit is $\BRinv <
\RESULTmodelBRUICVCFAlimitFC$ ($\RESULTmodelBRUICVCFBlimitFC$), when applying the constraint
$\Cc_V<1$ ($\Cc_{\rm on}=\Cc_{\rm off}$).  The corresponding expected limit on \BRinv, under
the hypothesis of the SM, is $\RESULTmodelBRUICVCFAlimitFCSM$ ($\RESULTmodelBRUICVCFBlimitFCSM$).

Also shown in Fig.~\ref{fig:bm:BRUICVCF} is
the uncertainty on the total width that the model variants allow,
expressed as the ratio $\Gamma_{\PH}/\Gamma_{\PH}^{\rm SM}$. These
estimates for the width are obtained from alternative
parameterisations of these benchmark models, where the
coupling-strength scale factor $\Cc_{F}$ is replaced by the expression that results
from solving Eq.~(\ref{eq:CH:1}) for $\Cc_{F}$, introducing
$\Gamma_{\PH}/\Gamma_{\PH}^{\rm SM}$ as a parameter of the
model. Figure~\ref{fig:bm:BRUICVCF} shows that the upper bound on the
Higgs boson width from the assumption $\Cc_{\rm off}=\Cc_{\rm on}$ is
substantially weaker than the bound from the assumption
$\Cc_{V}<1$. These choices of constraints on the Higgs boson width
complement each other in terms of explored parameter space: the
present limit of $\mu_{\rm off}<5.1$~\cite{Aad:2015xua} in the combined off-shell
measurement in the \hww\ and \hzz\ channels effectively constrains
$\Cc_{V}$ to be greater than one in the combined fit when exploiting the
assumption $\Cc_{\rm on}=\Cc_{\rm off}$.

The parameterisation of the off-shell signal strength $\mu_{\rm off}$ 
in terms of couplings implicitly requires that $\mu_{\rm
off}\ge 0$ (see Ref.~\cite{Aad:2015xua} for details). This boundary
condition causes the distribution of the test statistic to deviate
from its asymptotic form for low values of $\sigma_{\rm off}$, with
deviations in $p$-values of up to 10\% for $\sigma_{\rm off} \approx
2.5$, which corresponds to the value of $\sigma_{\rm off}$ at the upper
boundary of the 68\% asymptotic confidence interval of
$\Gamma_{\PH}/\Gamma_{\PH}^{\rm SM}$.  The upper bound of the 68\%~CL 
interval for the scenario $\Cc_{\rm off}=\Cc_{\rm on}$ shown in
Fig.~\ref{fig:bm:BRUICVCF} should therefore be considered to be only
approximate. Since the lower bound on
$\Gamma_{\PH}/\Gamma_{\PH}^{\rm SM}$ is always dominated by the
constraint $\BRinv\ge 0$, it is not affected by this
deviation from the asymptotic behaviour.

\subsubsection{No assumption about the total width}
\label{sec:bm:CF/CV,CV2/CH}
In the last benchmark model of this section no assumption about the total width is made.
In this model only ratios of coupling-strength scale factors are
measured, choosing as free parameters
\begin{eqnarray*}
  \Rr_{FV} &=& \Cc_F / \Cc_V \label{kFV}\\
  \Cc_{VV} &=& \Cc_V\cdot \Cc_V / \Cc_{\PH} \label{kVV}  ,
\end{eqnarray*}
where $\Rr_{FV}$ is the ratio of the fermion and vector boson coupling-strength
scale factors, $\Cc_{VV}$ is an overall scale that includes the total
width and applies to all rates, and $\Cc_{\PH}$ is defined in Table~\ref{tab:kexpr}.

Figure~\ref{fig:spbm:CF/CV,CV2/CH} shows the results of this fit. Both
ratio parameters are found to be consistent with the SM expectation and
the two-dimensional compatibility of the SM hypothesis with the
best-fit point is $\RESULTmodelRFVCVVpvalueSM$.
The best-fit values and uncertainties, when profiling the other parameter, are:
\begin{eqnarray*}
  \Rr_{FV} &=& \RESULTmodelRFVCVVvalueRFV \\
  \Cc_{VV} &=& \RESULTmodelRFVCVVvalueCVV .
\end{eqnarray*}
Similar to the model described in Section \ref{sec:bm:CF,CV},
Fig.~\ref{fig:spbm:CF/CV,CV2/CH,LFVCV2:CF/CV} shows the determination
of the sign of $\Rr_{FV}$ disfavouring $\Rr_{FV}=-1$ at approximately
$\RESULTNegativeCVCFZValue$, while Fig.~\ref
{fig:spbm:CF/CV,CV2/CH:LFVCV2} shows the two-dimensional likelihood
contour. The estimates of the two parameters are anticorrelated because
only their product appears in the model.

\begin{figure}[htbp!]
  \center
  \subfloat[]{
    \includegraphics[width=.45\textwidth]{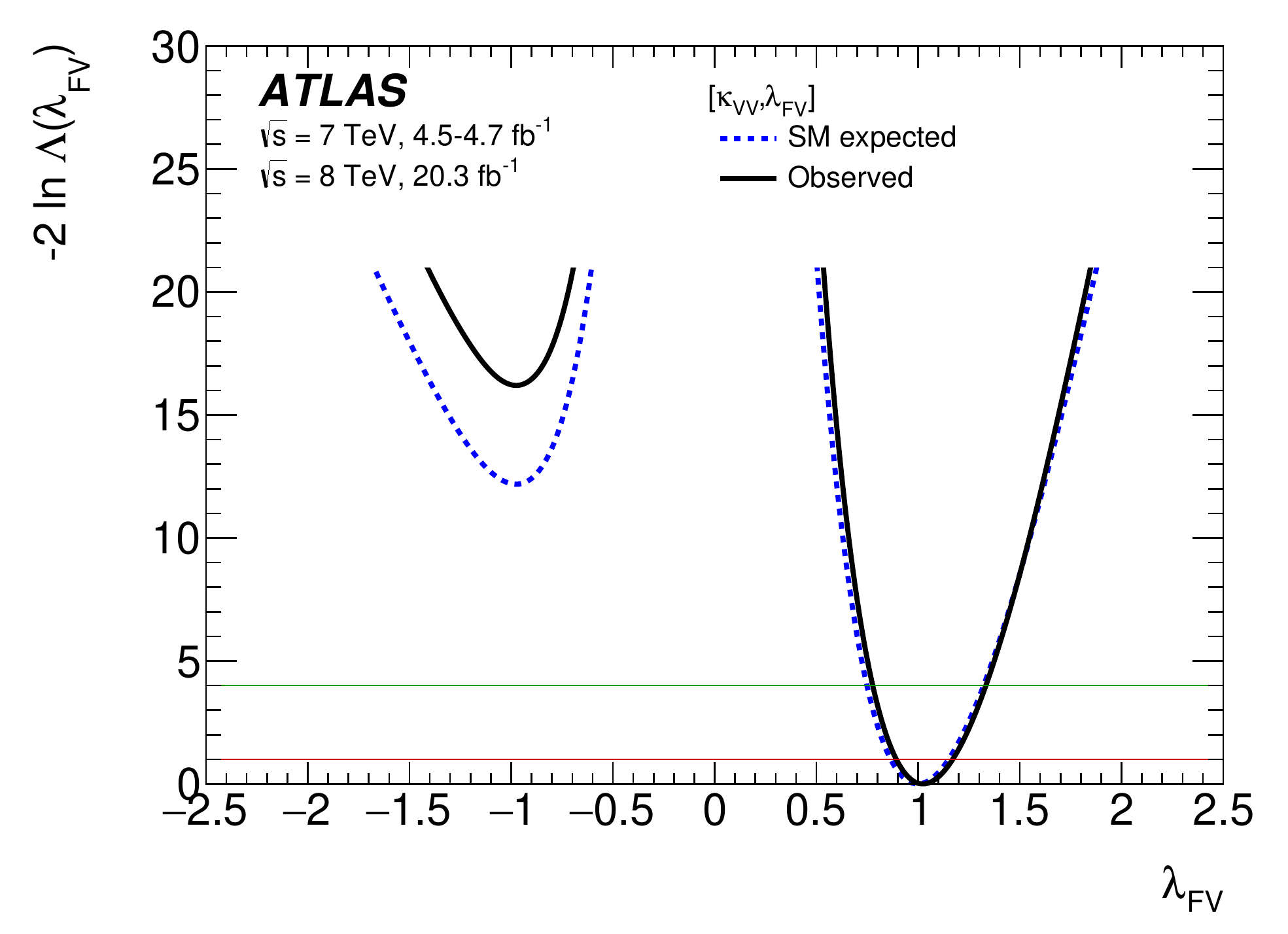} 
    \label{fig:spbm:CF/CV,CV2/CH,LFVCV2:CF/CV}
  }
  \subfloat[]{
    \includegraphics[width=.45\textwidth]{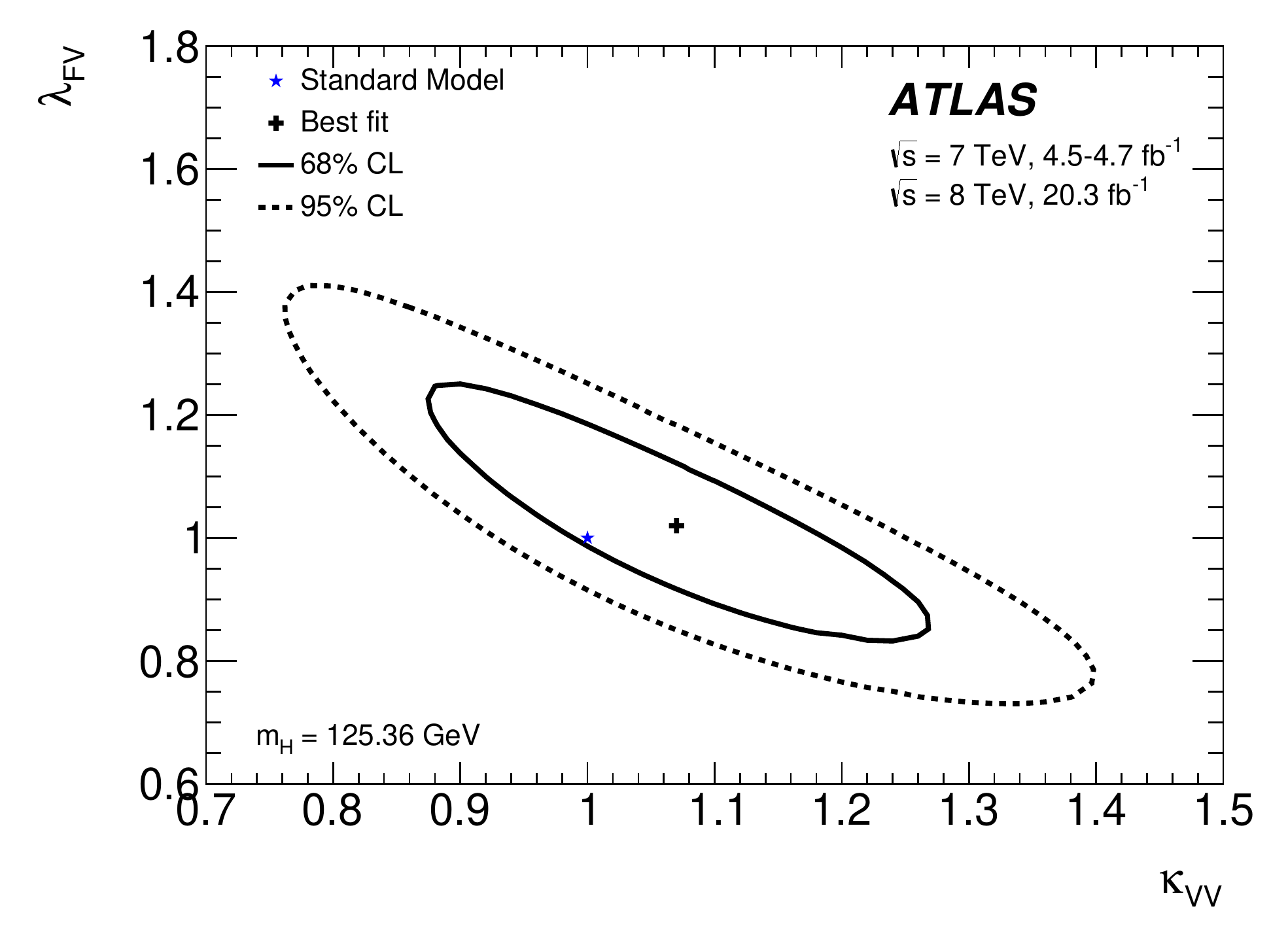}
    \label{fig:spbm:CF/CV,CV2/CH:LFVCV2}
  }

  \caption{ Results of fits for the two-parameter benchmark model defined
    in Section~\ref{sec:bm:CF/CV,CV2/CH} that probes different coupling-strength 
    scale factors for fermions and vector bosons without
    assumptions about the total width:
    (a)~profile likelihood ratio as a function of the coupling-strength scale factor ratio
    $\Rr_{FV}$ ($\Cc_{VV}$ is profiled). 
    The dashed curve shows the SM expectation.
    The red (green) horizontal line indicates the value of the profile likelihood ratio corresponding to a 68\% (95\%) confidence interval for the parameter of interest, assuming the asymptotic $\chi^2$ distribution for the test statistic.
    (b)~Results of the two-dimensional fit to $\Cc_{VV}$ and $\Rr_{FV}$, including $68\%$ and $95\%$ CL contours.
    \label{fig:spbm:CF/CV,CV2/CH}
  }

\end{figure}

\subsection{Probing relations within the fermion coupling sector}
\label{sec:fermion}
The previous sections assumed universal coupling-strength scale
factors for all fermions, while many extensions of the SM predict
deviations from universality within the fermion sector~~\cite{Heinemeyer:2013tqa}. In
this section, benchmark models are explored that probe the relations
between the up- and down-type fermions and between the lepton and
quark sectors, using the information in the currently accessible
channels, in particular in \hbb, \htt\ and \hmm\ decays and $ttH$
production.  The models considered assume that only SM particles
contribute to the {\ggF}, {\hgg}, {\Hzg} and {\ggZH} vertex loops, and
modifications of the coupling-strength scale factors are propagated through
the loop calculations. As only ratios of coupling-strength scale factors are
explored, no assumptions on the total width are made.

\subsubsection{Probing the up- and down-type fermion symmetry}
\label{sec:Cd/Cu,CV/Cu,Cu2/CH}
Many extensions of the SM contain different coupling strengths of the Higgs
boson to up-type and down-type fermions.  This is for instance the
case for certain Two-Higgs-Doublet
Models~(2HDM)~\cite{Lee:1973iz,Gunion:2002zf,Branco:2011iw}.
In this benchmark model the
ratio $\Rr_{du}$ of down- and up-type fermions coupling-strength scale factors is probed, while
vector boson coupling-strength scale factors are assumed to be unified and equal to $\Cc_{V}$. The indices
$u,d$ stand for all up- and down-type fermions, respectively.  The
free parameters are:
\begin{eqnarray*}
  \Rr_{du} &=& \Cc_{d} / \Cc_{u} \label{ldu}\\
  \Rr_{Vu} &=& \Cc_{V} / \Cc_{u} \label{kVu} \\
  \Cc_{uu} &=& \Cc_{u} \cdot \Cc_{u} / \Cc_{\PH} \label{kuu} .
\end{eqnarray*}

The up-type quark coupling-strength scale factor is mostly indirectly constrained
through the \ggF\ production channel, from the Higgs boson to top-quark
coupling strength, with an additional weak direct constraint from the \ttH\ production
channel, while the down-type coupling strength is constrained through the
{\hbb}, {\htt} and {\hmm} decays as well as weakly through the $\bb \to H$ production mode and the $b$-quark loop in the {\ggF} production mode.

The fit results for the parameters of interest in this benchmark model, when profiling the other parameters, are:
\begin{eqnarray*}
  \Rr_{du} &\in& \RESULTmodelUPDOWNvalueRDU~(68\%~ \rm{CL}) \\
  \Rr_{Vu} &=& \RESULTmodelUPDOWNvalueRVU \\
  \Cc_{uu} &=& \RESULTmodelUPDOWNvalueCUU .
\end{eqnarray*}
Near the SM prediction of $\Rr_{du}=1$, the best-fit value is 
$\Rr_{du} =\RESULTmodelUPDOWNvalueRDUupper$.
All parameters are measured to be
consistent with their SM expectation and the three-dimensional
compatibility of the SM hypothesis with the best-fit point is
$\RESULTmodelUPDOWNpvalueSM$.

The likelihood curves corresponding to these measurements are shown in
Figure~\ref{fig:bm:Rdu}. The likelihood curve of
Figure~\ref{fig:bm:RVuRduRuuh:Rdu} is nearly symmetric around
$\Rr_{du}=0$ as the model is almost insensitive to the relative sign
of $\Cc_u$ and $\Cc_d$. 
The interference of contributions from the $b$-quark and $t$-quark loops in the
\ggF\ production induces an observed asymmetry of about $0.6\sigma$
(no significant asymmetry is expected with the present sensitivity).
 The profile likelihood ratio value at $\Rr_{du}=0$ provides 
 $\RESULTmodelUPDOWNpvalueDownType$ evidence of the coupling of
 the Higgs boson to down-type fermions, mostly coming from the {\htt}
 measurement and to a lesser extent from the {\hbb} measurement.
 Vanishing coupling strengths of the Higgs boson to up-type fermions
 ($\Cc_{uu} = 0$) and vector bosons ($\Rr_{Vu} = 0$) are excluded at
a level of $>5\sigma$.

\begin{figure}[htbp!]
  \center

 \begin{minipage}{0.6\textwidth}
  \subfloat[]{
    \includegraphics[width=.99\textwidth]{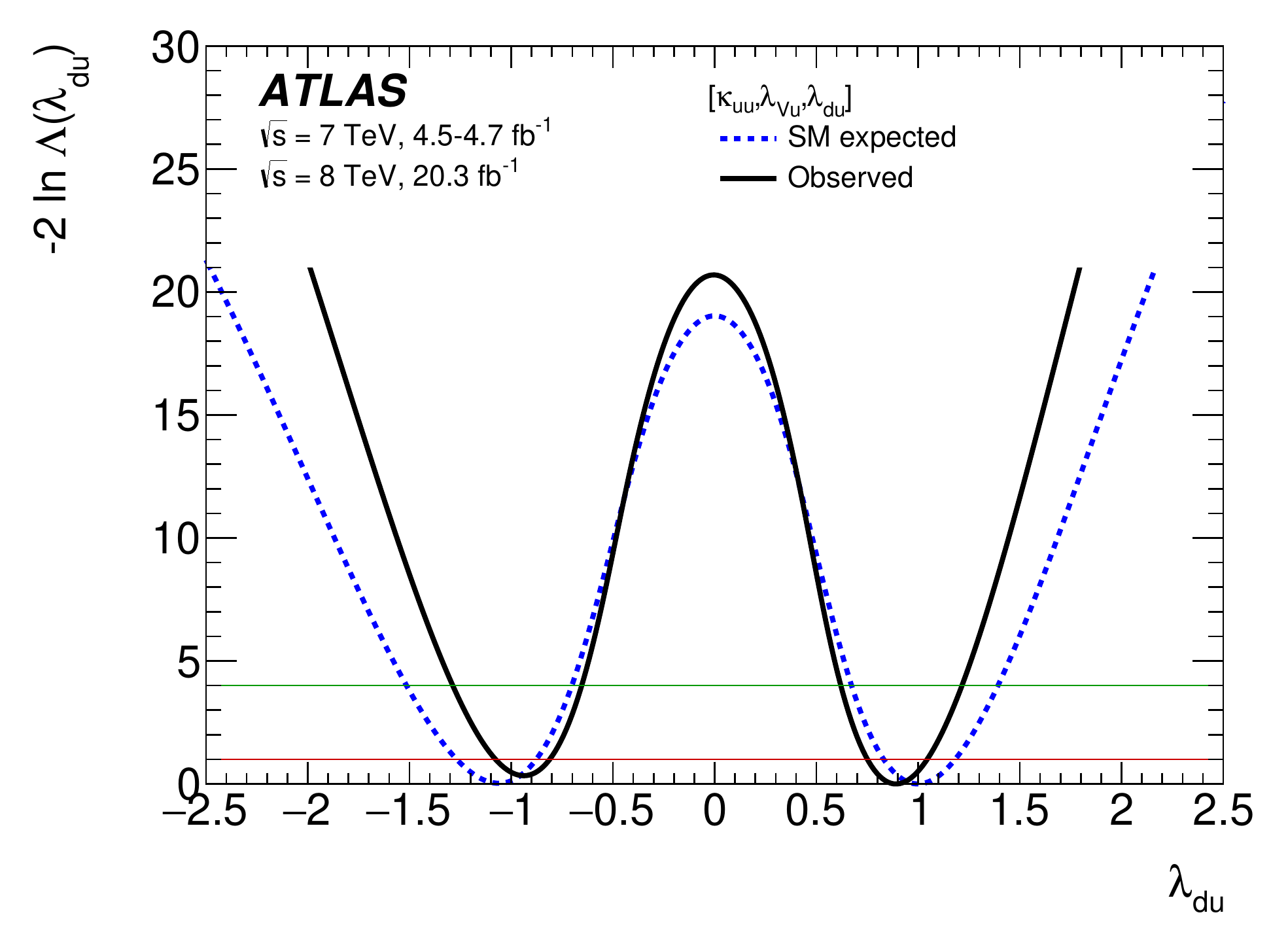}
    \label{fig:bm:RVuRduRuuh:Rdu}
    \vspace*{-0.5cm}
  }
 \end{minipage}
 \begin{minipage}{0.39\textwidth}
 \subfloat[]{
   \includegraphics[width=.99\textwidth,trim=0 1cm 0 0.5cm]{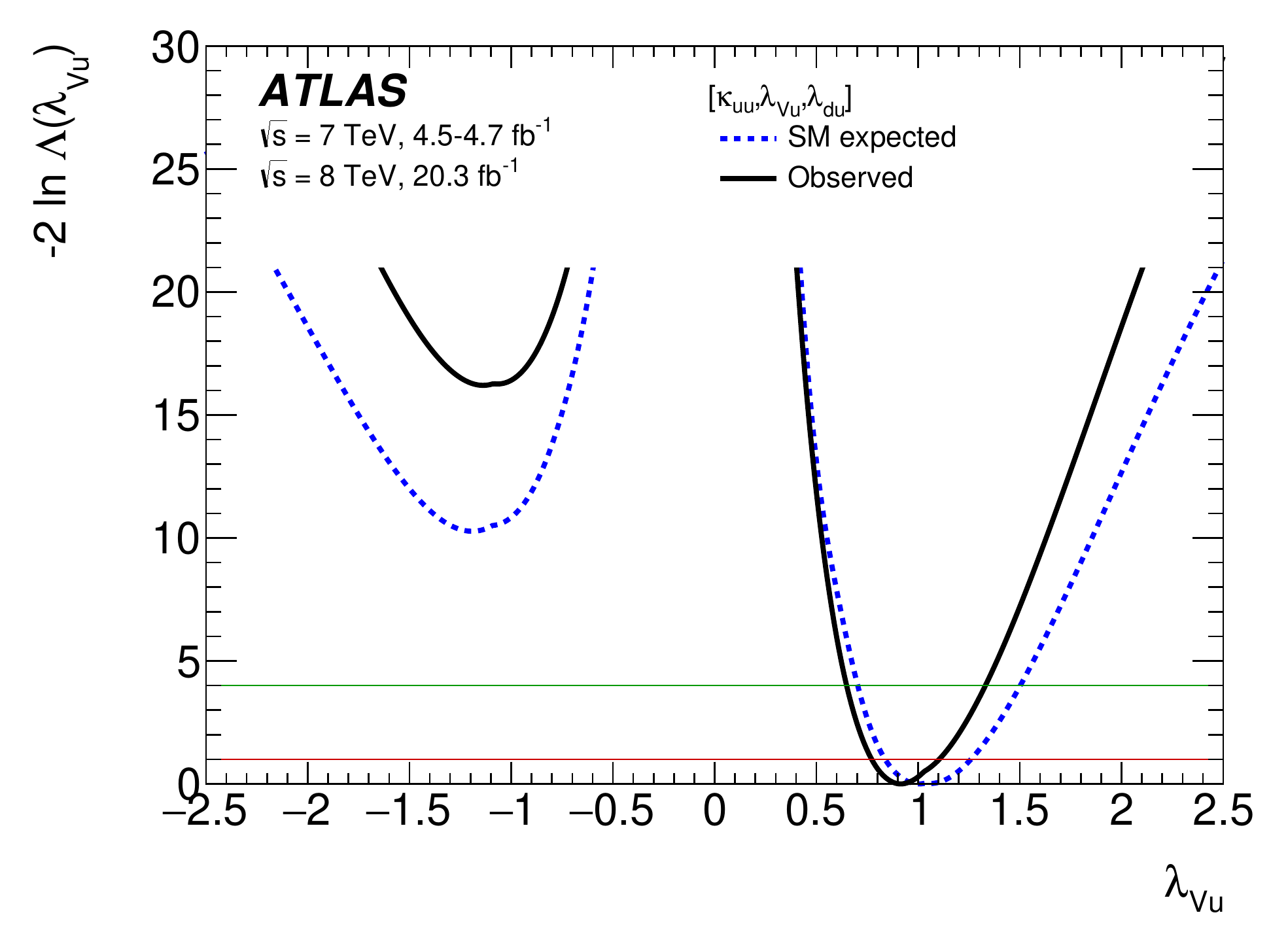}
   \label{fig:bm:RVuRduRuuh:RVu}
   \vspace*{-0.5cm}
 }

 \subfloat[]{
    \includegraphics[width=.99\textwidth,trim=0 1cm 0 0.5cm]{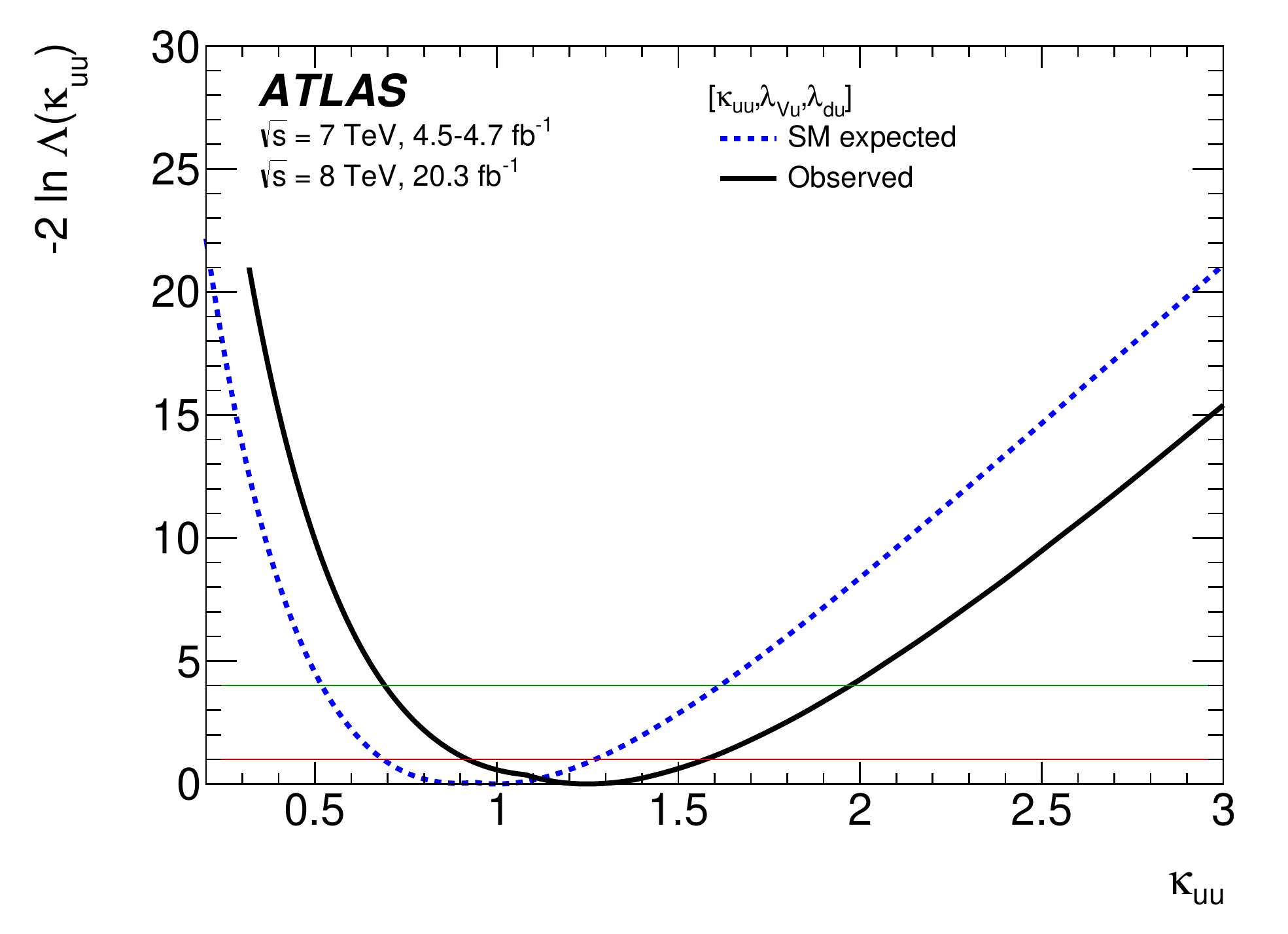}
    \label{fig:bm:RVuRduRuuh:Ruuh}
    \vspace*{-0.5cm}
  }
  \end{minipage}
  \vspace*{-0.2cm}

  \caption{Results of fits for the benchmark model described in Section~\ref{sec:Cd/Cu,CV/Cu,Cu2/CH} that probes the ratio of scale factors between down- and up-type fermions:
    profile likelihood ratios as functions of the coupling-strength scale factor ratios
    (a)~$\Rr_{du}$ ($\Rr_{Vu}$ and $\Cc_{uu}$ are profiled),
    (b)~$\Rr_{Vu}$ ($\Rr_{du}$ and $\Cc_{uu}$ are profiled),
    and (c)~the overall scale factor $\Cc_{uu}$ ($\Rr_{du}$ and $\Rr_{Vu}$ are profiled). The dashed curves show the SM expectations.
    The red (green) horizontal line indicates the value on the profile likelihood ratio corresponding to a 68\% (95\%) confidence interval for the parameter of interest, assuming the asymptotic $\chi^2$ distribution for the test statistic.
    }
    \label{fig:bm:Rdu}
\end{figure}

\subsubsection{Probing the quark and lepton symmetry}
\label{sec:Cl/Cq,CV/Cq,Cq2/CH}
Extensions of the SM can also contain different coupling strengths of the Higgs
boson to leptons and quarks, notably some variants of Two-Higgs-Doublet Models.
In this benchmark model the ratio $\Rr_{\ell q}$ of coupling-strength scale factors to leptons and quarks is probed, while
vector boson coupling-strength scale factors are assumed to be unified and equal to $\Cc_{V}$. The indices
$\ell$, $q$ stand for all leptons and quarks, respectively.  The free
parameters are:
\begin{eqnarray*}
  \Rr_{\ell q} &=& \Cc_{\ell} / \Cc_{q} \label{llq}\\
  \Rr_{Vq} &=& \Cc_{V} / \Cc_{q} \label{kVq} \\
  \Cc_{qq} &=& \Cc_{q} \cdot \Cc_{q} / \Cc_{\PH} \label{kqq} .
\end{eqnarray*}
The lepton coupling strength is constrained through the \htt\ and \hmm\ decays.
The fit results for the parameters of interest of this benchmark model, when profiling the other parameters, are:
\begin{eqnarray*}
  \Rr_{\ell q} &\in& \RESULTmodelQLvalueRLQ~(68\%~ \rm{CL})\\
  \Rr_{Vq} &=& \RESULTmodelQLvalueRVQ \\
  \Cc_{qq} &=& \RESULTmodelQLvalueCQQ  .
\end{eqnarray*}
Near the SM prediction of $\Rr_{\ell q}=1$, the best-fit value is 
 $\Rr_{\ell q} = \RESULTmodelQLvalueRLQupper$.
All parameters are measured to be consistent
with their SM expectation and the three-dimensional compatibility of the SM
hypothesis with the best-fit point is $\RESULTmodelQLpvalueSM$.

Figure~\ref{fig:bm:Rlq} shows the likelihood curves corresponding to the
fit results for this benchmark.  Similar to the model of
Section~\ref{sec:Cd/Cu,CV/Cu,Cu2/CH}, the likelihood curve is nearly symmetric around
$\Rr_{\ell q}=0$.  A vanishing coupling strength of the Higgs boson to
leptons, i.e. $\Rr_{\ell q}=0$, is excluded at the $\RESULTmodelQLpvalueLepton$
level due to the \htt\ measurement.
The profile likelihood ratio values at $\Cc_{qq}=0$ and $\Rr_{Vq}=0$ provide strong
confirmation of Higgs boson couplings to quarks and vector bosons with both significances of
$>5\sigma$.

\begin{figure}[htbp!]
  \center

  \begin{minipage}{0.6\textwidth}
  \subfloat[]{
    \includegraphics[width=.99\textwidth]{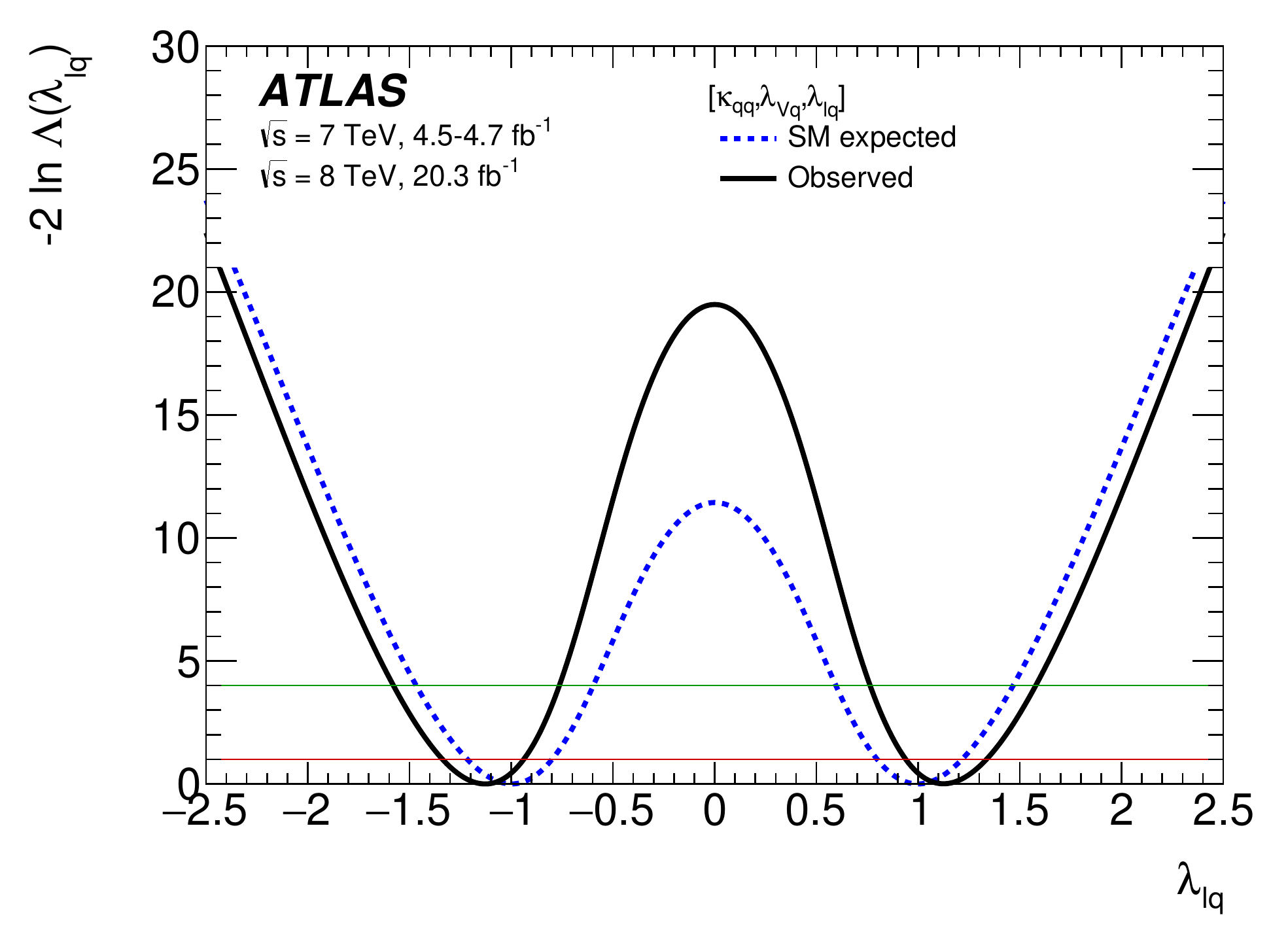}
    \label{fig:bm:RVqRlqRqqh:Rlq}
    \vspace*{-0.5cm}
  }
  \end{minipage}
  \begin{minipage}{0.39\textwidth}
  \subfloat[]{
    \includegraphics[width=.99\textwidth,trim=0 1cm 0 0.5cm]{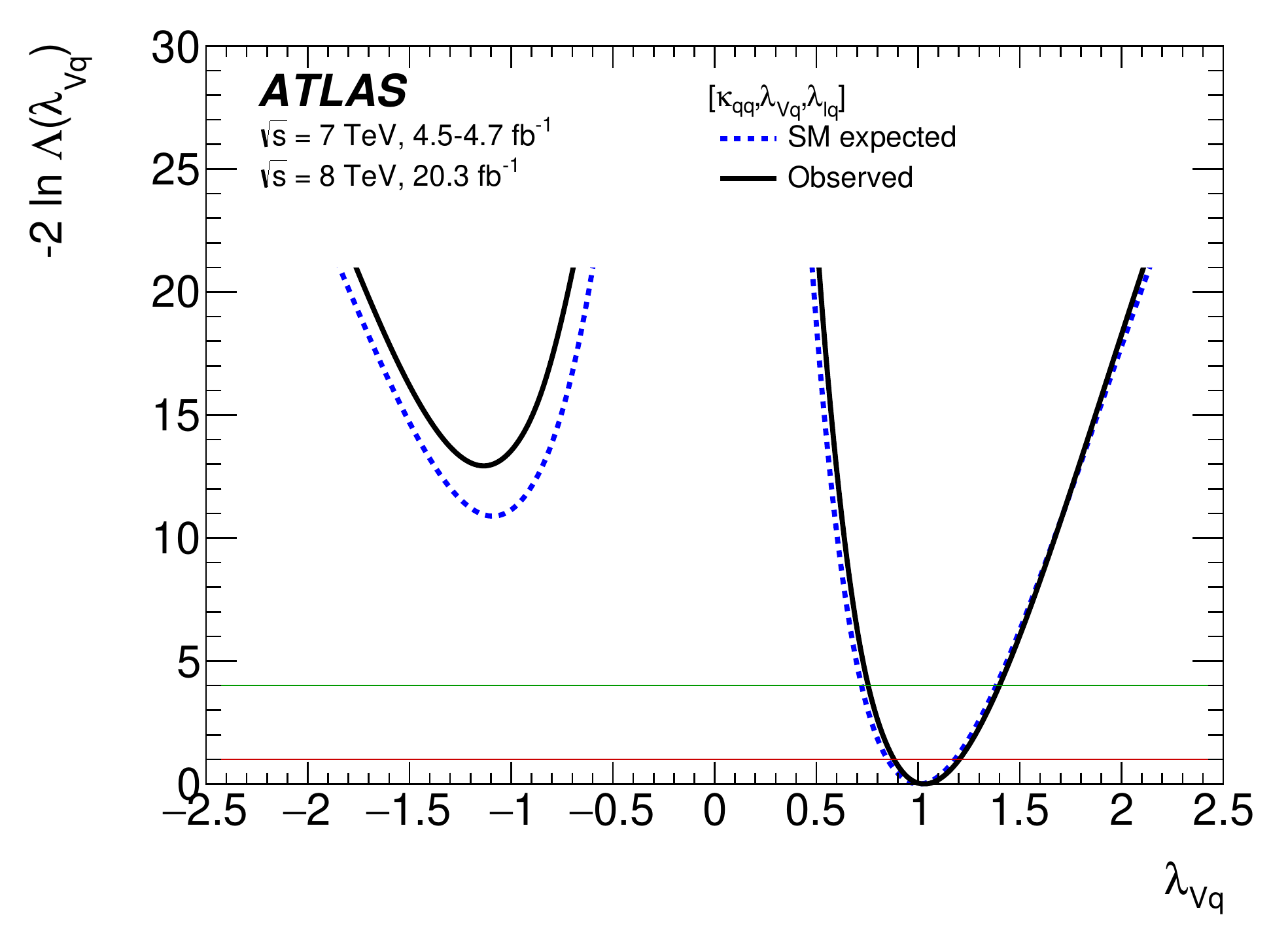}
    \label{fig:bm:RVqRlqRqqh:RVq}
    \vspace*{-0.5cm}
  }

  \subfloat[]{
    \includegraphics[width=.99\textwidth,trim=0 1cm 0 0.5cm]{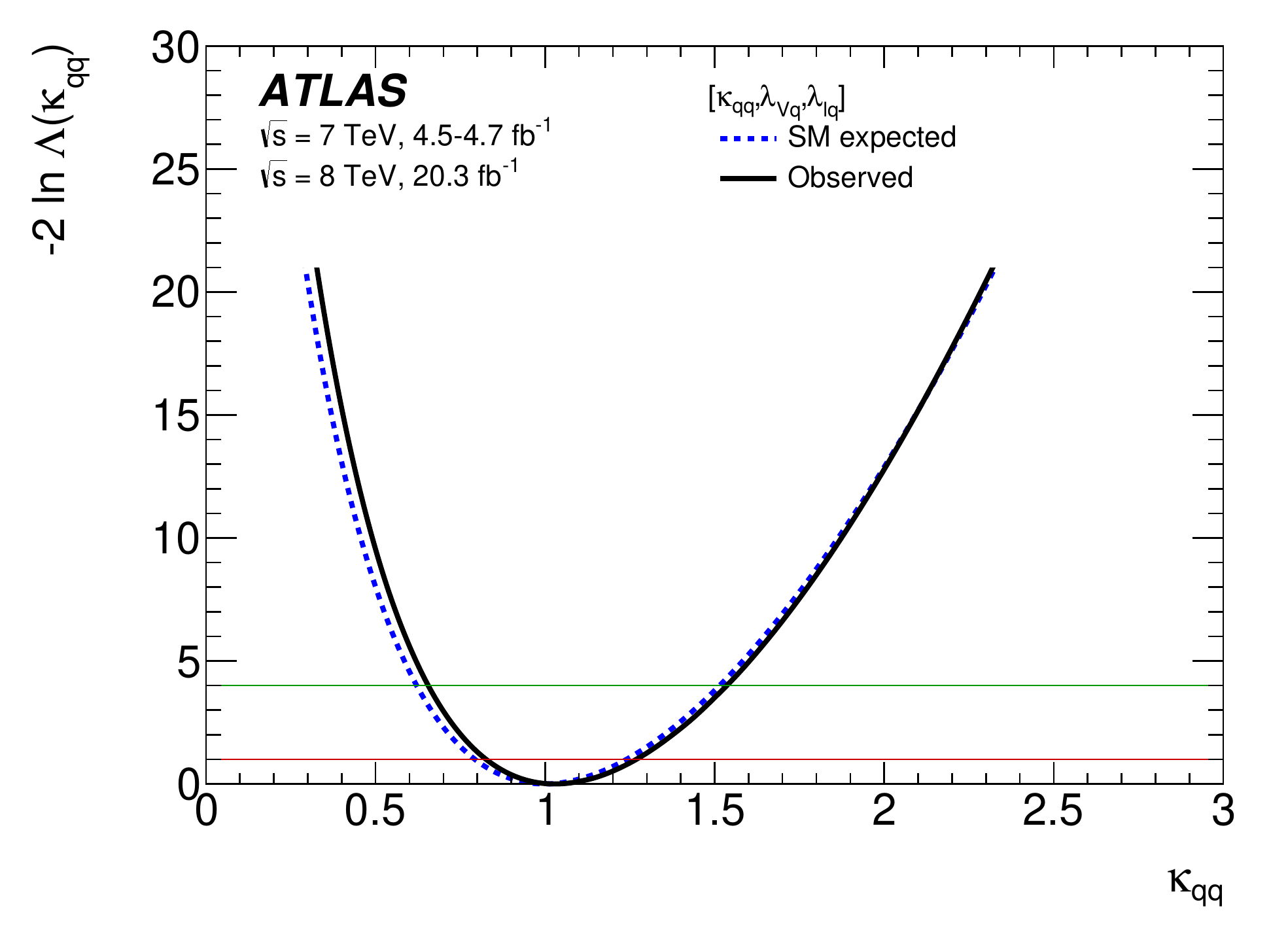}
    \label{fig:bm:RVqRlqRqqh:Rqqh}
    \vspace*{-0.5cm}
  }
  \end{minipage}

  \vspace*{-0.2cm}

  \caption{Results of fits for the benchmark model described in Section~\ref{sec:Cl/Cq,CV/Cq,Cq2/CH} that probes the symmetry between quarks and leptons:
    profile likelihood ratios as functions of the coupling-strength scale factor ratios
    (a)~$\Rr_{\ell q}$ ($\Rr_{Vq}$ and $\Cc_{qq}$ are profiled),
    (b)~$\Rr_{Vq}$ ($\Rr_{\ell q}$ and $\Cc_{qq}$ are profiled),
    and (c)~the overall scale factor $\Cc_{qq}$ ($\Rr_{\ell q}$ and $\Rr_{Vq}$ are profiled). The dashed curves show the SM expectations.
    The red (green) horizontal line indicates the value of the profile likelihood ratio corresponding to a 68\% (95\%) confidence interval for the parameter of interest, assuming the asymptotic $\chi^2$ distribution for the test statistic.
    }
  \label{fig:bm:Rlq}
\end{figure}

\subsection{Probing beyond the SM contributions in loops and decays}
\label{sec:bsm}
In this section, contributions from new particles either in loops or in
new final states are probed. 
For the {\hgg}, {\hzg} and {\ggF} vertices, effective scale factors
$\Cc_{\PGg}$, $\Cc_{\PZ\PGg}$ and $\Cc_{\Pg}$ are introduced that
allow for extra contributions from new particles. These effective
scale factors are defined to be positive as there is by construction
no sensitivity to the sign of these coupling strengths.  The potential
new particles contributing to these vertex loops may or may not
contribute to the total width of the observed state through direct
invisible or undetected decays. In the latter case the total width is
parameterised in terms of the additional branching ratio \BRinv into
invisible or undetected particles.

\subsubsection{Probing BSM contributions in loop vertices only} 
\label{sec:Cg,Cam}
In the first benchmark model of this section, BSM contributions can modify the loop
coupling strengths from their SM prediction, but it is assumed that
there are no extra contributions to the total width caused by non-SM
particles. Furthermore, all coupling-strength scale factors of known
SM particles are assumed to be as predicted by the SM,
i.e.\ $\Cc_{\PW} = \Cc_{\PZ} = \Cc_{\PQt} = \Cc_{\PQb} = \Cc_{\PGt} =
 \Cc_{\PGm} = 1$.  The free parameters  are thus
$\Cc_{\Pg}$, $\Cc_{\PGg}$ and $\Cc_{\PZ\PGg}$.

Figure~\ref{fig:bm:Cg,Cgamma:Cgamma,Cg:barchart} shows the results of
fits for this benchmark scenario and the best-fit values and
uncertainties, when profiling the other parameters. The effective
coupling-strength scale factors $\Cc_{\Pg}$ and $\Cc_{\PGg}$ are measured to be
consistent with the SM expectation, whereas a limit is set on the
effective coupling-strength scale factor
$\Cc_{\PZ\PGg}$. Figure~\ref{fig:bm:Cg,Cgamma:Cgamma,Cg} shows the
two-dimensional likelihood contour for $\Cc_{\Pg}$ vs. $\Cc_{\PGg}$,
where $\Cc_{\PZ\PGg}$ is profiled.  The three-dimensional
compatibility of the SM hypothesis with the best-fit point is
$\RESULTmodelELOOPpvalueSM$.

\begin{figure}[htbp!]
  \center

  \subfloat[]{
    \includegraphics[width=.80\textwidth]{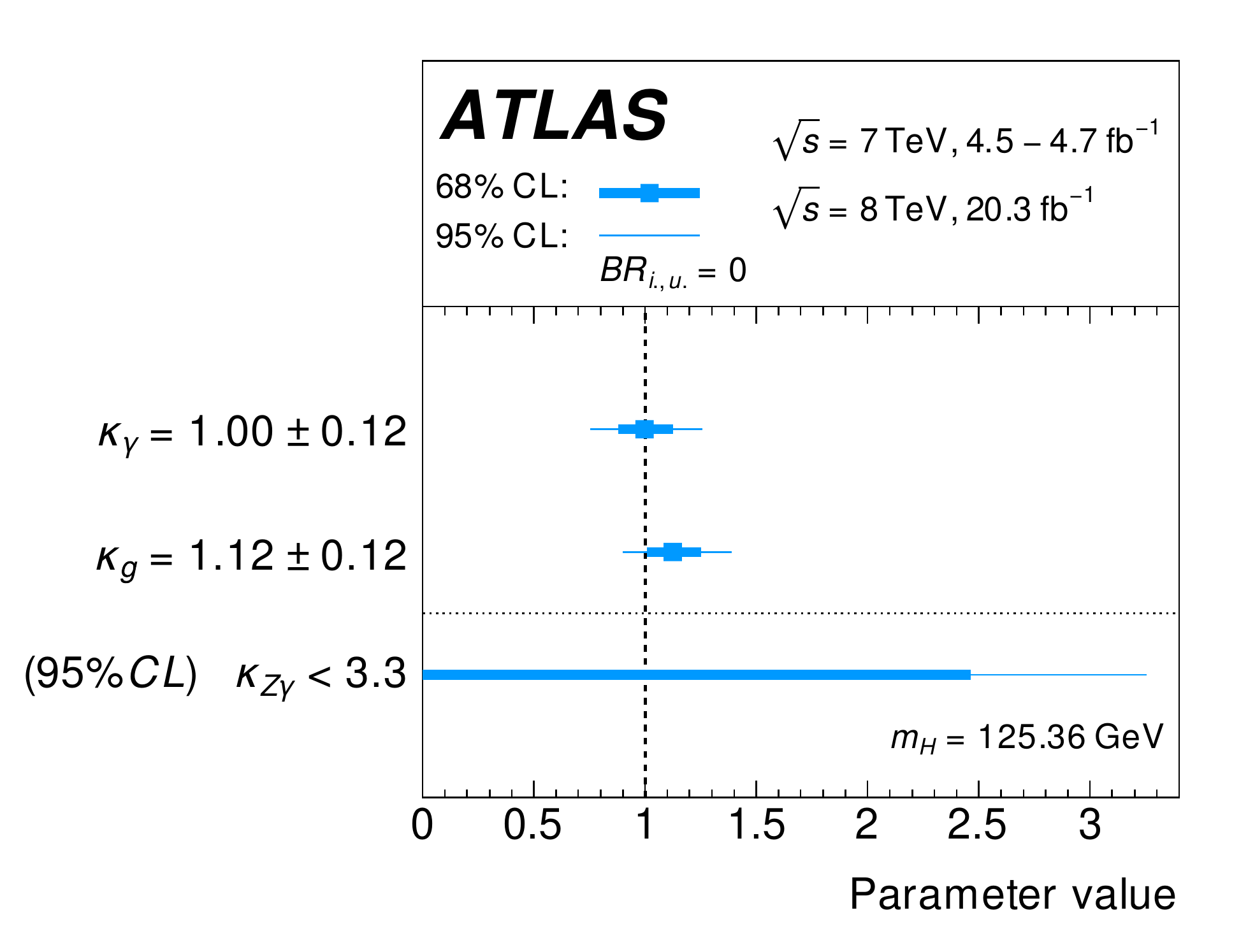}
    \label{fig:bm:Cg,Cgamma:Cgamma,Cg:barchart}
    \vspace*{-0.5cm}
  }

  \subfloat[]{
    \hspace{0.05\textwidth}
    \includegraphics[width=.70\textwidth]{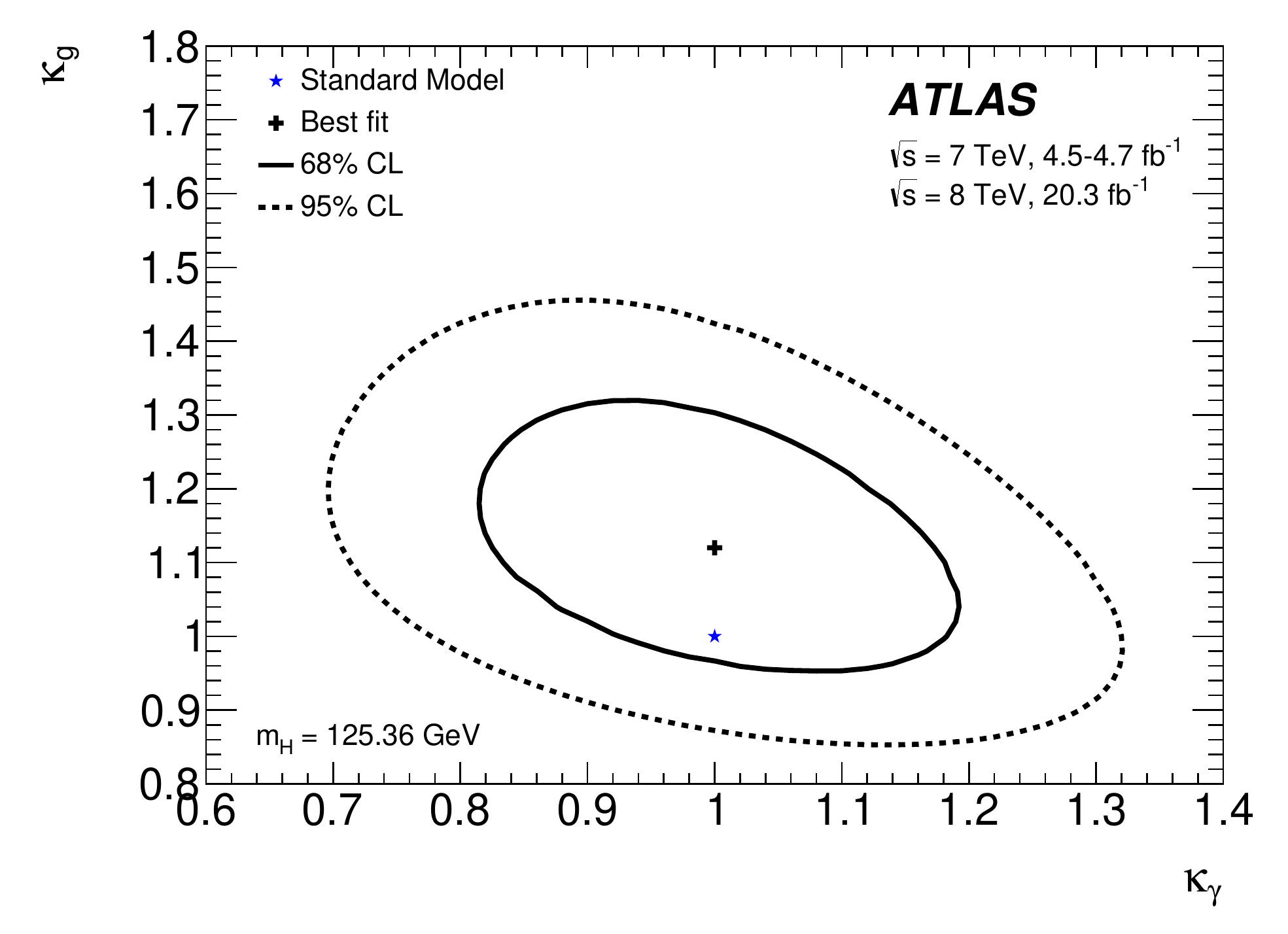}
    \label{fig:bm:Cg,Cgamma:Cgamma,Cg}
    \vspace*{-0.5cm}
  }

  \caption{Results of fits for the benchmark model that probes for
    contributions from non-SM particles in the \hgg, \hzg\ and \ggF\ loops,
    assuming no extra contributions to the total width:
    (a)~overview of fitted parameters, where the inner and outer bars correspond to 68\%~CL and 95\%~CL intervals,
    and (b)~results of the two-dimensional fit to $\Cc_{\PGg}$ and $\Cc_{\Pg}$, including $68\%$ and $95\%$ CL contours
    ($\Cc_{\PZ\PGg}$ is profiled).
    \label{fig:bm:Cg,Cgamma}
  }
\end{figure}
\begin{figure}[htbp!]

  \center
  \subfloat[]{
    \includegraphics[width=.70\textwidth]{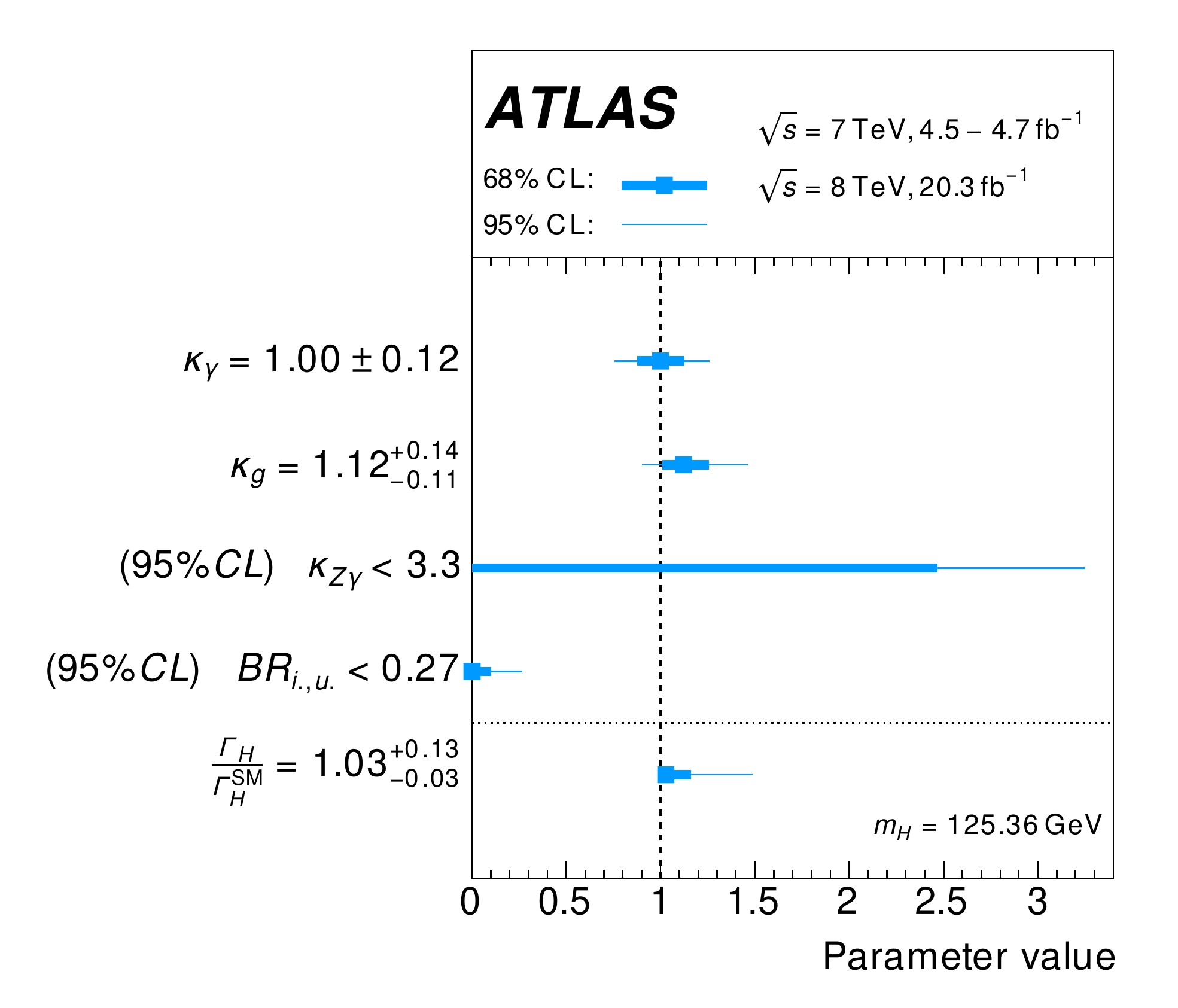}
    \label{fig:bm:Cg,Cgamma,BRinv:BRinv:barchart}
    \vspace*{-0.5cm}
  }%

  \subfloat[]{
    \hspace{0.20\textwidth}
    \includegraphics[width=.55\textwidth]{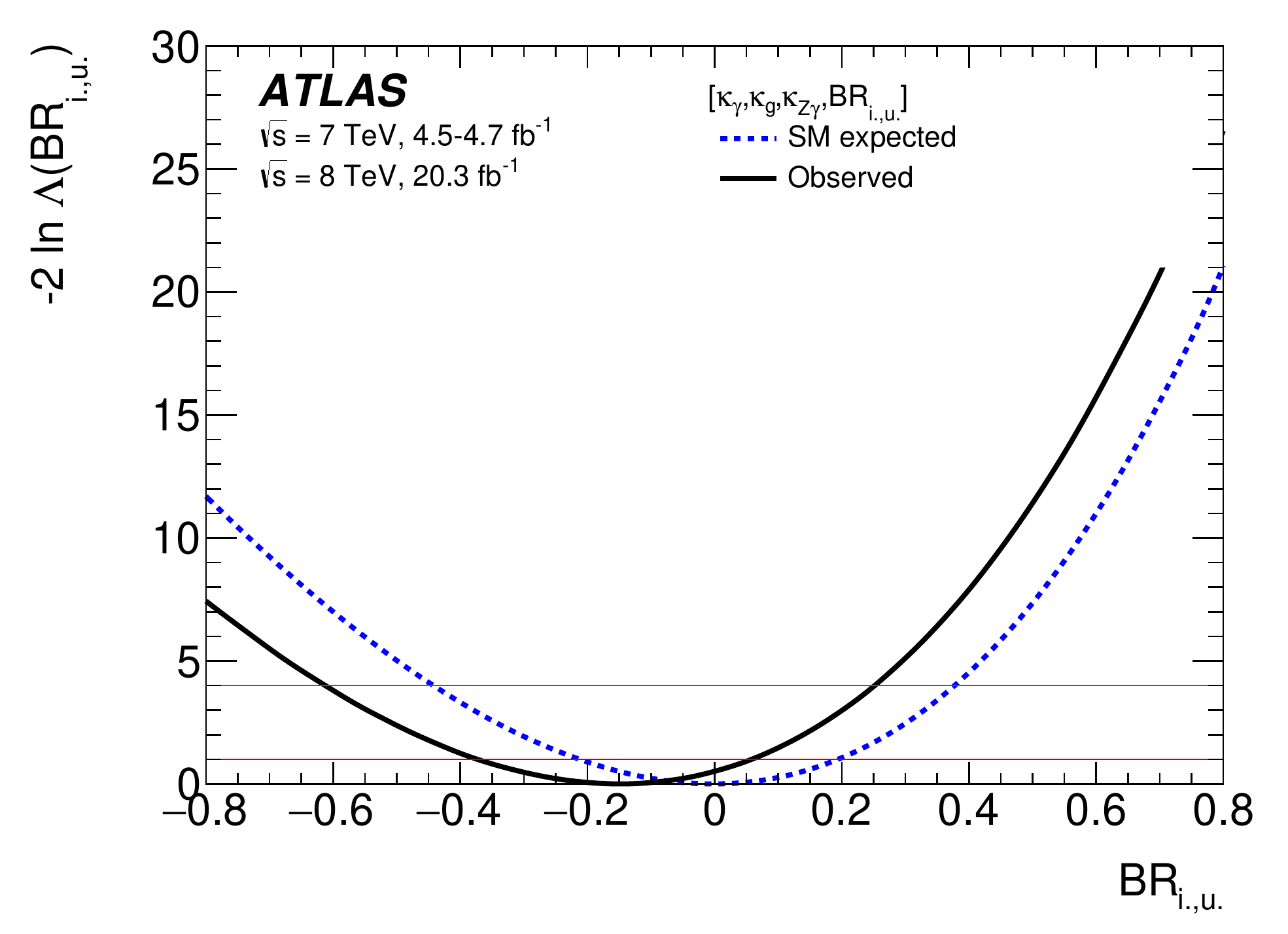}
    \hspace{0.05\textwidth}
    \label{fig:bm:Cg,Cgamma,BRinv:BRinv}
    \vspace*{-0.5cm}
  }%

  \caption{Results of fits for benchmark models that probe for
    contributions from non-SM particles in the \hgg, \hzg\ and \ggF\ loops,
    while allowing for potential extra contributions to the total
    width: (a)~overview of fitted parameters. The inner and outer bars correspond to 68\%~CL and 95\%~CL intervals.
    The confidence intervals for $\BRinv$ are estimated with respect to the physical bounds as described in the text.
    (b)~Profile likelihood ratio as a function of the branching fraction $\BRinv$ to invisible or undetected
    decay modes ($\Cc_{\PGg}$, $\Cc_{\Pg}$ and $\Cc_{\PZ\PGg}$ are profiled).
    The red (green) horizontal line indicates the value of the profile likelihood ratio corresponding to a 68\% (95\%) confidence interval for the parameter of interest, assuming the asymptotic $\chi^2$ distribution for the test statistic.
      \label{fig:bm:Cg,Cgamma,BRinv}
    }

\end{figure}

\subsubsection{Probing BSM contributions in loop vertices and to the total width}
\label{sec:Cg,Cam,BRinv}

The second benchmark model of this section removes the assumption of
no invisible or undetected Higgs boson decays, introducing \BRinv as
additional model parameter. The free parameters of this benchmark model are thus
$\Cc_{\Pg}$, $\Cc_{\PGg}$, $\Cc_{\PZ\PGg}$ and $\BRinv$.
The coupling-strength scale factors of
known SM particles are still assumed to be at their SM values of
1. Due to this assumption, the
parameterisation of Higgs boson channels that do not involve a
loop process, e.g. VBF production of \hww\ and associated production of \hbb, depends only on \BRinv in
this model, and not on $\Cc_{\Pg}$, $\Cc_{\PGg}$ or $\Cc_{\PZ\PGg}$, and can hence constrain \BRinv from the data.
Thus no additional constraints, beyond those introduced in the
benchmark model of Section~\ref{sec:CV,CF,BRinv}, are necessary in
this model.  

The results of fits to this benchmark model are shown in
Fig.~\ref{fig:bm:Cg,Cgamma,BRinv}, along with the uncertainty on the
total width that this model allows, obtained in the same fashion as
for the previous benchmark models.  The effective coupling-strength scale factors
$\Cc_{\Pg}$ and $\Cc_{\PGg}$ are measured to be consistent with the SM
expectation, whereas limits are set on the effective coupling-strength scale factor
$\Cc_{\PZ\PGg}$ and the branching fraction \BRinv.  By using the physical constraint 
$\BRinv>0$, the observed $95\%$~CL upper limit is $\BRinv<\RESULTmodelELOOPBRUIlimitFC$ 
compared with the expected limit of $\BRinv <\RESULTmodelELOOPBRUIlimitFCSM$ under the 
SM hypothesis. The four-dimensional compatibility of
the SM hypothesis with the best-fit point is
$\RESULTmodelELOOPBRUIpvalueSM$.  
The best-fit values of the
model parameters of interest and their uncertainties, when profiling
the other parameters, are

\begin{eqnarray*}
  \Cc_{\Pg} &=& \RESULTmodelELOOPBRUIvalueCGL\\
  \Cc_{\PGg} &=& \RESULTmodelELOOPBRUIvalueCGA\\
\end{eqnarray*}

In a variant of the fit where no limits are imposed on \BRinv its best-fit value is

\begin{eqnarray*}
 \BRinv &=& \RESULTmodelELOOPBRUIvalueBRINV\xspace ,
\end{eqnarray*}

corresponding to the likelihood curve shown in
Fig.~\ref{fig:bm:Cg,Cgamma,BRinv:BRinv}. Without the condition
$\BRinv\ge 0$, the best-fit value of \BRinv\ assumes a small
(unphysical) negative value that is consistent with zero within the
uncertainty. 

As the choice of free parameters in this model gives extra degrees of
freedom to  ggF production and \hgg\ and \hzg\ decays, the most
precise measurements based on ggF production or \hgg\ decays
(see Fig.~\ref{fig:muChannels}) do not give a sizeable contribution to
the determination of \BRinv.  Instead \BRinv\ is mostly constrained
by channels sensitive to VBF and $VH$ production, as the tree-level
couplings involved in these production modes are fixed to their SM
values within this model. The upward uncertainty on
$\Gamma_{\PH}/\Gamma_{\PH}^{\rm SM}$ is notably increased with respect to that
of the model in Section~\ref{sec:Cg,Cam} due to the removing the constraint
on \BRinv, whereas the downward uncertainty is identical due to the
 condition that $\BRinv \ge 0$.

\subsubsection{Probing BSM contributions in loop vertices and to the total width allowing modified couplings to SM particles}
\label{sec:CV,CF,Cg,Cam,BRinv}
The last benchmark model of this section removes the assumption of SM
couplings of the Higgs boson for non-loop vertices used so far in this
section, re-introducing the coupling-strength scale factors $\Cc_{F}$
and $\Cc_{V}$ defined in Section~\ref{sec:bm:CF,CV} to allow
deviations of the coupling strength of the Higgs boson to fermions and
gauge bosons, respectively.  As the expression for $\Cc_{\PH}$ is no
longer strongly constrained due to the newly introduced degrees of
freedom, the upper limit on $\Gamma_{\PH}$ is no longer bounded, and
an additional constraint on the total Higgs boson width must be
introduced. Similar to the model of Section~\ref{sec:CV,CF,BRinv} the
two choices of the constraints on the total width discussed in
Section~\ref{sec:framework} are studied: $\Cc_{V}<1$ and 
$\Cc_{\rm on}=\Cc_{\rm off}$. The free parameters of this model are $\Cc_F$,
$\Cc_V$, $\Cc_{\Pg}$, $\Cc_{\PGg}$, $\Cc_{\PZ\PGg}$ and \BRinv.

\begin{figure}[htbp!]
  \center
  \includegraphics[width=.85\textwidth]{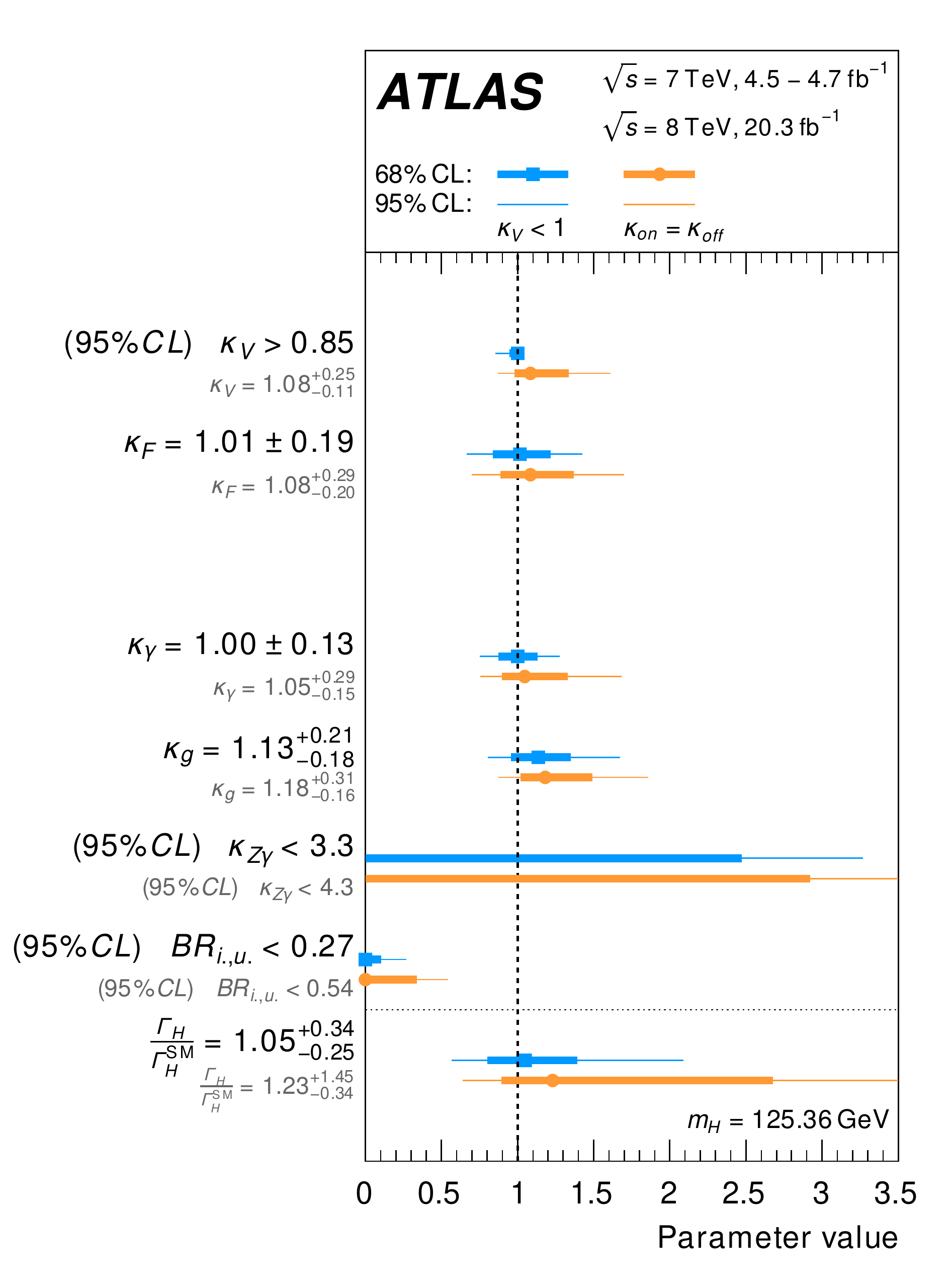}
  \vspace*{-0.3cm}

  \caption{Results of fits for benchmark models that probe for
    contributions from non-SM particles in the \hgg, \ggF\ and \hzg\ loops,
    with free gauge and fermion coupling-strength scale factors $\Cc_V,\Cc_F$,
    while allowing for potential extra contributions to the total
    width. The estimated values of each parameter under the constraint $\Cc_{V}<1$ or 
    $\Cc_{\rm on}=\Cc_{\rm off}$ are shown with markers in the shape of a box or a circle, 
    respectively. The inner and outer bars correspond to 68\%~CL and 95\%~CL intervals.
    The confidence intervals of $\BRinv$ and, in the benchmark model with the constraint 
    $\Cc_{V}<1$, also $\Cc_{V}$, are estimated with respect to their physical constraints 
    as described in the text. The numerical values of the fit under the constraint $\Cc_{V}<1$ 
    are shown on the left. Values for the alternative $\Cc_{\rm on}=\Cc_{\rm off}$ constraint are 
    also shown (in a reduced font size due to space constraints).
    \label{fig:bm:ELOOPBRUICVCF}
  }

\end{figure}

Figure~\ref{fig:bm:ELOOPBRUICVCF} shows the best-fit values and their
uncertainties. The coupling-strength scale factors $\Cc_{\Pg}$, $\Cc_{\PGg}$,
$\Cc_{V}$ and $\Cc_{F}$ are measured to be consistent with their SM
expectation, while limits are set on the coupling-strength scale factor
$\Cc_{\PZ\PGg}$ and the branching fraction \BRinv to invisible or
undetected decays. By using the physical constraint $\BRinv\ge 0$, the
$95\%$ CL upper limit is $\BRinv <
\RESULTmodelELOOPBRUICVCFAlimitFC$ ($\RESULTmodelELOOPBRUICVCFBlimitFC$)
when applying the constraint $\Cc_V<1$ ($\Cc_{\rm on}=\Cc_{\rm off}$).
The expected limit in case of the SM hypothesis is $\BRinv <
\RESULTmodelELOOPBRUICVCFAlimitFCSM$ ($\RESULTmodelELOOPBRUICVCFBlimitFCSM$).
The six-dimensional compatibility of the SM hypothesis with the
best-fit point is $\RESULTmodelELOOPBRUICVCFApvalueSM$ $(\RESULTmodelELOOPBRUICVCFBpvalueSM)$ when applying the
$\Cc_V<1$ ($\Cc_{\rm on}=\Cc_{\rm off}$) constraint, respectively.
The uncertainty on $\Gamma_{\PH}/\Gamma_{\PH}^{\rm SM}$ is significantly 
increased compared with models in Sections~\ref{sec:Cg,Cam} and~\ref{sec:Cg,Cam,BRinv} 
due to the further relaxed coupling constraints, in particular both the 68\% and 95\%~CL
intervals of $\Gamma_{\PH}/\Gamma_{\PH}^{\rm SM}$ extend below~1.

\subsection{Generic models}
\label{sec:gen}
In the benchmark models studied in Sections \ref{sec:CFCV},
\ref{sec:fermion} and \ref{sec:bsm}, specific aspects of the Higgs
sector are tested by combining coupling-strength scale factors into 
a minimum number of parameters under certain assumptions, thereby
maximising the sensitivity to the scenarios under study. In generic
models the scale factors for the coupling strengths to $W$, $Z$, $t$,
$b$, $\tau$ and $\mu$ are treated independently, while for the loop
vertices and the total width $\Gamma_{\PH}$, either the SM particle
content is assumed (Section~\ref{sec:gen1}) or no such assumption is
made (Sections~\ref{sec:gen2} and \ref{sec:gen3}).

\subsubsection{Generic model 1: no new particles in loops and in decays}
\label{sec:gen1}

In the first generic benchmark model all coupling-strength scale factors to SM
particles, relevant to the measured modes, are fitted
independently. The free parameters are: $\Cc_{\PW}$, $\Cc_{\PZ}$,
$\Cc_{\PQt}$, $\Cc_{\PQb}$, $\Cc_{\PGt}$, and $\Cc_{\PGm}$. It is
assumed that only SM particles contribute to Higgs boson vertices
involving loops, and modifications of the coupling-strength scale factors
for fermions and vector bosons are propagated through the loop
calculations. No invisible or undetected Higgs boson decays are assumed to
exist.  
Only the $W$ coupling-strength scale factor is assumed to be
positive without loss of generality: due to interference terms,
the fit is sensitive to the relative sign of the $W$ and $t$
couplings (through the $tH$, {\hgg}, {\hzg} processes) and the
relative sign of the $Z$ and $t$ coupling (through the {\ggZH}
process), providing indirect sensitivity to the relative sign of
the $W$ and $Z$ coupling.  Furthermore, the model has some
sensitivity to the relative sign of the $t$ and $b$ coupling
(through the ggF process).

\begin{figure}[hbt!]
  \center
    \includegraphics[width=0.9\textwidth]{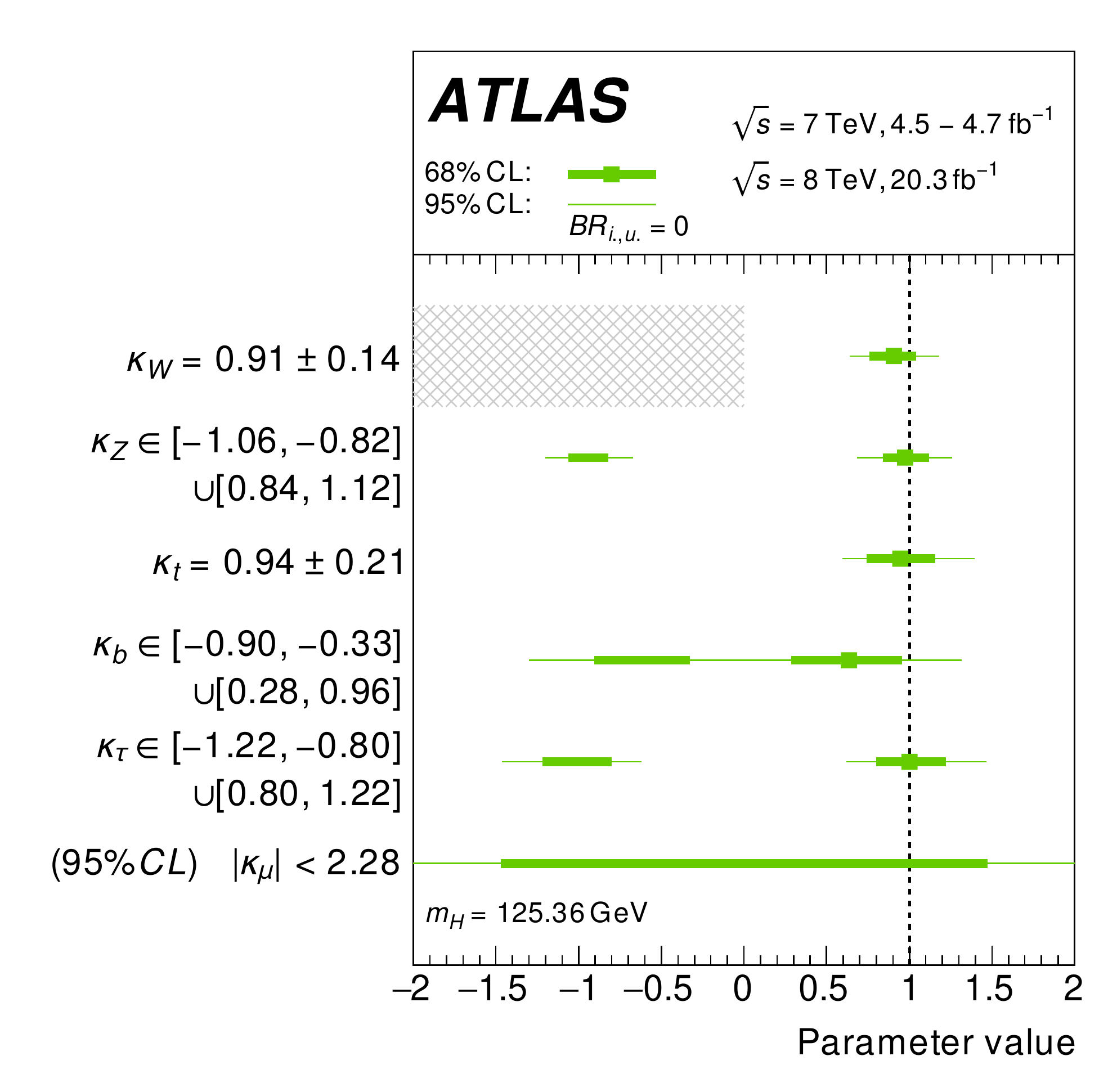}
  \caption{Overview of best-fit values of parameters with 68\% and 95\% CL intervals for generic model 1 (see text).
           In this model only SM particles are considered in loops and no invisible or undetected Higgs boson decays are allowed.
           The sign of $\Cc_{\PW}$ is assumed to be positive, as indicated by the hatched area, without loss of generality.
           The inner and outer bars correspond to 68\%~CL and 95\%~CL intervals.
 }
  \label{fig:bm:GenMod1:par}
\end{figure}

Figure~\ref{fig:bm:GenMod1:par} summarises the results of the fits for
this benchmark scenario. All measured coupling-strength scale factors in this
generic model are found to be compatible with their SM expectation,
and the six-dimensional compatibility of the SM hypothesis with the
best-fit point is $\RESULTmodelGENIpvalueSM$.  Illustrative likelihoods of the measurements
summarised in Fig.~\ref{fig:bm:GenMod1:par} are shown in Fig.~\ref{fig:bm:GenMod1:Cx}. As shown in
Figs. \ref{fig:bm:GenMod1:Cto} and \ref{fig:bm:GenMod1:Cb}, the
negative solution of $\Cc_{\PQt}$ is strongly disfavoured at
$3.1\sigma$ ($2.9\sigma$ expected), while the negative minimum of
$\Cc_{\PQb}$ is slightly disfavoured at $0.5\sigma$ (no sensitivity
expected).

\begin{figure}[hbt!]
  \center
  \subfloat[]{
    \includegraphics[width=.45\textwidth]{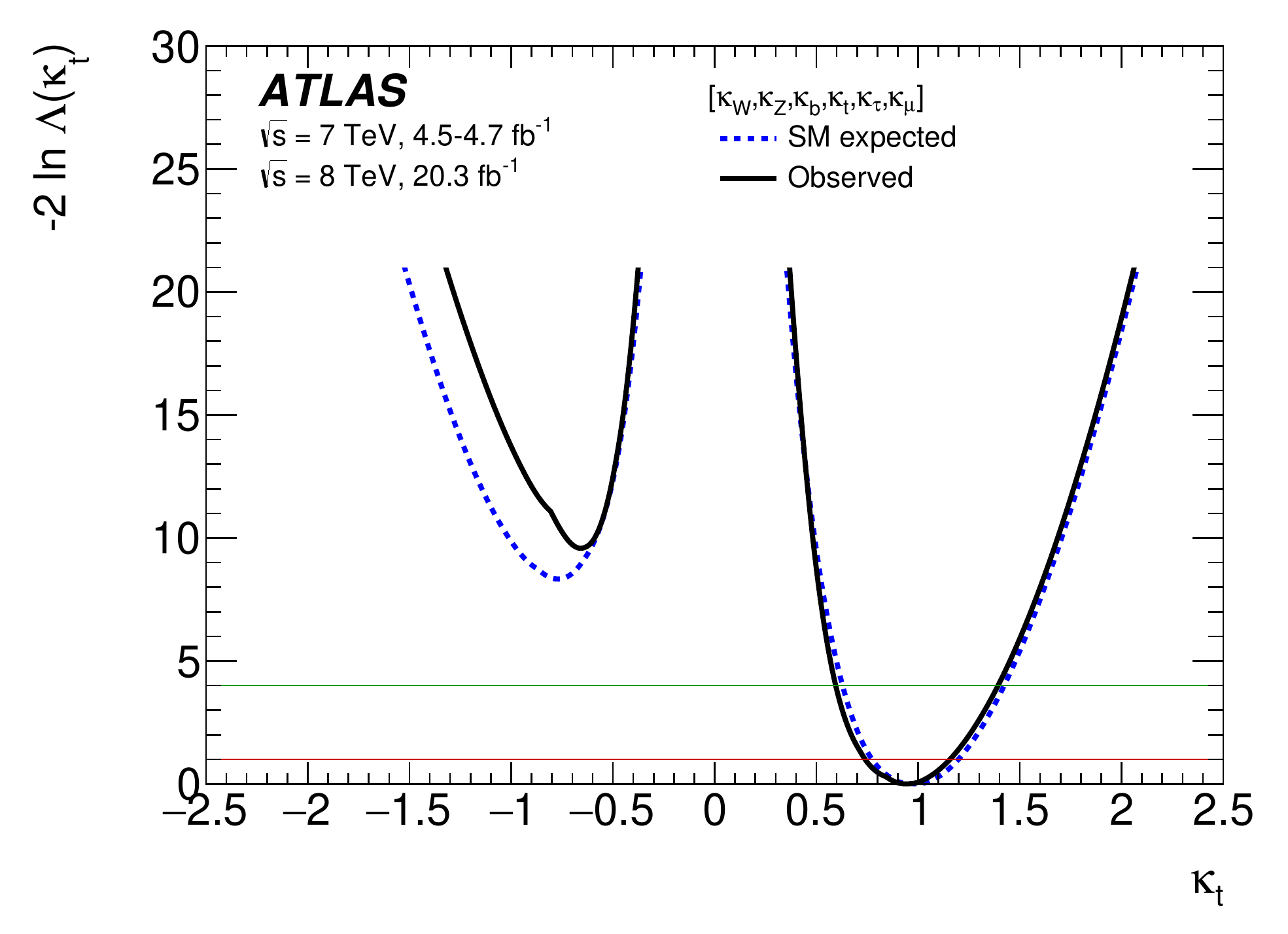}
    \label{fig:bm:GenMod1:Cto}
    \vspace*{-0.5cm}
  }
  \subfloat[]{
     \includegraphics[width=.45\textwidth]{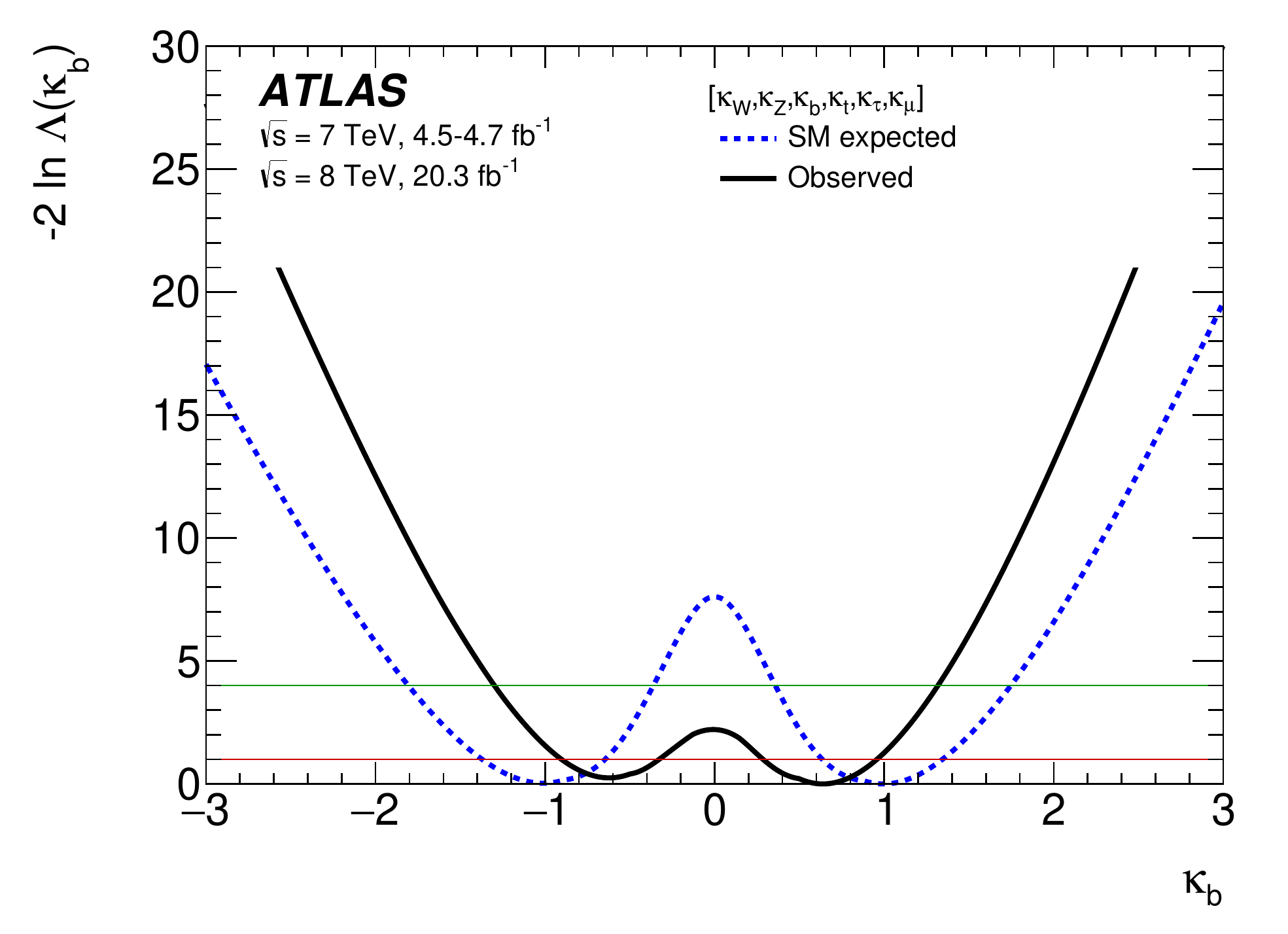}
    \label{fig:bm:GenMod1:Cb}
    \vspace*{-0.5cm}
  }

  \subfloat[]{
    \includegraphics[width=.45\textwidth]{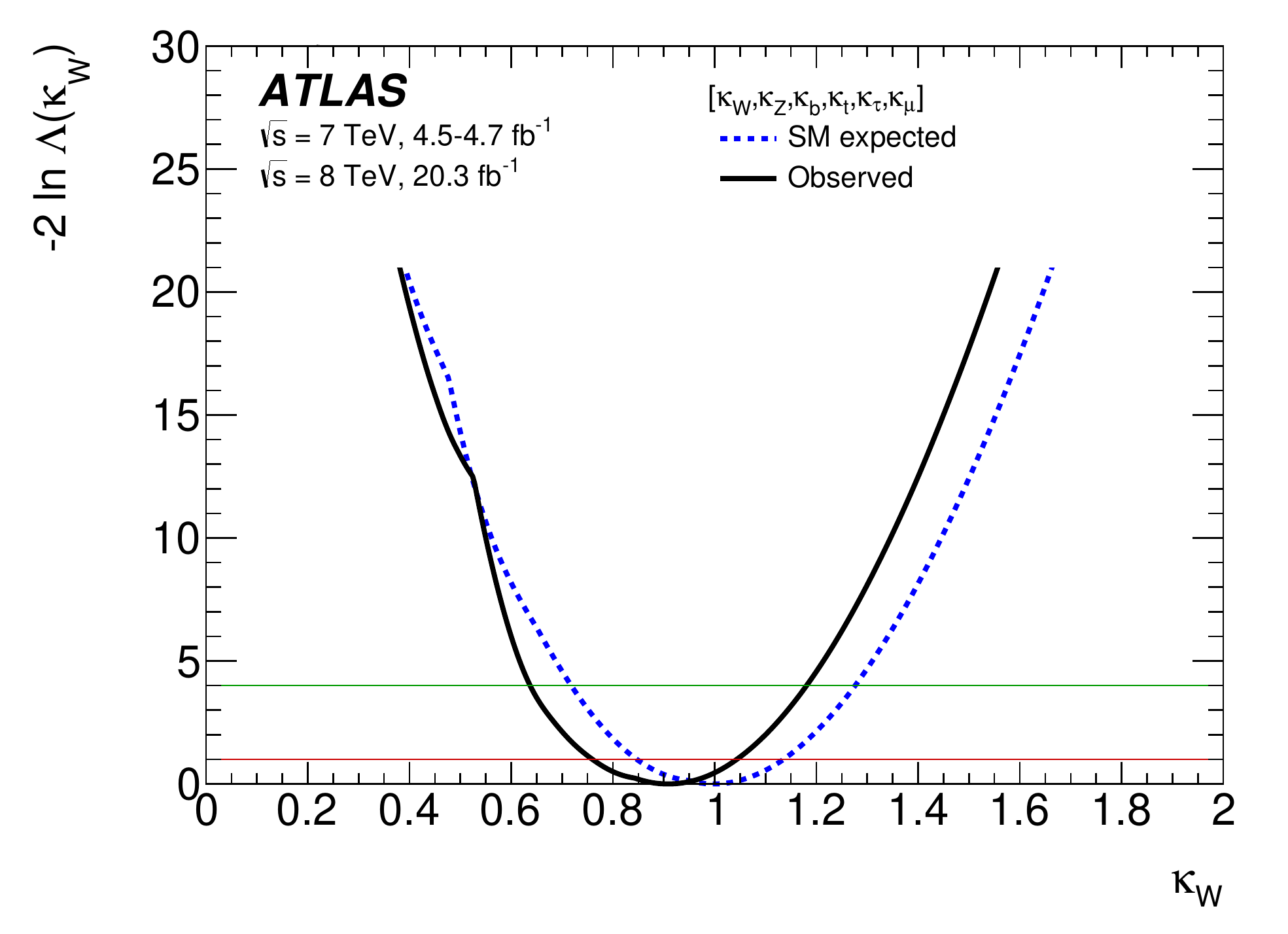}
    \label{fig:bm:GenMod1:CW}
    \vspace*{-0.5cm}
  }
  \subfloat[]{
     \includegraphics[width=.45\textwidth]{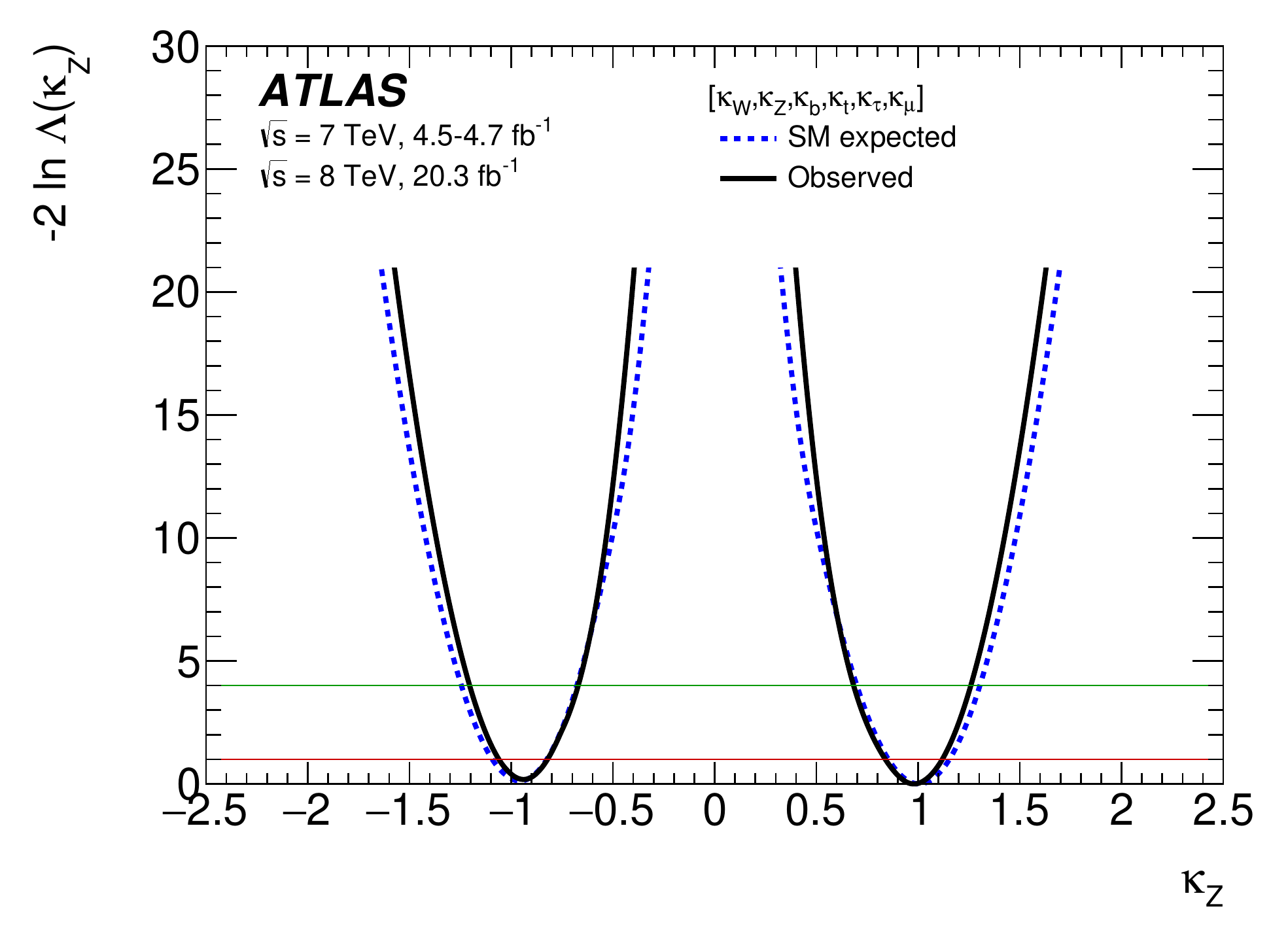}
    \label{fig:bm:GenMod1:CZ}
    \vspace*{-0.5cm}
  }

  \caption{Results of fits for generic model 1 (see text): 
    profile likelihood ratios as functions of the coupling-strength scale factors
    (a)~$\Cc_{\PQt}$,
    (b)~$\Cc_{\PQb}$,
    (c)~$\Cc_{\PW}$,
    and (d)~$\Cc_{\PZ}$. For each measurement, the other coupling-strength scale factors are profiled.
    The kinks in the curves of (a) and (c) are caused by transitions in solutions chosen by the profile likelihood for the relative sign between profiled couplings.
    The dashed curves show the SM expectations.
    The red (green) horizontal line indicates the value of the profile likelihood ratio corresponding to a 68\% (95\%) confidence interval for the parameter of interest, assuming the asymptotic $\chi^2$ distribution for the test statistic.
    }
  \label{fig:bm:GenMod1:Cx}
\end{figure}

For the measurements in this generic model, it should be noted that the
low fitted value of $\Cc_{\PQb}$ causes a reduction of the total width
$\Gamma_{\PH}$ by about 30\% compared to the SM expectation (see Table
\ref{tab:kexpr}), which in turn induces a reduction of all other
$\Cc$-values by about 20\%.

\begin{figure}[hbt!]
  \center
  \includegraphics[width=0.65\textwidth]{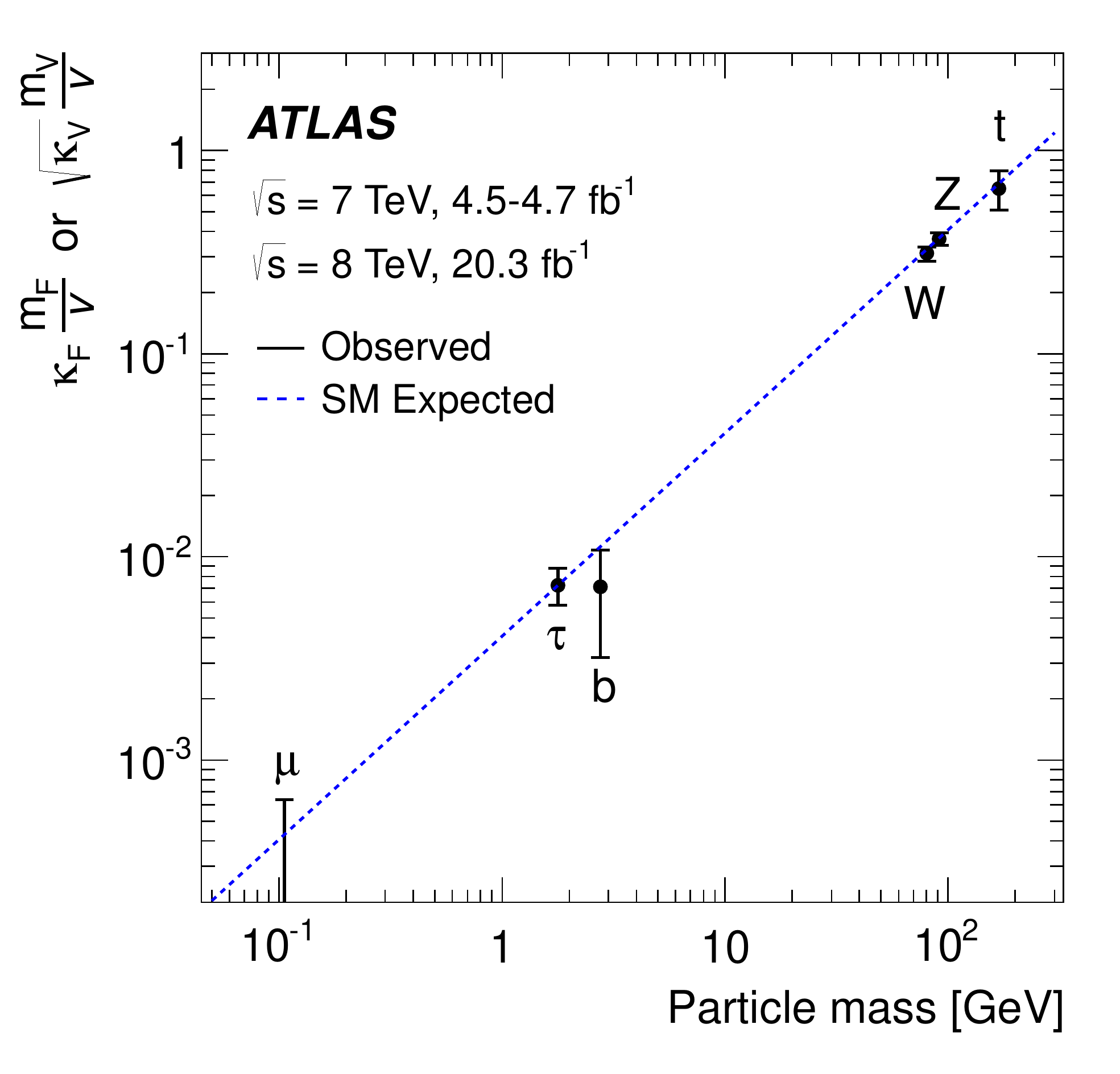}
  \caption{Fit results for the reduced coupling-strength scale factors $y_{V,i} = \sqrt{ \Cc_{V,i} \frac{g_{V,i}}{2v} } = \sqrt{\Cc_{V,i}}\frac{m_{V,i}}{v}$ for weak bosons and
          $y_{F,i} = \Cc_{F,i}\frac{g_{F,i}}{\sqrt{2}} = \Cc_{F,i}\frac{m_{F,i}}{v}$ for fermions as a
              function of the particle mass, assuming a SM Higgs boson with a mass of 125.36 GeV.
              The dashed line indicates the predicted mass dependence for the SM Higgs boson. }
            \label{fig:bm:GenMod1:prplot}
\end{figure}

Figure~\ref{fig:bm:GenMod1:prplot} shows the results of the fit for generic model 1 as reduced coupling-strength scale factors
\begin{equation}
  \label{eq:1}
  y_{V,i} = \sqrt{ \Cc_{V,i} \frac{g_{V,i}}{2v} } = \sqrt{\Cc_{V,i}}\frac{m_{V,i}}{v}
\end{equation}
for weak bosons with a mass $m_{V}$, where $g_{V,i}$ is the absolute Higgs boson coupling strength, $v$ is the vacuum expectation value of the Higgs field and
\begin{equation}
  \label{eq:2}
  y_{F,i} = \Cc_{F,i}\frac{g_{F,i}}{\sqrt{2}} = \Cc_{F,i}\frac{m_{F,i}}{v}
\end{equation}
for fermions as a function of the particle mass $m_{F}$, assuming a SM
Higgs boson with a mass of 125.36~GeV. For the $b$-quark mass in
Fig.~\ref{fig:bm:GenMod1:prplot} the $\overline{MS}$ running mass
evaluated at a scale of 125.36 GeV is assumed. 

\subsubsection{Generic model 2: allow new particles in loops and in decay}
\label{sec:gen2}
In the second generic benchmark model the six free parameters from
the first generic model are retained but the assumptions on the
absence of BSM contributions in loops and to the total width are
dropped. Effective coupling-strength scale factors for loop vertices
are introduced, and optionally a branching ratio \BRinv\ to new non-SM
decays that might yield invisible or undetected final states is
introduced, resulting in a total of 9~(10) free parameters. In the
variant where \BRinv\ is not fixed to zero, either the constraint
$\Cc_{V}<1$ is imposed, or the constraint on the total width from
off-shell measurements is included.

Figure~\ref{fig:bm:GenMod2:overview} summarises the results of the
fits for this benchmark scenario. The numerical results are shown in
Table~\ref{tab:bm:GenMod2:overview}. As an illustration of contributions
from different sources, the uncertainty components are shown for 
the case of $\BRinv=0$. All fundamental coupling-strength
scale factors, as well as the loop-coupling scale factors $\Cc_{\Pg}$
and $\Cc_{\PGg}$ are measured to be compatible with their SM
expectation under all explored assumptions, while limits are set on
the loop-coupling scale factor $\Cc_{\PZ\PGg}$ and the fraction of
Higgs boson decays to invisible or undetected decays.  When imposing the
physical constraint $\BRinv\ge 0$ in the inference on $\BRinv$, the
$95\%$ CL upper limit is $\BRinv < \RESULTmodelGENIIBRuiA$
($\RESULTmodelGENIIBRuiB$) under the constraint
$\Cc_{V}<1$ ($\Cc_{\rm on}=\Cc_{\rm off}$) on the Higgs boson total width.
The nine-dimensional compatibility of the SM hypothesis with the
best-fit point is $\RESULTmodelGENIIpvalueSMnoBRui$ when \BRinv is
fixed to zero.  The compatibilities for the fits with the conditions
$\Cc_{V}<1$ and $\Cc_{\rm on}=\Cc_{\rm off}$ imposed are
$\RESULTmodelGENIIpvalueSMkV$ and $\RESULTmodelGENIIpvalueSMoffshell$,
respectively.

\begin{figure}[hbtp!]
  \center
  \includegraphics[width=0.8\textwidth]{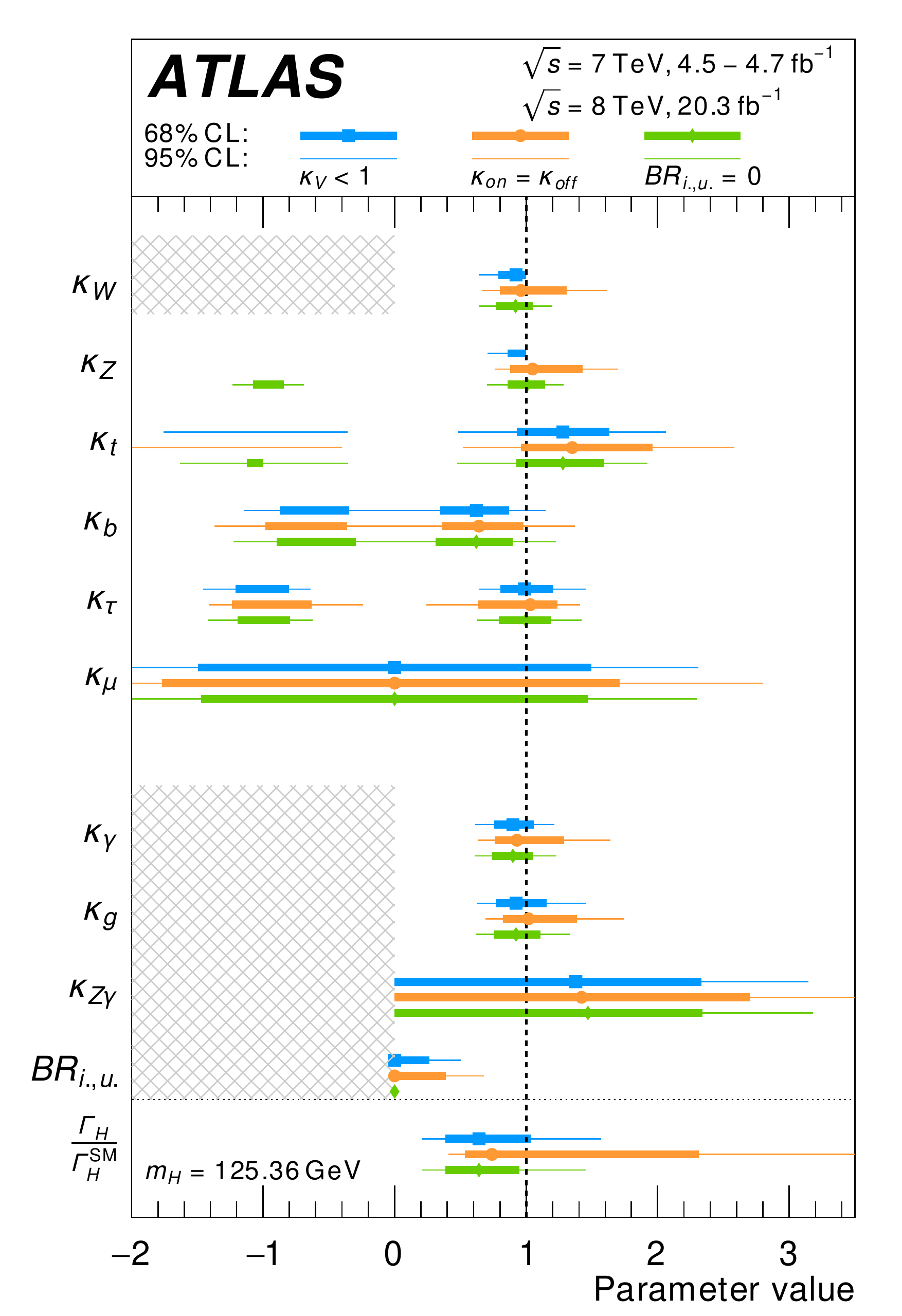}
  \caption{Results of fits for generic model 2 (see text): 
    the estimated values of each parameter under the constraint $\Cc_{V}<1$, $\Cc_{\rm on}=\Cc_{\rm off}$ or $\BRinv=0$ are shown with markers in the shape of a box, a circle, or a diamond, respectively.
    The hatched area indicates regions that are outside the defined parameter boundaries. 
    The inner and outer bars correspond to 68\%~CL and 95\%~CL intervals.
    The confidence intervals of $\BRinv$ and, in the benchmark model with the constraints $\Cc_{W}<1$ and $|\Cc_{Z}|<1$, also $\Cc_{W}$ and $\Cc_{Z}$, are estimated with respect to their physical bounds as described in the text. Numerical results are shown in Table~\ref{tab:bm:GenMod2:overview}.
  }
  \label{fig:bm:GenMod2:overview}
\end{figure}

\begin{sidewaystable}[htbp!]
  \caption{Numerical results of the fits to generic model 2 : effective coupling-strength 
    scale factors for loop processes allowing non-SM contributions with various assumptions 
    on the total Higgs boson width. These results are illustrated in Fig.~\ref{fig:bm:GenMod2:overview}.
    The confidence interval of $\BRinv$ in
    the benchmark model with the constraints $\Cc_{\PW}<1$ and
    $|\Cc_{\PZ}|<1$, and the confidence intervals $\Cc_{\PW}$ and
    $\Cc_{\PZ}$, are estimated with respect to their physical bounds,
    as described in the text. 
    Shown in square brackets are uncertainty components
    from different sources for the case of $\BRinv=0$ as an illustration.
    For $\Cc_{\PZ}$ and $\Cc_{\PQt}$, the uncertainty breakdowns are provided for the preferred
    positive solutions.  Also shown is the uncertainty on the
    total width that the model variants allow, expressed as the ratio
    $\Gamma_{\PH}/\Gamma_{\PH}^{\rm SM}$. These estimates for the
    width are obtained from alternative parameterisations of these
    benchmark models where the effective coupling-strength scale factor $\Cc_{\Pg}$
    is replaced by the expression that results from solving
    Eq.~(\ref{eq:CH:1}) for $\Cc_{\Pg}$, introducing
    $\Gamma_{\PH}/\Gamma_{\PH}^{\rm SM}$ as a parameter of the model.}
\label{tab:bm:GenMod2:overview}
\small
\begin{center}
\RESULTTableGenericModel
\end{center}
\end{sidewaystable}

Similar to the results of the benchmark model in
Section~\ref{sec:CV,CF,BRinv} the upper bound of the 68\%
CL interval for the scenario $\Cc_{\rm on}=\Cc_{\rm off}$ should
be considered to be only approximate due to  deviations of the test-statistic 
distribution from its asymptotic form. The deviation of the
asymptotic distribution was shown to be negligible for off-shell signal
strengths corresponding to the upper end of the 95\% asymptotic
confidence interval (Table \ref{tab:bm:GenMod2:overview}).  

Also shown in Fig~\ref{fig:bm:GenMod2:overview} are the resulting ranges of
the total width of the Higgs boson,  expressed as the ratio
$\Gamma_{\PH}/\Gamma_{\PH}^{\rm SM}$. 
These estimates
are obtained from alternative parameterisations of these benchmark
models, where the effective coupling-strength scale factor $\Cc_{g}$
is replaced by the expression that results from solving
Eq.~(\ref{eq:CH:1}) for $\Cc_{g}$, introducing
$\Gamma_{\PH}/\Gamma_{\PH}^{\rm SM}$ as a parameter of the model. 
 The figure shows that the upper bound on the
Higgs boson width from the assumption $\Cc_{\rm on}=\Cc_{\rm off}$ is
substantially weaker than the bound from the assumption $\Cc_{V}<1$.
These results on $\Gamma_{\PH}/\Gamma_{\PH}^{\rm SM}$ represent the
most model-independent measurements of the Higgs boson total width 
presented in this paper.

\begin{figure}[hbt!]
  \center
  \subfloat[]{
    \includegraphics[width=.45\textwidth]{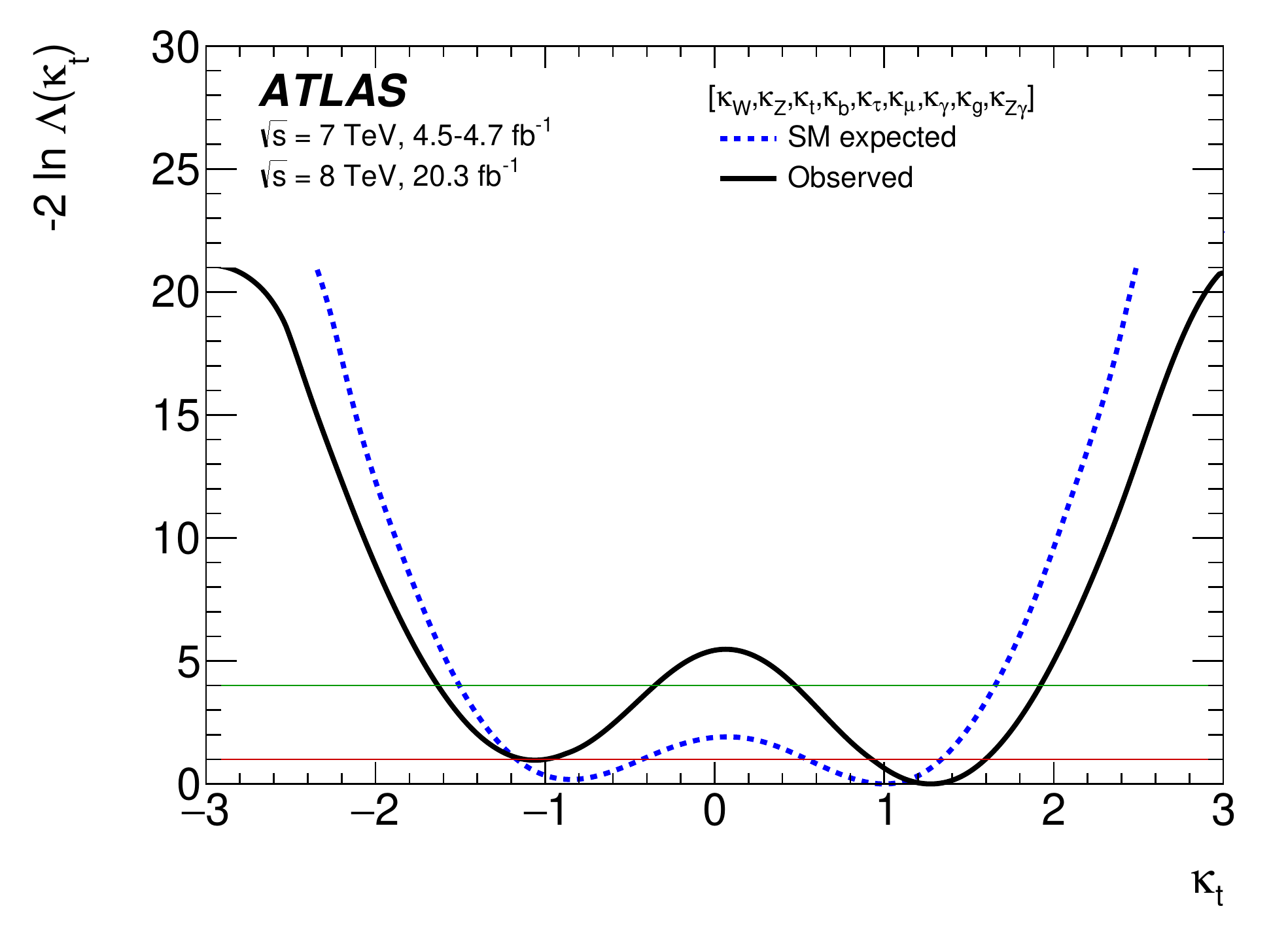}
    \label{fig:bm:GenMod2:Cto}
    \vspace*{-0.5cm}
  }
  \subfloat[]{
    \includegraphics[width=.45\textwidth]{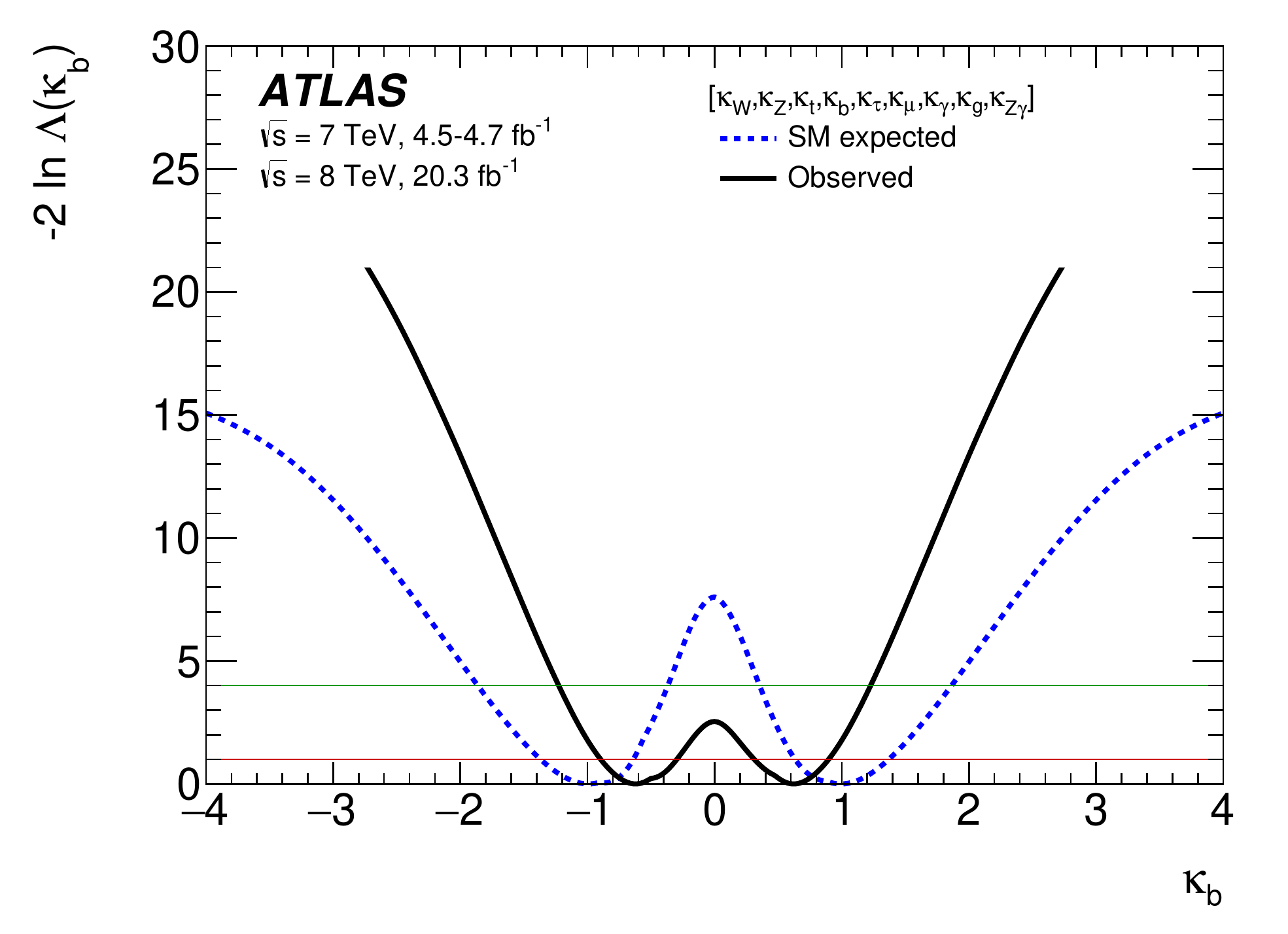}
    \label{fig:bm:GenMod2:Cb}
    \vspace*{-0.5cm}
  }

  \subfloat[]{
    \includegraphics[width=.45\textwidth]{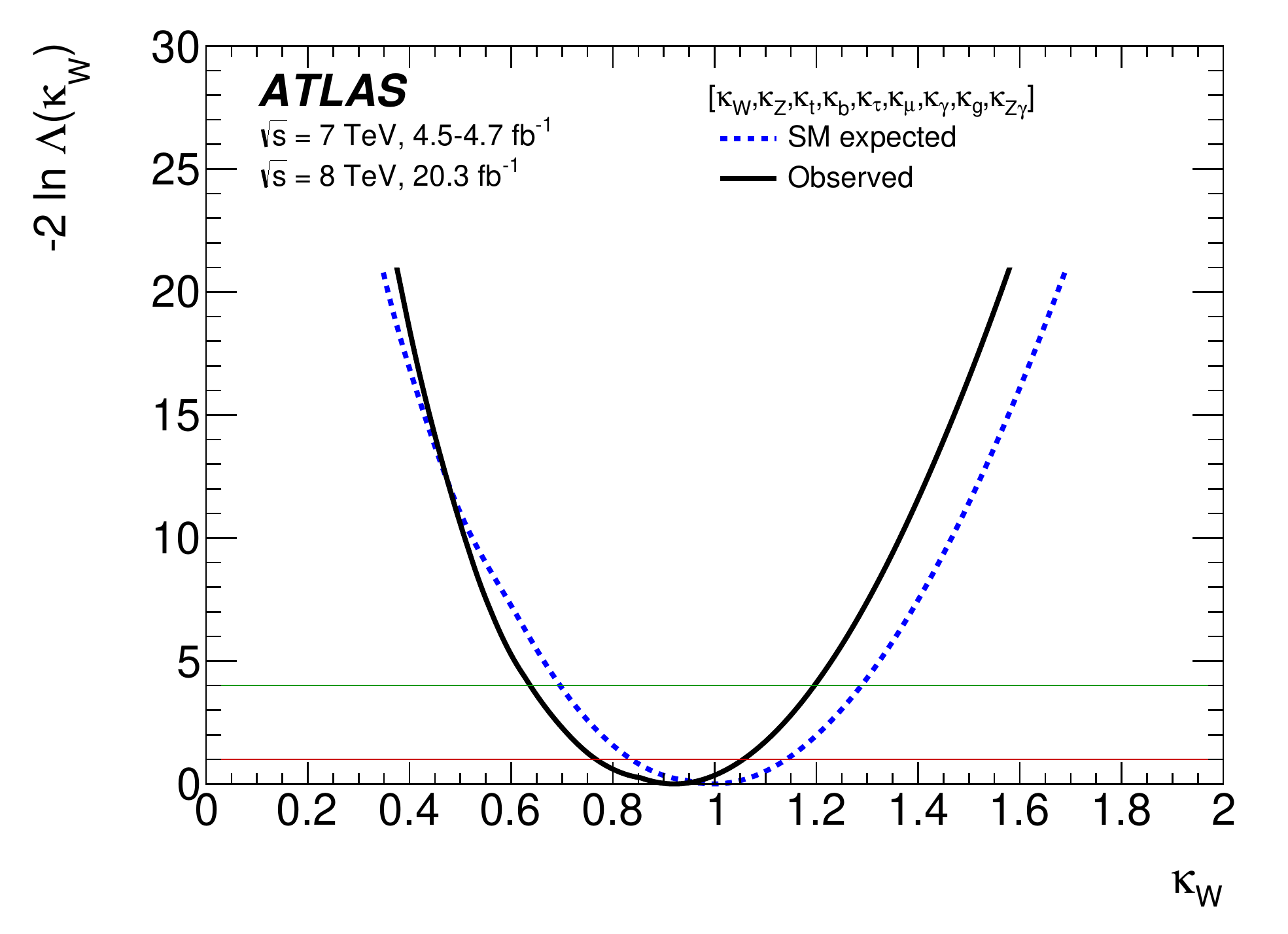}
    \label{fig:bm:GenMod2:CW}
    \vspace*{-0.5cm}
  }
  \subfloat[]{
    \includegraphics[width=.45\textwidth]{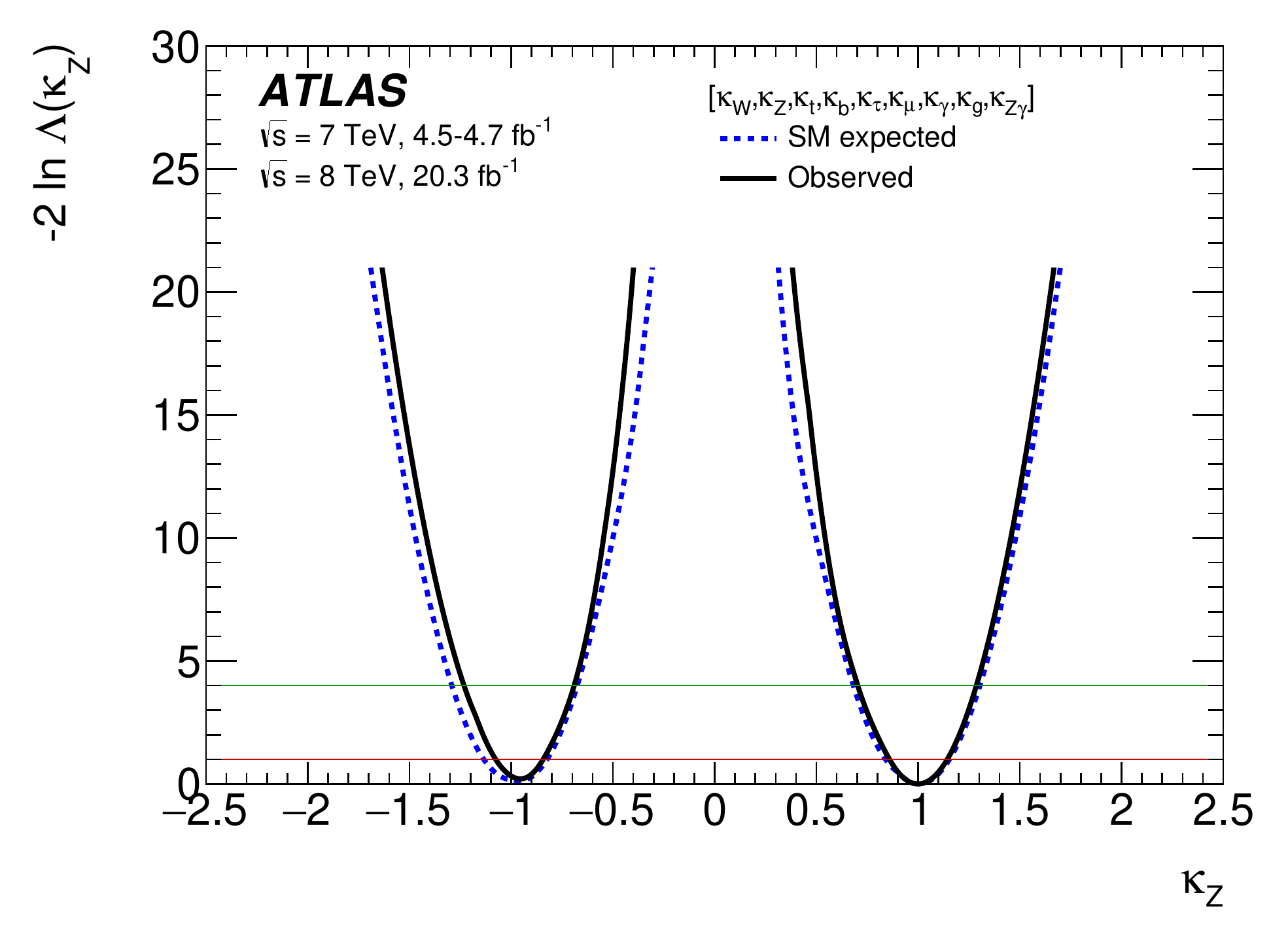}
    \label{fig:bm:GenMod2:CZ}
    \vspace*{-0.5cm}
  }

  \caption{Results of fits for generic model 2 (see text): 
    profile likelihood ratios as functions of the coupling-strength scale factors
    (a)~$\Cc_{\PQt}$,
    (b)~$\Cc_{\PQb}$,
    (c)~$\Cc_{\PW}$,
    and (d)~$\Cc_{\PZ}$.
    For each measurement, the other coupling-strength scale factors are profiled.
    The red (green) horizontal line indicates the value of the profile likelihood ratio corresponding to a 68\% (95\%) confidence interval for the parameter of interest, assuming the asymptotic $\chi^2$ distribution for the test statistic.
  }
  \label{fig:bm:GenMod2:Cx}
\end{figure}

Figure~\ref{fig:bm:GenMod2:Cx} shows profile likelihood ratios as a
function of selected coupling-strength scale factors.  In
Fig.~\ref{fig:bm:GenMod2:Cto}, the negative minimum of $\Cc_{\PQt}$ is
shown to be disfavoured at $1.0\sigma$.  The minimum corresponding to
the positive solution is found at $\Cc_{\PQt} = 1.28^{+0.32}_{-0.35}$.
The sensitivity to disfavour the negative solution of $\Cc_{\PQt}$ is
reduced with respect to generic model 1 as the interference in loop
couplings can no longer be exploited because effective coupling-strength scale factors
were introduced. The observed residual sensitivity to the sign of
$\Cc_{\PQt}$ is exclusively due to the tree-level interference effect
of the $tH$ background in the $ttH$ channel.  

The power of individual loop processes to measure the magnitude of
$\Cc_{\PQt}$ and resolve the sign of $\Cc_{\PQt}$ relative to
$\Cc_{\PW}$ is illustrated in more detail in
Fig.~\ref{fig:bm:GenMod2:loops}.  The blue curve shows the profile
likelihood ratio as a function of $\Cc_{\PQt}$ for a model with the
least sensitivity to the sign of $\Cc_{\PQt}$: all loop
processes are described with effective coupling parameters,
including the \ggZH\ loop process. Subsequently the red,
green and orange curves represent the profile likelihood ratios for
models that incrementally include information from loop processes by
resolving the \ggZH, ggF and $H\to\gamma\gamma,\,Z\gamma$ loop processes into their
expected SM content. Here the red curve corresponds to the
configuration of generic model 2, and the orange curve corresponds to
the configuration of generic model 1. As expected, resolving \ggZH\ process adds little information on
$\Cc_{\PQt}$. Additionally resolving the ggF loop process into its
SM content greatly improves the precision on $\Cc_{\PQt}$ (green
curve), but reduces the sensitivity to the relative sign of $\Cc_{\PQt}$
and $\Cc_{\PW}$. This reduction happens because on one hand
the ggF process yields no new information on this relative sign, as
it is dominated by  $t$--$b$ interference, and on the other hand because
it decreases the observed magnitude of $\Cc_{\PQt}$ to a more
SM-compatible level, thereby reducing the sensitivity of the $tH$
process to the relative sign. Further resolving the \hgg\ and
\hzg\ loop processes, which are dominated by $W$--$t$ interference,
greatly improves the measurement of the relative sign of $\Cc_{\PW}$
and $\Cc_{\PQt}$ (orange curve), but does not significantly contribute
to the precision of the magnitude of $\Cc_{\PQt}$.

\begin{figure}[hbt!]
  \center
  \includegraphics[width=0.65\textwidth]{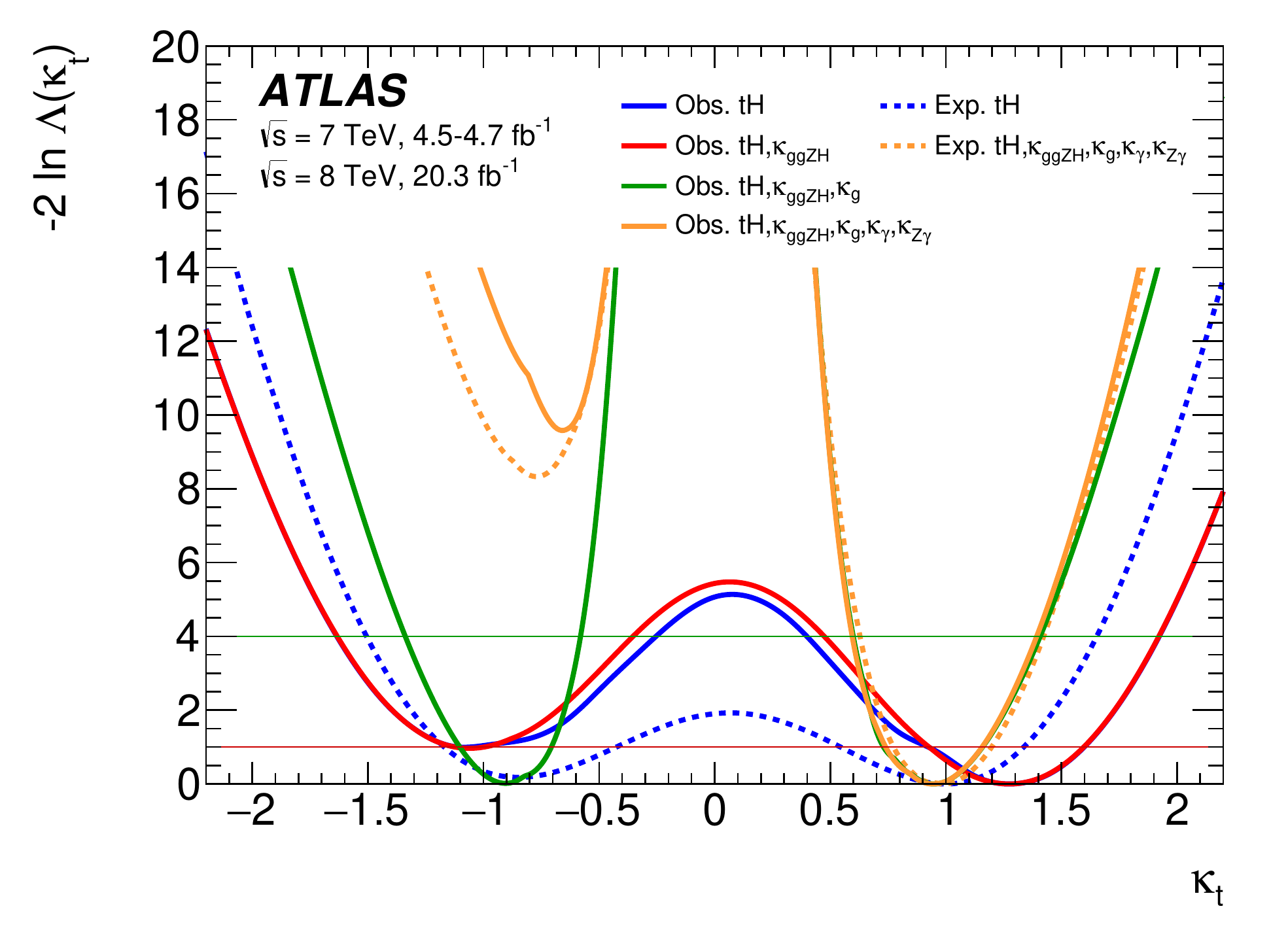}
  \caption{Profile likelihood ratio as a function of $\Cc_{\PQt}$ for models with and without resolved loop processes: shown are measurements of $\Cc_{\PQt}$ with no loop processes
           resolved (blue), only \ggZH\ resolved (red, generic model 2), \ggF\ additionally resolved (green), and \hgg\ and \hzg\ additionally resolved (orange, generic model 1).
           The dashed blue and orange curves correspond to the expected sensitivity for the no-loop and all-loop models.
All profile likelihood curves are drawn for the full range of $\Cc_{\PQt}$, however some curves are partially obscured 
when overlapping with another nearly identical curve.
           The red (green) horizontal line indicates the value of the profile likelihood ratio corresponding to a 68\% (95\%) confidence interval for the parameter of interest, assuming the asymptotic $\chi^2$ distribution for the test statistic.
  }
  \label{fig:bm:GenMod2:loops}
\end{figure}

\subsubsection{Generic model 3: allow new particles in loops, no assumptions on the total width}
\label{sec:gen3}
In the final benchmark model of this section, the six absolute
coupling-strength scale factors and three effective loop-coupling scale factors of
generic model 2 are expressed as ratios of scale factors that can be
measured independent of any assumptions on the Higgs boson total
width. The free parameters are chosen as:
\begin{center}
\begin{minipage}{0.3\textwidth}
\begin{eqnarray*}
  \Cc_{\Pg\PZ} &=& \Cc_{\Pg}\cdot\Cc_{\PZ} / \Cc_{\PH} \\
  \Rr_{\PZ\Pg} &=& \Cc_{\PZ} / \Cc_{\Pg} \label{GM3-kglZ}\\
  \Rr_{\PW\PZ} &=& \Cc_{\PW} / \Cc_{\PZ} \label{GM3-kWZ}\\
  \Rr_{\PQt\Pg} &=& \Cc_{\PQt} / \Cc_{\Pg} \label{GM3-ktg}\\
  \Rr_{\PQb\PZ} &=& \Cc_{\PQb} / \Cc_{\PZ} \label{GM3-kbZ}\\
\end{eqnarray*}
\end{minipage}
\begin{minipage}{0.3\textwidth}
\begin{eqnarray*}
  \Rr_{\PGt\PZ} &=& \Cc_{\tau} / \Cc_{\PZ} \label{GM3-ktaZ}\\
  \Rr_{\PGm\PZ} &=& \Cc_{\mu} / \Cc_{\PZ} \label{GM3-kmuZ}\\
  \Rr_{\PGg\PZ} &=& \Cc_{\PGg} / \Cc_{\PZ} \label{GM3-kgaZ}\\
  \Rr_{(\PZ\PGg)\PZ} &=& \Cc_{\PZ\PGg} / \Cc_{\PZ} .
\end{eqnarray*}
\end{minipage}
\end{center}
Figure~\ref{fig:bm:GenMod3:overview} shows the full set of results
obtained from the fit to this benchmark model. The fitted values and their
uncertainties are also shown in Table \ref{tab:gen3result}. As the
loop-induced processes are expressed by effective coupling-strength
scale factors, there is little sensitivity to the relative sign
of coupling-strength scale factors due to $tH$ and
\ggZH\ processes only. Hence only positive values for all
$\Cc$-factors except $\Cc_{\PQt}$ are shown without loss of
generality. The parameter
$\Cc_{\Pg\PZ},\Rr_{\PZ\Pg},\Rr_{\PW\PZ},\Rr_{\PQt\Pg},\Rr_{\PQb\PZ},\Rr_{\PGt\PZ}$
and $\Rr_{\PGg\PZ}$ are all measured to be compatible with their SM
expectation, while limits are set on the parameters $\Rr_{\PGm\PZ}$ and
$\Rr_{(\PZ\PGg)\PZ}$. The nine-dimensional compatibility of the SM
hypothesis with the best-fit point is $\RESULTmodelGENIIIpvalueSM$.

\begin{table}[htb]
\caption{Numerical results of the fits for generic model 3:
  measurements of ratios of coupling-strength scale factors in which assumptions on
  the Higgs boson total width cancel. These results are also shown in Fig.~\ref{fig:bm:GenMod3:overview}.
  Shown in square brackets are uncertainty components from different sources. 
  For $\Rr_{\PW\PZ}$ and $\Rr_{\PQt\Pg}$, the uncertainty breakdowns are provided for 
  the preferred positive solutions.}
\label{tab:gen3result}
\begin{center}
\RESULTTableGenericModelThree
\end{center}
\end{table}

The parameter $\Rr_{\PW\PZ} =\Cc_{\PW}/\Cc_{\PZ}$ in this model is of
particular interest: identical coupling-strength scale factors for the
$W$ and $Z$ bosons are required within tight bounds by the
$\mathrm{SU(2)}$ custodial symmetry and the $\rho$ parameter
measurements at LEP and at the Tevatron~\cite{ALEPH:2010aa}. This custodial
constraint is directly probed in the Higgs sector through the parameter $\Rr_{\PW\PZ}$.
The measured ratio $\Rr_{\PW\PZ}$ is in part directly constrained by the decays
in the \hWWlnln\ and \hZZllll\ channels and the $WH$ and $ZH$
production processes.  It is also indirectly constrained by the VBF
production process, which in the SM is $74\%$ $W$ fusion-mediated and $26\%$
$Z$ fusion-mediated (see Table \ref{tab:kexpr}).
Figure~\ref{fig:bm:GenMod3:rwz} shows the profile likelihood ratio as a
function of the coupling-strength scale factor ratio
$\Rr_{\PW\PZ}$. Due to the interference terms, the fit is sensitive to
the relative sign of the $W$ and $t$ coupling ($tH$) and the
relative sign of the $Z$ and $t$ coupling ({\ggZH}), providing
indirect sensitivity to the sign of $\Rr_{\PW\PZ}$. The negative
solution is disfavoured at $0.5\sigma$ ($0.3\sigma$ expected).  The
minimum corresponding to the positive solution is found at
$\Rr_{\PW\PZ} =0.92^{+0.14}_{-0.12}$, in excellent agreement with the
prediction of $\mathrm{SU(2)}$ custodial symmetry.

Also shown in Figs.~\ref{fig:bm:GenMod3:rtg},~\ref{fig:bm:GenMod3:rgz}
are the ratios $\Rr_{\PGg\PZ}$ and $\Rr_{\PQt\Pg}$.  The ratio
$\Rr_{\PGg\PZ}$ is sensitive to new charged particles contributing to
the {\hgg} loop in comparison to {\hzz} decays.  Similarly, the ratio
$\Rr_{\PQt\Pg}$ is sensitive to new coloured particles contributing
through the {\ggF} loop as compared to $ttH$. The minimum corresponding to
the positive solution is found at $\Rr_{\PQt\Pg} = 1.38 \pm 0.35$.
Both are observed to be compatible with the SM expectation.

\begin{figure}[hbt]
\center
\includegraphics[width=0.99\textwidth]{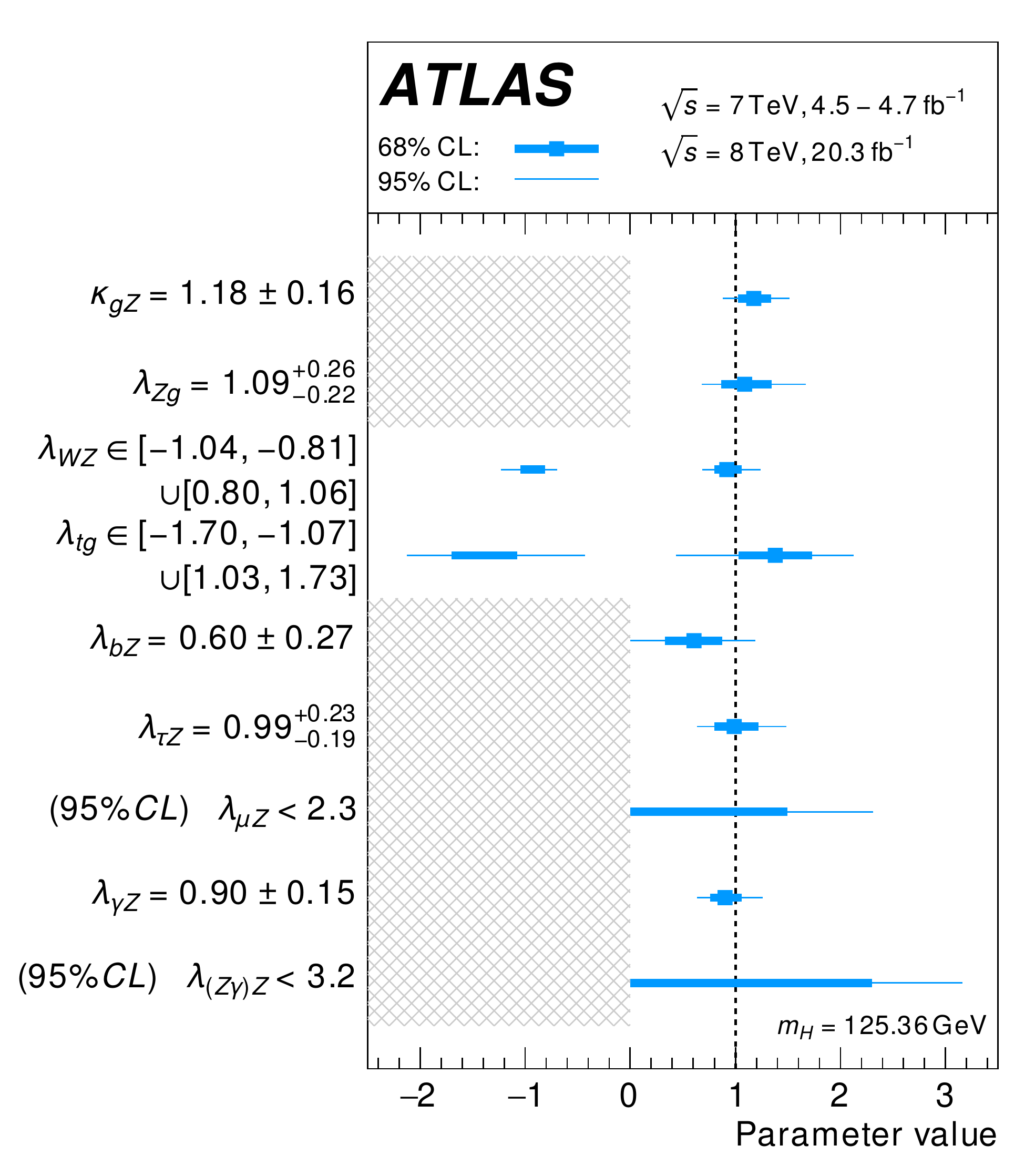}
\caption{Results of fits for generic model 3 (see text): allowing deviations in vertex loop-coupling scale factors and in the total width.
         Overview of best-fit values of parameters, where the inner and outer bars correspond to 68\%~CL and 95\%~CL intervals.
         The hatched areas indicate regions that are outside the defined parameter boundaries. }
\label{fig:bm:GenMod3:overview}
\end{figure}

\begin{figure}[hbt]
\center
  \subfloat[]{
    \includegraphics[width=0.80\textwidth]{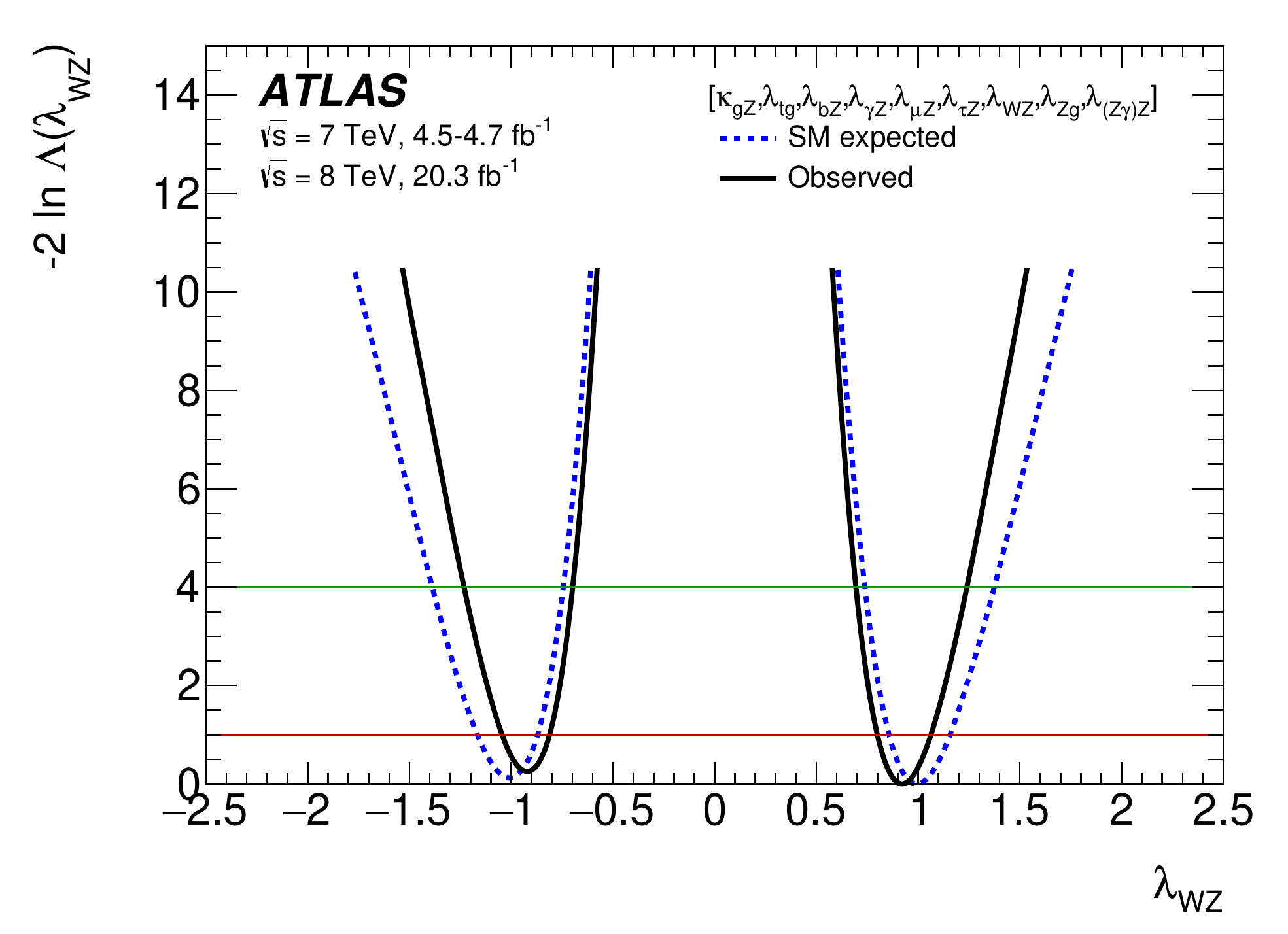}
    \label{fig:bm:GenMod3:rwz}
    \vspace*{-0.5cm}
  }

  \subfloat[]{
    \includegraphics[width=0.40\textwidth]{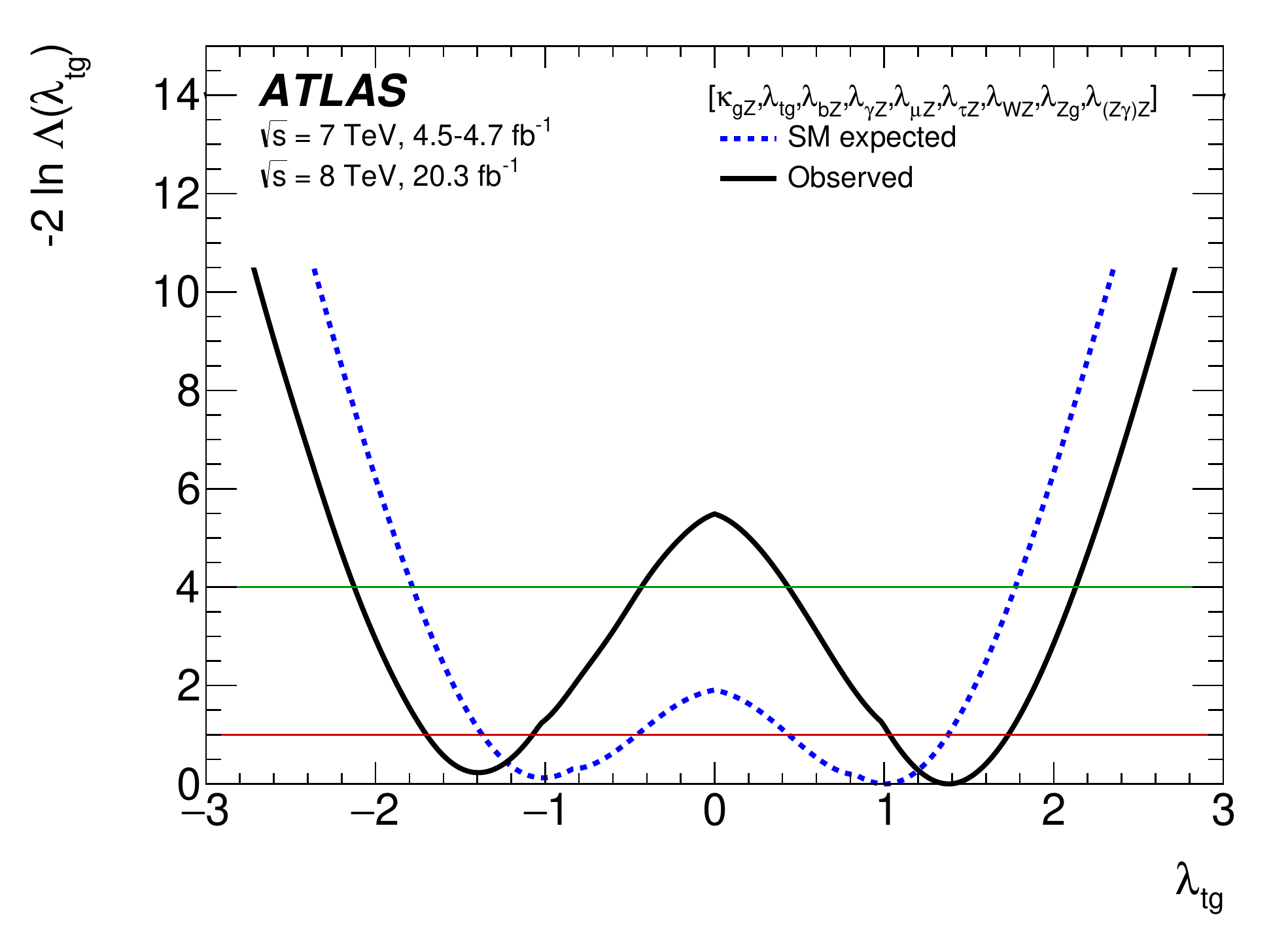}
    \label{fig:bm:GenMod3:rtg}
    \vspace*{-0.5cm}
  }
  \subfloat[]{
    \includegraphics[width=0.40\textwidth]{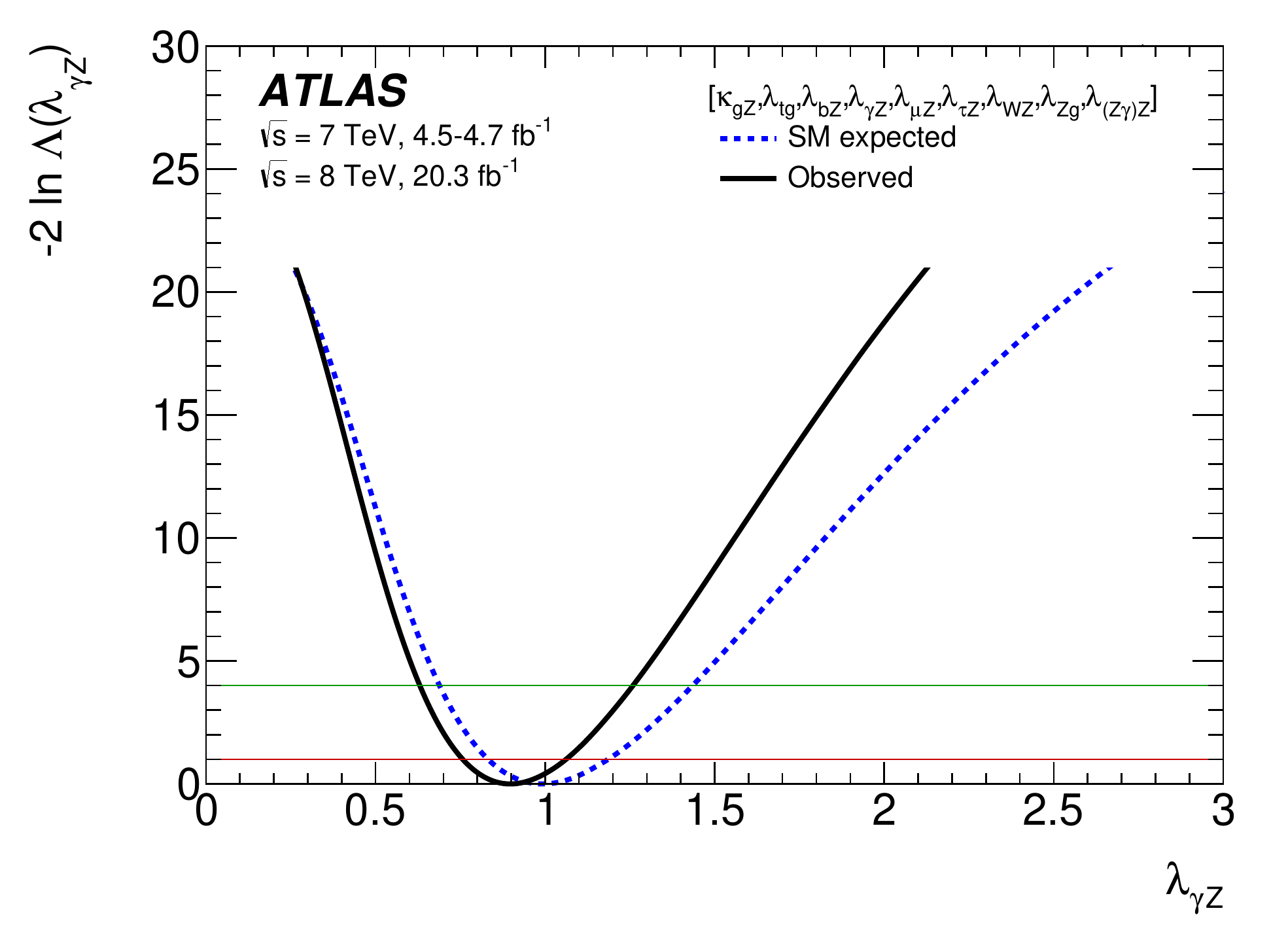}
    \label{fig:bm:GenMod3:rgz}
    \vspace*{-0.5cm}
  }
\caption{Results of fits for generic model 3 (see text): 
  profile likelihood ratios as functions of the coupling-strength scale 
  factor ratios (a)~$\Rr_{\PW\PZ}$, (b)~$\Rr_{\PQt\Pg}$ and (c)
  $\Rr_{\PGg\PZ}$.  In all cases, the other parameters are
  profiled. The dashed curves show the SM expectations.  The red
  (green) horizontal line indicates the cutoff value of the profile
  likelihood ratio corresponding to a 68\% (95\%) confidence interval
  for the parameter of interest, assuming the asymptotic $\chi^2$
  distribution for the test statistic.  }
\label{fig:bm:DetailGenMod3}
\end{figure}

The fit in the third generic benchmark model uses only the basic
assumptions, as stated at the beginning of this section, and hence
represents the most model-independent determination of coupling-strength scale
factors that is currently possible.

\clearpage

\mathversion{normal3}
\section{Conclusion}
\label{sec:Conclusion}

The Higgs boson production and decay properties are studied using 
proton--proton collision data collected by the ATLAS experiment at the Large Hadron Collider corresponding
to integrated luminosities of up to 4.7~\ifb\ at $\sqrt{s}=7$~TeV and
20.3~\ifb\ at $\sqrt{s}=8$~TeV. The study combines specific analyses
of the
$H\to\gamma\gamma,\,ZZ^*,\,WW^*,\,Z\gamma,\,b\bar{b},\,\tau\tau\,$ and
$\mu\mu$ decay channels, as well as searches for $ttH$ production and
measurements of off-shell Higgs boson production. It significantly
extends a previous combination of the $H\to\gamma\gamma,\,ZZ^*$ and
$WW^*$ decays~\cite{Aad:2013wqa}. In particular, the addition of the
fermionic decays of the Higgs boson in the combinations allows for
direct tests of the Yukawa interactions of the Higgs boson with
fermions.

The measured Higgs boson signal yields are compared with the SM
expectations at the fixed Higgs boson mass of $m_H=125.36$~GeV.  The
combined yield relative to its SM prediction is determined to be
$\RESULTGlobalMuLong$.
The combined analysis
provides unequivocal confirmation of gluon fusion production 
of the Higgs boson with a significance exceeding $5\sigma$ and
strong evidence of vector-boson fusion production with a significance of
$\RESULTNewVBFSigmaObs\sigma$. Furthermore, it supports the SM
predictions of Higgs boson production in association with a vector
boson or a pair of top quarks. Values for the total cross sections can
be obtained from the signal strength of each production
process within the uncertainties related to the modelling of Higgs boson
production and decay kinematics and assuming SM decay branching
ratios. The total cross sections at $\sqrt{s}=7$ and 8~TeV are
$\RESULTAXSAllLong$~pb and $\RESULTBXSAllLong$~pb, respectively.

The observed Higgs boson production and decay rates are also
interpreted in a leading-order coupling framework, exploring a wide
range of benchmark coupling models both with and without assumptions
about the Higgs boson width and the SM particle content of loop
processes. Higgs boson couplings to up-type fermions and vector bosons are found
with both significances above $5\sigma$ and to down-type fermions with a 
significance of $\RESULTmodelUPDOWNpvalueDownType$, under the assumption of unified coupling 
scale factors, one for each type of particles. In a different model with
separate unified coupling scale factors for leptons, quarks and vector bosons,
Higgs boson couplings to leptons are found with a significance of $\RESULTmodelQLpvalueLepton$.

The Higgs boson coupling strengths to fermions and bosons are measured with
a precision of $\pm 16$\% and $\pm 7$\% respectively, when assuming the SM Higgs boson
width, and are observed to be compatible with the SM expectations.
Coupling strengths of loop processes are measured with a precision of
$\pm 12$\% when assuming the SM expectations for non-loop Higgs boson coupling strengths and the
Higgs boson total width, increasing to about $\pm 20$\% when these
assumptions are removed.  No significant deviations from the SM expectations 
of Higgs boson coupling strengths in loop processes are observed.

Measurements of coupling strengths to
$\mu,\,\tau$ leptons, $b,\,t$ quarks and $W,\,Z$ bosons, or ratios of these
coupling strengths are presented in the context of generic Higgs boson coupling models. 
They can constrain the ratio of $W$
and $Z$ coupling strengths, a probe of custodial symmetry, with a
precision of $\pm 13$\%. For benchmark models that measure absolute coupling
strengths, a variety of physics-motivated constraints on the Higgs
boson total width have been explored. The measured Higgs boson
coupling strengths and their precision are found to depend only weakly on
the choice of these constraints. A third generic benchmark model
uses only the most basic assumptions and hence represents
the most model-independent determination of the coupling strength scale factors that
is currently possible. In this model ratios of couplings are constrained with a precision
of 15 -- 40\%.

The $p$-values expressing compatibility of the SM hypothesis with the best-fit point range 
between 29\% and 99\% for all considered benchmark models. The observed
data are thus very compatible with the SM expectation under a wide
range of assumptions.

\section*{Acknowledgements}

We thank CERN for the very successful operation of the LHC, as well as the
support staff from our institutions without whom ATLAS could not be
operated efficiently.

We acknowledge the support of ANPCyT, Argentina; YerPhI, Armenia; ARC, Australia; BMWFW and FWF, Austria; ANAS, Azerbaijan; SSTC, Belarus; CNPq and FAPESP, Brazil; NSERC, NRC and CFI, Canada; CERN; CONICYT, Chile; CAS, MOST and NSFC, China; COLCIENCIAS, Colombia; MSMT CR, MPO CR and VSC CR, Czech Republic; DNRF, DNSRC and Lundbeck Foundation, Denmark; IN2P3-CNRS, CEA-DSM/IRFU, France; GNSF, Georgia; BMBF, HGF, and MPG, Germany; GSRT, Greece; RGC, Hong Kong SAR, China; ISF, I-CORE and Benoziyo Center, Israel; INFN, Italy; MEXT and JSPS, Japan; CNRST, Morocco; FOM and NWO, Netherlands; RCN, Norway; MNiSW and NCN, Poland; FCT, Portugal; MNE/IFA, Romania; MES of Russia and NRC KI, Russian Federation; JINR; MESTD, Serbia; MSSR, Slovakia; ARRS and MIZ\v{S}, Slovenia; DST/NRF, South Africa; MINECO, Spain; SRC and Wallenberg Foundation, Sweden; SERI, SNSF and Cantons of Bern and Geneva, Switzerland; MOST, Taiwan; TAEK, Turkey; STFC, United Kingdom; DOE and NSF, United States of America. In addition, individual groups and members have received support from BCKDF, the Canada Council, CANARIE, CRC, Compute Canada, FQRNT, and the Ontario Innovation Trust, Canada; EPLANET, ERC, FP7, Horizon 2020 and Marie Skłodowska-Curie Actions, European Union; Investissements d'Avenir Labex and Idex, ANR, Region Auvergne and Fondation Partager le Savoir, France; DFG and AvH Foundation, Germany; Herakleitos, Thales and Aristeia programmes co-financed by EU-ESF and the Greek NSRF; BSF, GIF and Minerva, Israel; BRF, Norway; the Royal Society and Leverhulme Trust, United Kingdom.

The crucial computing support from all WLCG partners is acknowledged
gratefully, in particular from CERN and the ATLAS Tier-1 facilities at
TRIUMF (Canada), NDGF (Denmark, Norway, Sweden), CC-IN2P3 (France),
KIT/GridKA (Germany), INFN-CNAF (Italy), NL-T1 (Netherlands), PIC (Spain),
ASGC (Taiwan), RAL (UK) and BNL (USA) and in the Tier-2 facilities
worldwide.

\clearpage
\printbibliography
\clearpage
\appendix
\part*{Appendix}
\addcontentsline{toc}{part}{Appendix}

\section{Alternative parameterisation of ratios of cross sections and  of branching ratios}
\label{sec:oldRatio}

An alternative to the parameterisation of Section~\ref{sec:ratio} is to normalise the ratios of cross sections and of branching ratios to their SM values. Compared with Eq.~(\ref{eq:ratio}), the yield of the production and decay $i\to H\to f$ can be parameterised as
\begin{equation}
\sigma_i\cdot {\rm BR}_f = \mu_i^f \times \left[\sigma_i\cdot {\rm BR}_f\right]_{\rm SM} = \left(\mu_{\rm ggF}^{WW^*}\cdot  R_{i/{\rm ggF}}\cdot \rho_{f/WW^*}\right)\times \left[\sigma_i\cdot {\rm BR}_f\right]_{\rm SM}\,.
\end{equation}
Here $R$ and $\rho$ are ratios of cross sections and branching ratios relative to their SM expectations, respectively:
\begin{equation}
  R_{i/{\rm ggF}} =  \frac{\sigma_i/\sigma_{\rm ggF}}{\left[\sigma_i/\sigma_{\rm ggF}\right]_{\rm SM}}\hsb {\rm and}\hsb
  \rho_{f/WW^*}   =  \frac{{\rm BR}_f/{\rm BR}_{WW^*}}{\left[{\rm BR}_f/{\rm BR}_{WW^*}\right]_{\rm SM}}. 
\end{equation}

The data are fitted with $\mu_{\rm ggF}^{WW^*}$, four ratios of production cross sections and one ratio of branching ratios for each decay channel other than the $H\to WW^*$ decay. The results shown in Table~\ref{tab:oldRatio} are nearly identical to the best-fit values relative to their SM predictions shown in Table~\ref{tab:ratio}. The small differences are expected  from the inclusion of additional nuisance parameters of the SM predictions and from the precision of the fits. One clear advantage of the parameterisation of Section~\ref{sec:ratio} is that the results are independent of the SM predictions and are, therefore, not affected by the theoretical uncertainties of the predictions. Consequently, the fitted values of the ratios of cross sections and of partial decay widths shown in Table~\ref{tab:ratio} have significantly smaller theoretical uncertainties than their counterparts ($R_{i/{\rm ggF}}$ and $\rho_{r/WW^*}$) in Table~\ref{tab:oldRatio}. The former is only affected by the theoretical uncertainties in the modelling of Higgs boson production whereas the latter suffer from both the modelling uncertainties and the uncertainties of the SM predictions.

\begin{table}[htb!]
\caption{Best-fit values of $gg\to H\to WW^*$ signal strength $\mu_{\rm ggF}^{WW^*}$, ratios of cross sections $R_{i/{\rm ggF}}$ and of branching ratios $\rho_{f/WW^*}$.  All $R_{i/{\rm ggF}}$ and $\rho_{f/WW^*}$ are measured relative to their SM values for $m_H=125.36$~GeV from the combined analysis of the $\sqrt{s}=7$ and 8~TeV data. Shown in square brackets are uncertainty components: statistical (first), systematic (second) and signal theoretical (third) uncertainties.}
\label{tab:oldRatio}
\begin{center}
\RESULTRatioTable
\end{center}
\end{table}

\clearpage
\begin{flushleft}
{\Large The ATLAS Collaboration}

\bigskip

G.~Aad$^{\rm 85}$,
B.~Abbott$^{\rm 113}$,
J.~Abdallah$^{\rm 151}$,
O.~Abdinov$^{\rm 11}$,
R.~Aben$^{\rm 107}$,
M.~Abolins$^{\rm 90}$,
O.S.~AbouZeid$^{\rm 158}$,
H.~Abramowicz$^{\rm 153}$,
H.~Abreu$^{\rm 152}$,
R.~Abreu$^{\rm 30}$
% ...
\bigskip
\\
% ...
$^{11}$ Institute of Physics, Azerbaijan Academy of Sciences, Baku, Azerbaijan\\
% ...
$^{30}$ CERN, Geneva, Switzerland\\
% ...
$^{85}$ CPPM, Aix-Marseille Universit{\'e} and CNRS/IN2P3, Marseille, France\\
% ...
$^{90}$ Department of Physics and Astronomy, Michigan State University, East Lansing MI, United States of America\\
% ...
$^{107}$ Nikhef National Institute for Subatomic Physics and University of Amsterdam, Amsterdam, Netherlands\\
% ...
$^{113}$ Homer L. Dodge Department of Physics and Astronomy, University of Oklahoma, Norman OK, United States of America\\
% ...
$^{151}$ Institute of Physics, Academia Sinica, Taipei, Taiwan\\
$^{152}$ Department of Physics, Technion: Israel Institute of Technology, Haifa, Israel\\
$^{153}$ Raymond and Beverly Sackler School of Physics and Astronomy, Tel Aviv University, Tel Aviv, Israel\\
% ...
$^{158}$ Department of Physics, University of Toronto, Toronto ON, Canada\\
% ...
\end{flushleft}


\end{document}